\documentclass[a4paper,11pt]{article}
\usepackage{tikz}

\pdfoutput=1 
\usepackage{upgreek}
\usepackage{jheppub} 

\usepackage[T1]{fontenc} 

\usepackage{extarrows} 
\usepackage{amsmath}
\usepackage{graphicx}
\usepackage{dcolumn}

\usepackage{bm}
\usepackage{wasysym}
\usepackage{verbatim}
\usepackage{color}
\usepackage{mathtools,slashed}

\definecolor{bostonuniversityred}{rgb}{0.8, 0.0, 0.0}
\newcommand{\mpl}{M_{\mbox{\tiny{Pl}}}}
\newcommand{\Beq}{\begin{equation}\begin{aligned}}
\newcommand{\Eeq}{\end{aligned}\end{equation}}

\newcommand{\p}{\partial}

\newcommand{\bx}{{\textbf{x}}}
\newcommand{\bep}{{\boldsymbol{\varepsilon}}}

\newcommand{\mH}{\mathcal{H}}
\newcommand{\xa}{\xi_{A}}
\newcommand{\xpi}{\xi_{\varphi}}
\newcommand{\an}{\quad \textmd{and} \quad }

\newcommand{\I}{{\bf{I}}_2}
\newcommand{\II}{{\rm{I}_4}}
\newcommand{\III}{{\rm{I}_8}}
\newcommand{\bea}{\begin{eqnarray}}
\newcommand{\eea}{\end{eqnarray}}

\newcommand{\ga}{g_A}
\newcommand{\lvac}{\bigg\langle 0 \bigg\lvert}
\newcommand{\rvac}{\bigg\rvert 0 \bigg\rangle}

\newcommand{\xp}{\xi_{A}}
\newcommand{\sD}{\slashed{D}}

\newcommand{\cev}[1]{\reflectbox{\ensuremath{\vec{\reflectbox{\ensuremath{#1}}}}}}

\newcommand{\where}{\quad \textmd{where} \quad}
\newcommand{\x}{\tilde{\tau}}

\newcommand{\bk}{\boldsymbol{k}}

\newcommand{\reK}{{\rm{Re}}\kappa}

\title{\boldmath 
Dark Fermions and Spontaneous $CP$ violation in $SU(2)$-axion Inflation}

\author[a]{Azadeh Maleknejad}

\affiliation[a]{Max-Planck-Institute for Astrophysics, Karl-Schwarzschild-Str. 1, 85741 Garching, Germany}

\emailAdd{amalek@MPA-Garching.MPG.DE}

\abstract{
Remarkably, if $CP$ was spontaneously broken in the physics of inflation, fermions would notice and remember it. Based on that, we present a new (non-thermal) mechanism for generating self-interacting dark Dirac fermions prior to the Hot Big Bang. The non-Abelian gauge fields and axions are well-motivated matter contents for the particle physics of inflation. In this background, we analytical study Dirac fermion doublets charged under the $SU(2)$ gauge field and use point-splitting technique to regularize the currents. We show that the non-trivial $CP$-violating vacuum structure of $SU(2)$-axion models naturally leads to an efficient mechanism for generating massive fermions during inflation. The size of the fermionic backreaction and the density fraction of dark fermions put upper bounds on the fermion's mass. For a GUT scale inflation, the generated dark fermions, only gravitationally coupled to the visible sector, can be as heavy as $m\lesssim 10~TeV$.  }

\begin{document}
\maketitle
\flushbottom

\section{Introduction}\label{sec:Intro}
The current cosmological observations are in significant agreement with the general concept of inflation paradigm \cite{Guth:1980zm, Sato:1980yn, Linde:1981mu, Albrecht:1982wi} and several of its key predictions have been already confirmed by the cosmic microwave background (CMB) and large scale structure (LSS) data \cite{Ade:2015lrj}. Besides, the upcoming ambitious missions, e.g., LiteBIRD \cite{Matsumura:2013aja, Hazumi:2019lys} and CMB Stage-4  \cite{Abazajian:2019eic}, will soon provide us with even more data about the early Universe. However, our theoretical understanding of the particle content of inflation is still vague and incomplete in that sense. Given the fact that the energy scale of inflation can be as high as the GUT scale to a few TeV, it might provide an outstanding opportunity to explore the high energy physics beyond the standard model (BSM) and possibly breakthrough discoveries. In the past decade, several seminal and groundbreaking works have been done to systematically extract more information from the cosmological probes about the particle content of inflation through n-point functions, i.e. \textit{cosmological collider} physics \cite{Chen:2009zp, Baumann:2011nk, Noumi:2012vr, Arkani-Hamed:2015bza, Lee:2016vti, Kehagias:2017cym, Arkani-Hamed:2018kmz}. In these studies, the source of the particle production is either the expansion of the Universe or the decay of massive particles, while (up to this point) the vacuum is assumed to be topologically trivial and $CP$-\textit{preserving}. Recently, the $P$ and $CP$ violations on the cosmological collider for a scalar QED with an additional $\theta(\tau)F\tilde F$ operator is studied in \cite{Liu:2019fag}.

From the high-energy physics point of view, gauge fields and axions are very natural matter contents for high energy scales, e.g., the scale of inflation. Axion fields are abundant in theories BSM and hence well-motivated candidates
for the inflaton field. The first model of axion inflation, natural inflation, has been proposed almost 30 years ago in \cite{Freese:1990rb}. Enjoying an almost shift symmetry, the axion effective potential is protected from dangerous quantum corrections which guaranteed the flatness of the potential. 
Among the well-motivated axion inflation models, one can mention axion monodromy \cite{Easther:2013kla, McAllister:2014mpa}. See \cite{Pajer:2013fsa} for a review on axions in inflation, and see \cite{Marsh:2015xka} for a review on axions in cosmology. Besides, some theoretical issues of de sitter vacuum suggest not only the naturalness but the necessity of axions \cite{Obied:2018sgi, Dvali:2018dce}.

\begin{figure*}[!htb]
\centering
\includegraphics[scale=0.7]{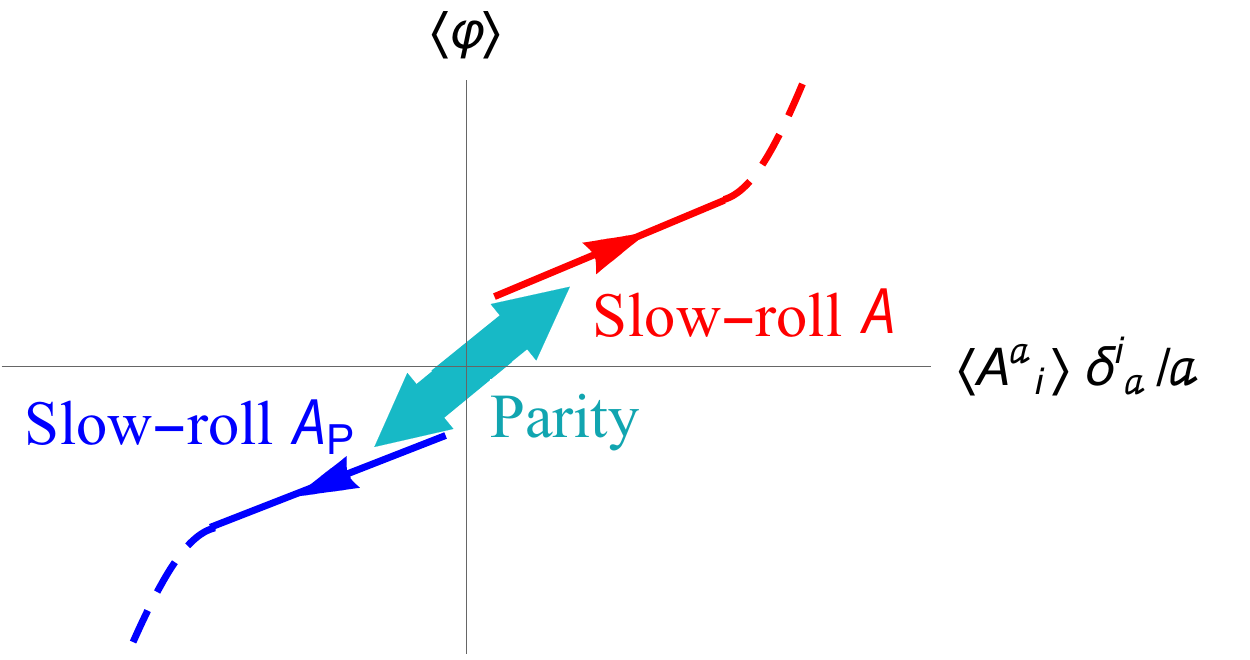} 
\caption{The $SU(2)$-axion vacuum structure. For each given parameter set and energy density, there are two $SU(2)$-axion field configuration vacuums, Eq.s \eqref{GF-SU2}-\eqref{axion}, which are related by the parity, i.e. slow-roll trajectories $A$ and $A_P$. The two have identical background cosmologies; however, vacuum spontaneously breaks P at the level of perturbations of the fields with spin, e.g., fermions and spin-2 fields. For more details about the vacuum see Sec. 2 of \cite{Maleknejad:2018nxz}.}\label{P-B}
\end{figure*}

Another natural matter content for inflation is non-Abelian gauge fields. Thanks to the isomorphy of the $su(2)$ and $so(3)$ algebras, any non-Abelian gauge field can acquire a homogeneous and isotropic vacuum field configuration (VEV) in its $SU(2)$ subsector. Moreover, their effective theory can break the conformal symmetry of the Yang-Mills theory, e.g., by coupling with the axion (to prevent its $a^{-4}$ decay) and let the gauge field enjoy a slow-roll dynamics with an almost constant energy density during inflation. Besides, the non-Abelian gauge field perturbed around its isotropic VEV has an extra spin-2 degree of freedom which is linearly coupled to the primordial gravitational waves. Such $SU(2)$-axion vacuums spontaneously break $P$ and $CP$ during inflation \cite{Maleknejad:2014wsa} (See Fig. \ref{P-B} for an illustration) and predict a $CP$\textit{-violating} pre-Hot Big Bang physics. Since the seminal paper \cite{Zeldovich:1974uw}, it is well known that broken discrete symmetries cause cosmological domain wall problem which can dominate the Universe if the symmetry is broken only spontaneously and happens after inflation \cite{Dvali:2018txx}. Given the fact that in $SU(2)$-axion models, the scale of symmetry breaking is high, and prior to the Hot Big Bang, they avoid the domain wall problem. The theoretically well-motivated and phenomenologically rich setting for $SU(2)$ inflation models have been discovered in \cite{Maleknejad:2011sq, Maleknejad:2011jw} which showed that non-Abelian gauge fields can contribute to the physics of inflation and the first realization of this class of models has been introduced, i.e., Gauge-flation. The next realization with the same phenomenology has been introduced as Chromo-natural inflation \cite{Adshead:2013nka}. In addition to the original models, several more inflationary models with the $SU(2)$ VEV have been proposed and studied which share the same key features. For a review on gauge fields in inflation see \cite{Maleknejad:2012fw} and for a recent overview on the $SU(2)$-inflation models so far in the literature and their classification see Sec. 2 of \cite{Maleknejad:2018nxz} and the references therein.

\begin{figure*}[!htb]
\centering
\includegraphics[scale=0.3]{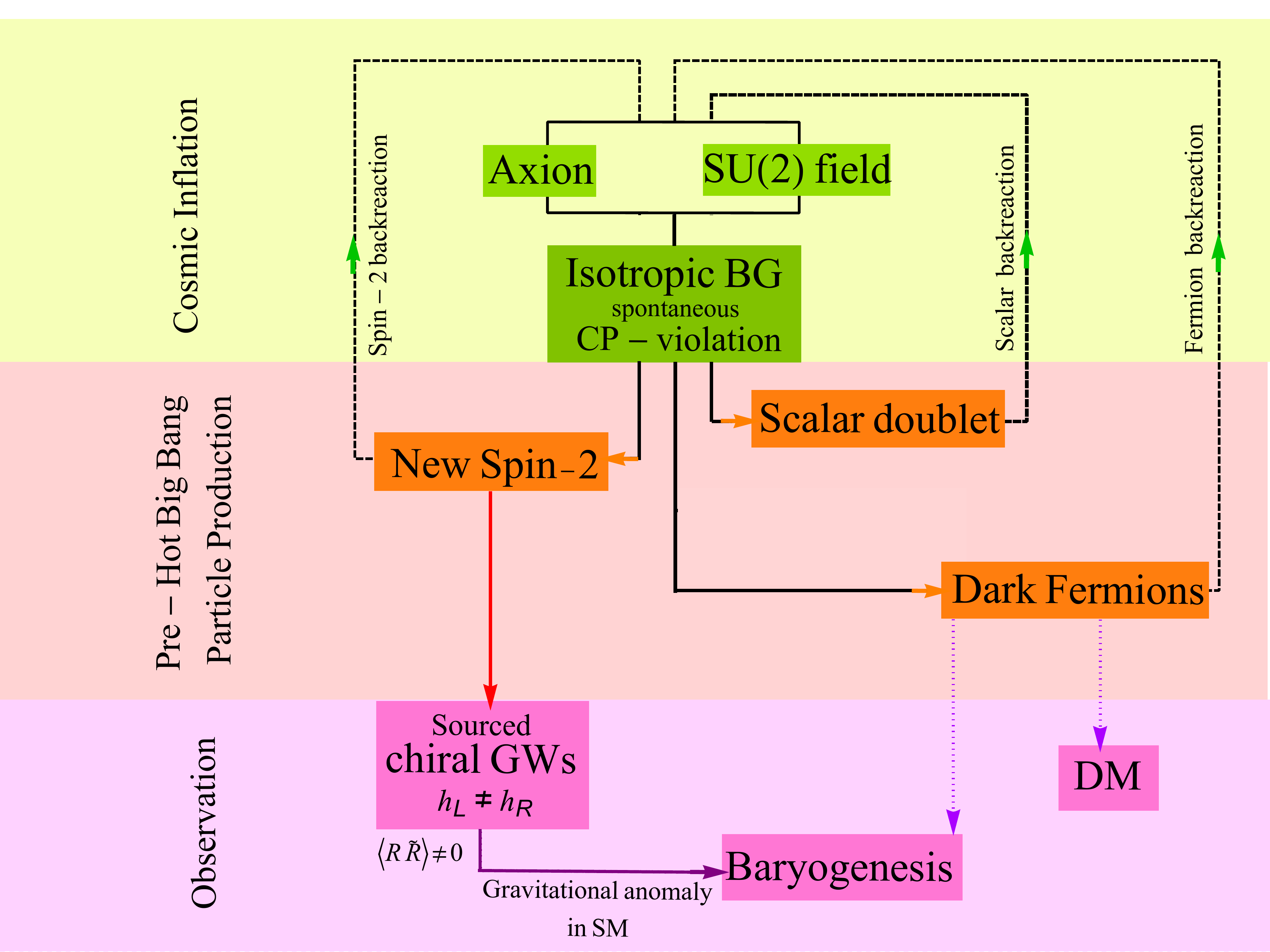}
\caption{
The $SU(2)$-axion inflation setup is a complete beyond standard model (BSM) which provides new mechanisms to explain and relate seemingly disparate phenomena in inflation and pre-Hot Big Bang (e.g. massive dark fermions, new spin-2 field and chiral gravity waves) with late time cosmological observations (dark matter, matter asymmetry, and primordial gravitational waves). Namely, the non-trivial $CP$-violating vacuum leads to distinctive features for the spin-2 and fermion fields coupled to it. The solid (dashed) back arrows show the particle production by (backreaction of the particles to) the vacuum. The colorful arrows show emerged phenomena due to the corresponding inflationary particle production. The dotted lines indicate unexplored directions. }\label{the-vac}
\end{figure*}

Despite their differences in details, all of the inflationary models with the $SU(2)$ VEV share several key features as robust consequences of having non-Abelian gauge fields in the physics of inflation. The illustration in Fig. \ref{the-vac} summarizes some of them.
The distinctive features of the non-trivial VEV are as follows. 1) The perturbed gauge field has a new \textit{chiral} tensorial mode (spin-2 field) \footnote{For details about the nature of this tensorial mode in the $SU(2)$ gauge field, perturbed around its isotropic and homogenous background, see App. B.1 of \cite{Maleknejad:2018nxz}.} which is linearly coupled to the gravitational waves, and generates chiral primordial gravitational waves \cite{Maleknejad:2011sq, Dimastrogiovanni:2012ew,Adshead:2013qp}, parity-odd correlations of CMB, i.e., non-zero $TB$ and $EB$ \cite{Thorne:2017jft}, and a possible large tensor bispectrum \cite{Agrawal:2017awz, Agrawal:2018mrg}. In \cite{Maleknejad:2018nxz}, the backreaction of the spin-2 field on the background fields has been studied analytically for all of the models so far in the literature. \footnote{At second order in perturbation, the spin-2 field couples to the curvature perturbation and contributes to the scalar power spectrum and bispectrum. For the chromo-natural model this effect has been studied in \cite{Papageorgiou:2018rfx}.} 2) Thanks to the $SU(2)$-axion vacuum which spontanously broke $CP$ in inflation and the new spin-2 field, this class of models has a non-trivial topology with $\langle R\tilde R\rangle \neq 0$ which through the gravitational anomaly in SM, $\nabla_{\mu}J^{\mu}_l = \frac{N_L-N_R}{384\pi^2} R\tilde R$, provides a natural leptogenesis mechanism during inflation \cite{Maleknejad:2014wsa, Maleknejad:2016dci, Noorbala:2012fh}. \footnote{The idea of inflationary leptogenesis through the gravitational anomaly in SM was first proposed within the context of modified gravity in the seminal work \cite{Alexander:2004us}.} Assuming type-I seesaw mechanism; the generated lepton number is proportional to the energy density of the gauge field during inflation, which provides the source of $CP$ violation and the mass of the heaviest right-handed neutrino \cite{Maleknejad:2016dci}. 3) It provides a new mechanism for the non-thermal generation of heavy, self-interacting dark fermions during inflation. The extensive analytical study of this \textit{pre-Hot Big Bang fermiogenesis} mechanism is the focus of the current work. 4) The produced heavy dark fermions might decay to the visible sector though some axion-mediated and/or direct feebly interactions which given the spontaneous $CP$-violation of the vacuum can provide a new mechanism for baryo- and leptogenesis.  This promising possibility is beyond the scope of the current work, and we leave for future study. It worth mentioning that the $SU(2)$ VEV can also produce scalar (Higgs) doublets coupled to it \cite{Lozanov:2018kpk}. For a comparison between the scalar and fermion production in this setup see Sec. \ref{scalar-fermion}.

\begin{figure*}[!htb]
\centering
\includegraphics[scale=0.25]{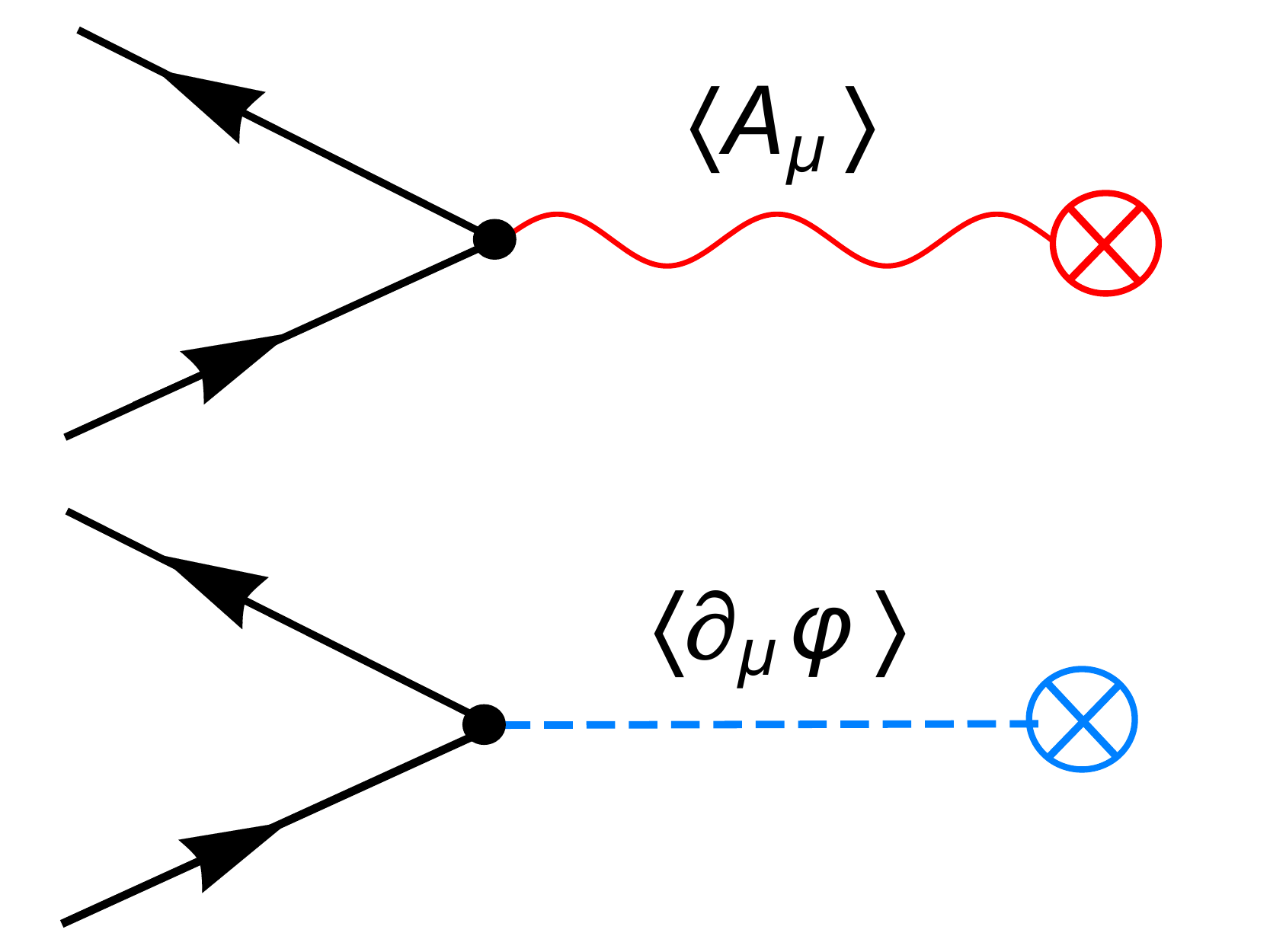}
\caption{The processes underlying inflationary fermion production studied in this work. The gauge field vacuum naturally induces fermionic currents through $i\bar{\Psi} \slashed{D} \Psi \supset \ga \bar{\Psi} \gamma^{\mu} A_{\mu} \Psi $ (top). Moreover, fermions can possibly be coupled to the axion by effective interaction $\p_{\mu}\varphi \bar{\Psi} \gamma^{\mu} \gamma^{5} \Psi$ (bottom).}
\label{fig:diag-F}
\end{figure*}

In de Sitter, gravitational production of massive fermions has been studied in \cite{Chung:2011ck}, their Schwinger production due to a constant and homogenous $U(1)$ field is studied in \cite{Hayashinaka:2016qqn}, and massive fermions coupled to slowly evolving axionic backgrounds was studied in \cite{Adshead:2015kza, Adshead:2018oaa} and the gravitational wave power spectrum induced by that fermions is studied in \cite{Adshead:2019aac}. Including the $SU(2)$ VEV in the physics of inflation, SM lepton and baryon production through gravitational anomaly has been studied in \cite{Maleknejad:2014wsa, Maleknejad:2016dci} (see also  \cite{Adshead:2017znw, Papageorgiou:2017yup, Alexander:2018fjp,Domcke:2018gfr}). Recently in \cite{Mirzagholi:2019jeb} by the author, a doublet of massive Dirac fermions in $SU(2)$-axion setting charged under the $SU(2)$ gauge field has been introduced, and the fermion backreaction to the background gauge field $J_a^{\mu}$, has been studied numerically for a small part of the parameter space, i.e. $m\sim H$. The current work presents an extensive study of the setup by computing all of the non-vanishing fermionic induced currents analytically and shows that this setup naturally leads to a new (non-thermal) mechanism for generating massive fermions during inflation. We regularize the UV divergent momentum integrals by the point-splitting regularization skim.

The paper is structured as follows. In Sec. \ref{sec:setup} we review the setup of a dark Dirac fermion charged under the gauge field in a generic $SU(2)$-axion inflation. In Sec. \ref{vacuum}, we discuss the symmetry structure of the $SU(2)$-axion vacuum and its consequences on the fermionic sector. In Sec. \ref{sec:PP}, we solve for the mode functions in (quasi) de Sitter. Next in Sec. \ref{gen-sol}, we take a closer look at the solutions and investigate their generic features. Sec. \ref{F-current} presents the analytical form of the fermionic currents after point-splitting regularization. Sec. \ref{fermi-de} discusses the main results and compares the inflationary dark fermion production in different limits. In Sec \ref{fermio-genesis}, we take a quick view on the (non-thermal) dark fermion production mechanism naturally arises in our non-trivial vacuum. We finally conclude in \ref{sec:Discussion}. Technical details of the computations as well as the underlying mathematical tools are provided in App.s \ref{Math}-\ref{sec:current-comp}.  Readers interested in the main results may proceed directly to Sec.s \ref{fermi-de} and \ref{fermio-genesis}.

\textbf{Notations and conventions:} ~  In this work, we adopt the notation introduced in \cite{Mirzagholi:2019jeb} as much as possible. Although in some cases we use a more convenient notation to better capture the novel aspects of the setup. Here we deal with $8$, $4$ and $2$-component spinors which are acted upon by $8\times8$, $4\times4$ and $2\times2$ matrices, respectively. We place a tilde $(\sim)$ on top of $8$ dimensional spinor and matrix. The $4$-dimensional spinor and matrix remains unchanged, while the $2$-dimensional ones are written in boldface. Moreover, $\rm{I}_n$ represents the $n\times n$ identity matrix and the gamma matrices are in the Dirac representation unless otherwise stated. We use the Einstein summation notation, i.e. repeated indices (one upper and one lower) are summed. Greek letters starting from the middle of the alphabet, i.e. $\mu$, $\nu$, $\dots$, are used for the space-time indices, whereas the starting ones, i.e.  $\alpha$, $\beta$, $\dots$, present the indices of the tangent space (non-coordinate) bases. In particular, the tetrads are defined as
$g^{\mu\nu}\equiv \eta^{\alpha\beta} {\bf{e}}^{\mu}_{~\alpha} {\bf{e}}^{\nu}_{~\beta}$. As a result, $\gamma^\mu = {\bf{e}}^\mu_{~\alpha} \gamma^{\alpha}$. 
 We recall that the FRW tetrads can be defined as ${\bf{e}}^{\mu}_{~\alpha}= a^{-1}(\tau)\delta^{\mu}_{\alpha}$ and $\gamma^{\alpha}$ are the flat space Dirac matrices 
\bea
\gamma^0=\begin{pmatrix*}[c] \I &\boldsymbol{0} \\ \boldsymbol{0}& -\I\end{pmatrix*}\, \an \gamma^i=\begin{pmatrix*}[c] \boldsymbol{0} & \boldsymbol{\sigma}^i \\ -\boldsymbol{\sigma}^i & \boldsymbol{0}\end{pmatrix*} \,, \nonumber
\eea 
which obey the Clifford algebra
$$  \{\gamma^\alpha,\gamma^\beta\}=-2\eta^{\alpha\beta}\II,$$
where $\eta^{\alpha\beta}$ is the Minkowski metric with signature $(-+++)$. The indices of $\gamma^{\alpha}$ and $\boldsymbol{\sigma}^i$ are lowered as $\gamma_{\alpha}=\eta_{\alpha\beta}\gamma^{\beta}$ and $\boldsymbol{\sigma}_i=\delta_{ij}\boldsymbol{\sigma}^j$. The gamma matrices in the Weyl representation are presented as $\gamma^{\alpha}_{_{W}}$. The 4-dimensional Hermitian adjoint row spinor $\bar\Psi$ is given as $\bar\Psi \equiv \Psi^{\dag}\gamma^0$. Besides, $\gamma^5=i\gamma^0\gamma^1\gamma^2\gamma^3$. The $4\times 4$ helicity operator, $h(k^{\alpha})$, is defined as
\bea
h(k^{\alpha}) \equiv \I \otimes k^j.\boldsymbol{\sigma}_j.
\eea

 ${\bf{T}}_a$ are the generators of the $SU(2)$ group in the fundamental representation $
[  {\bf{T}}_a , {\bf{T}}_b ] = i \epsilon^c_{~ab} {\bf{T}}_c$, and we call the $SU(2)$ index \textit{color}. The generators of the $SU(2)$ and the Pauli matrices are related as ${\bf{T}}_a=\frac12 \boldsymbol{\tau}^a$. As a mathematical tool to take care of the entanglement of the color and Lorentz indices in our setup, we define the color-spin helicity operator (c-helicity) as
\bea\label{h4}
\mathfrak{h}(k^{\alpha}) \equiv \frac{\delta_{ai}}{k^2}~k^i.\boldsymbol{\tau}^a\otimes k^j.\boldsymbol{\sigma}_j,
\eea
in which $k^{\alpha}$ is the momentum of the mode with $k=\lvert \vec{k}\rvert$. Both $\boldsymbol{\sigma}^i$ and $\boldsymbol{\tau}^a$ are the Pauli matrices. However, $\boldsymbol{\sigma}^i$ represents the spin operator which carries a Lorentz index while $\boldsymbol{\tau}^a$ represents the generators of the gauge group and acts on color indices. Whenever convenient, we also use $\boldsymbol{\tau}^i = \delta_a^i \boldsymbol{\tau}^a$.

 Throughout this work we will use three different representations of the doublet fermions, i.e. isospin, Weyl, and c-helicity frames which are related by two unitary operators, $\tilde{T}_1$ and $\tilde{T}_2$. We present these mappings in the following table. 
\begin{table}[h!]
\begin{center}
\begin{tabular}{ | m{3cm} | m{2cm} | m{0.5cm} | m{2cm} | m{0.5cm}| m{3.1cm}|} 
\hline
~ Representation  & ~~~ Isospin  & $\rightarrow$ & ~~~~~Weyl & $\rightarrow$ & Color-spin helicity   \\ \hline    
~~Transformation & & \vskip 1mm  $\tilde{T}_1$ & & \vskip 1mm  $\tilde{T}_2$ & \\
~~~~  $\tilde\Uppsi \equiv a^{\frac32}\tilde{\Psi}$  & ~~~  $\begin{pmatrix}
\Uppsi^1 \\
\Uppsi^2
\end{pmatrix}$  &   & ~~~ $\begin{pmatrix}
\Uppsi_L \\
\Uppsi_R
\end{pmatrix}$ & & ~~~~~~ $\begin{pmatrix}
\Uppsi^+ \\
\Uppsi^-
\end{pmatrix}$  \\ 
& & & & & \\
 \hline
\end{tabular}
\end{center}
\label{T0}
\end{table}

Here, $\Uppsi^1$ and $\Uppsi^2$ are the eigenstates of the isopin, while $\Uppsi_L$ and $\Uppsi_R$ are chiral states which are the eigenstates of $\gamma^5$, i.e. $\gamma^5\Uppsi_{L,R}=\mp \Uppsi_{L,R}$. Finally, $\Uppsi^{\pm}$ are the eigenstates of the c-helicity operator defined in Eq. \eqref{h4}. More precisely, we have $ \mathfrak{h}(k^{\alpha}) \Uppsi^{\pm}=\pm \Uppsi^{\pm}$. Moreover, the subscripts $s$ and superscripts $p$ are labels of the c-helicity frame. Throughout this work, the superscript $\pm$ denotes the plus/minus subspace, while the subscript $\pm$ denotes the corresponding 2d helicity polarization states, i.e.
\bea\label{E+-E-}
{\bf{E}}_+ = \begin{pmatrix}
1 \\ 0
\end{pmatrix} \an {\bf{E}}_- = \begin{pmatrix}
0 \\ 1
\end{pmatrix}.
\eea
besides, the subscripts $C$ and $P$ denote the charge conjugated and parity transformed quantities respectively.

In constructing physical observables associated with Grassmann variables and computing the expectation values, we use the antinormal ordering on the creation and annihilation operators of the spinors, $\vdots$, defined as
\bea\label{anti-normal}
\vdots b^{~}_{\bf{k}} b^{\dag}_{\bf{k}} \vdots \equiv b^{~}_{\bf{k}} b^{\dag}_{\bf{k}} \an \vdots b^{\dag}_{\bf{k}} b^{~}_{\bf{k}} \vdots \equiv -b^{~}_{\bf{k}} b^{\dag}_{\bf{k}}.
\eea
Finally, the point-splitting 4-vector which splits each point in the space-time into a forward and backward point is denoted by $\varepsilon^{\mu}=(\varepsilon, {\boldsymbol{\varepsilon}}^i)$. The forward/backward (physical) coordinates corresponding to $X^{\mu}$ and the splitting 4-vector $\varepsilon^{\mu}$, are $X^{\mu}_f$/~$X^{\mu}_b$ defined as 
\begin{center}
\centering{
\begin{tikzpicture}
  \draw[line width=2pt,blue,-stealth](3,0)--(4,0) node[anchor=south west]{(forward) $X^{\mu}_f=X^{\mu}+\frac12\varepsilon^{\mu}$};
   \draw[black,thick] (3,0) circle (0.1cm);
  \draw[bostonuniversityred,thick,dotted] (2,0) circle (0.1cm);
  \draw[line width=2pt,bostonuniversityred,-stealth](3,0)--(2,0) node[anchor=north east]{$X^{\mu}_b=X^{\mu}-\frac12\varepsilon^{\mu}$ (backward)};
   \draw[blue,thick,dotted] (4,0) circle (0.1cm);
   \draw[thick](3.1,-0.2)--(3.1,-0.2) node[below]{$X^{\mu}$};
\end{tikzpicture}
}
\end{center}
and we will eventually take $\varepsilon^{\mu}$ to zero. Notice that the slow-roll parameter is shown by $\epsilon =-\frac{\dot H}{H^2}$.

\section{Setup}\label{sec:setup}
In this work, we consider an $SU(2)$ doublet fermion in an $SU(2)$-axion inflation setting. This setup has been recently introduced in \cite{Mirzagholi:2019jeb}. We consider a doublet of Dirac fermions 
\Beq
\label{eq:Higgs}
\tilde{\Psi}=\begin{pmatrix}\Psi^{1}\\ \Psi^{2}\end{pmatrix}\,,
\Eeq
charged under the gauge field of the $SU(2)$-axion model
\Beq
\label{eq:FullAction-}
S=\int\text{d}^4x\sqrt{-g}\left[i \bar{\tilde{\Psi}} {\sD} \tilde{\Psi} - m\bar{\tilde{\Psi}}\tilde{\Psi} \right]\,,
\Eeq
with mass $m$ and $\slashed{D}$ as 
\Beq\label{slashD}
\slashed{D} \equiv {D}_{\mu} \otimes \gamma^{\mu} = {\bf{e}}^{\mu}_{~\alpha} \big[ \I \nabla_{\mu}  - i\ga A^a_{\mu} {\bf{T}}_a\big] \otimes \gamma^\alpha.
\Eeq
For the sake of completeness, we also consider the possible effective interaction of the axion with the fermion as
\bea
S_{\rm{int}}=\int\text{d}^4x\sqrt{-g} \beta \frac{\lambda \varphi}{f}  \nabla_{\mu} J^{\mu}_5,
\eea
where $\varphi$ is the axion field, $f$ is the axion decay constant, $\lambda$ is the dimensionless coefficient of the Chern-Simons interaction term of the axion, and $\beta$ is a dimensionless parameter. Here, $\otimes$ is the Kronecker product. For definition of $\otimes$ see App. \ref{Math}.
Moreover, the quantity $\beta$ can be of order unity or lower. Here, $J^{\mu}_5$ is the chiral current given as
\bea\label{axial}
J^{\mu}_5 \equiv  \bar{\tilde{\Psi}} \I\otimes (\gamma^\mu \gamma^5)\tilde{\Psi}.
\eea
We assume slow-roll inflation with (quasi) de-Sitter metric
\bea
ds^2= a^2(\tau) ( - d\tau^2 + \delta_{ij}dx^i dx^j),
\eea
where $\tau$ is the conformal time, $H$ is the (almost) constant Hubble parameter during inflation, $a(\tau)$ is the scale factor, and $$\mH\equiv aH.$$
 Moreover, we assume a slowly-evolving homogeneous and isotropic $SU(2)$ gauge field in the temporal gauge as \cite{Maleknejad:2011sq} \footnote{ The stability of this homogeneous-isotropic gauge field solution against initial stochastical anisotropies of the Bianchi cosmology has been studied in \cite{Maleknejad:2011jr, Maleknejad:2013npa}. Interestingly, although such systems can in principle  acquires anisotropic hair \cite{Maleknejad:2012as, Watanabe:2009ct}, the homogeneous-isotropic gauge field ansatz is the attractor solution (For the massive gauge field case see \cite{Adshead:2018emn}).}
\bea\label{GF-SU2}
A_0(\tau,{\bf{x}})=A^a_0 {\bf{T}}_a =0  \an  {\bf{A}}_{i}(\tau,{\bf{x}}) = A^a_{i} {\bf{T}}_a = \frac12 a(t) \psi(\tau) \boldsymbol{\tau}^i,
\eea 
and a homogeneous axion field     
\bea\label{axion}
\varphi(\tau,{\bf{x}}) = \varphi(\tau),
\eea
where $\psi(\tau)$ and $\varphi(\tau)$ are two slow-varying pseudo-scalars during slow-roll inflation. 
Combination of the slow-varying $SU(2)$ gauge field Eq. \eqref{GF-SU2} and axion Eq. \eqref{axion} make our \textit{$SU(2)$-axion vacuum} (See Fig. \ref{P-B}). 
The class of inflationary models with such $SU(2)$ VEV has several different realizations. See section 2 of \cite{Maleknejad:2018nxz} and the references therein for a recent comprehensive review on the models. In the perturbation sector, the $SU(2)$-axion vacuum appears in terms of the following dimensionless quantities \footnote{Notice that $\xpi$ is different from $\xi \equiv \frac{\lambda\p_{\tau}\varphi}{2afH}$ in the axion inflation backgrounds. In fact, $\xi$ and $\xpi$ are related as
$\xpi = \beta \xi$, where $\beta$ can be of order unity.}
\bea\label{def:xa-xphi}
\xa \equiv \frac{\ga \psi}{H}, \quad \xi_{\varphi} \equiv \beta \frac{\lambda\p_{\tau}\varphi}{2afH}.
\eea
It is also useful to define a dimensionless parameter from the mass of the fermion as
\bea
\mu_{{\rm m}} \equiv \frac{m}{H}.
\eea
Moreover, it is more convenient to work with the canonically normalized field
\bea
\tilde\Uppsi \equiv a^{\frac32}\tilde\Psi,
\eea
which is the comoving fermionic field. In the Fourier space, we can expand $\tilde\Uppsi(\tau,{\bf{x}})$ as
\bea
\tilde\Uppsi(\tau,{\bf{x}}) = \int {\rm{d}}^3k e^{i{\bf{k}}.{\bf{x}}} \tilde\Uppsi_{{\bf{k}}},
\eea
and the fermion theory reads as
\bea\label{theory}
\mathcal{L}_{\Psi} =  \bar{\tilde\Uppsi}_{{\bf{k}}} \bigg[ i\I   \otimes \gamma^0 \p_{\tau} -  \bigg( k^i \I  - \frac{1}{2}  \xa \mH \boldsymbol{\tau}^i \bigg) \otimes \gamma_{i} -  \mu_{{\rm{m}}} \mH \III  - 2 \xi_{\varphi} \mH \I \otimes (\gamma^0 \gamma^5) \bigg] \tilde\Uppsi_{{\bf{k}}}.\,\nonumber\\
\eea
The above theory is in terms of $8$-spinors wherein the color and spin indices are entangled. The reason is the homogenous-isotropic gauge field's VEV in Eq. \eqref{GF-SU2} in which the gauge group index associated with $su(2)$ and spatial Lorentz index associated with $so(3)$ has been identified to restore the rotational symmetry in the background (For a detailed discussion see \cite{Maleknejad:2012fw}). The next step would be to find a frame in which the system is possibly reducible to lower dimension subspinors.

\subsection{$\Uppsi^{+}$ and $\Uppsi^{-}$ decomposition }\label{decompose-sec}

Up to this point, we were in the isospin frame. However, the ideal frame, would be the one in which the three of $\I\otimes (\gamma^0\gamma^i)$, $\I\otimes k^i\gamma_i$, and $\boldsymbol{\tau}^i\otimes\gamma_i$ are diagonal. If exists, such a frame is made of two copies of the common eigenstates of $  \I\otimes k^i.\boldsymbol{\sigma}_i $, and $ \boldsymbol{\tau}^i \otimes \boldsymbol{\sigma}_i$. However, these two operators have only two common eigenstates. Thus, the theory is only block-diagonalizable and can be decomposed into two subspaces. 

As a mathematical tool for any given 4-momentum $k^{\alpha}$, we define the color-spin helicity operator (c-helicity) as
\bea
\mathfrak{h}(k^{\alpha}) \equiv \frac{1}{k^2}~k^i.\boldsymbol{\tau}_i\otimes k^j.\boldsymbol{\sigma}_j,
\eea
in which the first Pauli matrix is the $su(2)$ generator while the second one is the spin operator. As we will show shortly, it is easier to expand the system in terms of the eigenstates of c-helicity, which automatically takes care of the entangled color and spin indices in the action \eqref{theory}.
The orthonormal eigenstates of c-helicity satisfy in
\Beq
\mathfrak{h}(k^{\alpha})~ E^{p}_{~s}(k^{\alpha}) = sp ~ E^{p}_{~s}(k^{\alpha})\,,
\Eeq
where $p=\pm1$, $s=\pm1$, and $E^{p}_{~s}(k^{\alpha})$ are
\Beq
E^{+}_{~+}(k^{\alpha})&=\frac{\check{k}^{\alpha}\bar{\boldsymbol{\tau}}_{\alpha}\otimes\check{k}^{\beta}\bar{\boldsymbol{\sigma}}_{\beta}}{2k(k+k^3)}\begin{pmatrix}1 \\ 0\\ 0 \\ 0\end{pmatrix}\,,\quad E^{+}_{~-}(k^{\alpha})=-\frac{\check{k}^{\alpha}\boldsymbol{\tau}_{\alpha}\otimes\check{k}^{\beta}\boldsymbol{\sigma}_{\beta}}{2k(k+k^3)}\begin{pmatrix}0 \\ 0\\ 0 \\ 1\end{pmatrix}\,,\\
E^{-}_{~+}(k^{\alpha})&=-\frac{\check{k}^{\alpha}\boldsymbol{\tau}_{\alpha}\otimes\check{k}^{\beta}\bar{\boldsymbol{\sigma}}_{\beta}}{2k(k+k^3)}\begin{pmatrix}0 \\ 0\\ 1 \\ 0\end{pmatrix}, \, \quad E^{-}_{~-}(k^{\alpha})=-\frac{\check{k}^{\alpha}\bar{\boldsymbol{\tau}}_{\alpha}\otimes\check{k}^{\beta}\boldsymbol{\sigma}_{\beta}}{2k(k+k^3)}\begin{pmatrix}0 \\ 1\\ 0 \\ 0\end{pmatrix}\,,
\Eeq
in which $\boldsymbol{\sigma}^{\alpha}$ and $\bar{\boldsymbol{\sigma}}^{\alpha}$ are 
\bea
\boldsymbol{\sigma}^{\alpha}= (\I, \boldsymbol{\sigma}^i)    \an \bar{\boldsymbol{\sigma}}^{\alpha}=(\I, -\boldsymbol{\sigma}^i),
\eea
and their indices are lowered with the 
Minkowski metric, i.e. $\boldsymbol{\sigma}_{\alpha}= \eta_{\alpha\beta} \boldsymbol{\sigma}^{\beta}$. In a similar way, we define $\boldsymbol{\tau}^{\alpha}$ and $\bar{\boldsymbol{\tau}}^{\alpha}$ from $\boldsymbol{\tau}^i$. Besides, 
 $\check{k}^{\alpha}$ is a four vector given as
\bea
\check{k}^{\alpha} \equiv (k,\bk),
\eea
where $k=\sqrt{{\bf{k}}. {\bf{k}}}$. Notice that $\check{k}^{\alpha}$ is the four momentum of the massless field, but for the massive cases it is just a mathematical tool. It is straightforward to see that $ E^{p}_{~s}(k^{\alpha})$ are eigenstates of helicity operator as well
\Beq
 \I\otimes k^i.\boldsymbol{\sigma}_i  E^{p}_{~s}(k^{\alpha}) = s ~ k E^{p}_{~s}(k^{\alpha})\,,
\Eeq
and satisfy the orthonormality condition
\bea
 E^{p\dagger}_{~s}(k^{\alpha})\cdot E^{p'}_{~s'}(k^{\alpha})=\delta_{ss'}\delta^{pp'}\,.
\eea
The $p=+1$ elements are also eigenvectors of $\boldsymbol{\tau}^i\otimes \boldsymbol{\sigma}_i$ in Eq. \eqref{Sigma4-}
\Beq
\boldsymbol{\tau}^i\otimes \boldsymbol{\sigma}_i E^{+}_{~s}(k^{\alpha})= E^{+}_{~s}(k^{\alpha})\,.
\Eeq
However, the $p=-1$ elements does not satisfy eigenstate equations with $\boldsymbol{\tau}^i\otimes \boldsymbol{\sigma}_i$ operator. Thus, the system is not fully, but only block diagonalizable. 

The $E^{p}_{~s}(k^{\alpha})$ 4-vectors make an orthonormal basis and two copies of them define a $8\times 8$ frame, i.e c-helicity frame. As it is implies by the form of $E^{p}_{~s}(k^{\alpha})$s, the color and spin are totally mixed in this frame. It is shown in \cite{Mirzagholi:2019jeb} by the author that the system is decomposed into two independent subsystems. For self-sufficiency, we provide the details in App. \ref{subsec:b1} and here we only present the final forms. The plus subspace is made of two copies of $E^+_{~s}(k^{\alpha})$, and the minus subspace is made of two copies of $E^-_{~s}(k^{\alpha})$, i.e.
\bea
\tilde\Uppsi^{\pm}_{{\bf{k}}} = \Uppsi^+_{{\bf{k}}} \oplus \Uppsi^-_{{\bf{k}}} = \begin{pmatrix}
\Uppsi^+_{{\bf{k}}} \\
\Uppsi^-_{{\bf{k}}}
\end{pmatrix},
\eea
 such that the theory is given as
$$S[\tilde\Uppsi] = S^+[\Uppsi^+] + S^-[\Uppsi^-],$$
where $S^+[\Uppsi^+]$ and $S^-[\Uppsi^-]$ are \footnote{Notice that our notation in defining the c-helicity frame is slightly different than \cite{Mirzagholi:2019jeb} so that the sign of $k\gamma^3$ term be the same in both subspaces.}
\bea    
\label{eq:SLSR-P}
&S^+&=\int \text{d}\tau\text{d}^3k \bar{\Uppsi}^+_{\bk} \left[i\gamma^0 \partial_{\tau}-\gamma^3k-\big(2\xpi -\frac{\xi_A}{2}\big)\mathcal{H}\gamma^0\gamma^5 - \mu_{{\rm m}}\mH \II \right]\Uppsi^+_{\bk}\,,\\
\label{eq:SLSR-M}
&S^-&=\int \text{d}\tau\text{d}^3k \bar{\Uppsi}^-_{\bk} \left[i\gamma^0\partial_{\tau}-\gamma^3k-\big(2\xpi +\frac{\xi_A}{2}\big)\mathcal{H}\gamma^0\gamma^5 -\mu_{{\rm m}} \mH \II +\gamma^1\xi_A\mathcal{H}\right] \Uppsi^-_{\bk} \,.
\eea

The system is reducible into two 4-Dirac spinors, $\Uppsi^+$ and $\Uppsi^-$, thanks to the isospin symmetry as well as the isotropic-homogeneous configuration of the gauge field's VEV which identifies the $su(2)$ gauge index and $so(3)$ rotation index. In particular, in case that the mass term breaks the isospin symmetry, i.e., the isospinors have different masses, the $\Uppsi^{+}_{\bf{k}}$ and $\Uppsi^{-}_{\bf{k}}$ fields would be coupled by terms proportional to the difference of the isospinor masses. Hence the reduction is not possible in the absence of the isospin symmetry. 

Our Dirac fields can be expanded as
\bea
\Uppsi^{\pm}_{\bf{k}} = \sum_{s=\pm} \begin{pmatrix}
~\uppsi^{\pm\uparrow}_{s}(\tau,k) {\bf{E}}_s \\
\\
s \uppsi^{\pm\downarrow}_{s}(\tau,k)  {\bf{E}}_s
\end{pmatrix},
\eea
where $\uppsi^{\pm\uparrow}_{s}(\tau,k)$ and $\uppsi^{\pm\downarrow}_{s}(\tau,k)$ are mode functions and ${\bf{E}}_s$ with $s=\pm 1$ are the two-spinor polarization states
\bea\label{E+-E-}
{\bf{E}}_+ = \begin{pmatrix}
1 \\ 0
\end{pmatrix} \an {\bf{E}}_- = \begin{pmatrix}
0 \\ 1
\end{pmatrix}.
\eea
Note 1: the superscript $\pm$ denotes the plus/minus subspace, while the subscript $\pm$ denotes the corresponding 2d helicity polarization states, ${\bf{E}}_{\pm}$. Note 2: being already in the helicity frame for the given momentum, the 2-spinor polarization states are $k$-independent. As a result, the plus spinor is decoupled in terms of the polarization spinor ${\bf{E}}_{s}$ and it is makes two pairs of coupled field equations for each polarization. On the other hand, the minus spinor is not diagonalizable due to the extra (time dependent) term proportional to $\gamma^1$ in the action \eqref{eq:SLSR-M}. As a result, the minus subsystem includes four coupled field equations. In the limit of either well inside the horizon, i.e. $k\gg \mH$, or $\xa/\mu_{\rm{m}}\ll 1$, this term is negligible and the minus system approximately decoupled into two pairs of coupled field equations.

Our fermionic sector generates the fermion current and isospin current as
\bea\label{fermion-number}
J^{\mu} = \bar{\tilde{\Psi}} \I \otimes\gamma^{\mu} \tilde{\Psi} \an  J^{\mu a} = \ga \bar{\tilde{\Psi}}  {\bf{T}}^a\otimes \gamma^{\mu} \tilde{\Psi},
\eea
as well as axial current, $J^{\mu}_5$, in Eq \ref{axial} and axial isospin current
\bea\label{axial-isospin}
J^{\mu a}_5 = \ga \bar{\tilde{\Psi}} {\bf{T}}^a\otimes\gamma^{\mu}\gamma^{5} \tilde{\Psi}.
\eea
Among them, $J^{\mu a}$ backreacts on the background equation of the gauge field as \footnote{For the explicit form of the background equations in different realizations of $SU(2)$-axion inflationary models see Sec. 2 of \cite{Maleknejad:2018nxz}.}
\Beq
\label{eq:CurlyJ}
 \frac{1}{3a} \delta^a_i \frac{\delta\mathcal{L}_{\rm{inf}}}{\delta A^a_i} \bigg\rvert_{BG}= \mathcal{J}  \equiv \frac{1}{3a}\delta_{a}^{i}\langle J_{i}^{a}\rangle\,,
\Eeq
where the left hand side is the background field equation of the $SU(2)$ gauge field and $\mathcal{L}_{\rm{inf}}$ is the specific theory of the $SU(2)$-axion inflation model.
Moreover, the axial current backreacts on the axion background equation as
\Beq\label{axion-BR}
-\frac{\delta\mathcal{L}}{\delta \varphi} \bigg\rvert_{BG}  =  \beta \frac{\lambda}{2f}\nabla_{\mu}J^{\mu}_5 \equiv \mathcal{B} \,,
\Eeq
where the left hand side is the background field equation of the axion, and the $\nabla_{\mu}J^{\mu}_5$ is given as
\Beq
\label{eq:BAxionBack}
 \nabla_{\mu}J^{\mu}_5 =  -  \frac{2im}{a^3}\bar{\tilde{\Uppsi}}\gamma_5 \tilde{\Uppsi} - \frac{2\ga^2}{16\pi^2}F^{a\mu\nu}\tilde{F}_{a\mu\nu} \,.
\Eeq
In the right-hand side, the first term is the tree-level effect of explicite breaking of the chiral symmetry by the mass term, while the second term is the loop effect due to the well-known chiral (Adler-Bell-Jackiw) anomaly \cite{Adler:1969gk, Bell:1969ts}.

\section{Non-trivial vacuum structure and fermions}\label{vacuum}
To gain more insight on the nature of our dark fermions and capture some of the novel aspects of the vacuum, here we explore the action of the discrete symmetries $C$, $P$ and $CP$, as well as the continuous chiral symmetry on fermions. Here, just by using the symmetry structure of the system, we extract the general features of the observable quantities.  Later in Sec. \ref{F-current}, we explicitly compute these quantities and confirm this qualitative analysis.
The chiral charge and the fermion backreaction to the background gauge field equation, Eq.s \eqref{axial} and \eqref{eq:CurlyJ}, in the Weyl frame can be written as \footnote{Notice that the chiral charge is defined in terms of $J^{t}_5=\frac{\p t}{\p \tau}J^{0}_5$ which justifies the extra factor of $a$ inside the integral Eq. \eqref{Q5}.}
\bea\label{Q5}
Q_5 &\equiv & \langle aJ^0_{5} \rangle = Q_{R}-Q_{L},\\ \label{cJ}
\mathcal{J} & = & \frac{1}{3a} \delta^i_{a} \langle J^a_{iL} + J^a_{iR} \rangle,
\eea 
in which the R/L subscription denotes the contribution of the right-/left-handed fields to that quantity. The net fermion number Eq \eqref{fermion-number} is given as
\bea
Q \equiv  \langle aJ^0 \rangle  =   Q_{R}+Q_{L}, 
\eea
and the axial isospin current can be written as
\bea
J^{\mu a}_5 =  J^{\mu a}_R  - J^{\mu a}_L. 
\eea
The left- and right-handed isospin currents are
\bea\label{Isospin-L-R}
J^{\mu a}_L \equiv \ga \Uppsi^{\dag}_L \textbf{T}^a \otimes  \bar{\boldsymbol{\sigma}}^{\mu} \Uppsi_L \an J^{\mu a}_R \equiv \ga \Uppsi^{\dag}_R  \textbf{T}^a  \otimes\boldsymbol{\sigma}^{\mu} \Uppsi_R.
\eea

\subsection{C-symmetry, P- and CP-violation}\label{CP-v}

As showed in Eq. \ref{CC}, our fermion theory in Eq. \eqref{theory} is invariant under the action of the charge conjugation operator, $\tilde{C}$, defined as
\bea
\tilde{C} = \I\otimes C \where C=i\gamma^2\gamma^0.
\eea
In other words, both $\tilde{\Uppsi}$, and its charge conjugated field $$\tilde{\Uppsi}_c\equiv \tilde{C} \bar{\tilde{\Uppsi}},$$ obey the same theory (See Eq. \eqref{Psi-Psi-c}) and the theory is $C$-symmetric. The $4\times 4$ operator $C$ is the charge conjugation operator of each plus and minus subsystems. 

However, parity is different since both $\varphi$ (axion) and $\psi$ (the effective field value of the $SU(2)$ gauge field) in Eq.s \eqref{GF-SU2} and \eqref{axion}, are pseudo-scalars. As a result, parity is spontaneously broken in the action \eqref{theory}. Under the action of parity, this theory goes to its parity conjugate with (see Fig. \ref{P-B})
\bea
\xp \xrightarrow{P} -\xp \an \xpi \xrightarrow{P} -\xpi.
\eea
 As a result, both $P$ and $CP$ are spontaneously broken by the vacuum of the $SU(2)$-axion, i.e., the vacuum is $CP$-violating with non-vanishing $\tilde R R$ and $F\tilde F$ \cite{Maleknejad:2014wsa, Maleknejad:2016dci}. Violation of $CP$ is the essential condition in any matter-antimatter asymmetric scenario. Then the out of thermal equilibrium state which is guaranteed by inflation ensures that the production of matter by one mechanism is not immediately compensated by its disappearance through the inverse reaction. This robust aspect of $SU(2)$-axion models make them natural settings for inflationary leptogenesis \cite{Maleknejad:2014wsa, Maleknejad:2016dci}. In the current fermionic sector with $C$-symmetry and $CP$-violation, we expect a vanishing net fermion number, $J^0=0$ and a non-zero chiral charge, $J^0_5\neq0$ which will be confirmed by the exact computation in section \ref{F-current}. There we will show that both $\mathcal{J}$ and $Q_5$ are directly proportional to the sources of the $CP$-violation, i.e., $\xp$ and $\xpi$.

\subsection{Chiral anomaly and conserved currents}\label{CSym}

In the massless limit, the fermionic sector enjoys the chiral symmetry in the classical level under the unitary chiral transformation
\bea
\tilde\Psi  \rightarrow e^{i\alpha^a(x) {\bf{t}}_a \otimes\gamma_5} ~ \tilde\Psi,
\eea
where ${\bf{t}}_a$ is the isospin matrix (equals to ${\bf{T}}_a$ for isospin doublet and $\I$ for isospin singlet), and $\alpha^a(x)$ is the parameter of the transformation.
All four fermionic currents are classically conserved under the above unitary transformation.
At the quantum level, the left- and right-handed isospin currents in Eq \eqref{Isospin-L-R} satisfy the following equalities by the non-Abelian chiral anomaly
\bea
D_{\mu} J^{\mu a}_{L,R} = \pm \frac{\ga^2}{24\pi^2} \nabla_{\mu} \bigg(  \epsilon^{\mu\nu\lambda\sigma} {\rm{tr}} \bigg[ \boldsymbol{T}^a \big( A_{\nu} \p_{\lambda} A_{\sigma} - \frac{\ga}{2} A_{\nu} A_{\lambda} A_{\sigma} \big)
\bigg]\bigg),
\eea
which implies that the (vector) isospin current satisfies 
\bea\label{anomaly-free}
D_{\mu} J^{\mu a} = 0,
\eea
and the axial isospin current satisfies 
\bea\label{DJa5}
D_{\mu} J^{\mu a}_5 = - \frac{\ga^2}{12\pi^2} \nabla_{\mu} \bigg(  \epsilon^{\mu\nu\lambda\sigma} {\rm{tr}} \bigg[ \boldsymbol{T}^a \big( A_{\nu} \p_{\lambda} A_{\sigma} - \frac{\ga}{2} A_{\nu} A_{\lambda} A_{\sigma} \big)
\bigg]\bigg).
\eea
Thus, the fermion vector current and isospin current remain anomaly free at the quantum level.
In particular, Eq. \eqref{anomaly-free} consistency condition for the gauge theory \cite{Peskin:1995ev}. 
However, the axial vector and axial isospin currents receives corrections according to the Adler-Bell-Jackiw anomaly \cite{Bell:1969ts, Adler:1969gk} respectively as
\bea\label{Abelian-CA}
\nabla_{\mu} J^{\mu}_{5} =  - \frac{\ga^2}{16\pi^2} \epsilon^{\mu\nu\lambda\sigma} F^b_{\mu\nu} F^c_{\lambda\sigma} {\rm{tr}}\big[ \{{\bf{T}}_b,{\bf{T}}_c\}\big] =  - \frac{2\ga^2}{16\pi^2} F\tilde F,
\eea
and the chiral isospin current anomaly in Eq. \eqref{DJa5} can be written as
\bea
D_{\mu} J^{\mu a}_{5} =  - \frac{\ga^2}{16\pi^2} \epsilon^{\mu\nu\lambda\sigma} F^b_{\mu\nu} F^c_{\lambda\sigma} {\rm{tr}} \big[ {\bf{T}}_a \{{\bf{T}}_b,{\bf{T}}_c\}\big] =0.
\eea
Therefore, in this setup the axial isospin current is not affected by the chiral anomaly and remains conserved. 

In the zero mass limit, the only fermionic effect is the Abelian chiral anomaly in Eq. \eqref{Abelian-CA}, $\frac{\ga^2}{16\pi^2}F\tilde{F}$ which backreacts to the axion field as well as the spatial component of the chiral isospin current, $J^{ai}_5\simeq  \frac{\ga^2}{6\pi^2}\psi^2H = \frac{\xp^2}{6\pi^2}H^3 $. Massive fermions, however, break chiral symmetry explicitly and have non-zero $\mathcal{J}$ as well. Recalling \eqref{cJ}, it should be directly proportional to the mass of the fermions, i.e.
\bea
\mathcal{J} \propto \mu_{\rm{m}},
\eea
since $\mu_{\rm{m}}=\frac{m}{H}$ quantifies the deviation from chiral symmetry.

\section{Fermions in SU(2)-axion inflation}\label{sec:PP}
Now we turn to solve the field equations and find $\Uppsi^{\pm}_{\bk}$ specified by the actions in the Eq.s \eqref{eq:SLSR-P} and \eqref{eq:SLSR-M}. 
The $\Uppsi^{p}_{\bk}$ spinors with $p=\pm 1$ can be decomposed as
\Beq
\label{eq:psipls}
\Uppsi^{p}_{\bk}=\sum_{s=\pm}\left[U_{s,{\bf{k}}}^{p}(\tau)a^{p}_{s,\bk}+V_{s,-{\bf{k}}}^{p}(\tau)b^{p\dagger}_{s,-\bk}\right]\,,
\Eeq
where $a^{p}_{s,\bk}$ and $b^{p}_{s,\bk}$ are creation and annihilation operators obey
\Beq
\label{eq:Anticommab-}
\{a^{p}_{s,\bk},a^{p\dagger}_{s',\bk'}\}=\delta_{ss'}\delta^{(3)}(\bk-\bk') \an \{b^{p}_{s,\bk},b^{p\dagger}_{s',\bk'}\}=\delta_{ss'}\delta^{(3)}(\bk-\bk')\,.
\Eeq
Note that the superscript $p=\pm$ denotes the plus/minus subspace, while the subscript $s=\pm$ denotes the corresponding 2d helicity polarization states, ${\bf{E}}_{\pm}$.
Moreover, the charge conjugation relates the field to its charge conjugated field as (see Sec. \ref{CC})
\bea
\Uppsi^{p}_{C\bf{k}} = i\gamma^2 \Uppsi^{p*}_{-\bf{k}},
\eea
which gives 
\bea\label{V-U}
V^{p}_{s,\bf{k}} =  i \gamma^2 U^{p*}_{s,\bf{k}}.
\eea
The 4-spinor fields, $\Uppsi^{p}_{\bk}$, are govern by first order differential equations. Therefore, they are each made of four linearly independent solutions. We will solve them by setting the Bunch-Davies vacuum as the initial condition for the fields
\Beq
\label{eq:BD-}
\lim_{\tau\rightarrow -\infty} U_{s,{\bf{k}}}^p(\tau)=\frac{e^{-ik\tau}}{(2\pi)^3}\begin{pmatrix} ~{\bf E}_s \\ s{\bf E}_s\end{pmatrix}.
\Eeq
Here, we define the rescaled physical momentum as
\bea
\x \equiv \frac{k}{aH} = - k\tau.
\eea
Since the plus and minus spinors are decoupled, we study them individually.

\subsection{$\Uppsi^+$ fermions}

As we discussed in Sec. \ref{decompose-sec}, the $\Uppsi^+_{\bf{k}}$ spinors are defined to be diagonalizable in terms of the c-helicity. The $\Uppsi^+_{\bk}$ spinors are described by the linear differential equation specified by the action \eqref{eq:SLSR-P}. Therefore, $U_{s,{\bf{k}}}^+(\tau)$ and $V_{s,{\bf{k}}}^+(\tau)$ in Eq. \eqref{eq:psipls} can be decomposed in terms of the 2d helicity polarization states, ${\bf{E}}_{\pm}$, defined in Eq. \eqref{E+-E-} as
\Beq
\label{eq:UVpsipls}
U_{s,{\bf{k}}}^+(\tau)=\frac{1}{\sqrt{2}}\begin{pmatrix} ~{\bf E}_su^{\uparrow}_s(k,\tau) \\ s{\bf E}_su^{\downarrow}_s(k,\tau)\end{pmatrix} \quad {\rm and} \quad V_{s,-{\bf{k}}}^+(\tau)=\frac{1}{\sqrt{2}}\begin{pmatrix} ~{\bf E}_sv^{\uparrow}_s(k,\tau) \\ s{\bf E}_sv^{\downarrow}_s(k,\tau)\end{pmatrix}\,.
\Eeq
Moreover, using Eq. \eqref{V-U}, we can read $V^{+}_{s,-{\bf{k}}}(\tau)$ as
\bea\label{V-U+}
V^{+}_{s,-{\bf{k}}}(\tau) =  - \frac{1}{\sqrt{2}}\begin{pmatrix} ~~{\bf E}_s u^{\downarrow*}_{-s}(k,\tau) \\ s{\bf E}_s u^{\uparrow*}_{-s}(k,\tau)\end{pmatrix}.
\eea
Using \eqref{eq:UVpsipls} in the action \eqref{eq:SLSR-P}, we arrive at two sets of coupled field equations for $u^{\uparrow}_s$ and $u^{\downarrow}_s$ as
\bea \label{u-ups-eq}
&(i\partial_{\tau}-\mu_{{\rm{m}}}\mathcal{H})u^{\uparrow}_s-\left[k+s\left(2\xpi-\frac{\xi_A}{2}\right)\mathcal{H}\right]u^{\downarrow}_s=0\,,\\ \label{u-downs-eq}
&(i\partial_{\tau}+\mu_{{\rm{m}}}\mathcal{H})u^{\downarrow}_s-\left[k+s\left(2\xpi-\frac{\xi_A}{2}\right)\mathcal{H}\right]u^{\uparrow}_s=0\,.
\eea
The above coupled set of equations can be decoupled in terms of the following decomposition
\Beq
\label{eq:uupdowngen}
u^{\uparrow}_s=\frac{1}{\sqrt{2\x}}\left(Y_s+Z_s\right)\quad{\rm and}\quad u^{\downarrow}_s=\frac{1}{\sqrt{2\x}}\left(Y_s-Z_s\right)\,.
\Eeq
The coupled set of first order differential equations \eqref{u-ups-eq} and \eqref{u-downs-eq} can be decoupled into two second order differential equations for $Y_s$ and $Z_s$ as
\bea\label{Y-eq-+}
&&\partial_{\x}^2 Y_s+\left[1-\frac{2i\kappa^+_s}{\x}+\frac{1/4-(\mu^{+})^2}{\x^2}\right]Y_s=0\,,\\ \label{Z-eq-+}
&&\partial_{\x}^2 Z_s+\left[1-\frac{2i\tilde\kappa^+_s}{\x}+\frac{1/4-(\mu^{+})^2}{\x^2}\right]Z_s=0\,,
\eea
where $\kappa^+_s$, and $\tilde{\kappa}^+_s$ are
\bea\label{kappa-s+}
\kappa^+_s=\frac{1}{2}+is\left(2\xpi-\frac{\xi_A}{2}\right) \an
\tilde\kappa^+_s=-\frac{1}{2}+is\left(2\xpi-\frac{\xi_A}{2}\right),
\eea
and $\mu^{+}$ is 
\bea\label{mu+}
\mu^{+}= i \bigg[ \mu_{{\rm{m}}}^2 + \big(2\xpi-\frac{\xi_A}{2}\big)^2 \bigg]^{\frac12}\,.
\eea 
The general solutions for Eq.s \eqref{Y-eq-+} and \eqref{Z-eq-+} are $W$ and $M$ Whittaker functions. Setting the Bunch-Davies vacuum as the initial condition for $u^{\uparrow}_s$ and $u^{\downarrow}_s$ and using Eq.s \eqref{WM-asymp} and \eqref{WM-asymp-2}, we find $Y_s$ and $Z_s$ as
\Beq\label{Ys-Zs}
Y_s &= \frac{1}{(2\pi)^{\frac32}} e^{s(\xp/4-\xpi)\pi} W_{\kappa_s^+,\mu^+}(-2i\x),\\
Z_s &= - \frac{i\mu_{_{\rm{m}}}}{(2\pi)^{\frac32}}  e^{s(\xp/4-\xpi)\pi} W_{\tilde\kappa_s^+,\mu^+}(-2i\x).
\Eeq

\subsection{$\Uppsi^-$ fermions}

Next, we turn to $\Uppsi^-_{\bk}$ modes, described by  the action \eqref{eq:SLSR-M}. The theory of the minus subspace is not diagonalizable in a time independent frame. Hence, the helicity eigenstates are only the eigenstates of the Lagrangian in the asymptotic past limit. Therefore, we use the ansatz
\Beq
\label{eq:UVpsimns}
U_{s,k}^-(\tau)=\frac{1}{\sqrt{2}}\begin{pmatrix} ~ {\bf E}_su^{\uparrow}_{s,+}(k,\tau) \\ s{\bf E}_su^{\downarrow}_{s,+}(k,\tau)\end{pmatrix}+\frac{1}{\sqrt{2}}\begin{pmatrix} ~~{\bf E}_{-s}u^{\uparrow}_{s,-}(k,\tau) \\ -s{\bf E}_{-s}u^{\downarrow}_{s,-}(k,\tau)\end{pmatrix}\,,
\Eeq
where ${\bf{E}}_s$ are the 2-spinor polarization states defined in Eq. \eqref{E+-E-}. Since in the asymptotic past limit the system is diagonalized in this particular basis, we have 
\bea
 \lim_{\x \rightarrow \infty} u^{\uparrow}_{s,-}(k\tau) &=& \lim_{\x \rightarrow \infty} u^{\downarrow}_{s,-}(k\tau) = 0.
\eea
Moreover, using the charge conjugation relation, we can read $V^{-}_{s,-{\bf{k}}}(\tau)$ as
\bea\label{V-U-}
V_{s,k}^-(\tau)= -\frac{1}{\sqrt{2}}\begin{pmatrix} ~{\bf E}_s u^{\downarrow*}_{-s,+}(k,\tau) \\ s{\bf E}_s u^{\uparrow*}_{-s,+}(k,\tau)\end{pmatrix} -\frac{1}{\sqrt{2}}\begin{pmatrix} ~~{\bf E}_{-s}u^{\downarrow*}_{-s,-}(k,\tau) \\ -s{\bf E}_{-s}u^{\uparrow*}_{-s,-}(k,\tau)\end{pmatrix}\,.
\eea
Using the ansatz Eq. \eqref{eq:UVpsimns} in the action \eqref{eq:SLSR-M}, we arrive at the field equations below
\bea
&&(i\partial_{\tau}-\mu_{\rm{m}}\mathcal{H})u^{\uparrow}_{s,s'} - \left[k+ss'\left(2\xpi+\frac{\xi_A}{2}\right)\mathcal{H}\right]u^{\downarrow}_{s,s'}-ss'\xi_{A}\mathcal{H}u^{\downarrow}_{s,-s'}=0\,,\\
&&(i\partial_{\tau}+\mu_{\rm{m}}\mathcal{H})u^{\downarrow}_{s,s'} - \left[k+ss'\left(2\xpi+\frac{\xi_A}{2}\right)\mathcal{H}\right]u^{\uparrow}_{s,s'}+ss'\xi_{A}\mathcal{H}u^{\uparrow}_{s,-s'}=0\,.
\eea
Note that for each given $s$ and $s'$, the equations with $s'=+$ and $s'=-$ are coupled. We make the decompositions
\Beq
\label{eq:uupdowngenm}
u^{\uparrow}_{s,s'}=\frac{1}{\sqrt{2\tilde{\tau}}}\left(Y_{s,s'}+Z_{s,s'}\right)\quad{\rm and}\quad u^{\downarrow}_{s,s'}=\frac{1}{\sqrt{2\tilde{\tau}}}\left(Y_{s,s'}-Z_{s,s'}\right)\,,
\Eeq
which gives
\Beq
\label{eq:YZ1storderm-2}
&i\partial_{\x} Y_{s,s'} + \left[1 + \frac{ss'(4\xpi+\xi_A)-i}{2\x}\right]Y_{s,s'}+\frac{\mu_{\rm{m}}}{\x}Z_{s,s'}-\frac{ss'\xi_{A}}{\x}Z_{s,-s'}=0\,,\\
&i\partial_{\x} Z_{s,s'} - \left[1 + \frac{ss'(4\xpi+\xi_A)+i}{2\x}\right]Z_{s,s'}+\frac{\mu_{\rm{m}}}{\x}Y_{s,s'}+\frac{sp\xi_{A}}{\x}Y_{s,-s'}=0\,.
\Eeq

We can reduce the above coupled first order equations to pairs of coupled second order equations
\Beq
\partial_{\x}^2 Y_{s,s'}+\Bigg[1 - \frac{2is'[\kappa_s^- + \frac{1}{2} (s'-1)]}{\x}+&\frac{1/4 - (\mu^{-})^2}{\x^2}\Bigg]Y_{s,s'} -\frac{2\xi_A(2\xpi+\xi_A/2)}{\x^2}Z_{s,-s'}=0\,,\\
\partial_{\x}^2 Z_{s,s'}+\Bigg[1 - \frac{2is'[\tilde{\kappa}_s^- - \frac{1}{2} (s'-1)]}{\x}+&\frac{1/4-(\mu^{-})^2}{\x^2}\Bigg]Z_{s,s'}-\frac{2\xi_A(2\xpi+\xi_A/2)}{\x^2}Y_{s,-s'}=0\,,
\Eeq
where $\kappa^-_s$, and $\tilde{\kappa}^-_s$ are
\bea\label{kappa-s-}
\kappa^-_s= \frac{1}{2}+is(2\xpi+\frac{\xi_A}{2}) \an
\tilde\kappa^-_s= - \frac{1}{2}+is(2\xpi+\frac{\xi_A}{2}),
\eea
while $\mu^{-}$ is 
\bea\label{mu-}
\mu^{-}= i \bigg[ \mu_{{\rm{m}}}^2 + \big(2\xpi+\frac{\xi_A}{2}\big)^2 + \xp^2 \bigg]^{\frac12}\,.
\eea

Unlike the plus spinors, the set of equations in minus system is still coupled and cannot be solved analytically. However, since the coupling term becomes relevant around and after horizon crossing, it is possible to solve the system analytically in the limit that the coupling is negligible, i.e. $k\tau\gg 1$ and/or $\xp/\mu_{\rm{m}}\ll 1$. More precisely, we can decompose the fields as \footnote{The prefactor $-\frac{s\xp}{\mu_{\rm{m}}}$ of $Z_{s,-}(\x)$ in Eq. \eqref{ansatz-YZ} is chosen such that the initial value of $z_{s,-}$ is one. }
\Beq\label{ansatz-YZ}
Y_{s,+}(\x)  =  y_{s,+}(\x) Y_{s}(\x) & \an  Y_{s,-}(\x)  =  y_{s,-}(\x) Y_{-s}(\x),\\   
Z_{s,+}(\x)  =  z_{s,+}(\x) Z_{s}(\x) & \an 
Z_{s,-}(\x)  =  -\frac{s\xp}{\mu_{\rm{m}}} z_{s,-}(\x) Z_{-s}(\x),
\Eeq
where $Y_{s}$, and $Z_{s}$ are analytically solvable while the other four functions, $y_{s,p}(\x)$ and $z_{s,p}(\x)$, should be solved numerically. In the asymptotic past, we have
\bea\label{tf-tg-in}
\lim_{\x\rightarrow \infty} y_{s,+}(\x) &=& 1 \\\lim_{\x\rightarrow \infty} z_{s,+}(\x) &=& 1,\\ \label{tf-tg-inI}
\lim_{\x\rightarrow \infty} y_{s,-}(\x) &=&  0,
\eea
and $z_{s,-}(\x)$ up to a phase, $(2\x_0)^{2i s\kappa_{I}}$, is
\bea\label{tf-tg-in-II}
\lim_{\x\rightarrow \infty} z_{s,-}(\x) = 1.
\eea
Notice that at $\x\rightarrow \infty$, $Z_s(\x)\sim \x^{-1} Y_s(\x)$ (See Eq. \eqref{Y-Z-minus-}).
The fields $Y_{s}$ and $Z_{s}$ satisfy 
\Beq
&\partial_{\x}^2 Y_{s}+\Bigg[1 - \frac{2i\kappa_s^-}{\x}+&\frac{1/4 - \mu^{2}_{-}}{\x^2}\Bigg]Y_{s} =0\,,\\
&\partial_{\x}^2 Z_{s}+\Bigg[1-\frac{2i\tilde\kappa_s^{-}}{\x}+&\frac{1/4-\mu^{2}_{-}}{\x^2}\Bigg]Z_{s} =0\,.
\Eeq

The above differential equations can be solved analytically in terms of Whittaker functions and setting the Bunch-Davies vacuum as the initial condition for $u^{\uparrow}_{s,+}$ and $u^{\downarrow}_{s,+}$, $Y_s$ and $Z_s$ are given as
\bea\label{Y-Z-minus-}
Y_s = e^{-s(\xp/4+\xpi)\pi} W_{\kappa_s^-,\mu^-}(-2i\x) \an Z_s = -i \mu_{_{\rm{m}}} e^{-s(\xp/4+\xpi)\pi} W_{\tilde\kappa_s^-,\mu^-}(-2i\x).
\eea 
Next, using the ansatz Eq. \eqref{ansatz-YZ}, as well as the first order differential equations of $Y_{s}$ and $Z_{s}$ in Eq. \eqref{eq:YZ1storderm-2}, we find a set of coupled first order differential equations for $y_{s,s'}$ and $z_{s,s'}$. Given the analytical forms of $Y_s$ and $Z_s$ in Eq. \eqref{Y-Z-minus-} and the initial condition Eq.s \eqref{tf-tg-in}-\eqref{tf-tg-in-II}, we solve the above coupled linear differential equations numerically. The results are presented in Sec. \ref{sec:numerics} (Fig.s \ref{fig:ysp-zsp-44}-\ref{fig:ysp-ysp-1}). In order to have a better insightful vision of the minus subsystem, next we discuss the qualitative behavior of $\Uppsi^{-}_{\bf{k}}$ and the $y_{s,s'}$ and $z_{s,s'}$ functions.

\subsubsection*{Geometry of the minus subspace}

Unlike the plus subspace, helicity is not a good quantum number in the minus subspace. That is because of the off-diagonal term proportional to $\xp$ in its action in Eq. \eqref{eq:SLSR-M}.  Therefore, although both $\Uppsi^{+}$ and $\Uppsi^{-}$ spinors have similar asymptotic past limits, but the more minus spinors are stretched and approach the  horizon, the more they experience the off diagonal time-dependent term. As a result, a minus fermion with a given initial polarization state ${\bf{E}}_s$, gradually rotates in the helicity space and turns to a mixed state of ${\bf{E}}_s$ and ${\bf{E}}_{-s}$, i.e.
\bea
Y_s(\x) \begin{pmatrix} ~~{\bf{E}}_s \\
~s {\bf{E}}_s \end{pmatrix} &\longrightarrow & \big( \lvert Y_{s,+}(\x)\rvert^2 + \lvert Z_{s,-}(\x)\rvert^2 \big)^{\frac12} \begin{pmatrix} ~~{\bf{E}}_s^{\rm{Y}}(\x) \\
~s {\bf{E}}^{\rm{Y}}_s(\x) \end{pmatrix},\\
Z_s(\x) \begin{pmatrix} ~~{\bf{E}}_s \\
-s {\bf{E}}_s \end{pmatrix} &\longrightarrow & \big(\lvert Z_{s,+}(\x)\rvert^2 + \lvert Y_{s,-}(\x)\rvert^2\big)^{\frac12} \begin{pmatrix} ~~{\bf{E}}^{\rm{Z}}_s(\x) \\
-s {\bf{E}}^{\rm{Z}}_s(\x) \end{pmatrix},
\eea
where the time dependent polarization states, ${\bf{E}}^{{\rm{Y}}}_s(\x)$ and ${\bf{E}}^{{\rm{Z}}}_s(\x)$, can be read off in terms of the helicity eigenstates, ${\bf{E}}_{\pm}$, as
\bea
{\bf{E}}^{{\rm{Y}}}_s(\x) &=& \frac{Y_{s,+}(\x)}{ \big( \lvert Y_{s,+}(\x)\rvert^2 + \lvert Z_{s,-}(\x)\rvert^2 \big)^{\frac12}} {\bf{E}}_s +  \frac{Z_{s,-}(\x)}{\big( \lvert Y_{s,+}(\x)\rvert^2 + \lvert Z_{s,-}(\x)\rvert^2 \big)^{\frac12}} {\bf{E}}_{-s},\\
{\bf{E}}^{{\rm{Z}}}_s(\x) &=& \frac{Z_{s,+}(\x)}{\big( \lvert Z_{s,+}(\x)\rvert^2 + \lvert Y_{s,-}(\x)\rvert^2 \big)^{\frac12}} {\bf{E}}_s +  \frac{Y_{s,-}(\x)}{\big( \lvert Z_{s,+}(\x)\rvert^2 + \lvert Y_{s,-}(\x)\rvert^2 \big)^{\frac12}} {\bf{E}}_{-s}.
\eea

\begin{figure*}[!htb]
\centering
\includegraphics[scale=0.5]{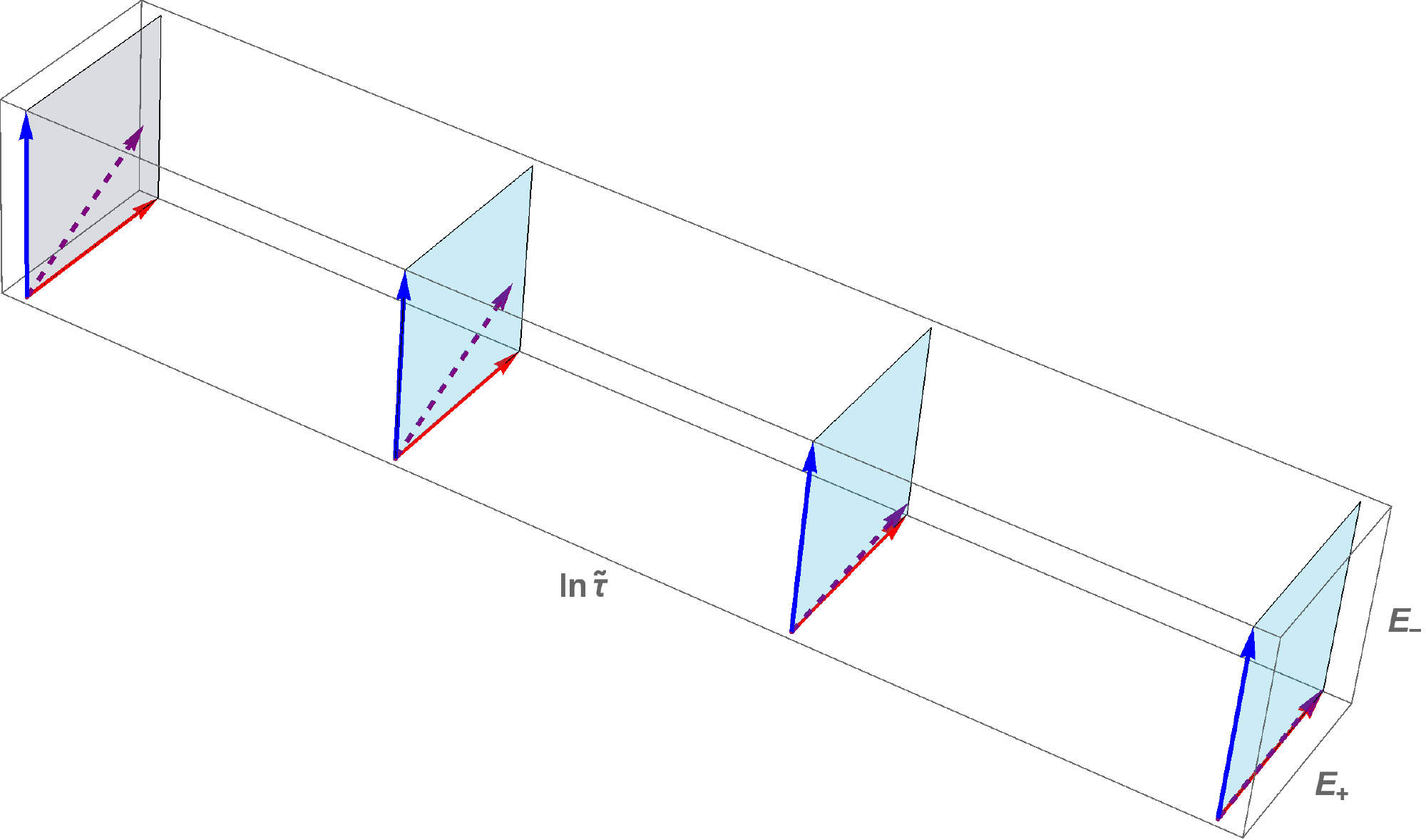}\\ 
\caption{Schematic time evolution of the $\Uppsi^{-}$ polarization states, ${\bf{E}}^{Y}_{+}(\x)$ (and ${\bf{E}}^{Z}_{+}(\x)$). The (purple) dashed vectors show ${\bf{E}}^{Y}_{+}(\x)$ (and ${\bf{E}}^{Z}_{+}(\x)$) while the helicity polarization vectors, ${\bf{E}}_{+}$ and ${\bf{E}}_{-}$ are shown in red and blue respectively. The direction of time is from right to left, and the gray plane presents the horizon crossing, i.e. $\x=1$.}
\label{fig:EsYZ}   
\end{figure*}
All ${\bf{E}}^{Y}_{s}(\x)$ and ${\bf{E}}^{Z}_{s}(\x)$ polarization states start from ${\bf{E}}_{s}$ at asymptotic past. Then, around and at horizon crossing, they make most of their rotation in the ${\bf{E}}_+-{\bf{E}}_-$ plane. Here, for more insight, Fig. \ref{fig:EsYZ} shows schematic rotation of ${\bf{E}}^{Y}_{+}(\x)$ (and ${\bf{E}}^{Z}_{+}(\x)$) in ${\bf{E}}_+-{\bf{E}}_-$ plane.

Before going any further and numerically solve the minus subsystem in Sec. \ref{sec:numerics} (Fig.s \ref{fig:ysp-zsp-44}-\ref{fig:ysp-ysp-1}), let us discuss the qualitative behavior of these functions. 
\begin{itemize}
\item{The $y_{s,+}(\x)$, and $z_{s,+}(\x)$ start from one at asymptotic past and after a fast transition phase each reduces to another constant number after horizon crossing, i.e. $y_{0s+}$ and $z_{0s+}$. Their behavior can be roughly described as $\big(\tanh[\alpha_s\x-\x_{0s}]+\beta_s\big)/(1+\beta_s)$ with $\x_{0s}\gtrsim 1$, $\alpha_s\lesssim 1$, while for $y_{s,+}(\x)$ we have $\beta_s=\frac{1+y_{0s+}}{1-y_{0s+}}$, and for $z_{s,+}(\x)$ we get $\beta_s=\frac{1+z_{0s+}}{1-z_{0s+}}$.}
\item{The $y_{s,-}(\x)$ starts from zero at asymptotic past and after a quick transition phase increases to a constant number less than one after horizon crossing, i.e. $y_{0s-}$. Its behavior can be roughly described by $\frac12y_{0s-}\big(-\tanh[\alpha_s\x-\x_{0s}]+1\big)$ with $y_{0s-}\lesssim 1$.}
\item{The $z_{s,-}(\x)$ starts from one at asymptotic past and after horizon crossing approach a constant value either less or more (but of the order of) one. Yet it also follows a $\tanh$-type behavior.}
\end{itemize}

\section{Mode functions, a closer look}\label{gen-sol}
In this part, first, we discuss the generic features of the analytical solutions $Y_s$ and $Z_s$, which are in common between the plus and minus subsystems. Next, we present the numerical study of the four functions, $y_{s,p}$ and $z_{s,p}$, which we introduced in the minus subspace to take care of the rotation of its mode functions in the helicity plane.

\subsection{$\Uppsi^{\pm}$ spinors: $Y_s$ and $Z_s$}

\begin{figure*}[!htb]
\centering
\includegraphics[scale=0.49]{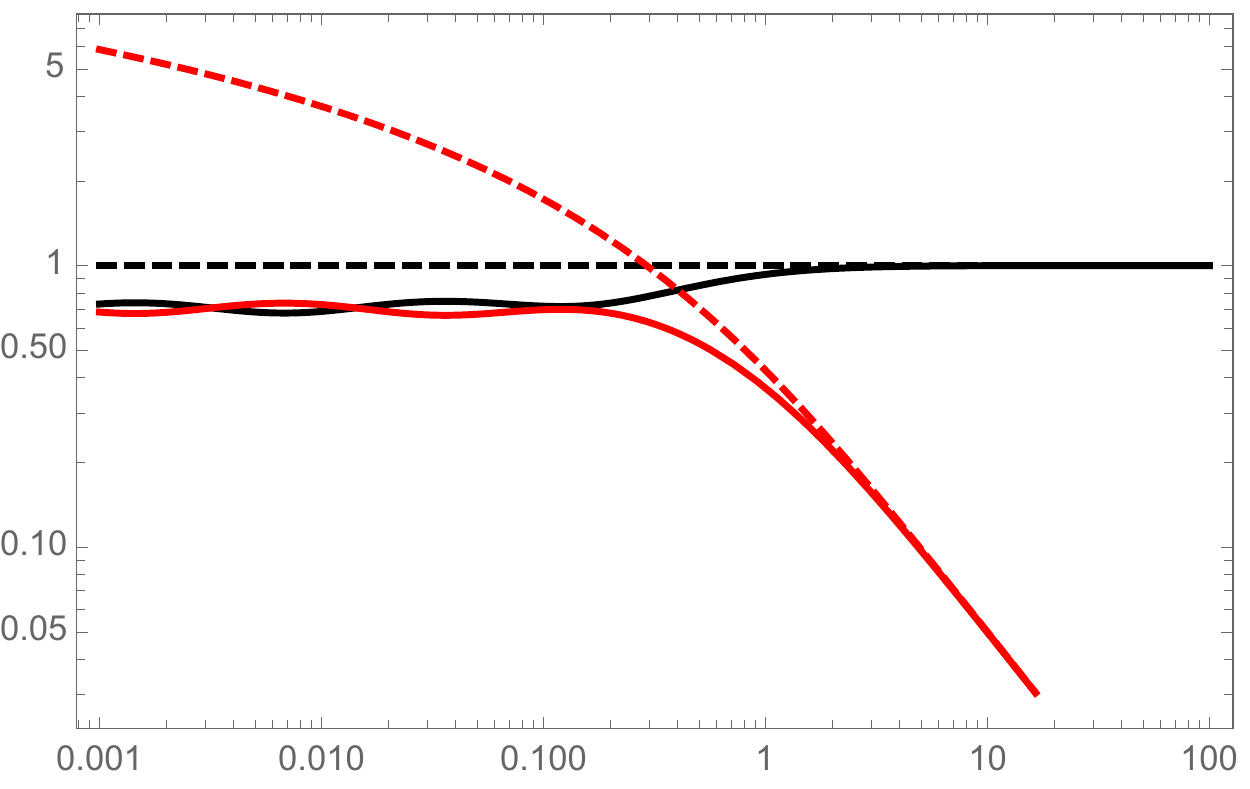} 
\caption{The evolution of absolute values of $(2\pi)^{\frac32}\frac{Y_{\pm}(\x)}{\sqrt{2\x}}$ (black) and $(2\pi)^{\frac32}\frac{Z_{\pm}(\x)}{\sqrt{2\x}}$ (red) for trivial vacuum with $\kappa_I=0$ with respect to $\x$. The dashed lines shows $\mu_{m}=0$, and the solid lines presents $\mu_m=1$.}
\label{fig:free-Kappa}   
\end{figure*}

\begin{figure*}[!htb]
\centering
\includegraphics[scale=0.49]{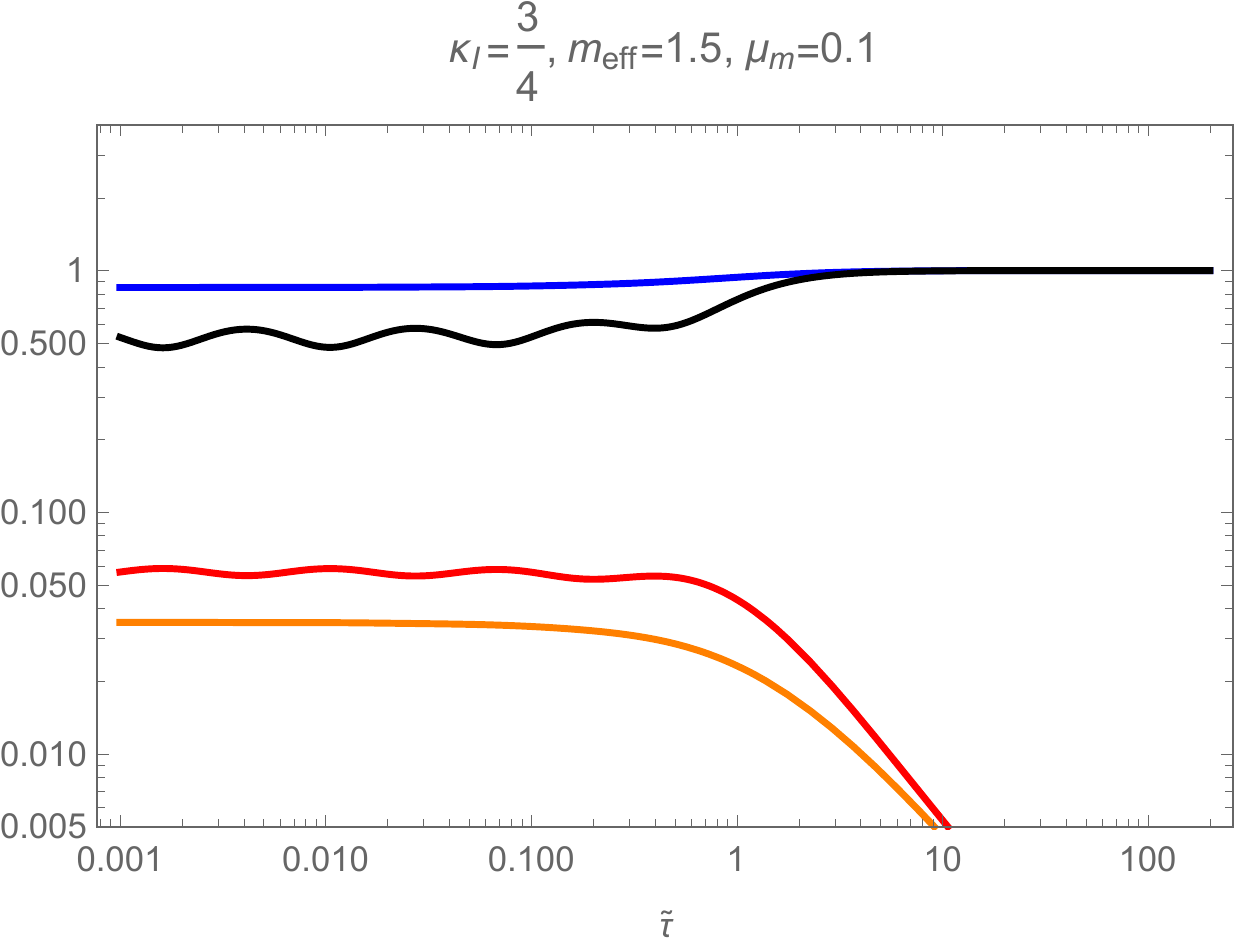} 
\includegraphics[scale=0.49]{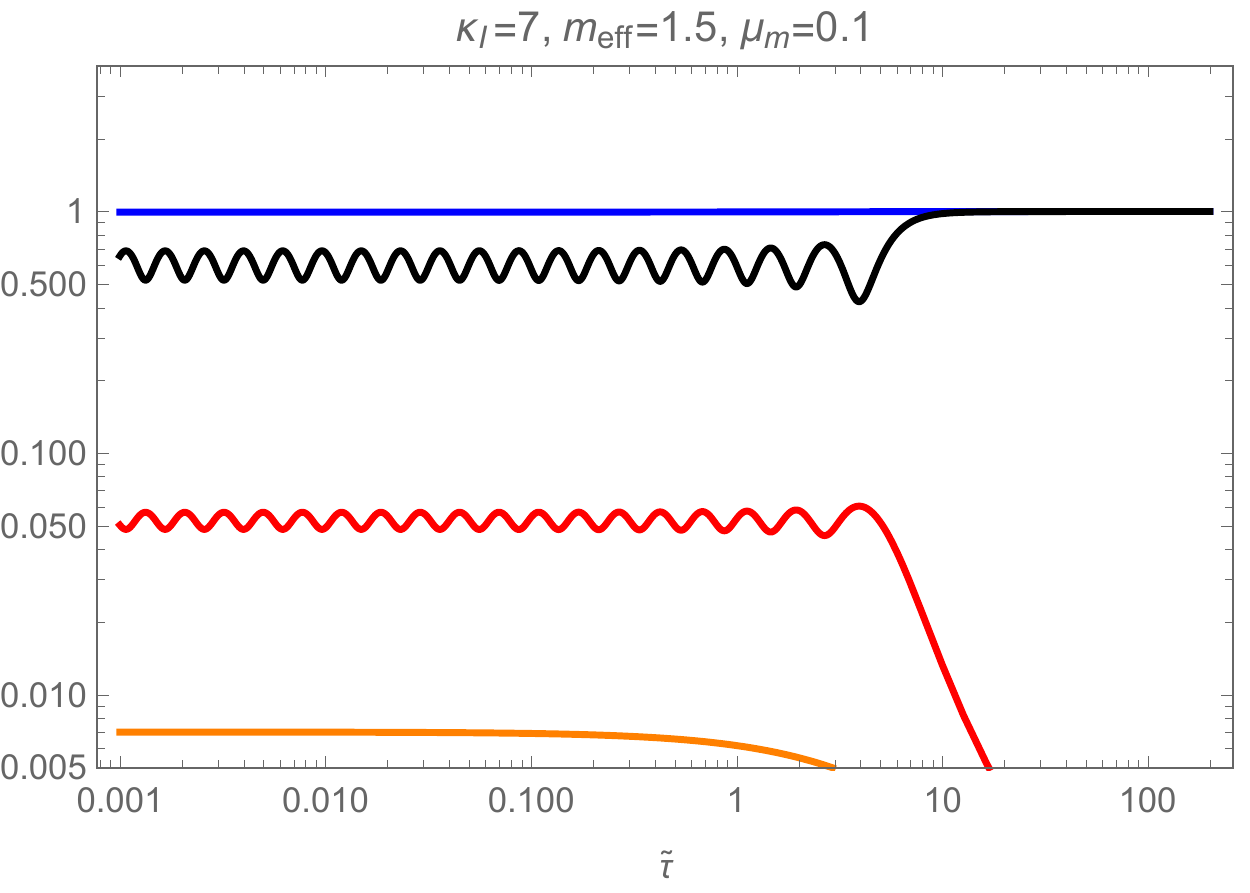}\\
\includegraphics[scale=0.49]{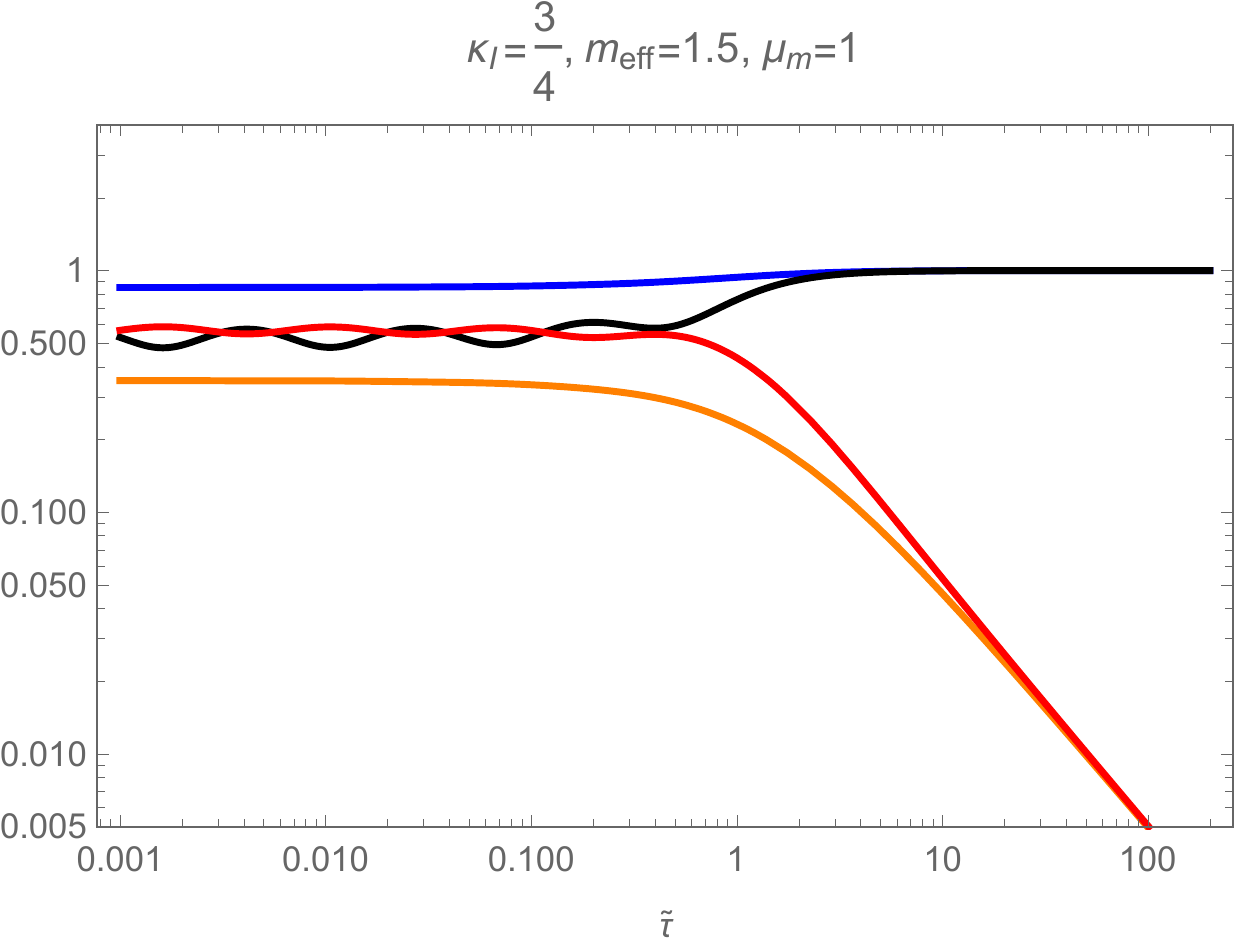} 
\includegraphics[scale=0.49]{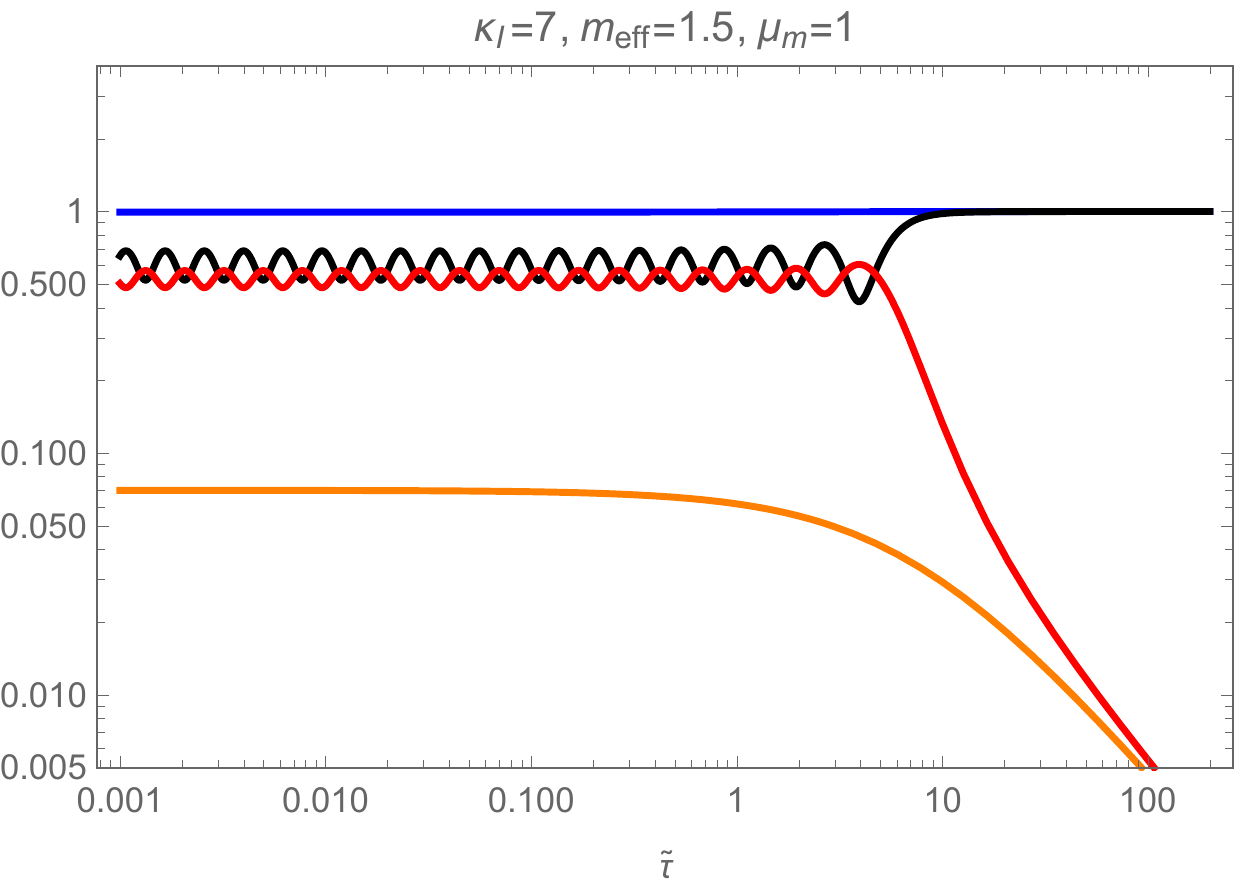}
\caption{The evolution of absolute values of $(2\pi)^{\frac32}\frac{Y_{\pm}(\x)}{\sqrt{2\x}}$ and $(2\pi)^{\frac32}\frac{Z_{\pm}(\x)}{\sqrt{2\x}}$ with respect to $\x$ for different values of $\kappa_I$, $m_{\rm{eff}}$, and $\mu_m$. The blue shows $(2\pi)^{\frac32}\frac{Y_+(\x)}{\sqrt{2\x}}$, orange is $(2\pi)^{\frac32}\frac{Z_+(\x)}{\sqrt{2\x}}$, black is $(2\pi)^{\frac32}\frac{Y_-(\x)}{\sqrt{2\x}}$, and red shows $(2\pi)^{\frac32}\frac{Z_-(\x)}{\sqrt{2\x}}$.}
\label{fig:Ys-Zs+}   
\end{figure*}

\begin{figure*}[!htb]
\centering
\includegraphics[scale=0.52]{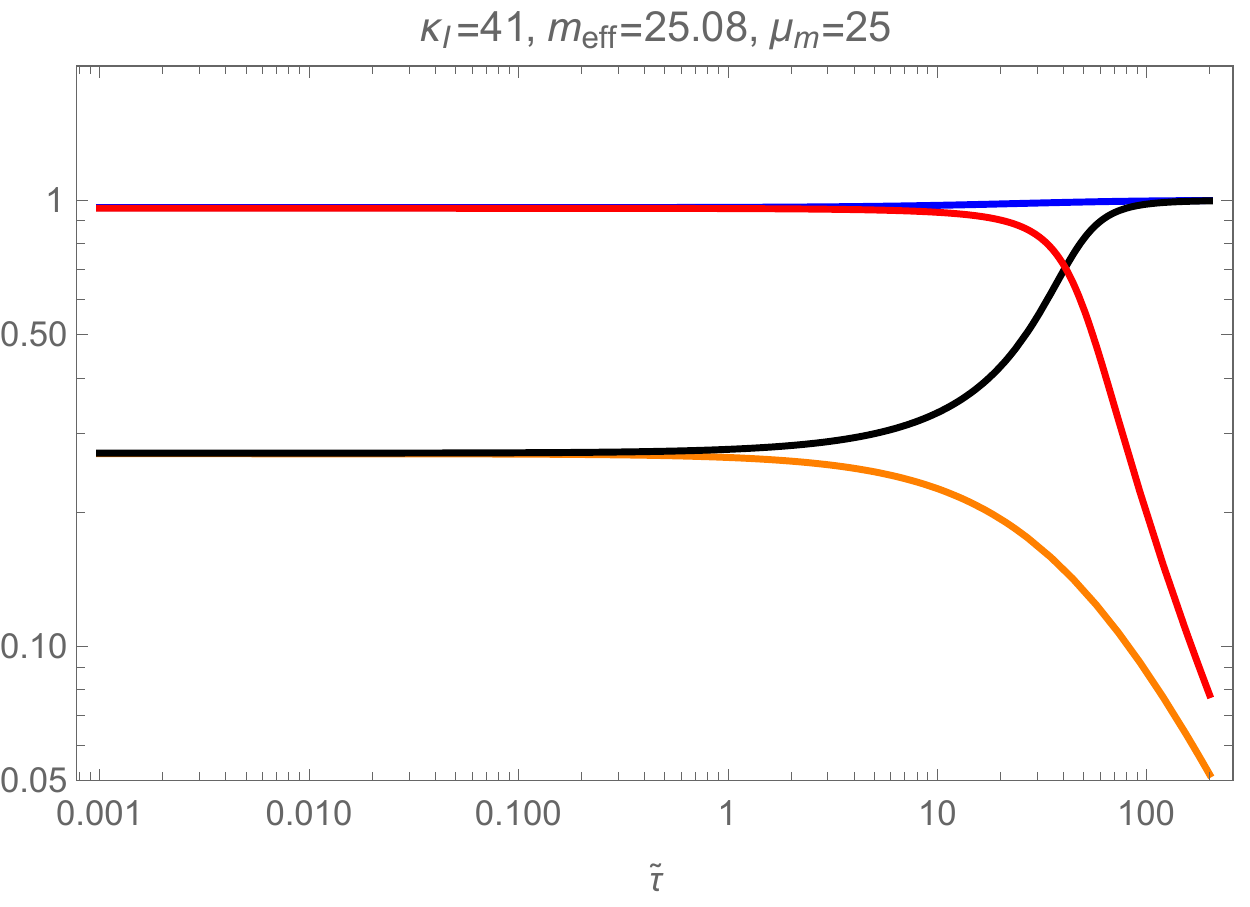} 
\includegraphics[scale=0.52]{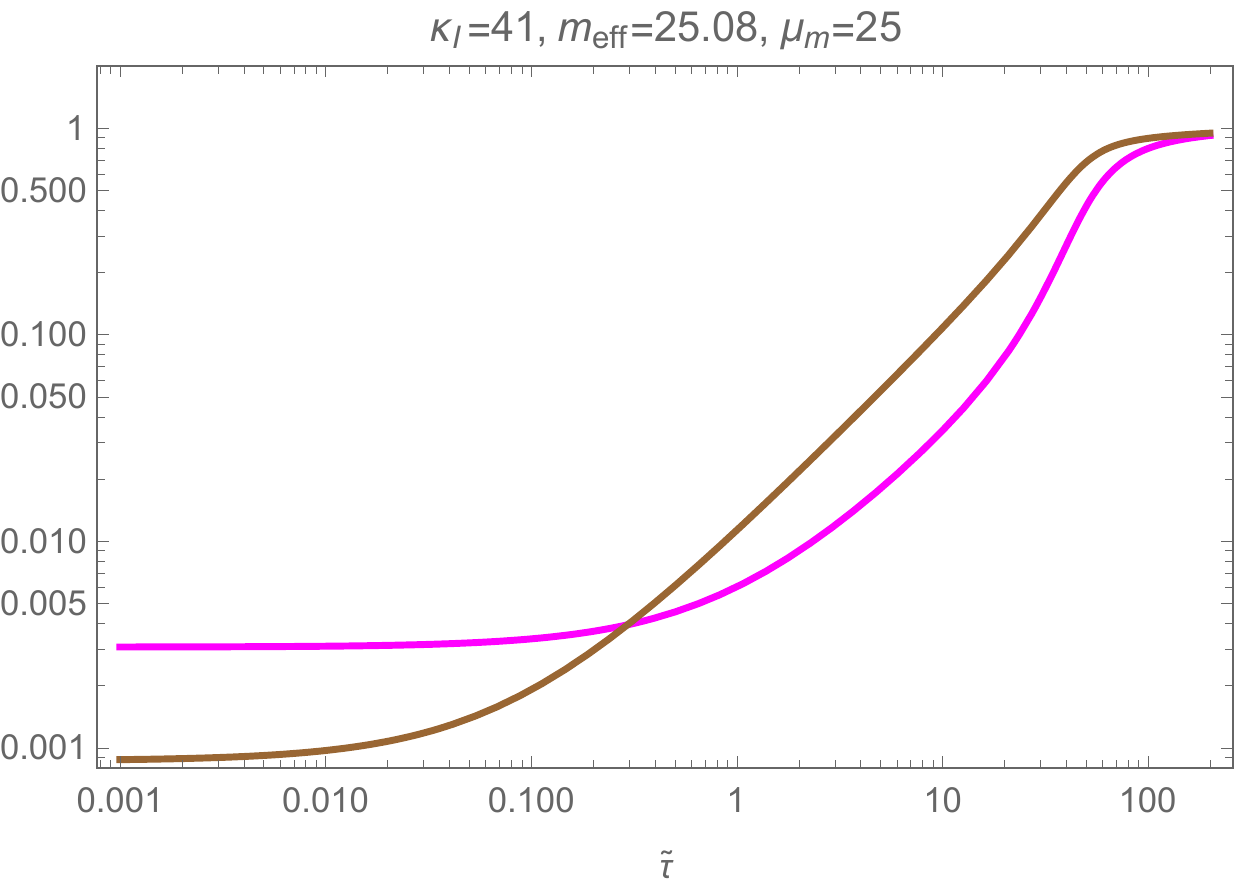}\\
\includegraphics[scale=0.52]{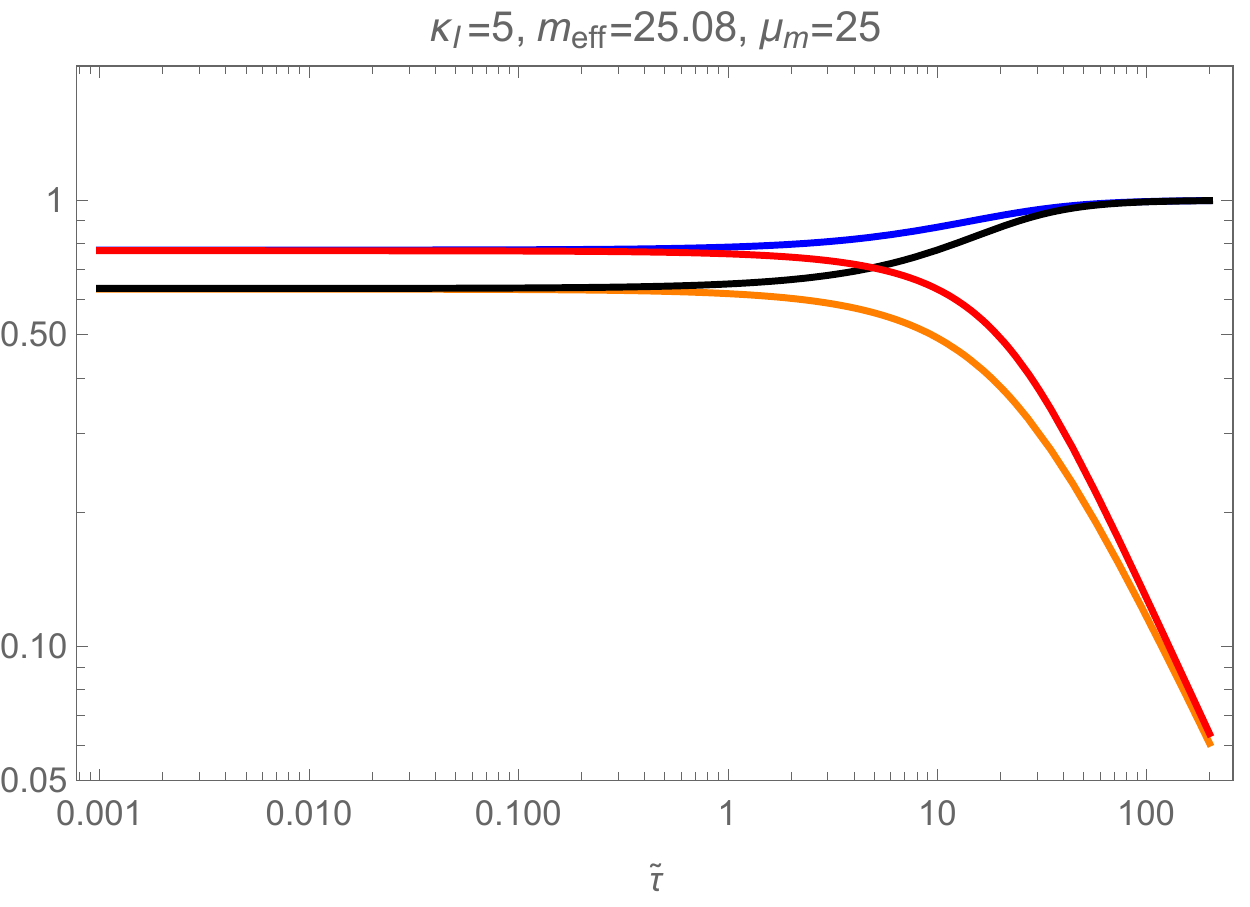} 
\includegraphics[scale=0.52]{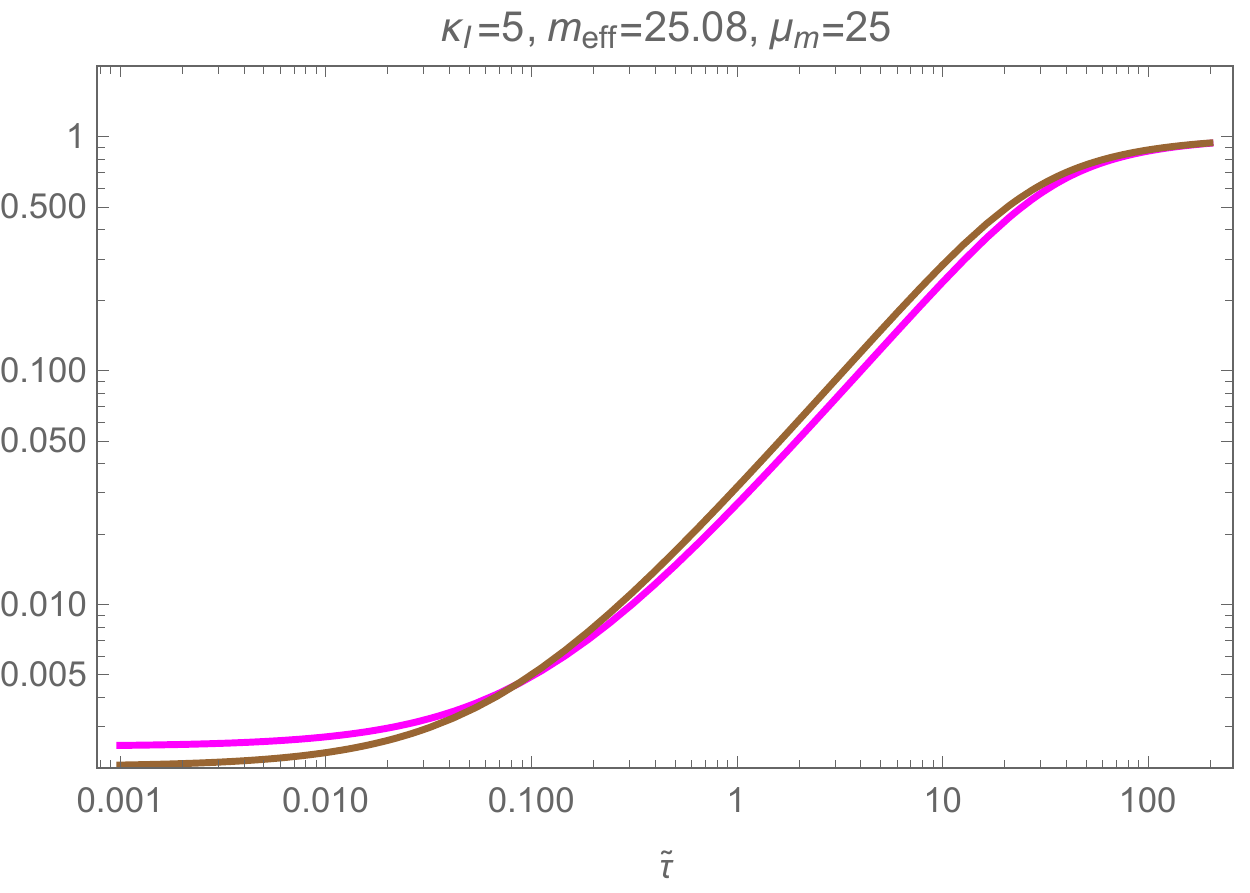}
\caption{Large mass limit behavior of $(2\pi)^{\frac32}\frac{Y_{\pm}(\x)}{\sqrt{2\x}}$ and $(2\pi)^{\frac32}\frac{Z_{\pm}(\x)}{\sqrt{2\x}}$. The left panels show absolute values of $(2\pi)^{\frac32}\frac{Y_{+}(\x)}{\sqrt{2\x}}$ (blue), $(2\pi)^{\frac32}\frac{Z_{+}(\x)}{\sqrt{2\x}}$ (orange), $(2\pi)^{\frac32}\frac{Y_{-}(\x)}{\sqrt{2\x}}$ (black), and $(2\pi)^{\frac32}\frac{Z_{-}(\x)}{\sqrt{2\x}}$ (red) respectively with respect to $\x$. The right panels show the absolute values of $(2\pi)^{\frac32}\frac{Y_{+}(\x)-Z_{-}(\x)}{\sqrt{2\x}}$ (magenta) and $(2\pi)^{\frac32}\frac{Y_{-}(\x)-Z_{+}(\x)}{\sqrt{2\x}}$ (brown) respectively with respect to $\x$.}
\label{fig:Ys-Zs++}   
\end{figure*}

Here, we consider $Y_s$ and $Z_s$ mode functions which are in common between the plus and minus sub-sectors. In fact, in both subsystems, $\mu$ is pure imaginary, i.e. (see Eq.s \eqref{mu+} and \eqref{mu-})
\bea
\mu^{\pm} = i \lvert \mu^{\pm}\rvert,
\eea
while $\kappa^{\pm}_s$ and $\tilde\kappa^{\pm}_s$ in Eq.s \eqref{kappa-s+} and \eqref{kappa-s-}, can be written as
\bea
\kappa^{\pm}_s = \frac12 + is\kappa_I^{\pm} \an \tilde\kappa^{\pm}_s = -\frac12 + is\kappa_I^{\pm},
\eea
where $\kappa^{\pm}_I$ is the absolute value of the imaginary parts of $\kappa^{\pm}_s$ (and $\tilde\kappa^{\pm}_s$) and $\mu^{\pm}$ can be decomposed as
\bea
\mu^{\pm}= i [(m^{\pm}_{\rm{eff}})^2+(\kappa_I^{\pm})^2]^{\frac12},
\eea
where $m^{\pm}_{\rm{eff}}$ is
\bea
m^{+}_{\rm{eff}} = \mu_m \an m^{-}_{\rm{eff}} = (\mu_m^2+\xp^2)^{\frac12}.
\eea
Here, $\lvert\mu^{\pm}\rvert$, $m^{\pm}_{\rm{eff}}$, and $\kappa_I^{\pm}$ are real parameters in terms of $\xp$, $\xpi$, and $\frac{m}{H}$. 
As a result, we can present the analytical solutions in both subspaces, Eq.s \eqref{Ys-Zs} and \eqref{Y-Z-minus-}, in a unified form and in terms of three independent parameters, $\xp$, $\kappa_I$, and $m_{\rm{eff}}$ as
\Beq\label{Ys-Zs}
Y_s(\x) &= \frac{1}{(2\pi)^{\frac32}} e^{-s \kappa_I \pi/2} W_{\frac12+ i s\kappa_I,\mu}(-2i\x),\\
Z_s(\x) &= - \frac{i\mu_{_{\rm{m}}}}{(2\pi)^{\frac32}}  e^{-s \kappa_I \pi/2} W_{-\frac12+ i s\kappa_I,\mu}(-2i\x).
\Eeq
Fig. \ref{fig:free-Kappa} shows the mode functions for trivial vacuum with $\kappa_I=0$. As we see, in the absence of the source of CP violation, the mode functions are parity symmetric and their after horizon amplitude decreases with the mass of the fermion. In Fig. \ref{fig:Ys-Zs+}, we present $\frac{Y_s}{\sqrt{x}}$ and $\frac{Z_s}{\sqrt{x}}$ with $s=\pm$ in terms of $\x$ for several values of parameters. Moreover, Fig. \ref{fig:Ys-Zs++} shows the same quantities as well as $\frac{Y_{+}(\x)-Z_{-}(\x)}{\sqrt{2\x}}$ and $\frac{Y_{-}(\x)-Z_{+}(\x)}{\sqrt{2\x}}$ in the large mass limit.

Here we summarize the generic behavior of these mode functions.

\begin{itemize}
\item{After horizon crossing, the mode functions $Y_{\pm}/\sqrt{\x}$ and $Z_{\pm}/\sqrt{\x}$, oscillate like $\x^{i\lvert \mu\rvert}$ around a constant value.}
\item{The frequency of these oscillations as well as the amplitude of $Y_{\pm}$ increase with the increase of $\kappa_I$. However, the amplitude of $Z_{\pm}$ decreases with $\kappa_{I}$.}
\item{Increasing $m_{\rm{eff}}$ without changing $\mu_m$, which is possible in the minus subspace, decreases $Y_{-}$ and $Z_{-}$, while has negligible effect on $Y_{+}$ and $Z_{+}$.}
\item{Increasing $\mu_{m}$ has negligible effect on $Y_s$ while it linearly increases $Z_s$, i.e. $Z_s\propto \mu_m$. }
\item{In the large mass limit, $\mu_m\gg \xi_{\phi},\xp$ and for each $s=\pm$, $Z_{s}$ increases to a constant value very close to (but less than) $Y_{-s}$. See the right panels of Fig. \ref{fig:Ys-Zs++}. }
\item{The mode functions of the comoving plus fermion, i.e. $u^{\uparrow}_{\pm}(\x)$ and $u^{\downarrow}_{\pm}(\x)$, and hence $\Uppsi^+_{{\bf{k}}}$ are freezed out after horizon crossing. }
\item{For the $\Uppsi^{-}_{\bf{k}}$, one needs $y_{s,s'}(\x)$ and $z_{s,s'}(\x)$ functions in addition to the $Y_s$ and $Z_s$ to fully identify this subsystem. That is the task of Sec. \ref{sec:numerics}.}
\end{itemize}

\subsection{$\Uppsi^{-}$ spinors: $y_{s,s'}$ and $z_{s,s'}$}\label{sec:numerics}

In Eq \eqref{ansatz-YZ} we decomposed the fermion mode functions of the minus subspace in terms of the analytical functions, $Y_s$ and $Z_s$, as well as four unknown functions, $y_{s,\pm}$ and $z_{s,\pm}$ which should be solved numerically. 
In this part, we present the numerical solutions of these functions in terms of $\x$ and for several values of $\mu_{\rm{m}}$, $\xp$, and $\xpi$ parameters. Moreover, for later use, here we also present $\lvert y_{s,+}\rvert^2+\lvert y_{-s,-}\rvert^2$, and $\lvert z_{s,+}\rvert^2+\frac{\xp^2}{\mu^2_{\rm{m}}} \lvert z_{-s,-}\rvert^2$. Later we will see that these are the combinations which appearers in the currents of the minus spinors. 

\begin{figure*}[!htb]
\centering
\includegraphics[scale=0.272]{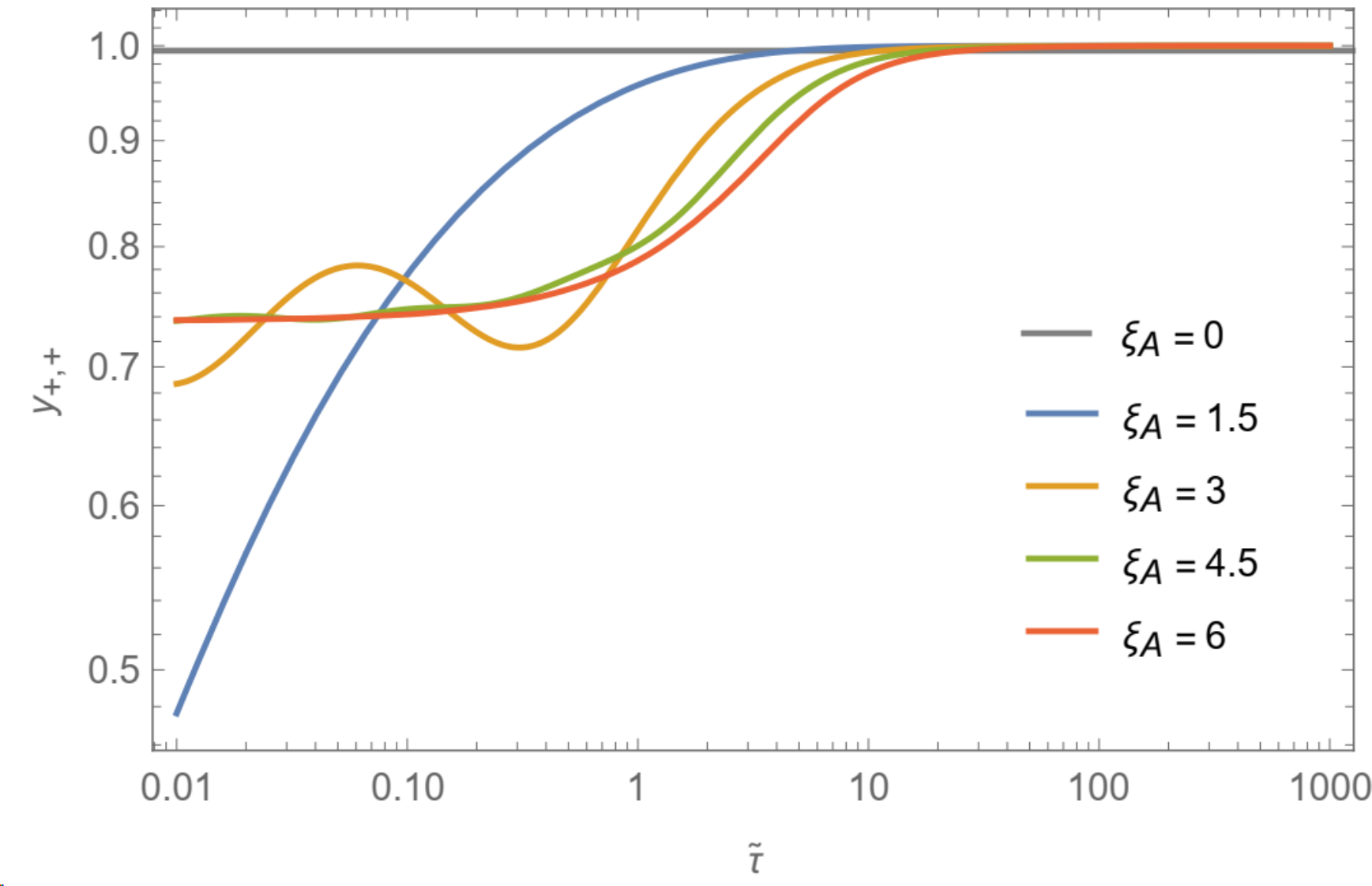} 
\includegraphics[scale=0.272]{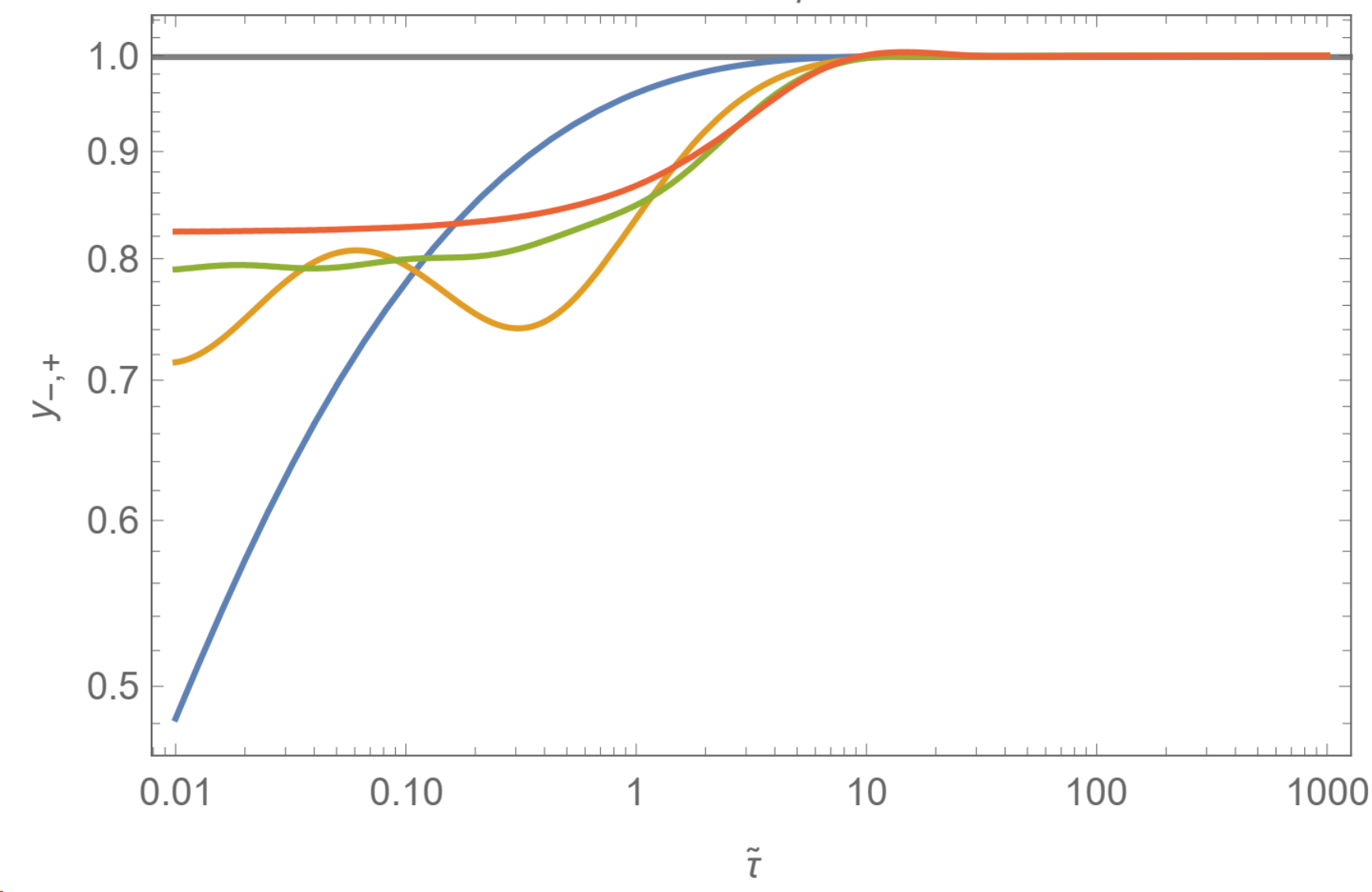}\\
\includegraphics[scale=0.272]{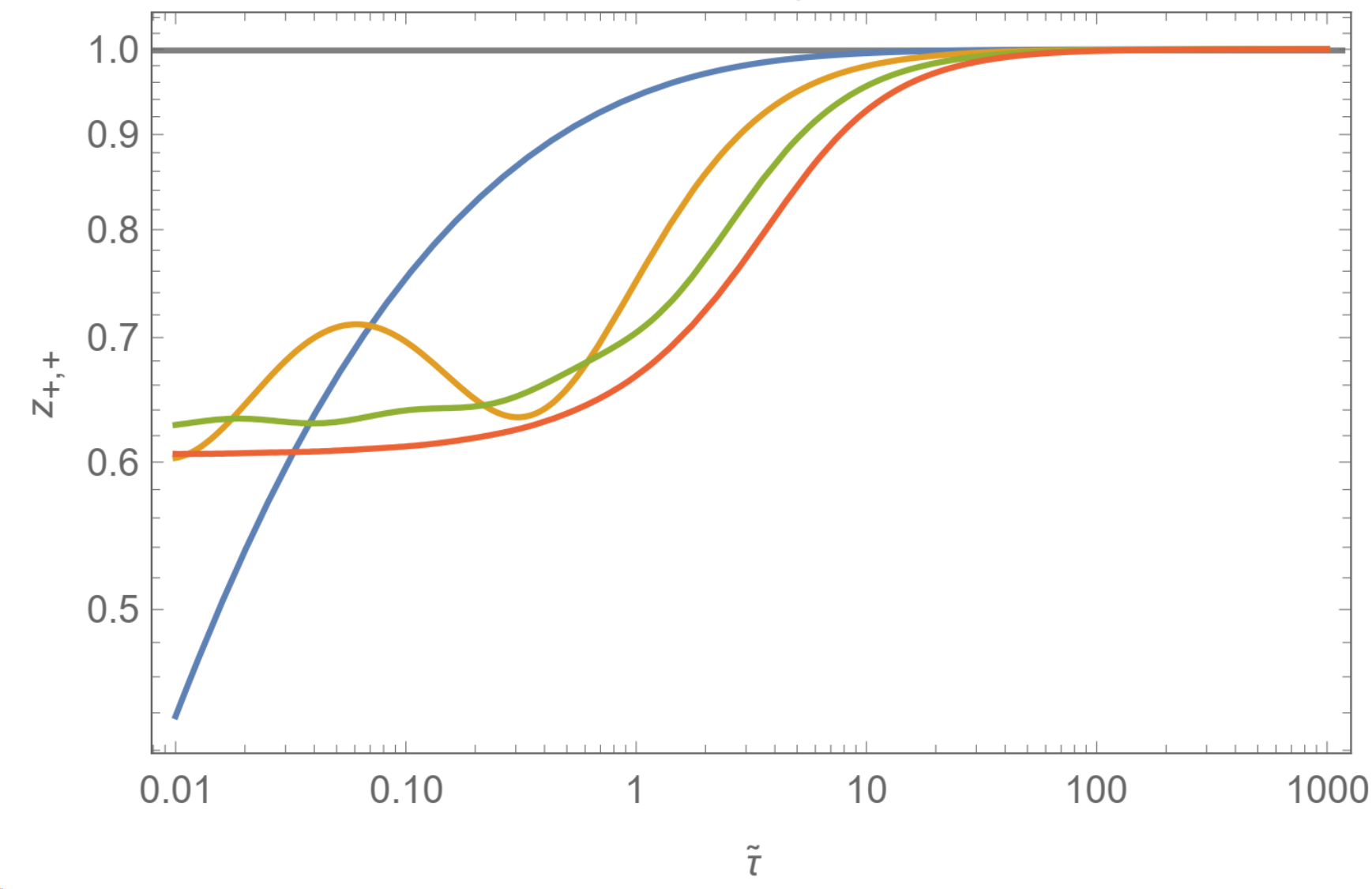} 
\includegraphics[scale=0.272]{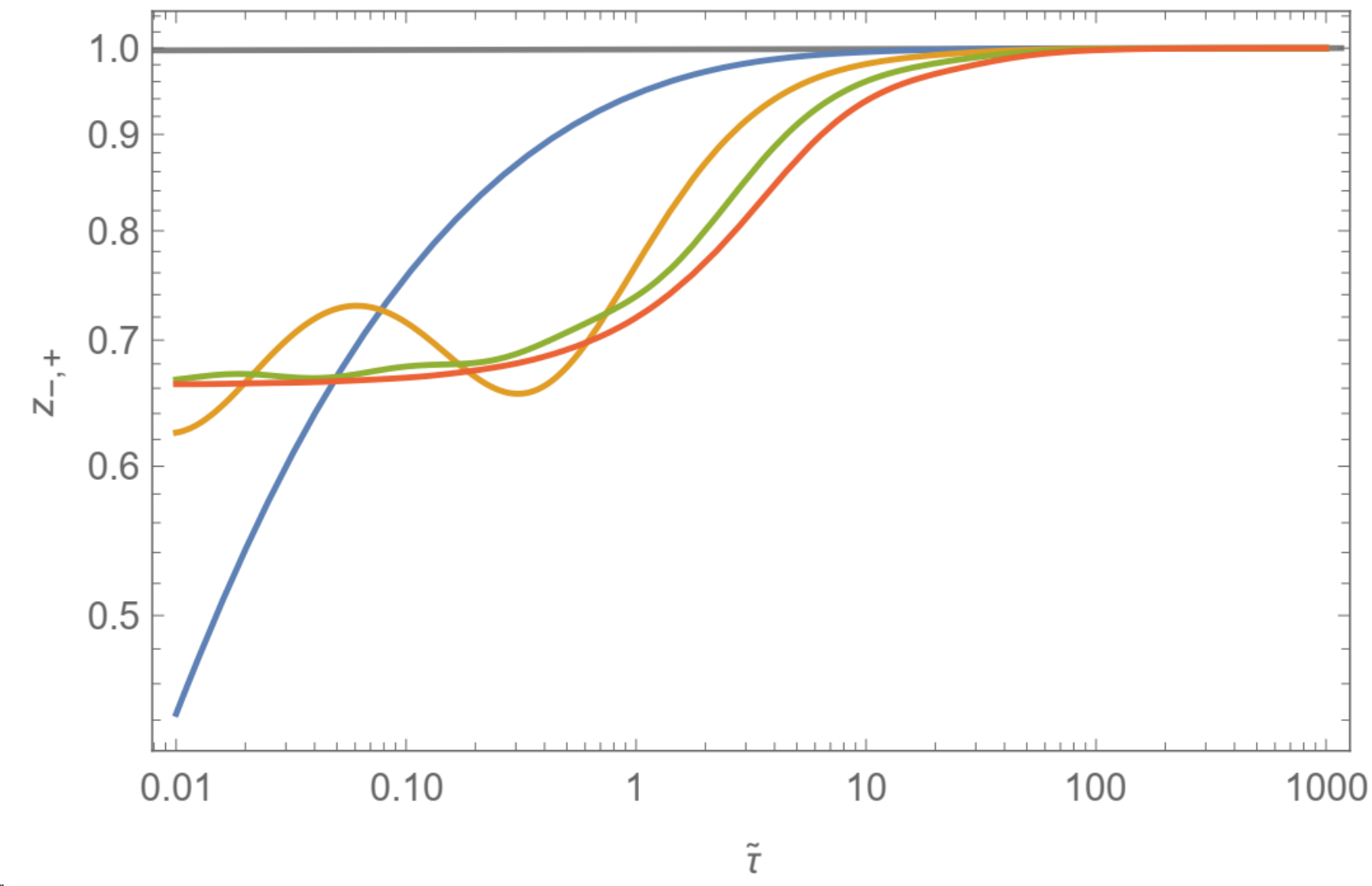}\\
\includegraphics[scale=0.272]{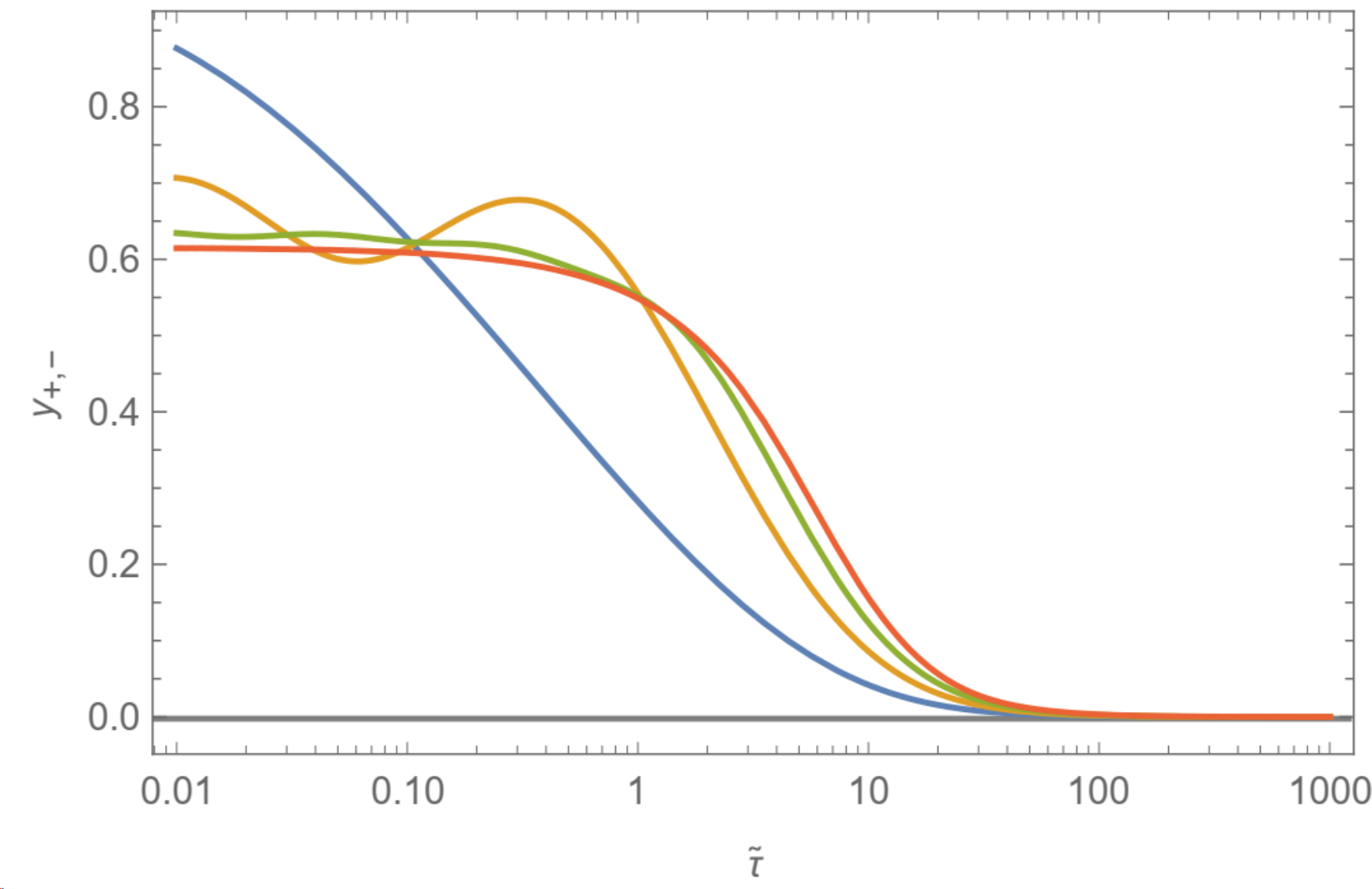} 
\includegraphics[scale=0.272]{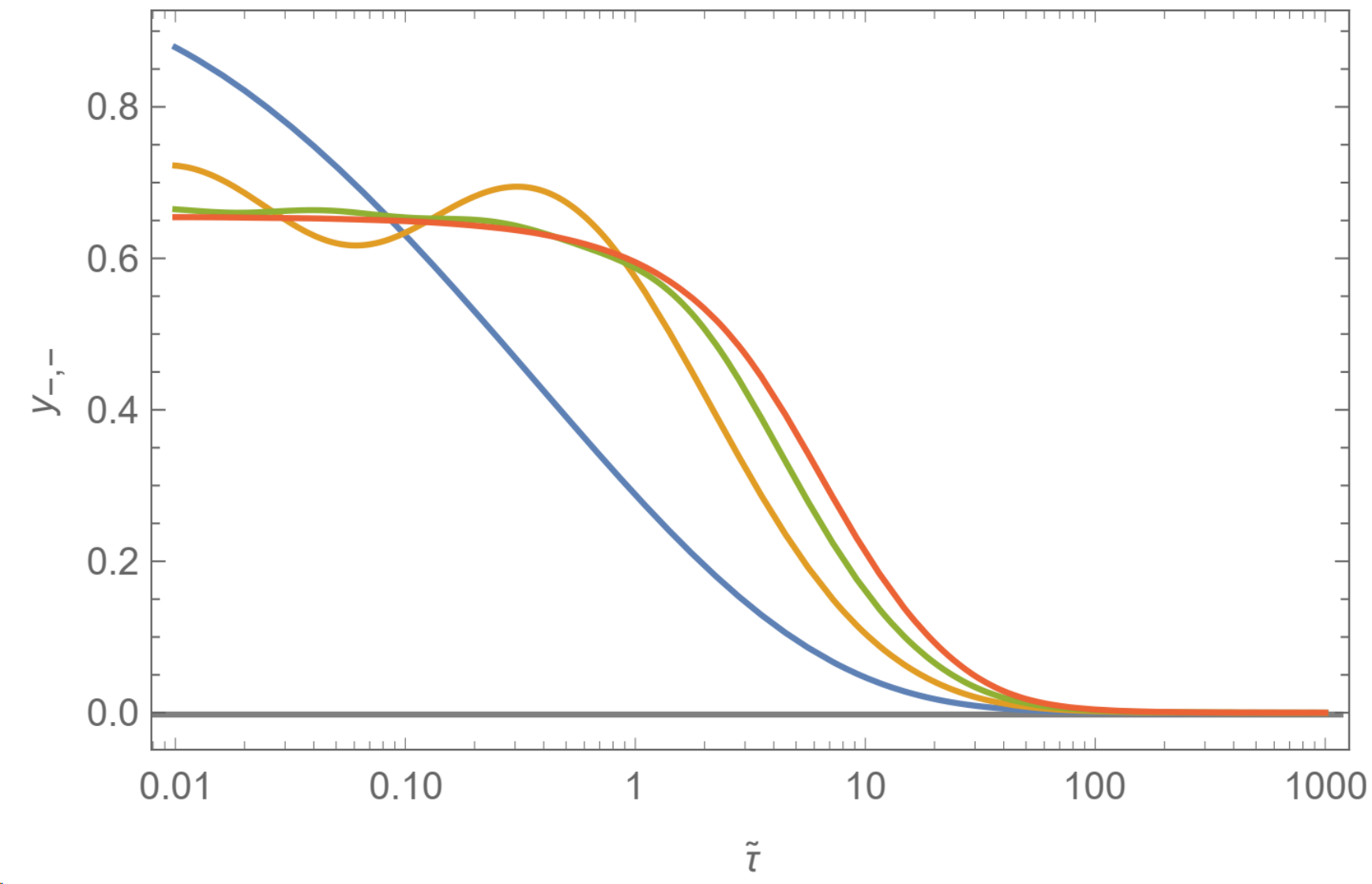}\\
\includegraphics[scale=0.272]{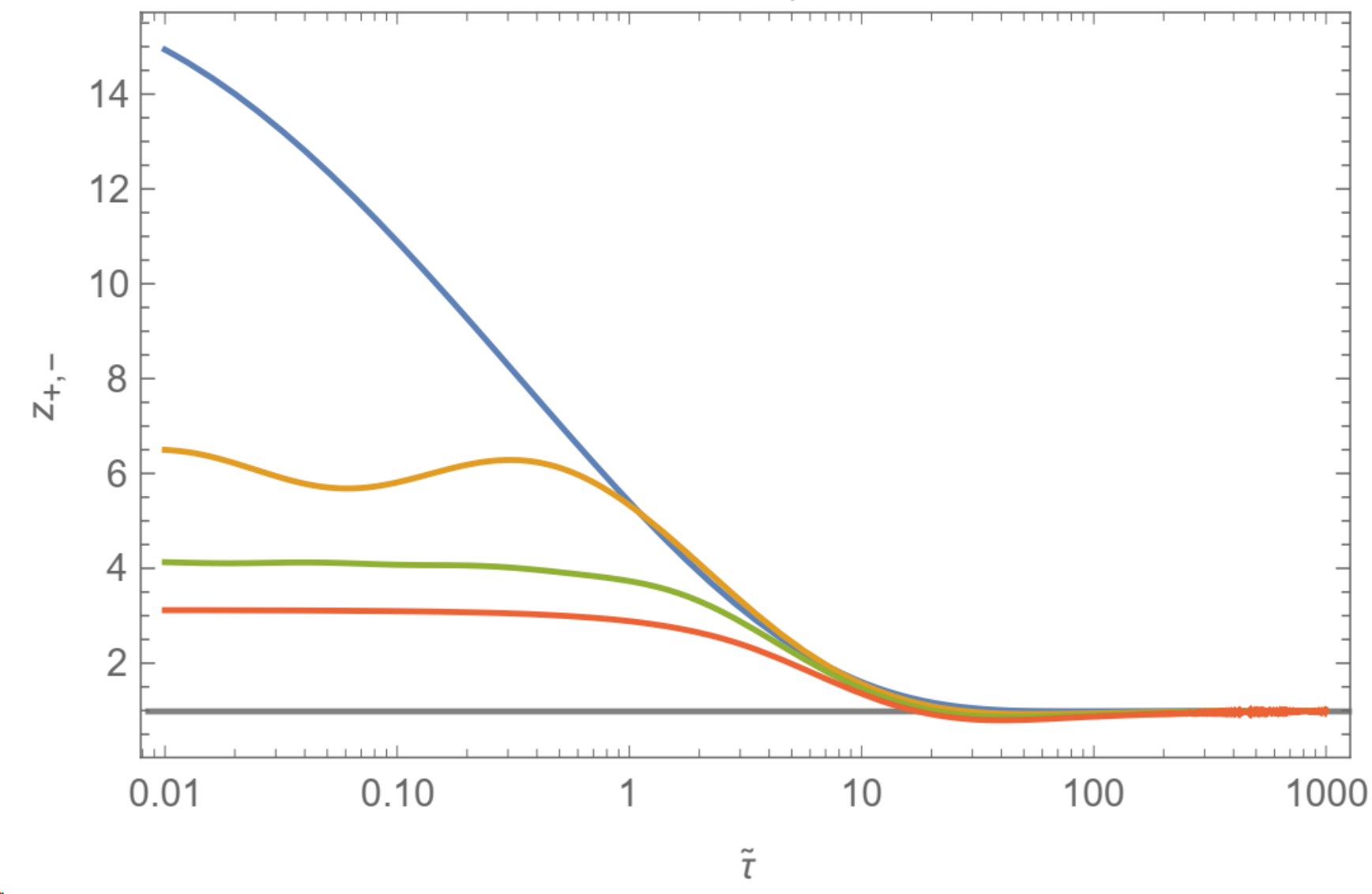} 
\includegraphics[scale=0.272]{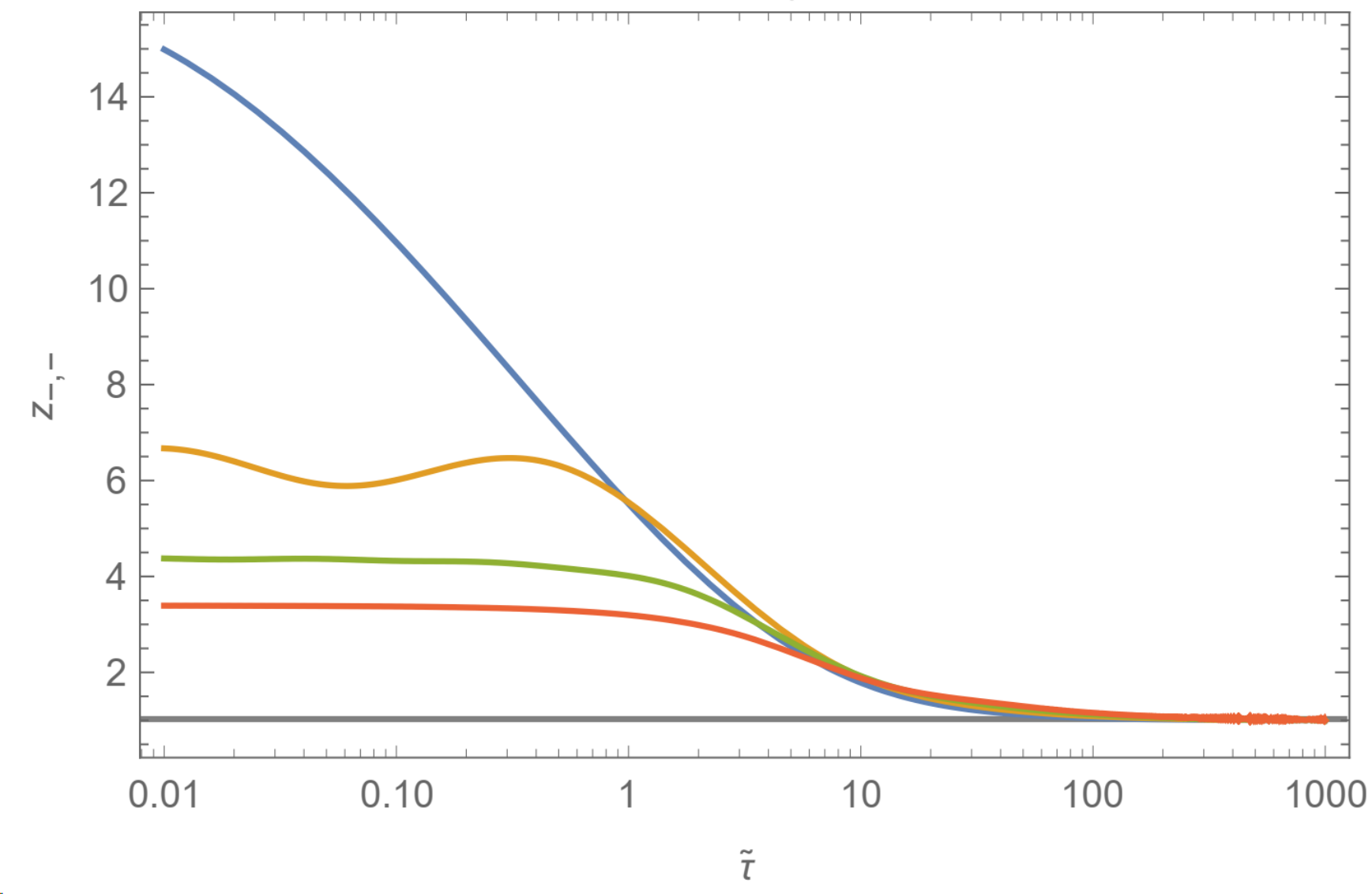}
\caption{$y_{s,s'}(\x)$ and $z_{s,s'}(\x)$ with respect to $\x$ for $\mu_m=25$, $\xpi=\xp$, and different values of $\xp$.}
\label{fig:ysp-zsp-44}  
\bigskip
\hspace{0.09cm}
\includegraphics[scale=0.39]{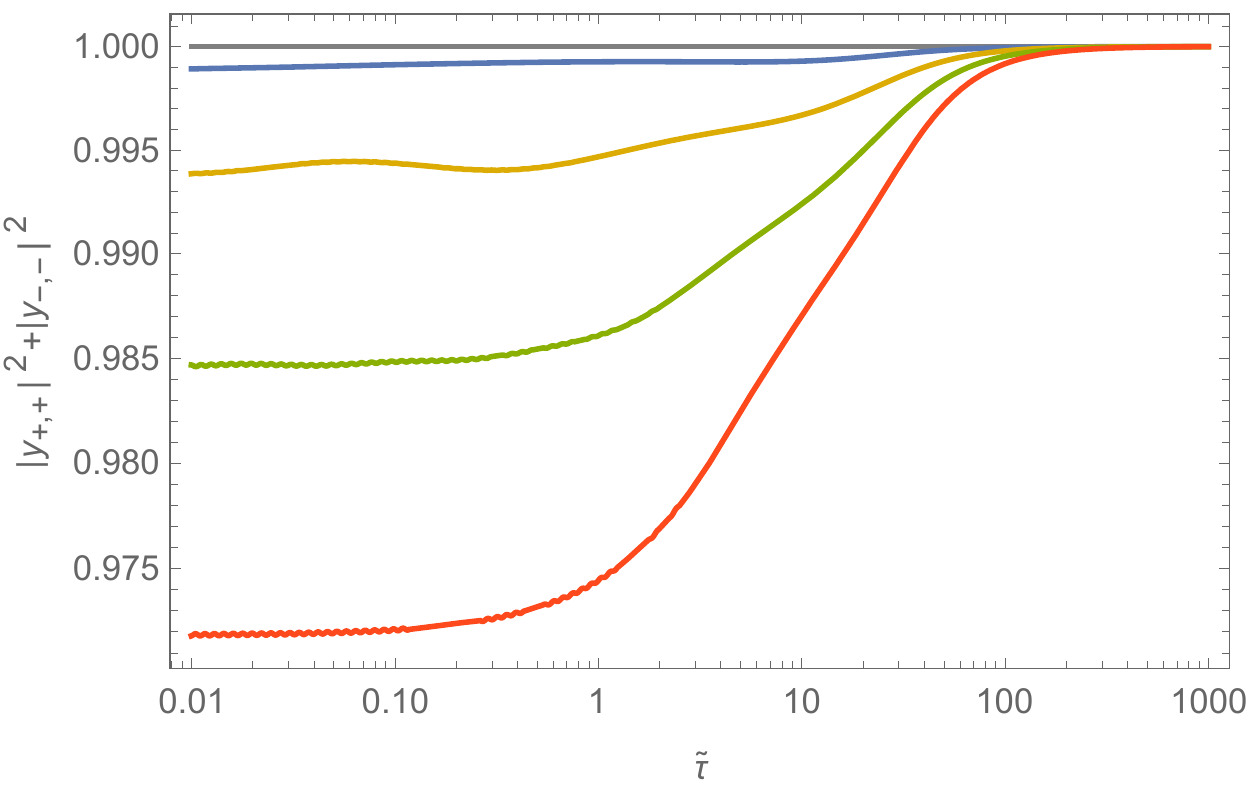}\hspace{0.11cm} 
\includegraphics[scale=0.38]{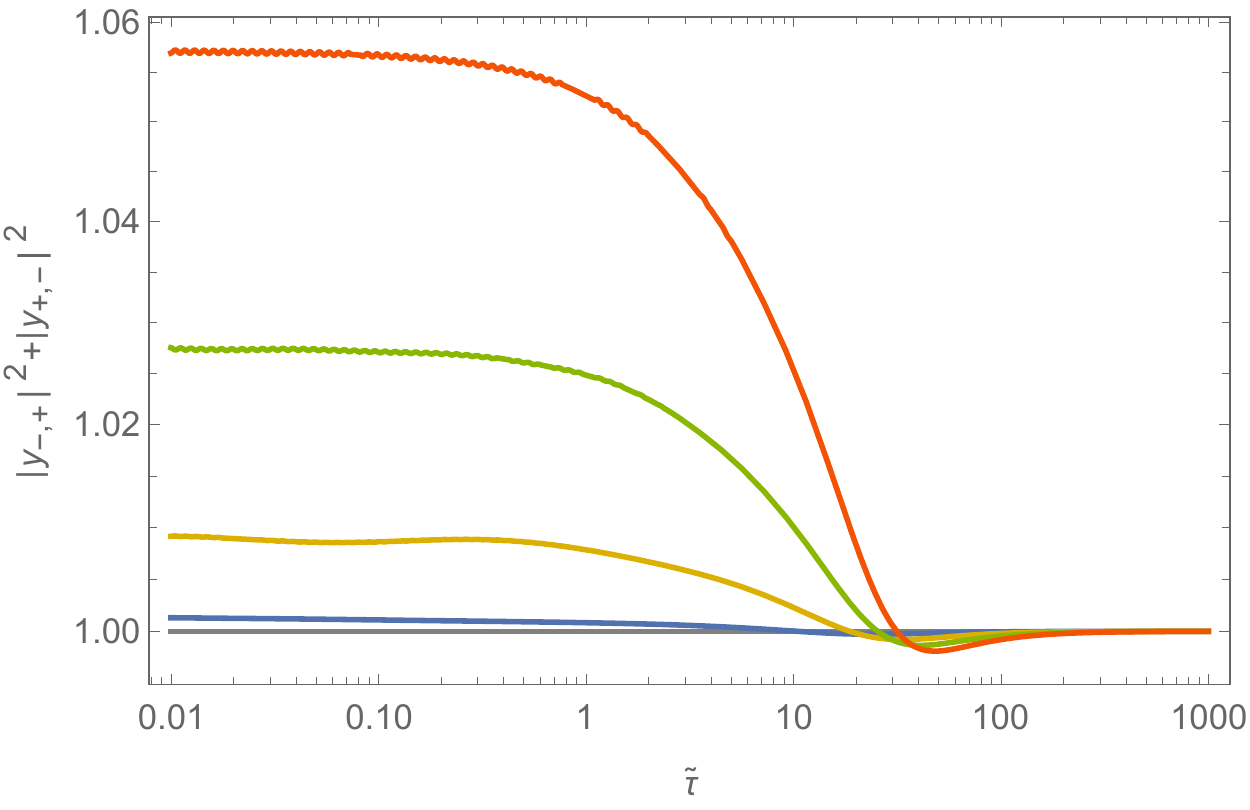}\\
\includegraphics[scale=0.4]{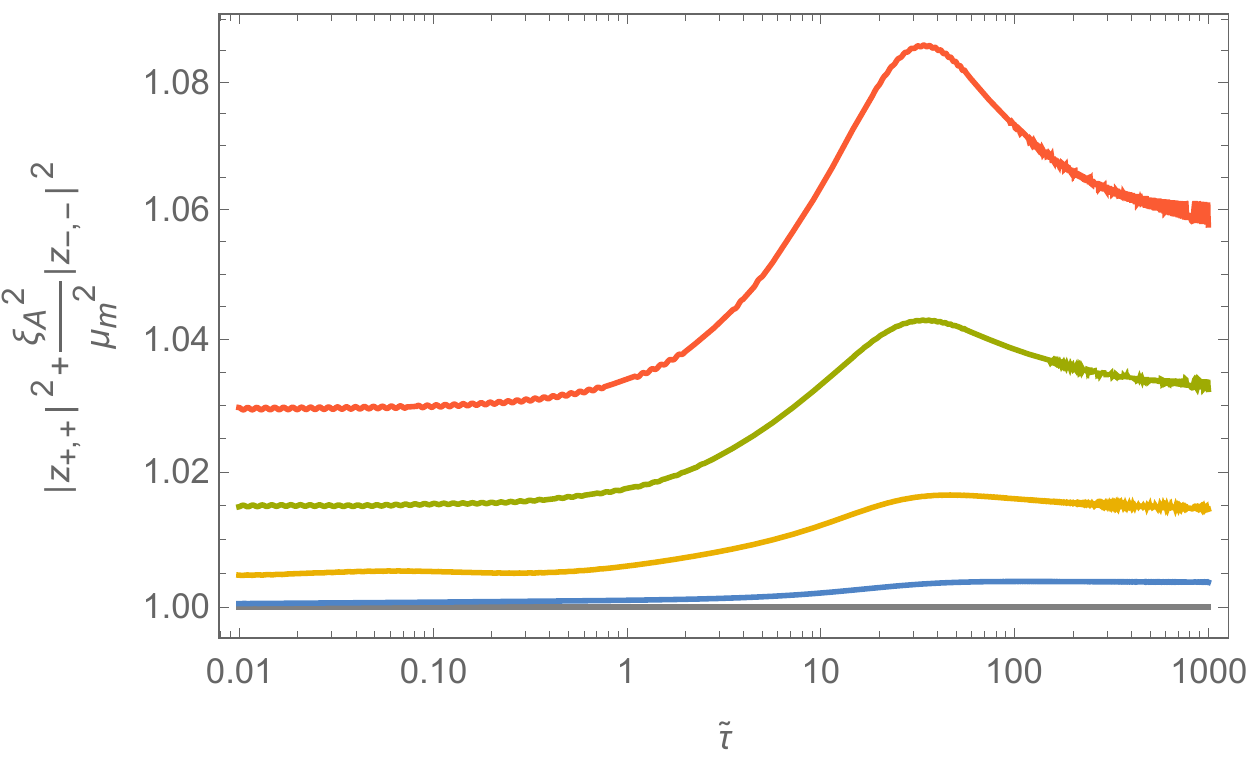} 
\includegraphics[scale=0.4]{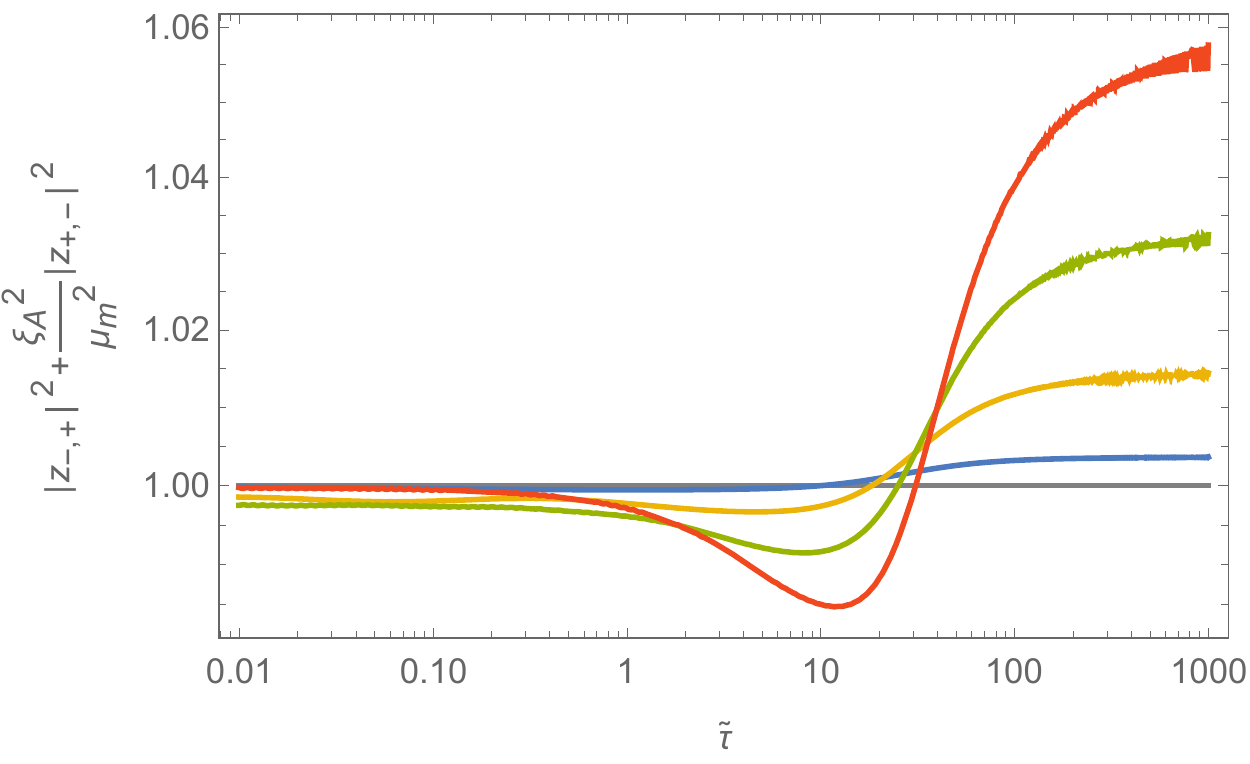}
\caption{$\lvert y_{s,+}\rvert^2+\lvert y_{-s,-}\rvert^2$, and $\lvert Z_{s,+}\rvert^2+\frac{\xp^2}{\mu^2_{\rm{m}}} \lvert Z_{-s,-}\rvert^2$ for the same parameters of Fig. \ref{fig:ysp-zsp-44}. }
\label{fig:ysp-ysp-44}   
\end{figure*}

\begin{figure*}[!htb]
\centering
\includegraphics[scale=0.28]{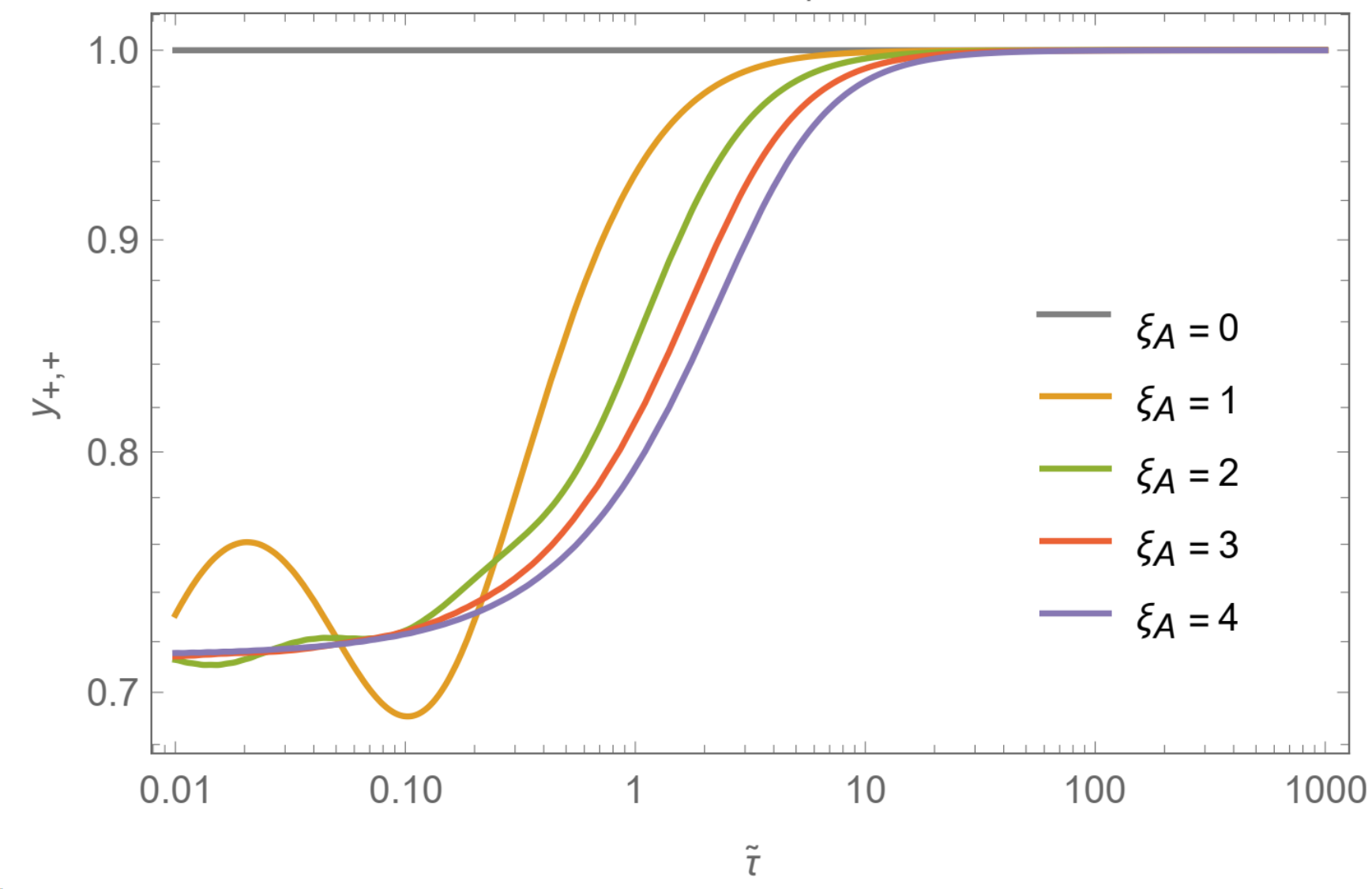} 
\includegraphics[scale=0.385]{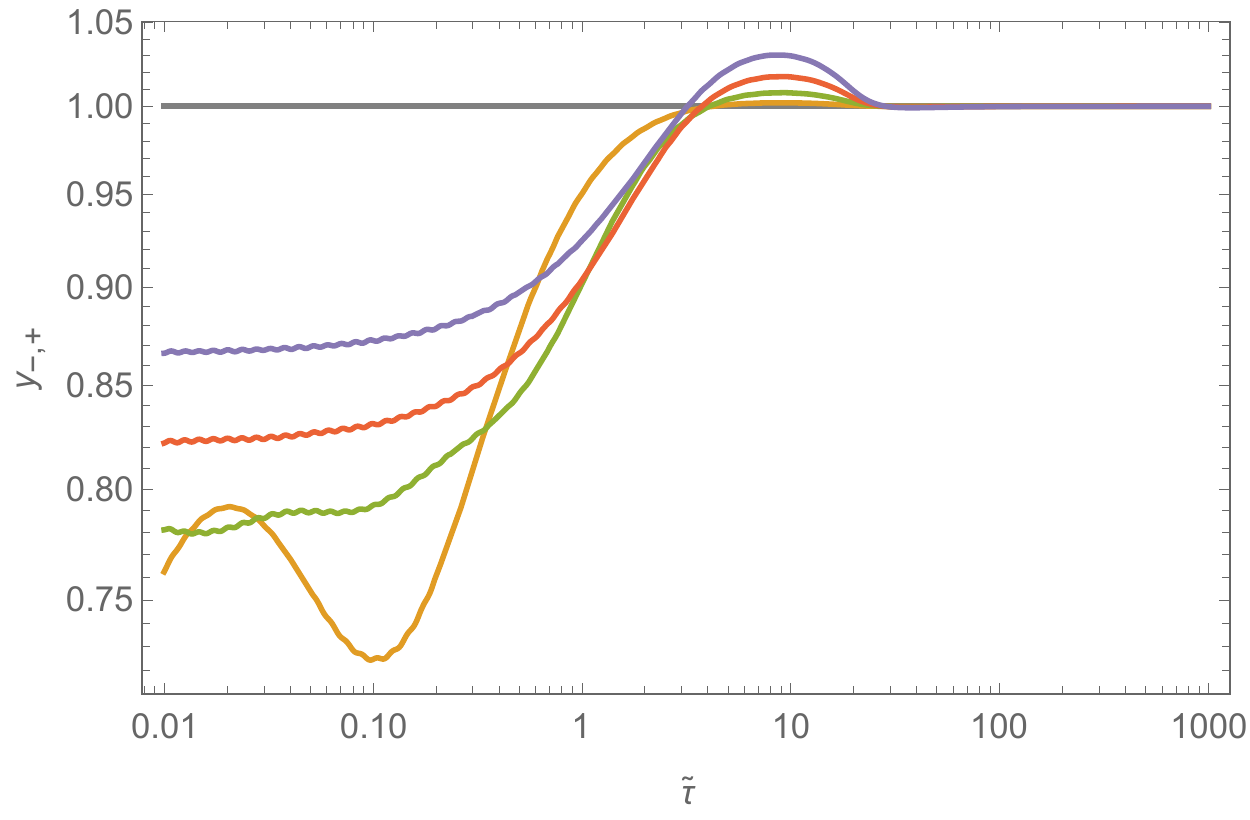}\\
\includegraphics[scale=0.385]{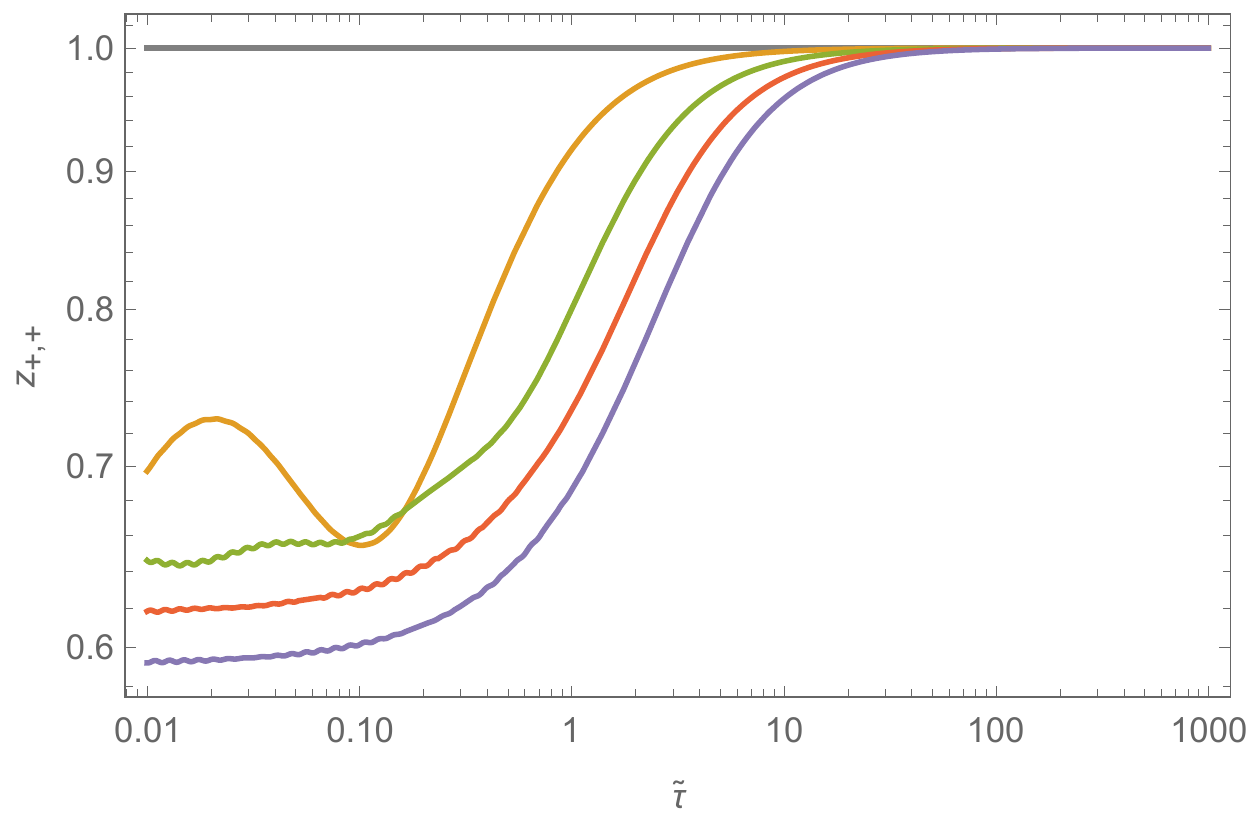} 
\includegraphics[scale=0.385]{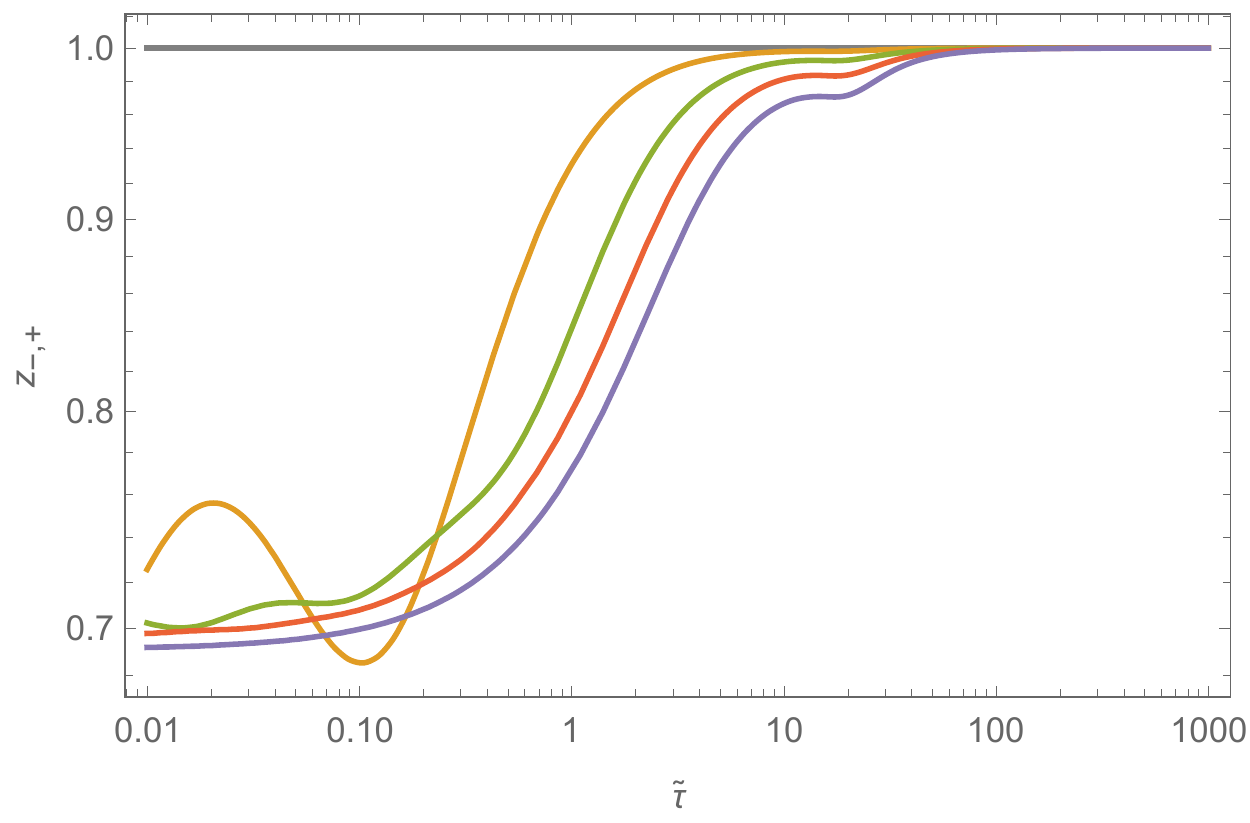}\\
\includegraphics[scale=0.28]{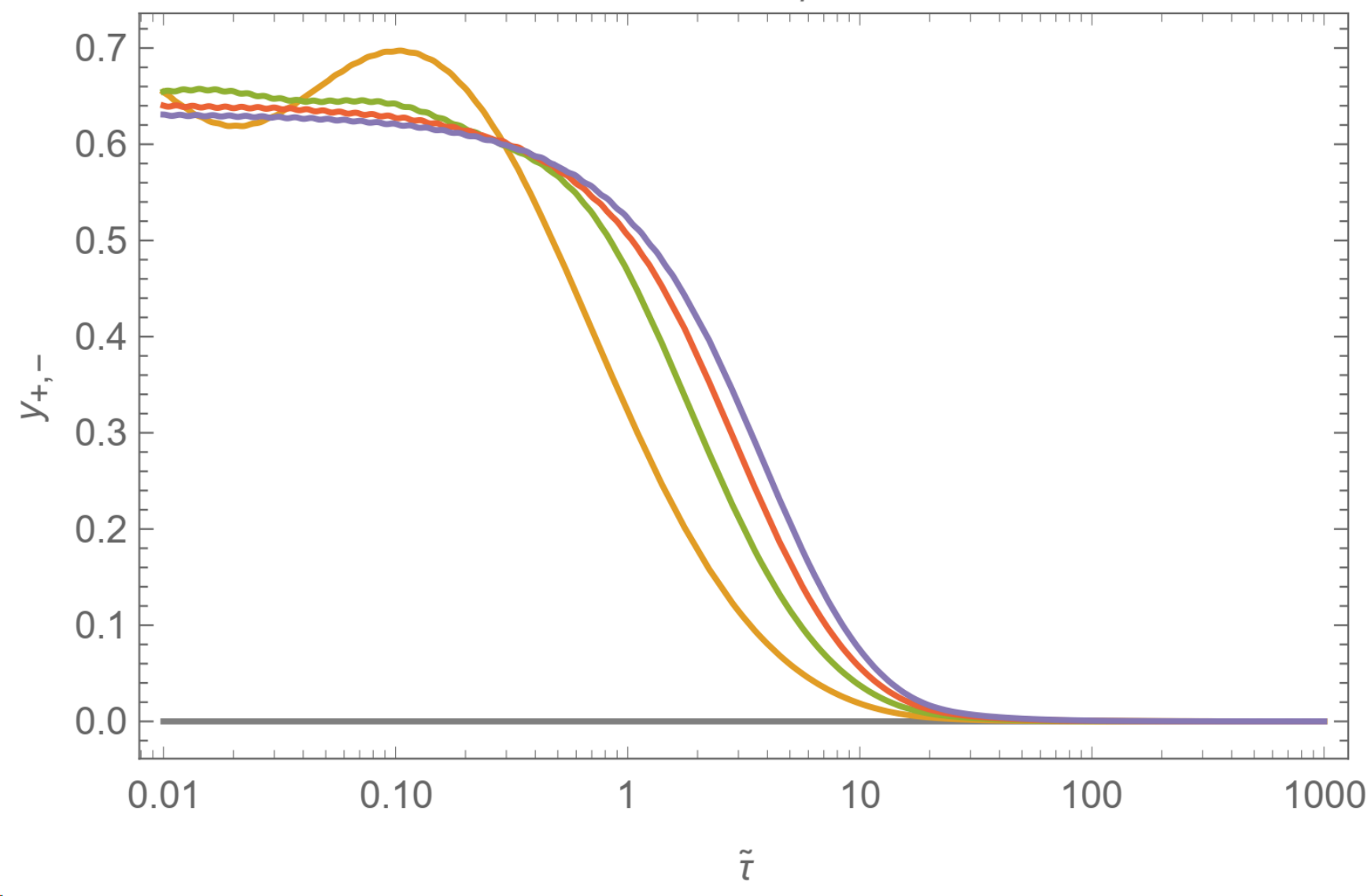} 
\includegraphics[scale=0.385]{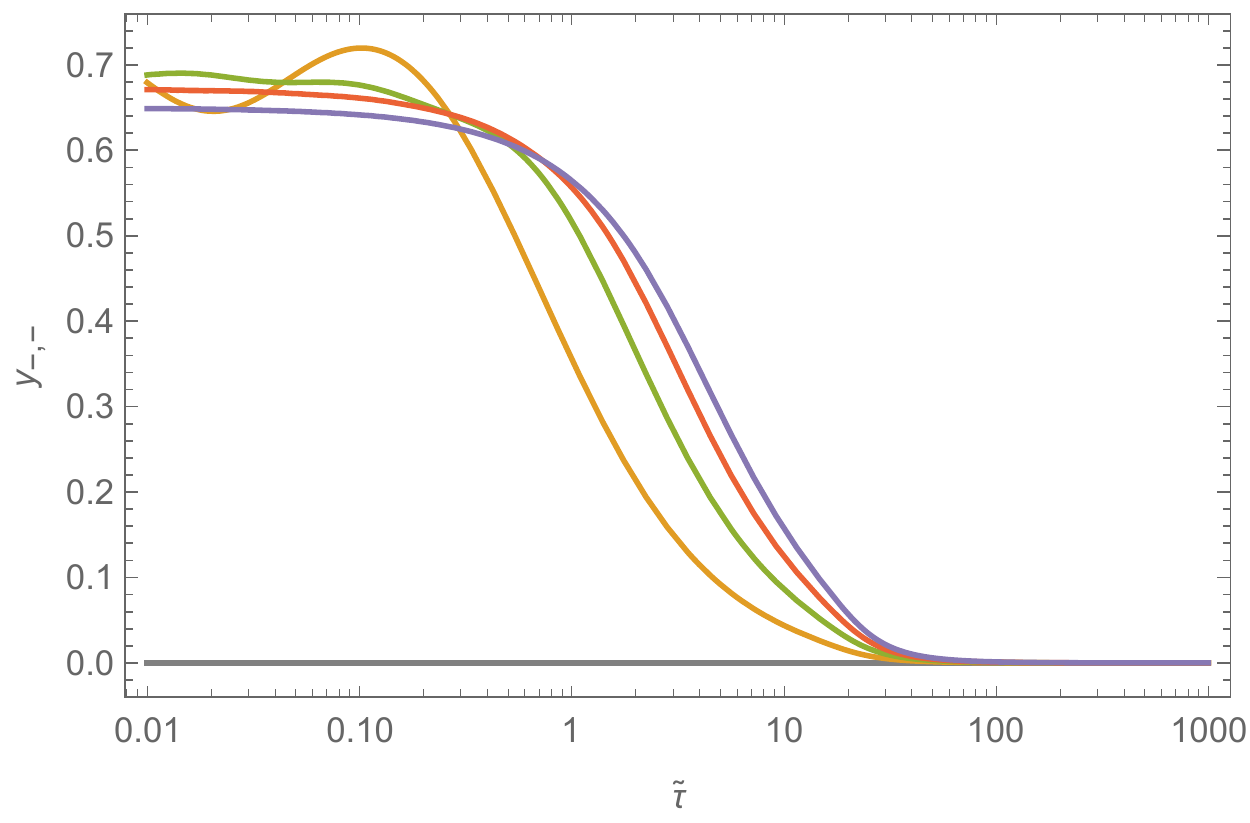}\\
\includegraphics[scale=0.385]{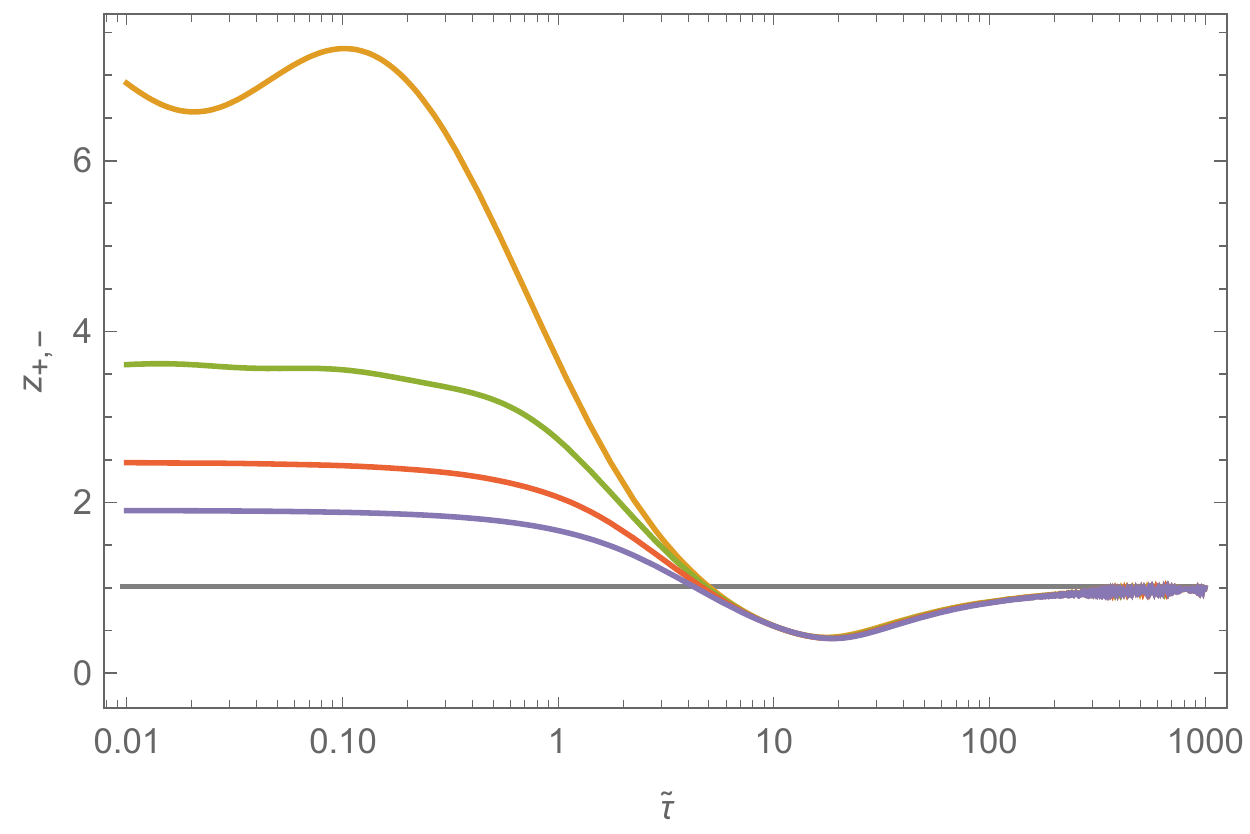} 
\includegraphics[scale=0.385]{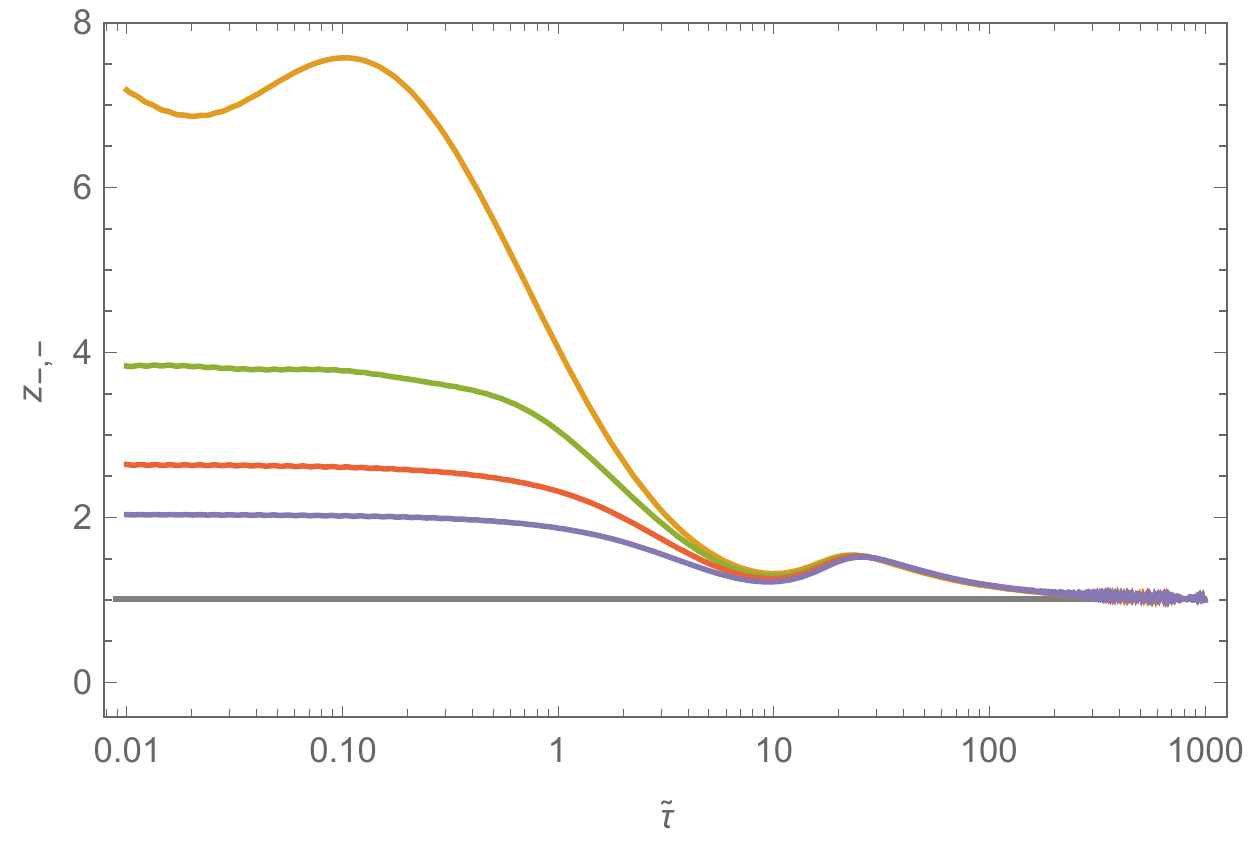}
\caption{$y_{s,s'}(\x)$ and $z_{s,s'}(\x)$ with respect to $\x$ for $\mu_m=10$, $\xpi=8$, and different values of $\xp$.}
\label{fig:ysp-zsp-4}  
\hspace{0.2cm}
\includegraphics[scale=0.28]{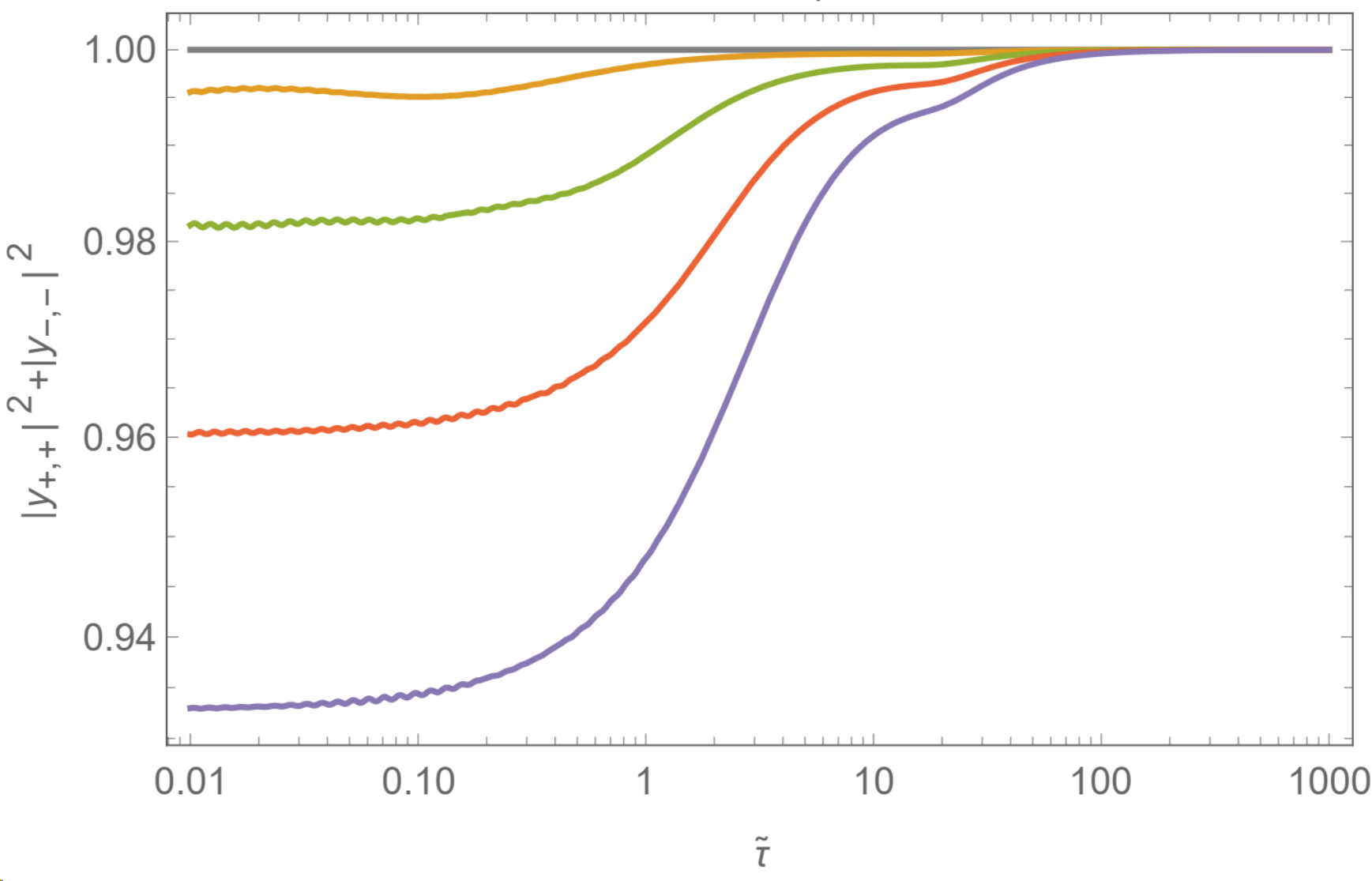}\hspace{0.46cm} 
\includegraphics[scale=0.28]{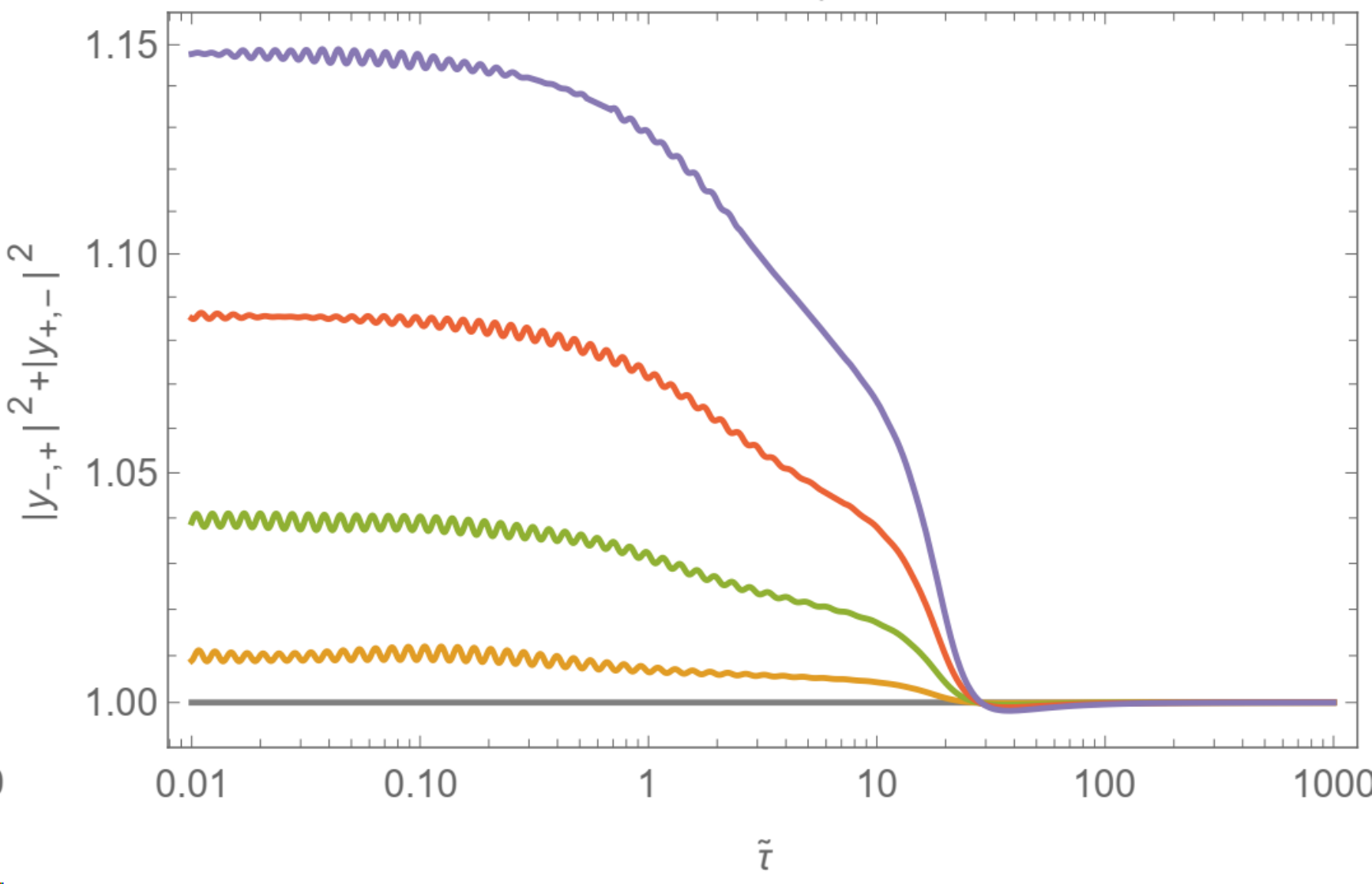}\\
\includegraphics[scale=0.3]{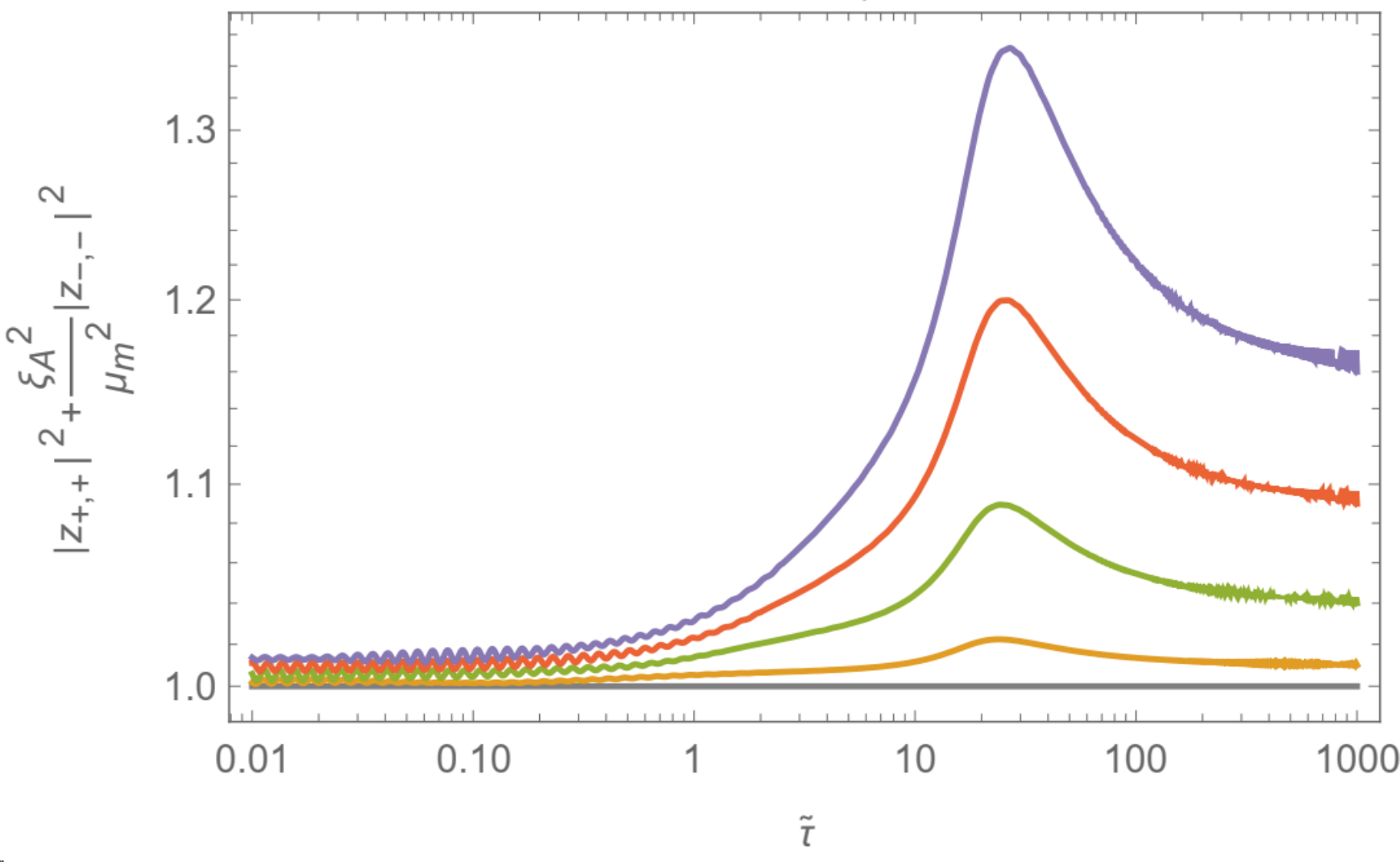} 
\includegraphics[scale=0.3]{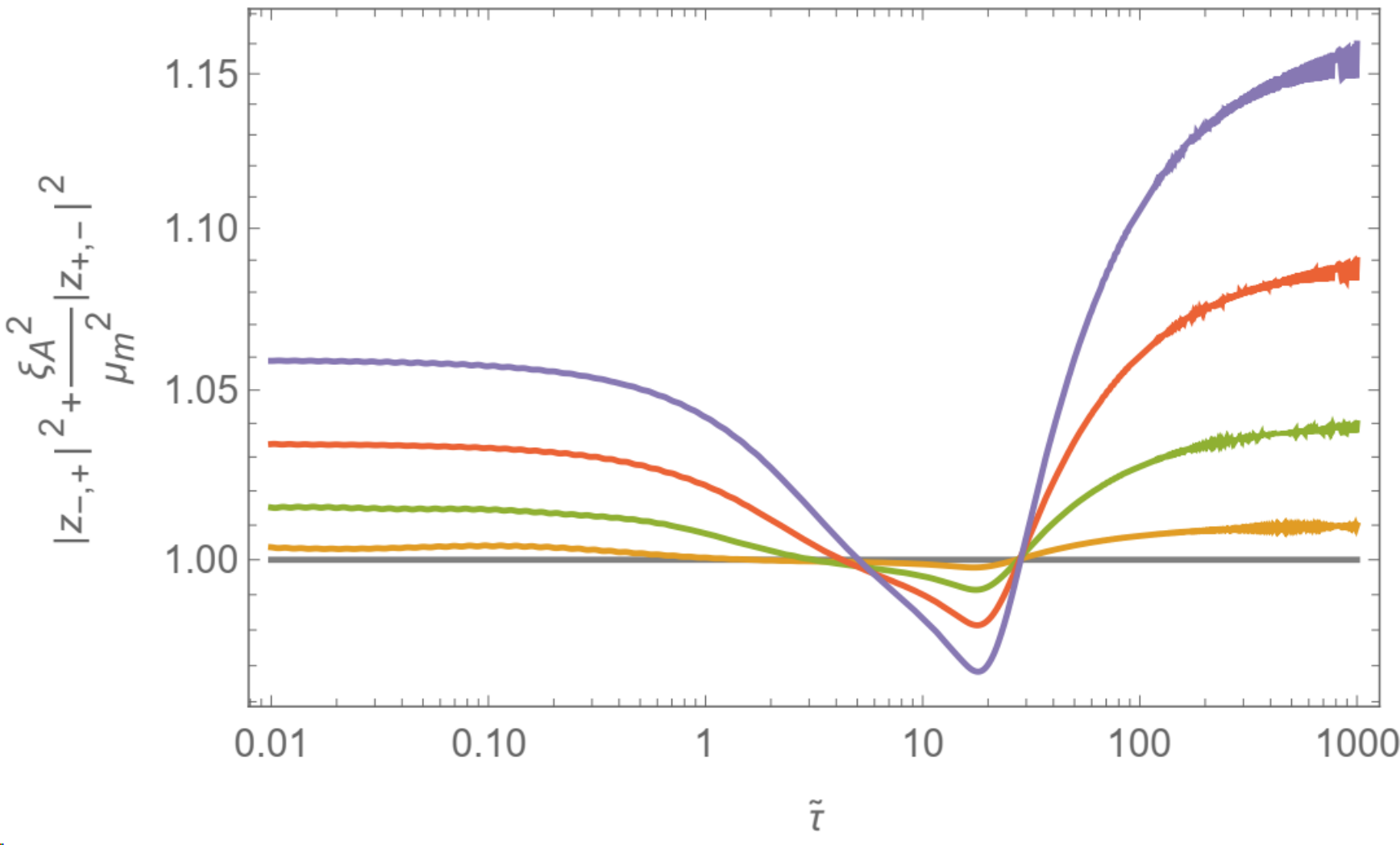}
\caption{$\lvert y_{s,+}\rvert^2+\lvert y_{-s,-}\rvert^2$, and $\lvert Z_{s,+}\rvert^2+\frac{\xp^2}{\mu^2_{\rm{m}}} \lvert Z_{-s,-}\rvert^2$ for the same parameters of Fig. \ref{fig:ysp-zsp-4}. }
\label{fig:ysp-ysp-4}   
\end{figure*}

\begin{figure*}[!htb]
\centering
\includegraphics[scale=0.272]{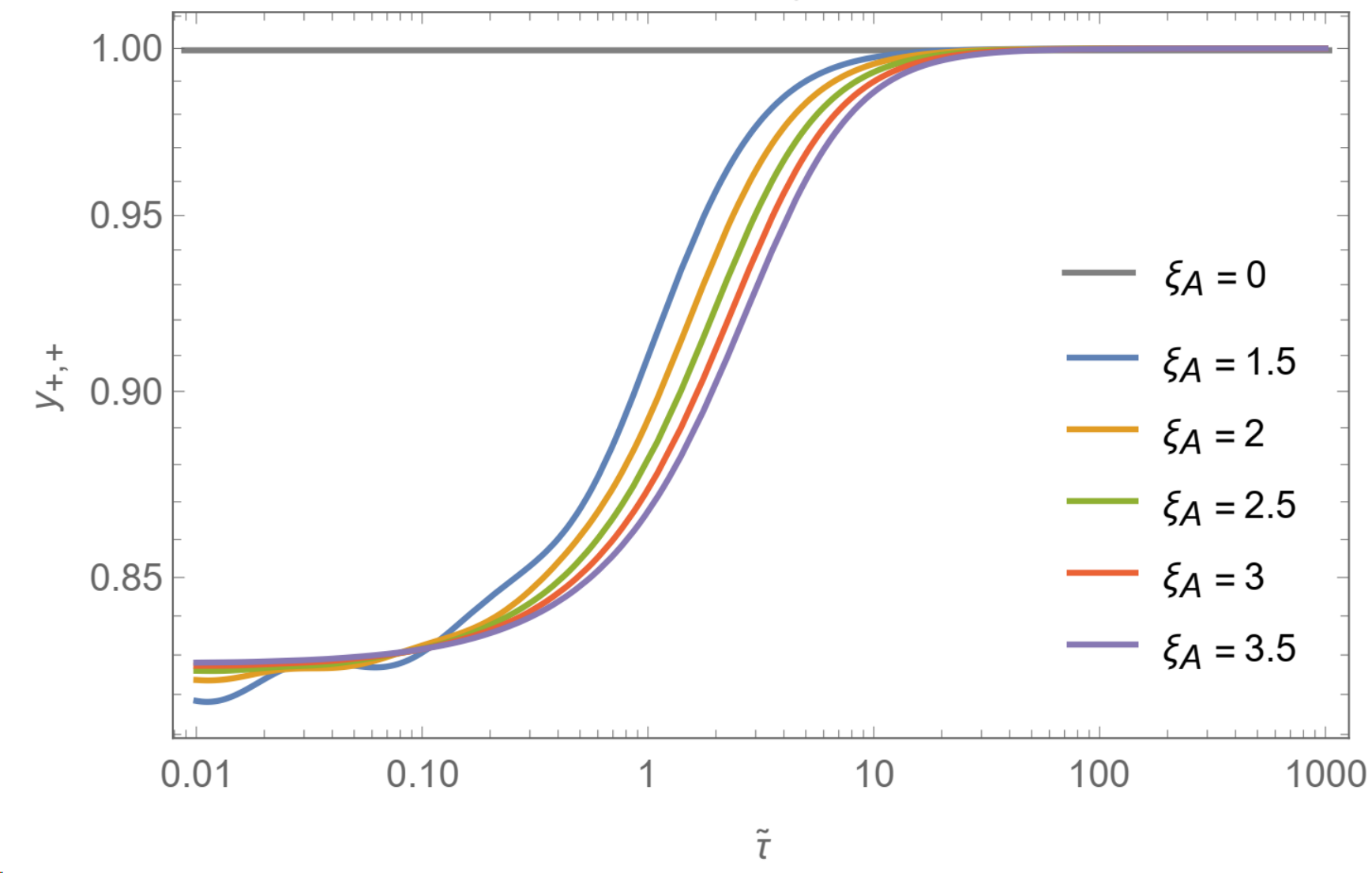} 
\includegraphics[scale=0.272]{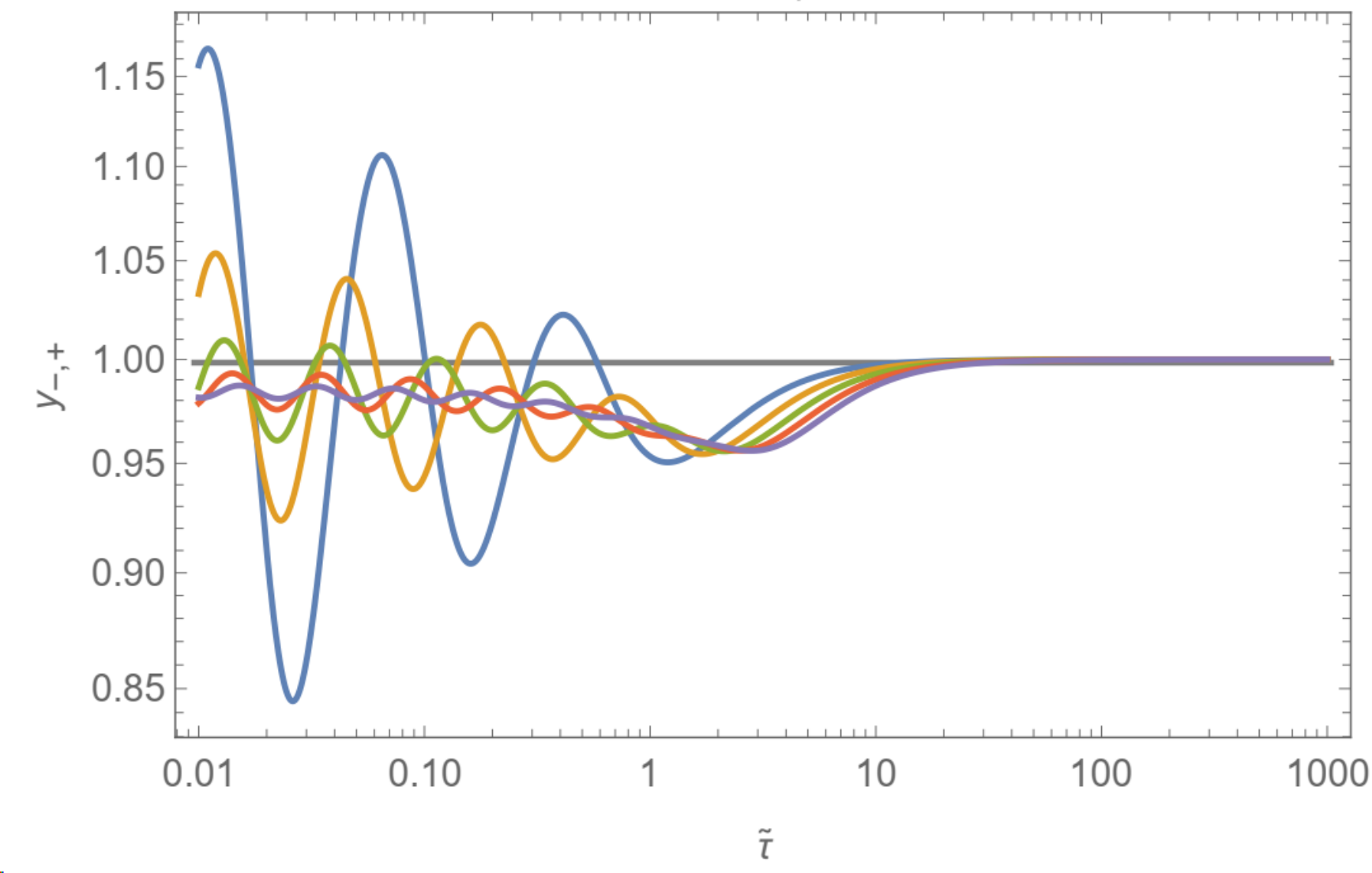}\\
\includegraphics[scale=0.272]{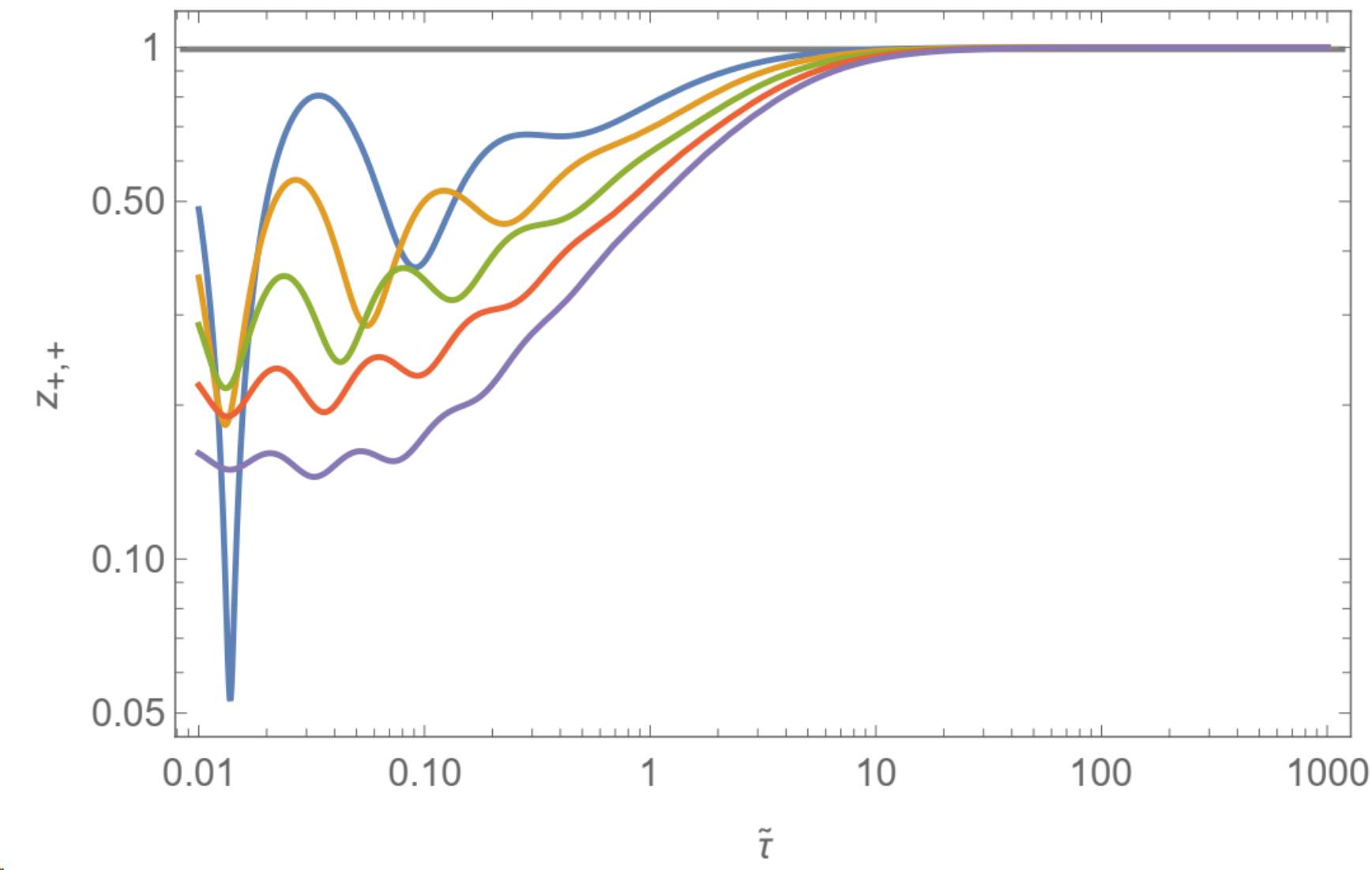} 
\includegraphics[scale=0.272]{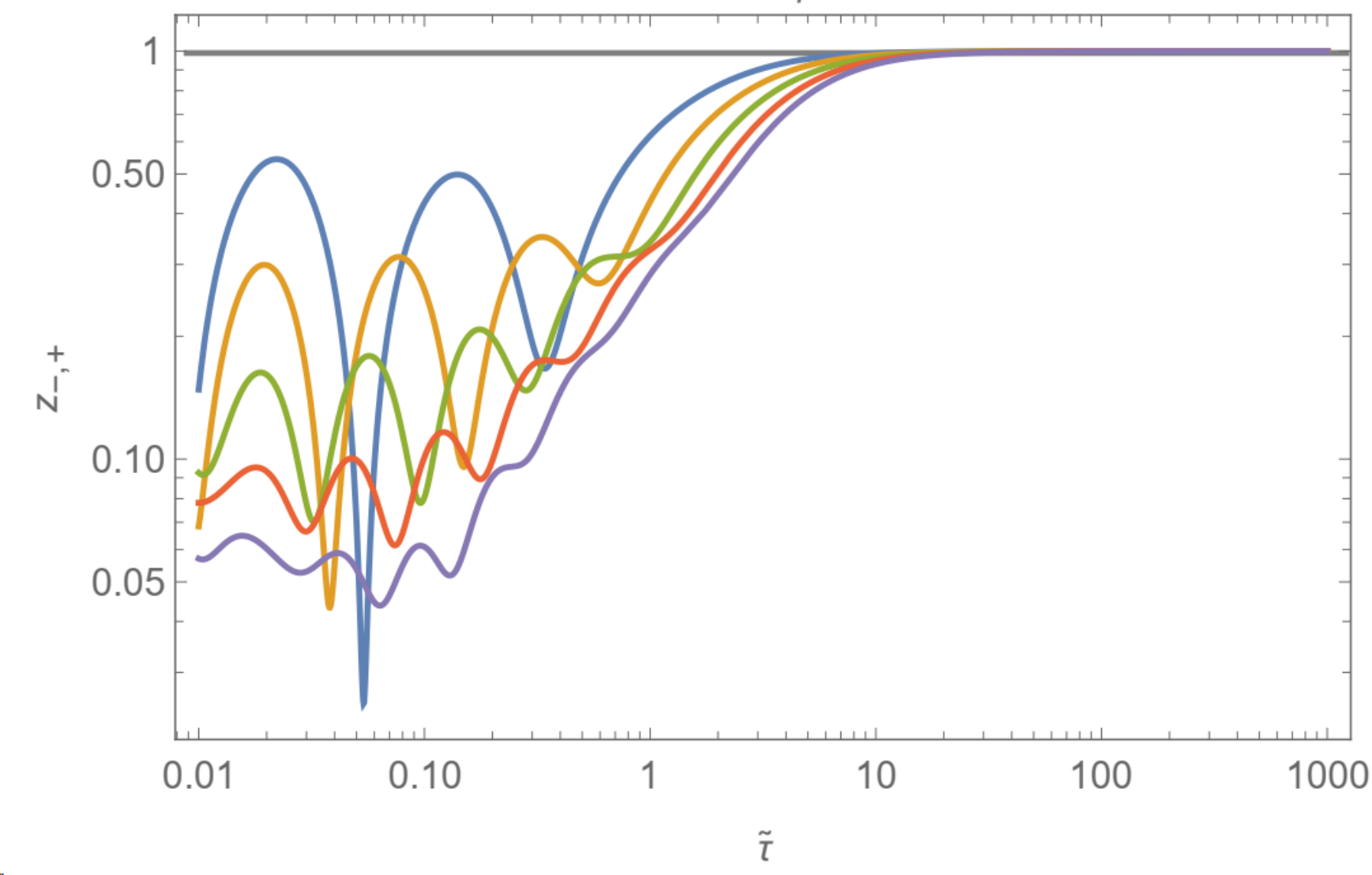}\\
\includegraphics[scale=0.272]{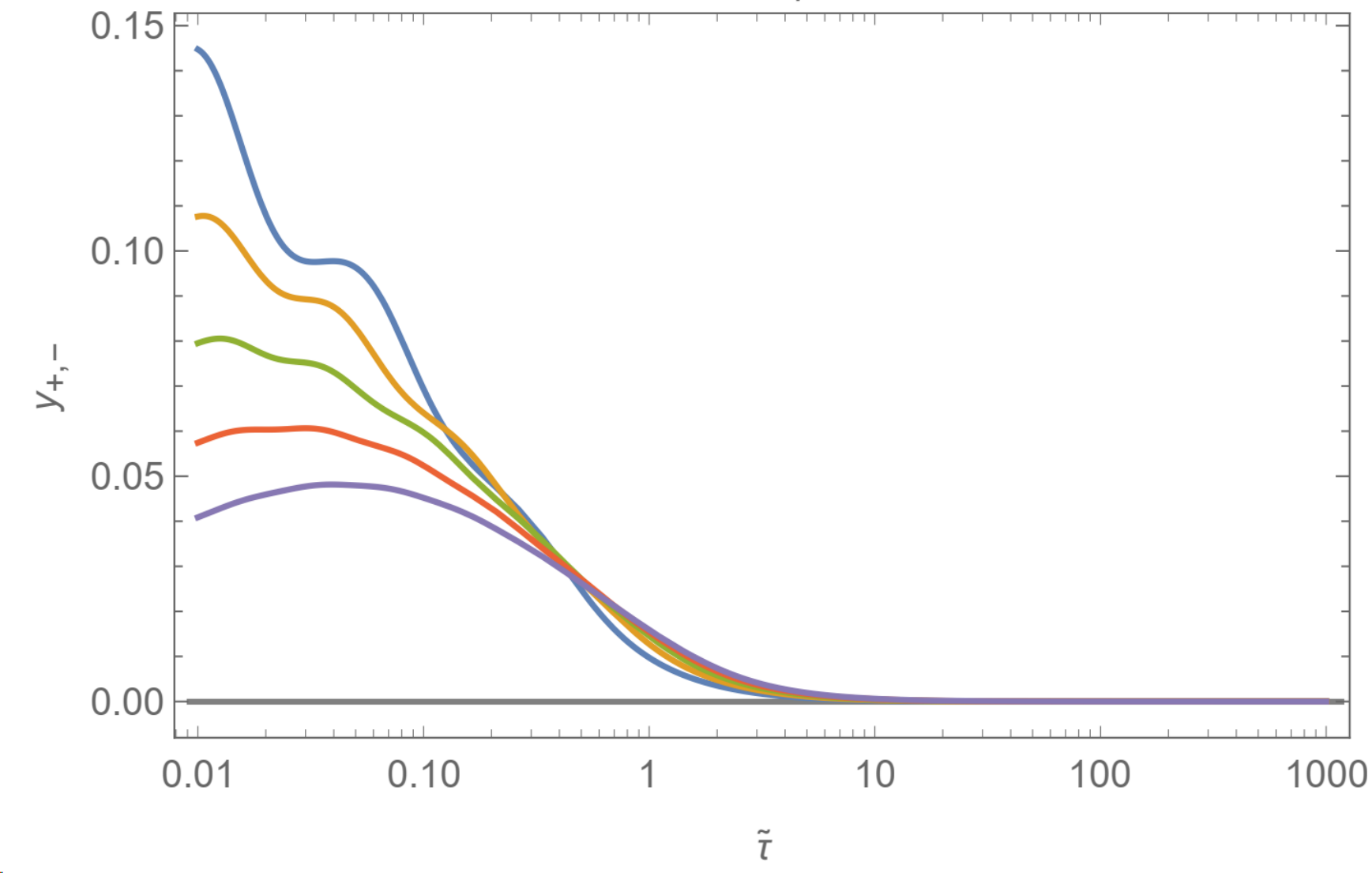} 
\includegraphics[scale=0.272]{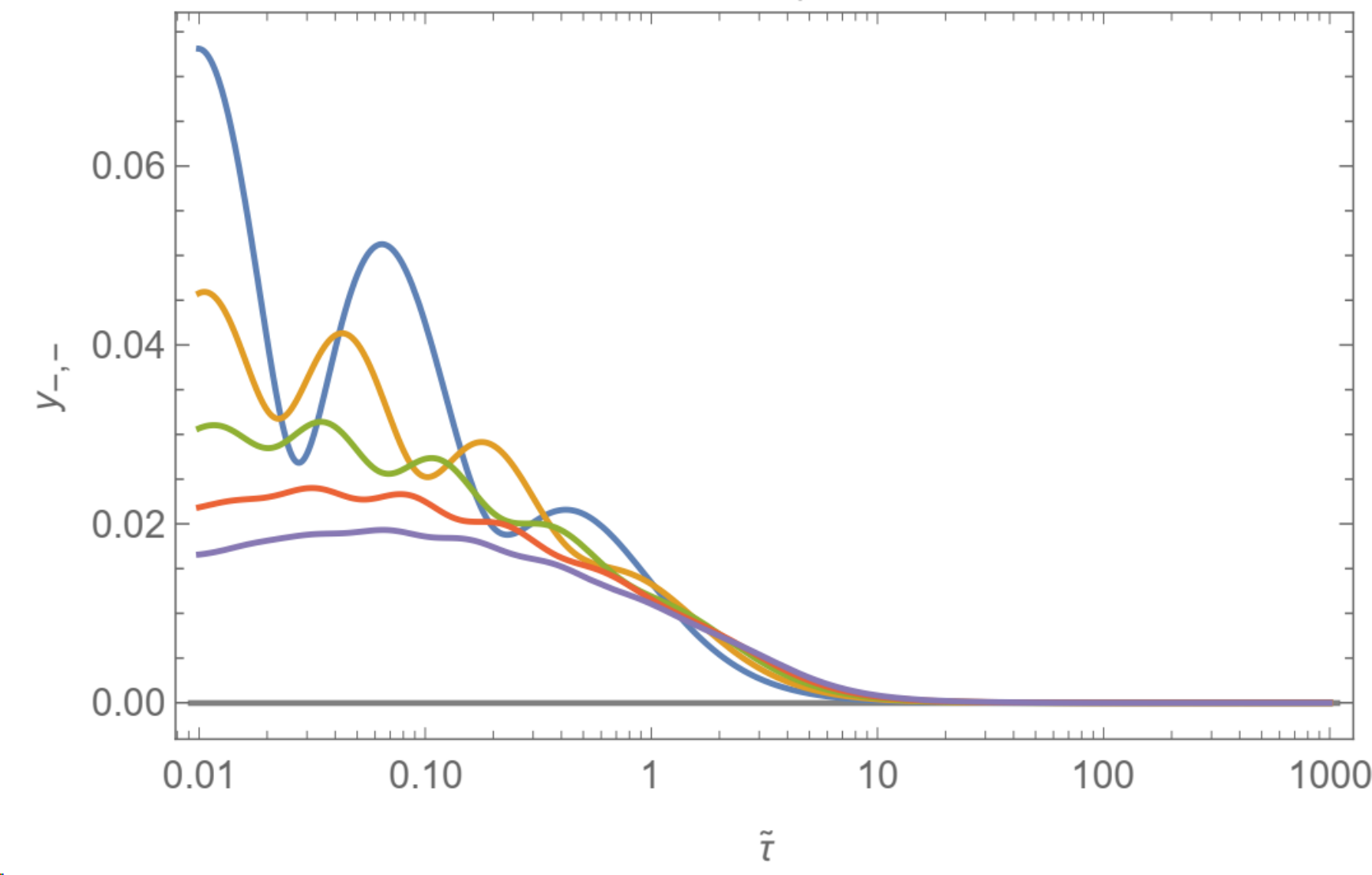}\\
\includegraphics[scale=0.272]{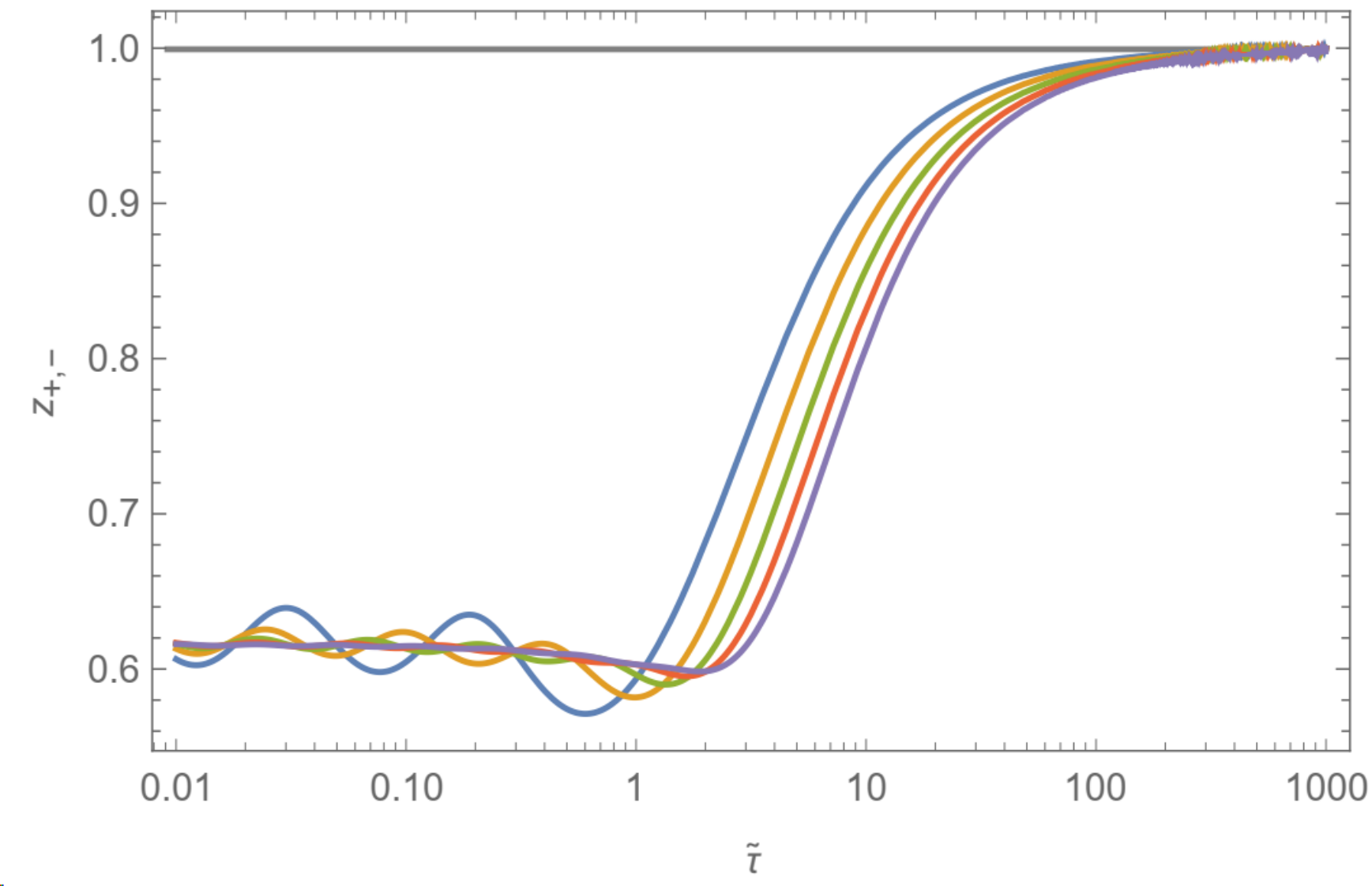} 
\includegraphics[scale=0.272]{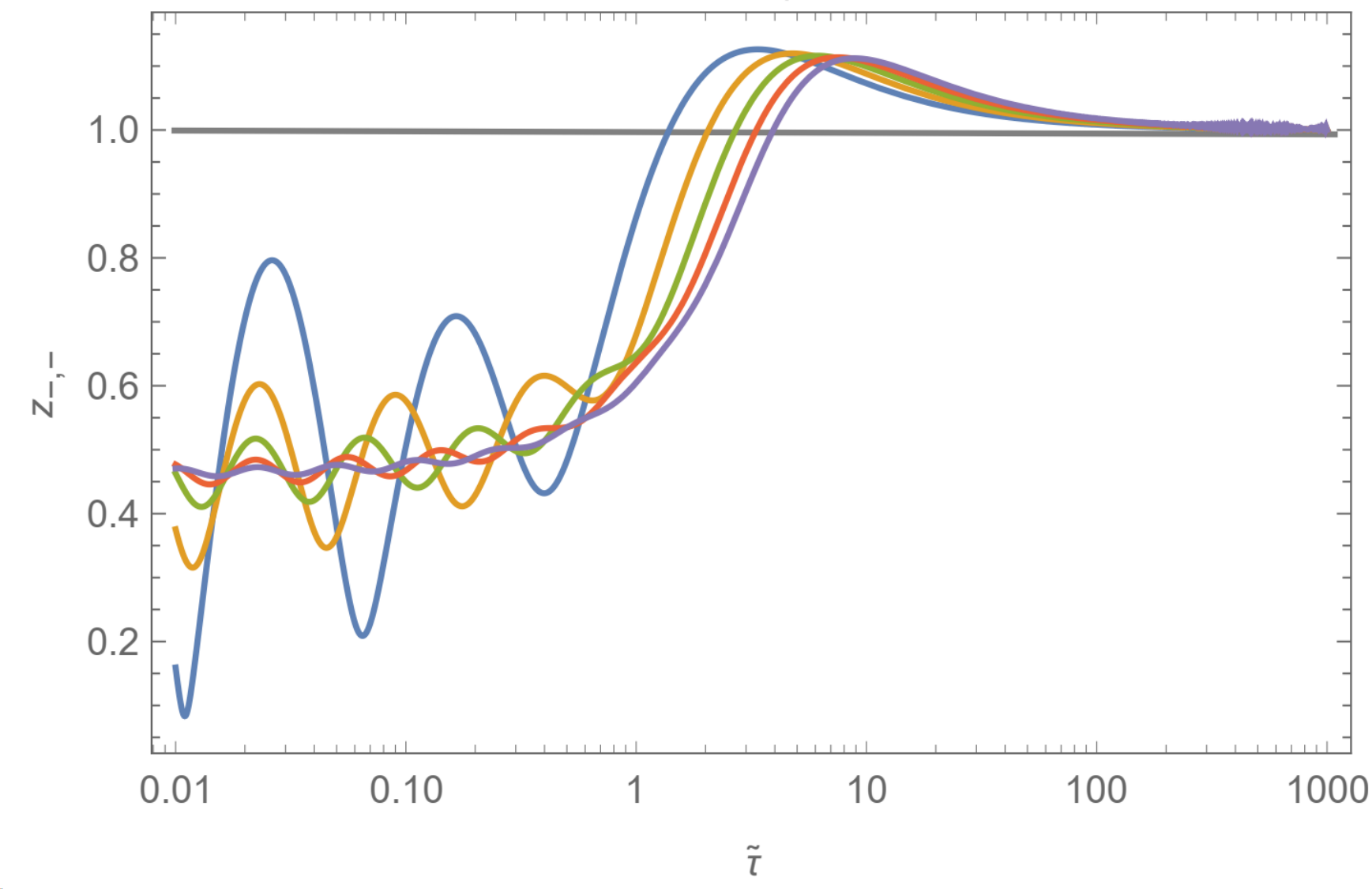}
\caption{$y_{s,s'}(\x)$ and $z_{s,s'}(\x)$ with respect to $\x$ for $\mu_m=0.1$, $\xpi=0.05$.}
\label{fig:ysp-zsp-5}   
\bigskip
\hspace{0.2cm}
\includegraphics[scale=0.39]{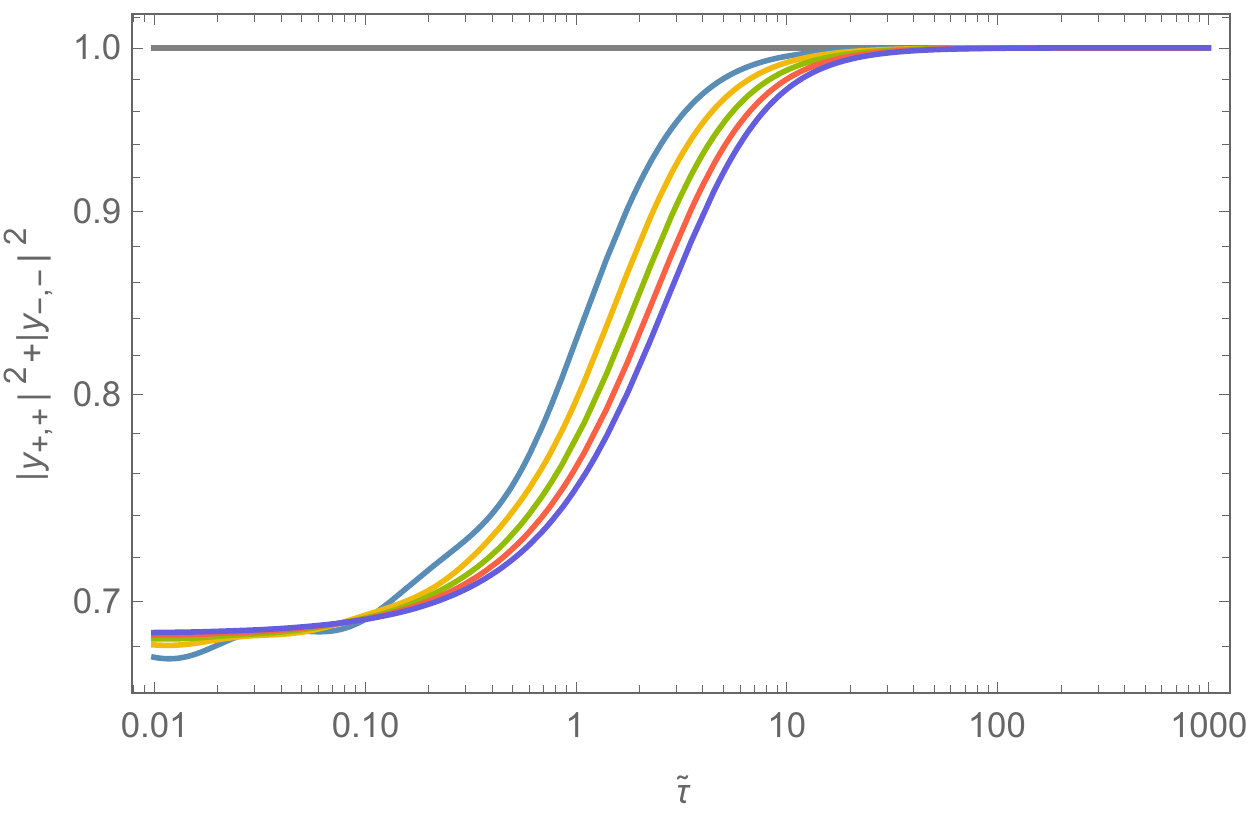}\hspace{0.55cm} 
\includegraphics[scale=0.38]{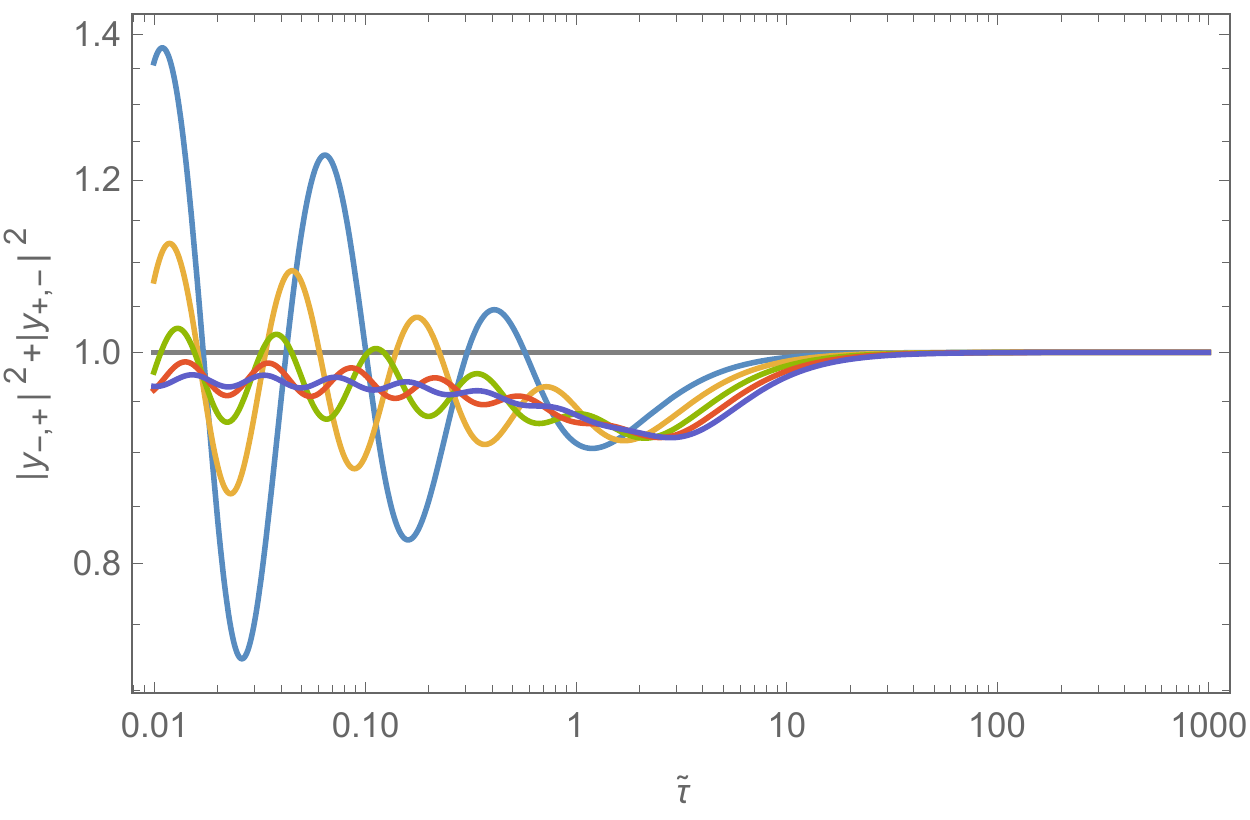}\\
\includegraphics[scale=0.43]{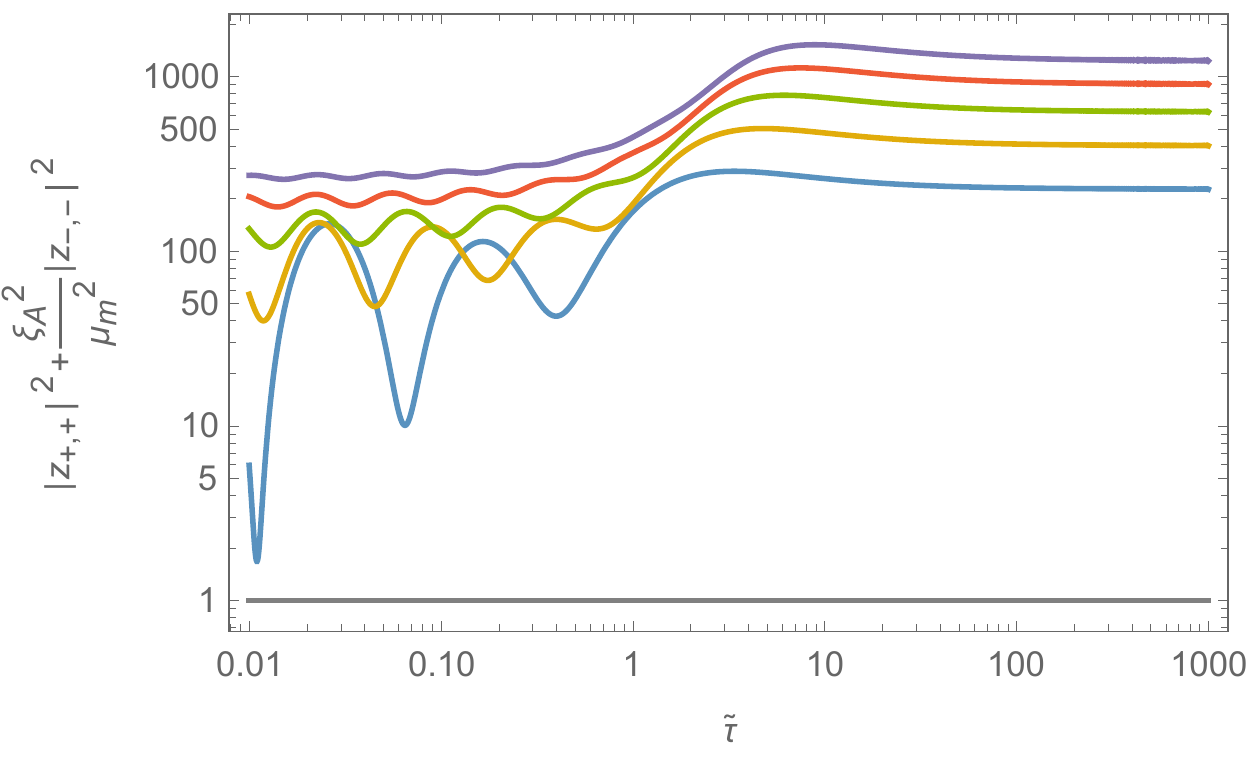} 
\includegraphics[scale=0.43]{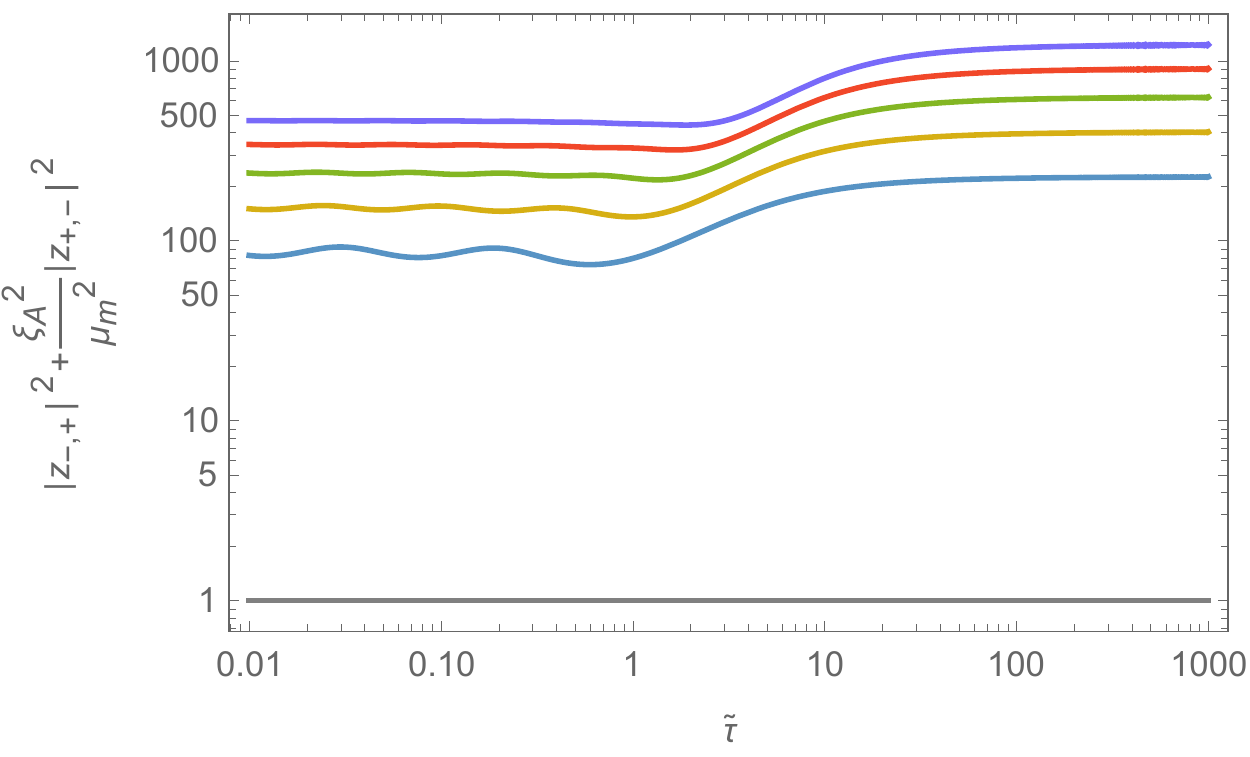}
\caption{$\lvert y_{s,+}\rvert^2+\lvert y_{-s,-}\rvert^2$, and $\lvert Z_{s,+}\rvert^2+\frac{\xp^2}{\mu^2_{\rm{m}}} \lvert Z_{-s,-}\rvert^2$ for the same parameters of Fig. \ref{fig:ysp-zsp-5}. }
\label{fig:ysp-ysp-5}   
\end{figure*}

\begin{figure*}[!htbp] 
\centering
\includegraphics[scale=0.28]{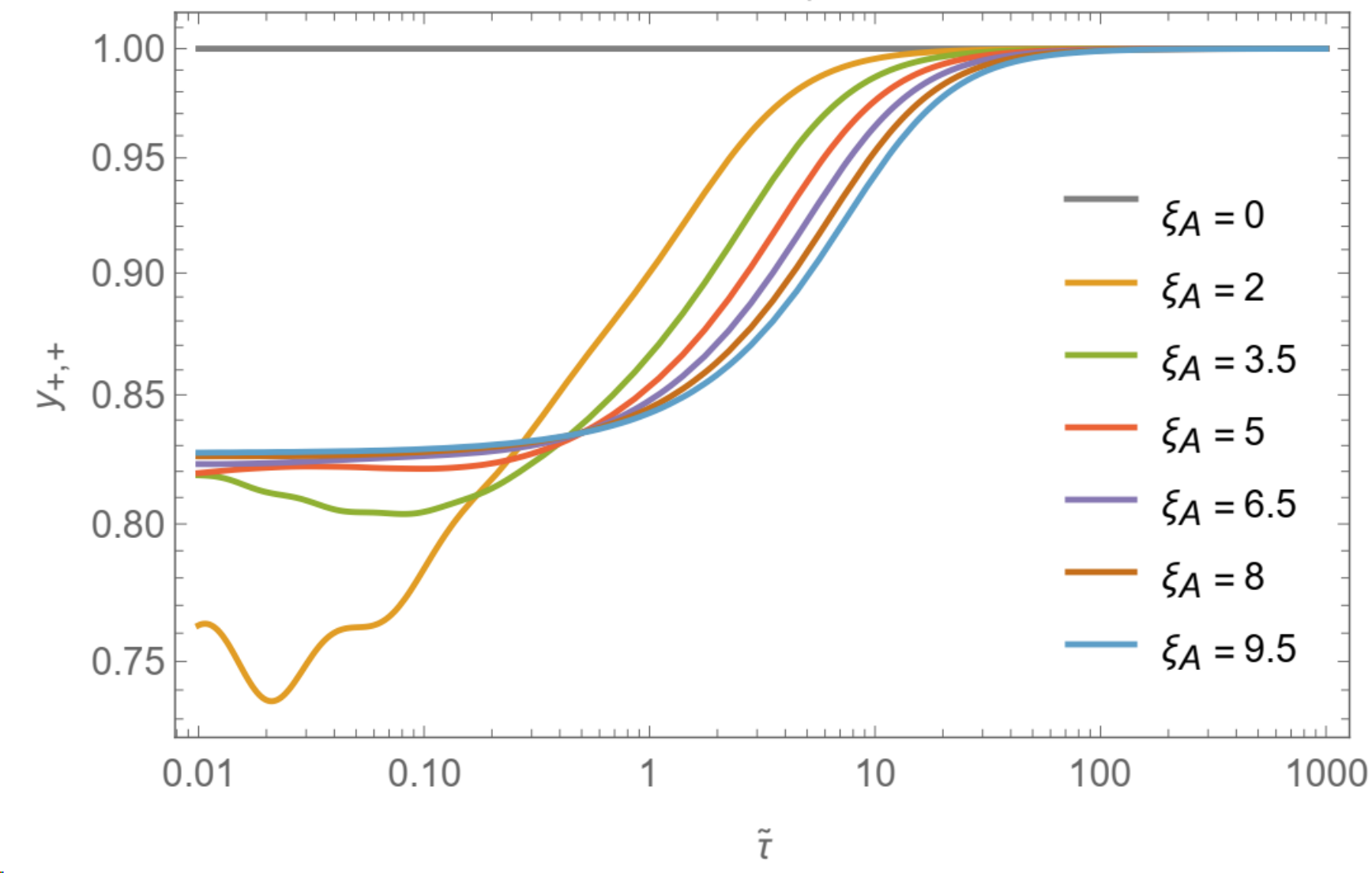} 
\includegraphics[scale=0.39]{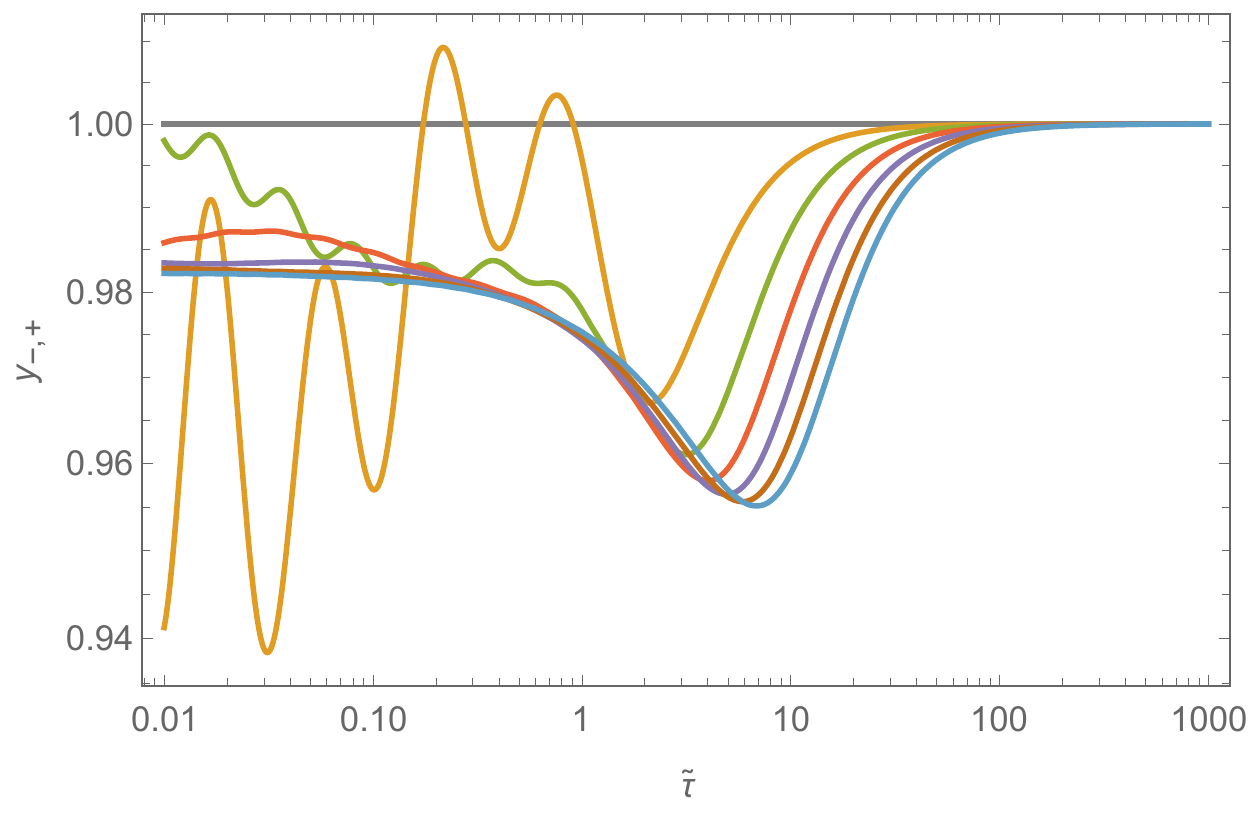}\\
\includegraphics[scale=0.395]{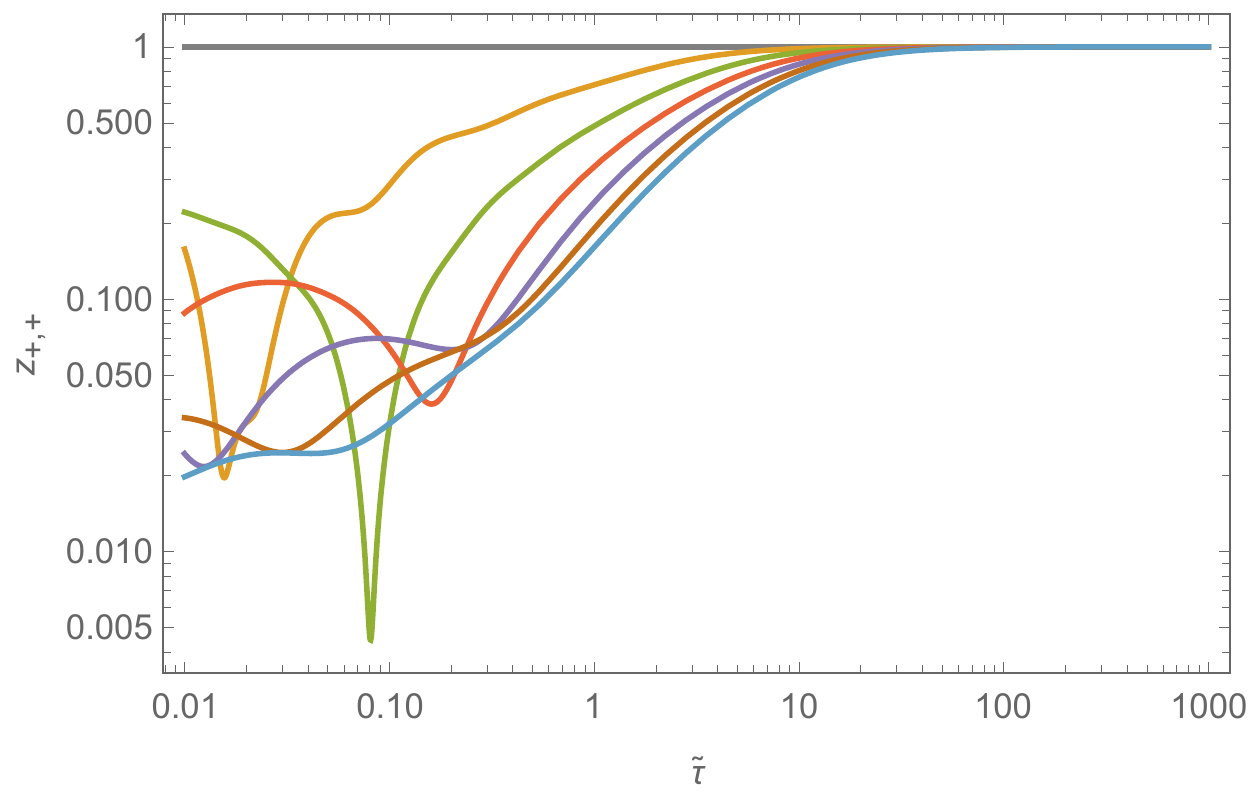} 
\includegraphics[scale=0.39]{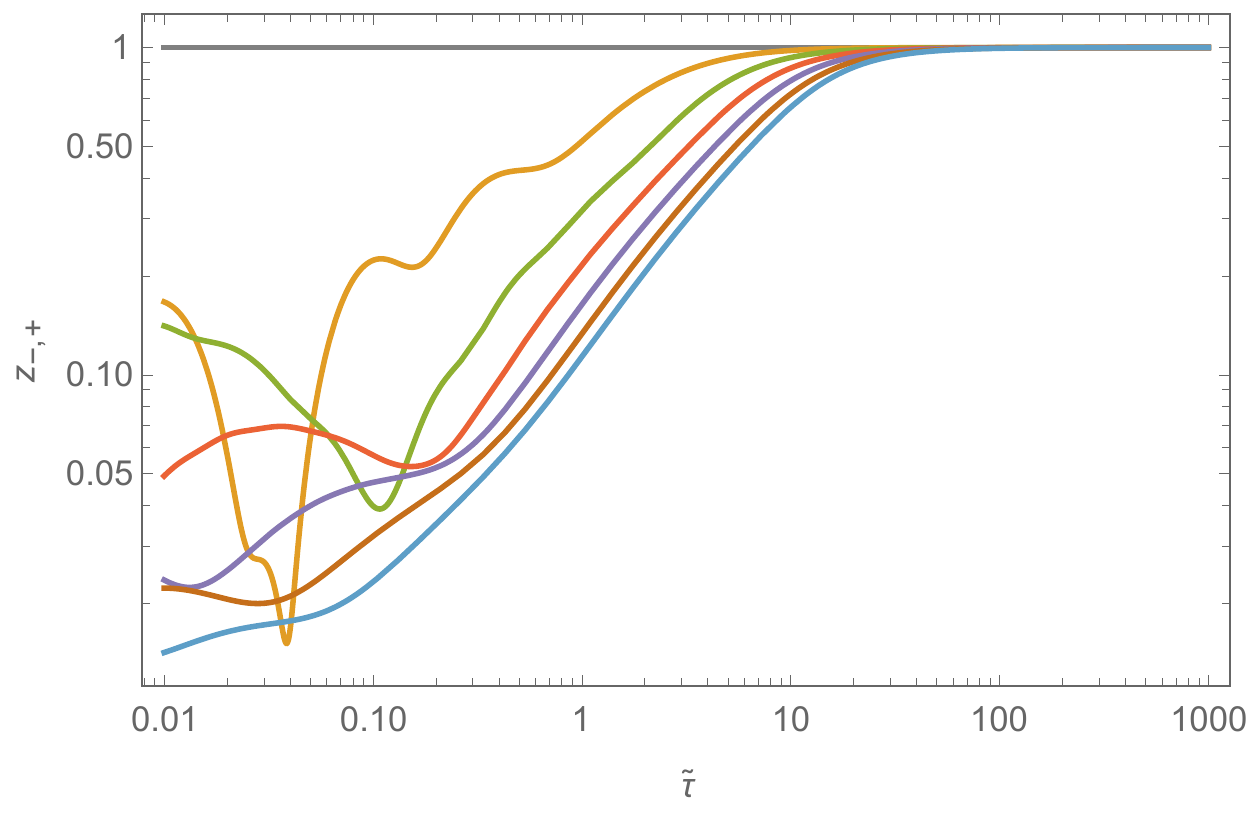}\\
\includegraphics[scale=0.275]{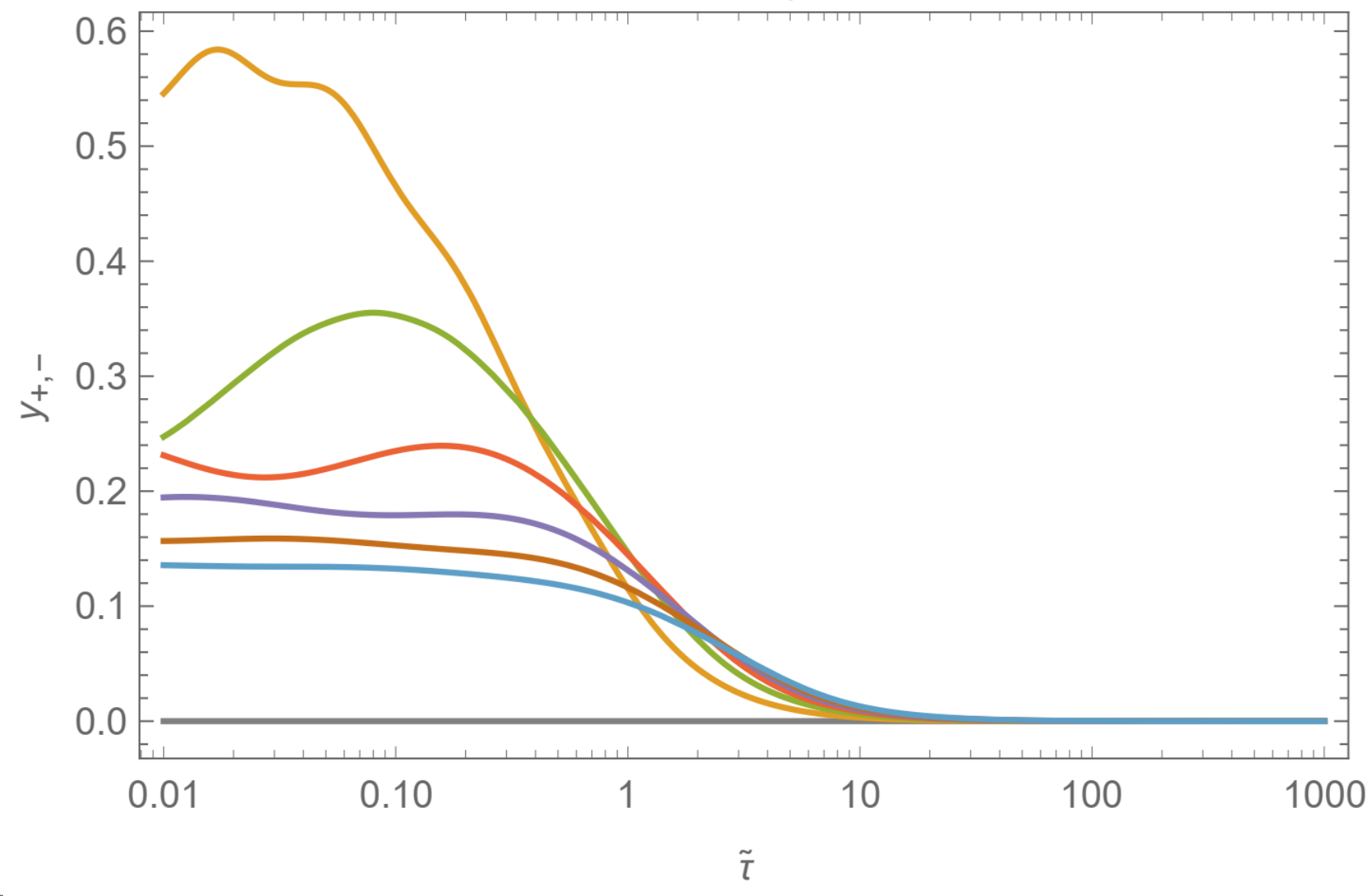} 
\includegraphics[scale=0.39]{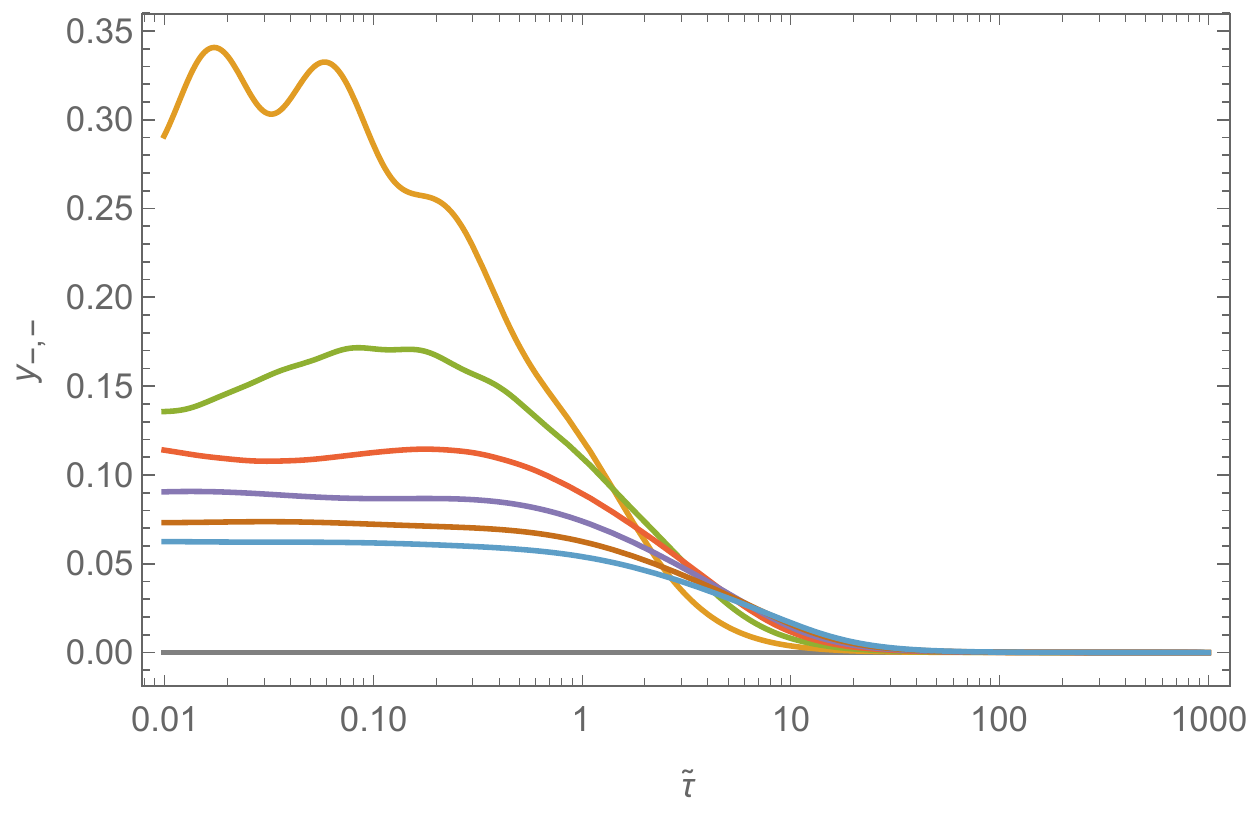}\\
\includegraphics[scale=0.39]{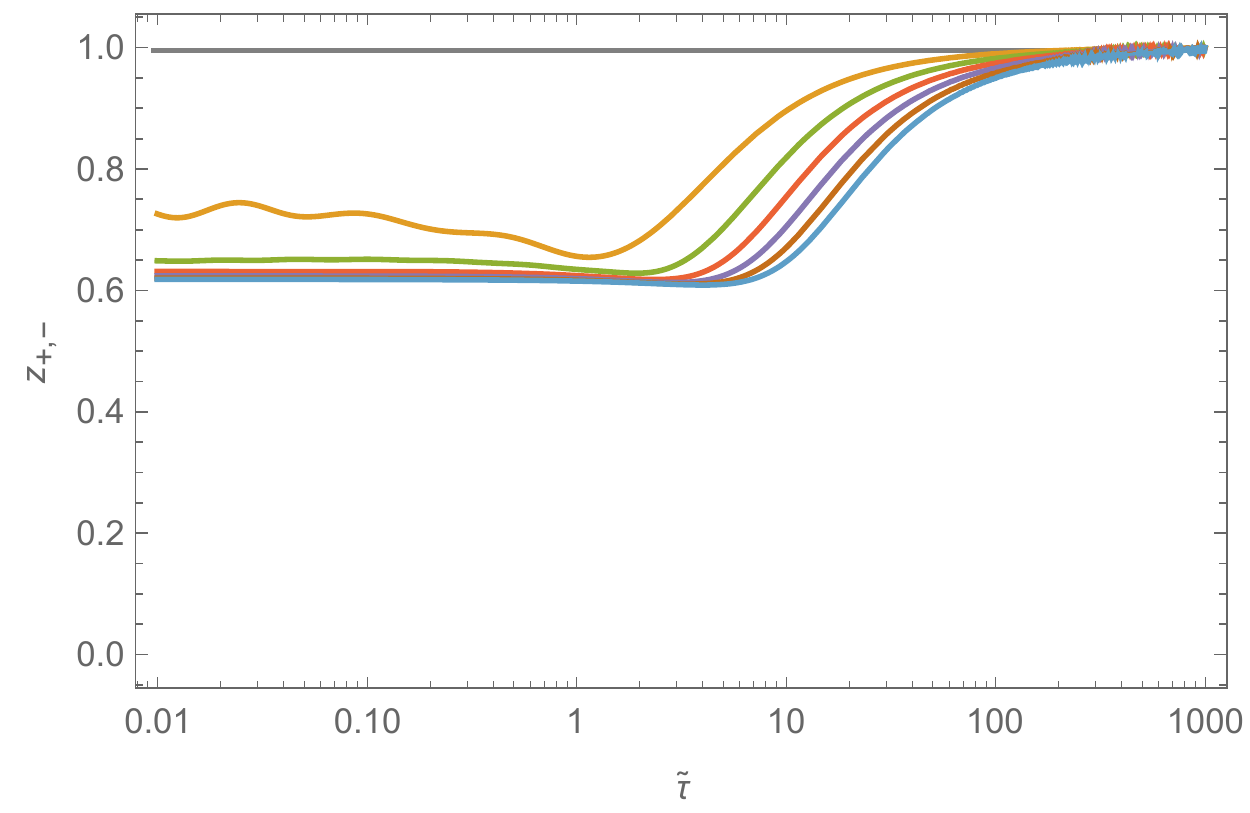} 
\includegraphics[scale=0.39]{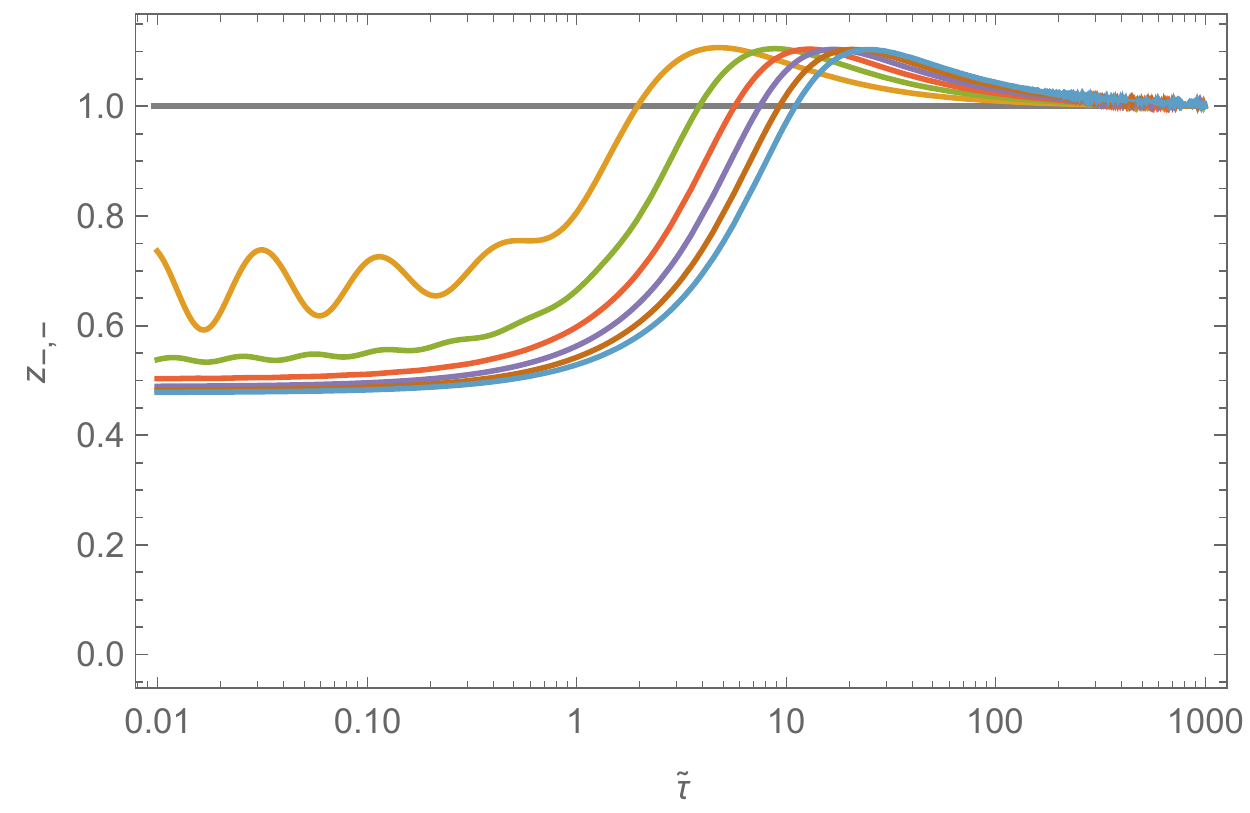}
\caption{$y_{s,s'}(\x)$ and $z_{s,s'}(\x)$ with respect to $\x$ for $\mu_m=1$, $\xpi=0$, and different values of $\xp$.}
\label{fig:ysp-zsp-1}   
\hspace{0.2cm}
\includegraphics[scale=0.275]{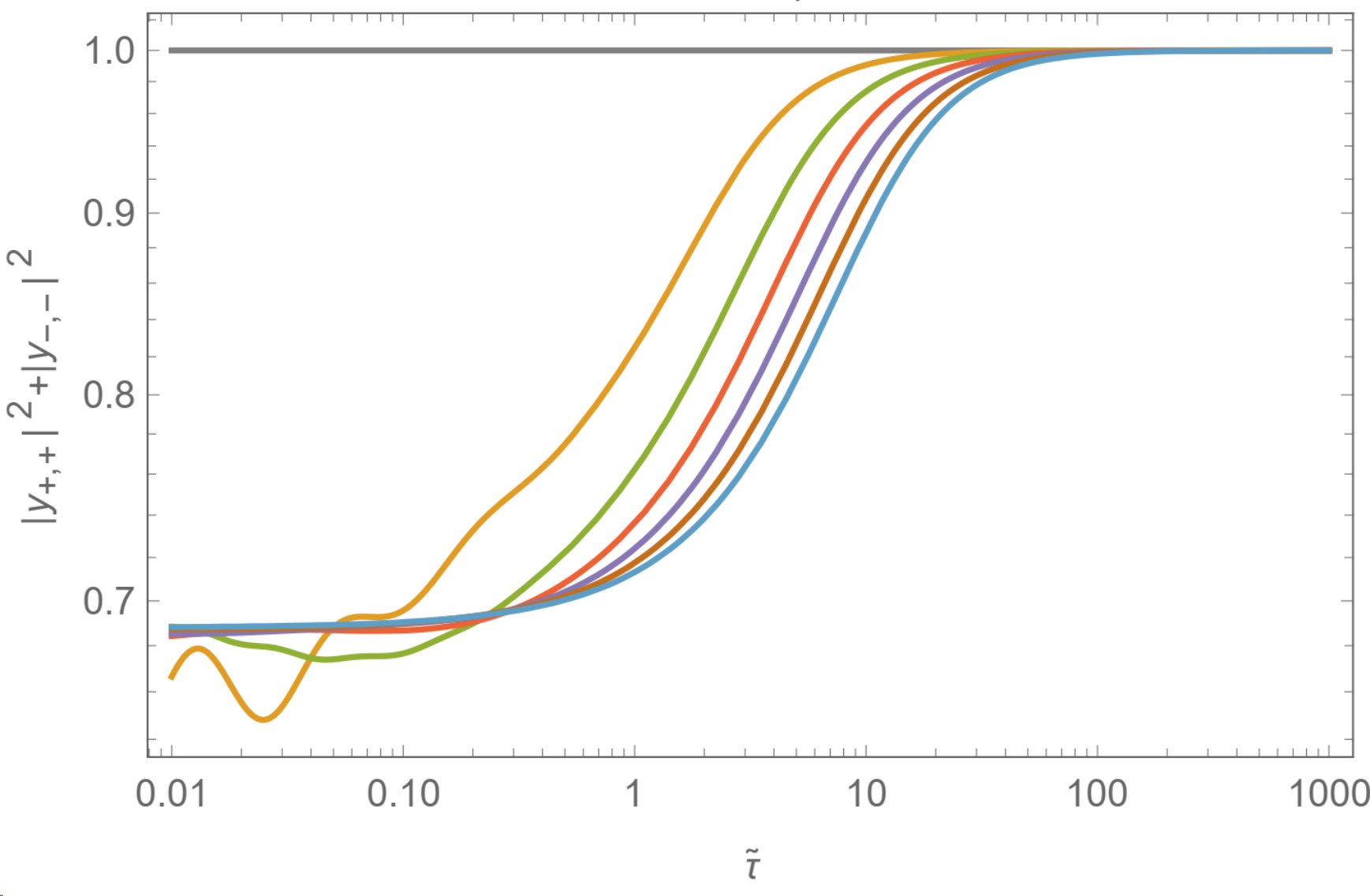}\hspace{0.44cm} 
\includegraphics[scale=0.275]{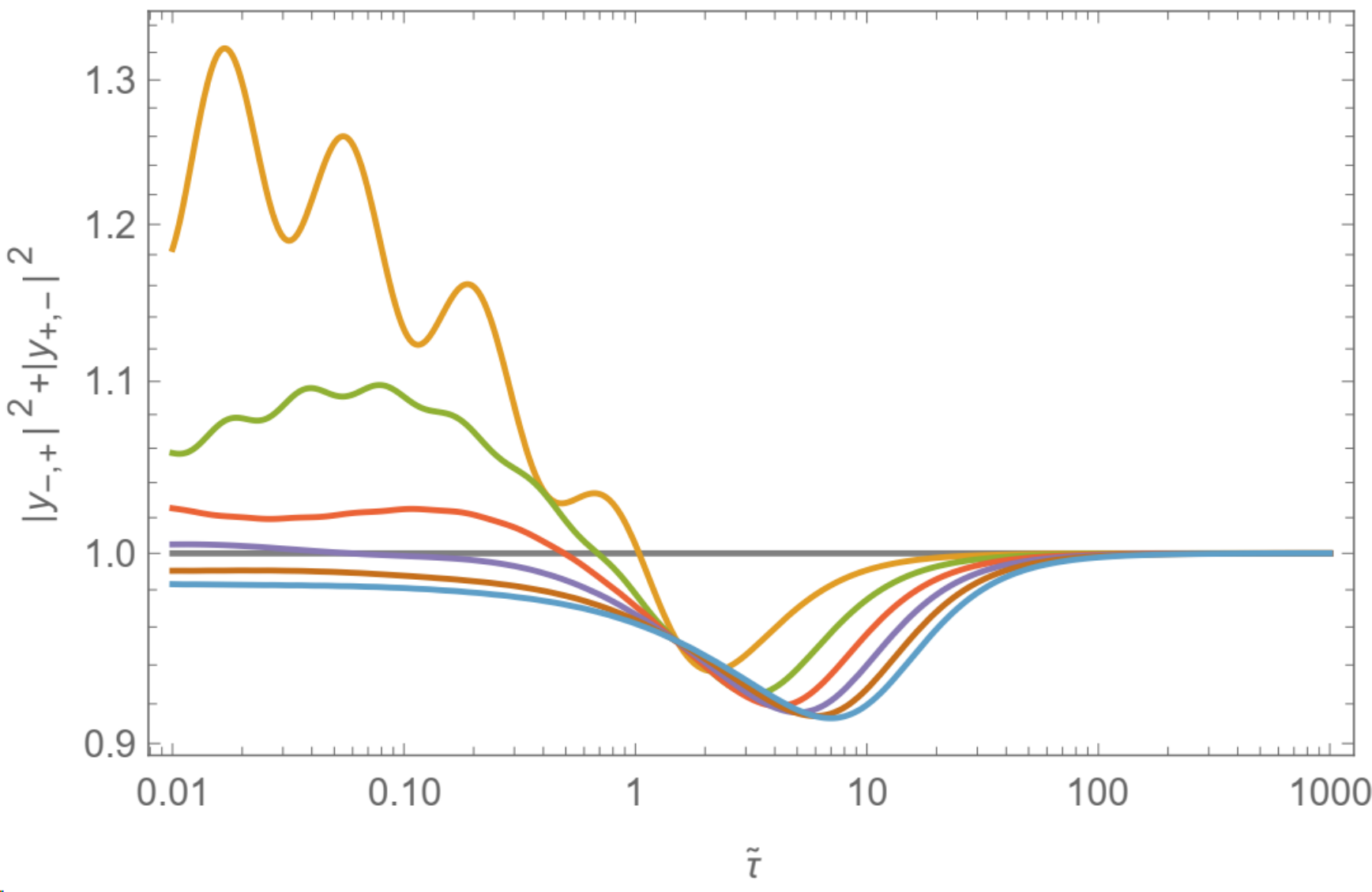}\\
\includegraphics[scale=0.295]{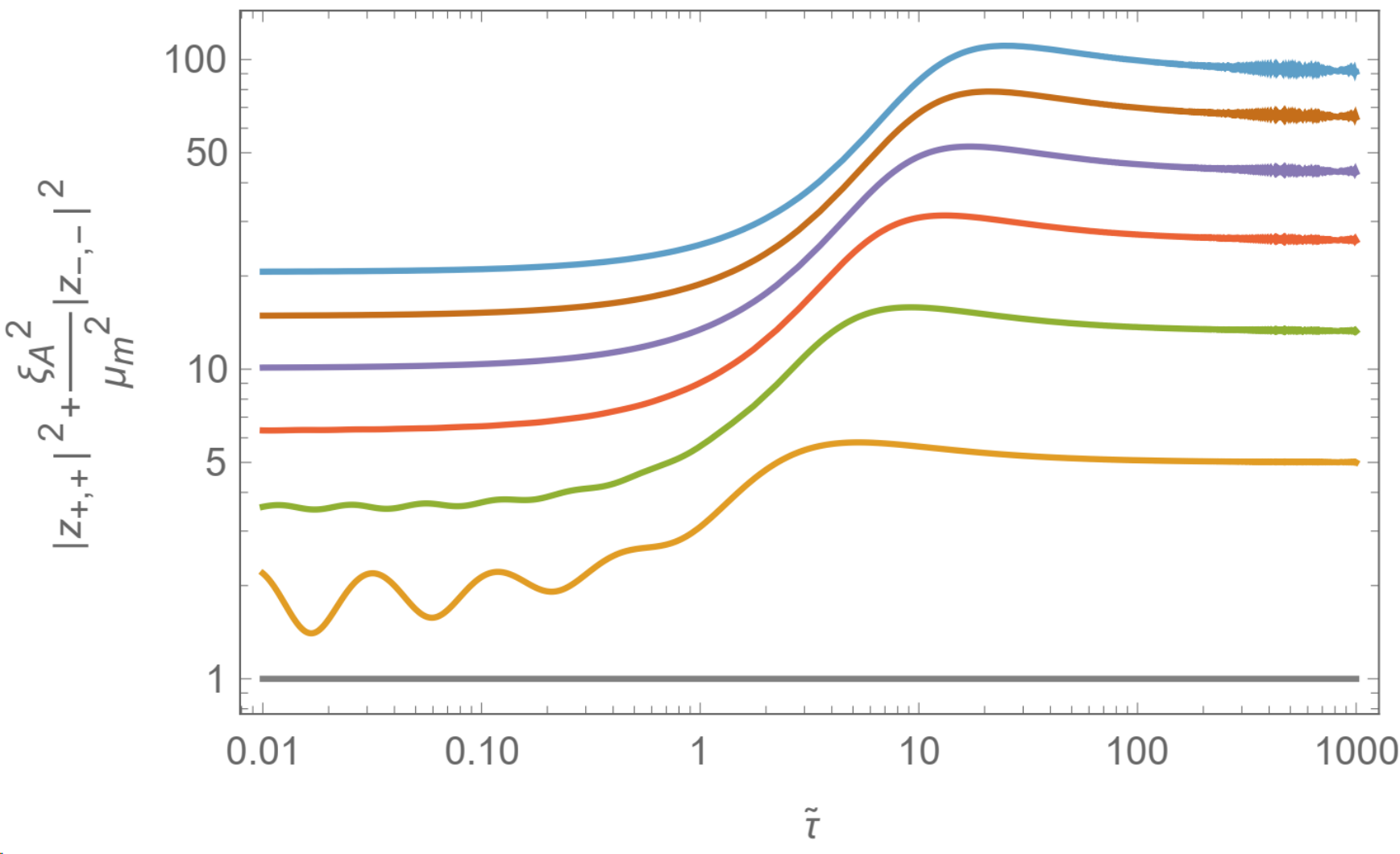} 
\includegraphics[scale=0.295]{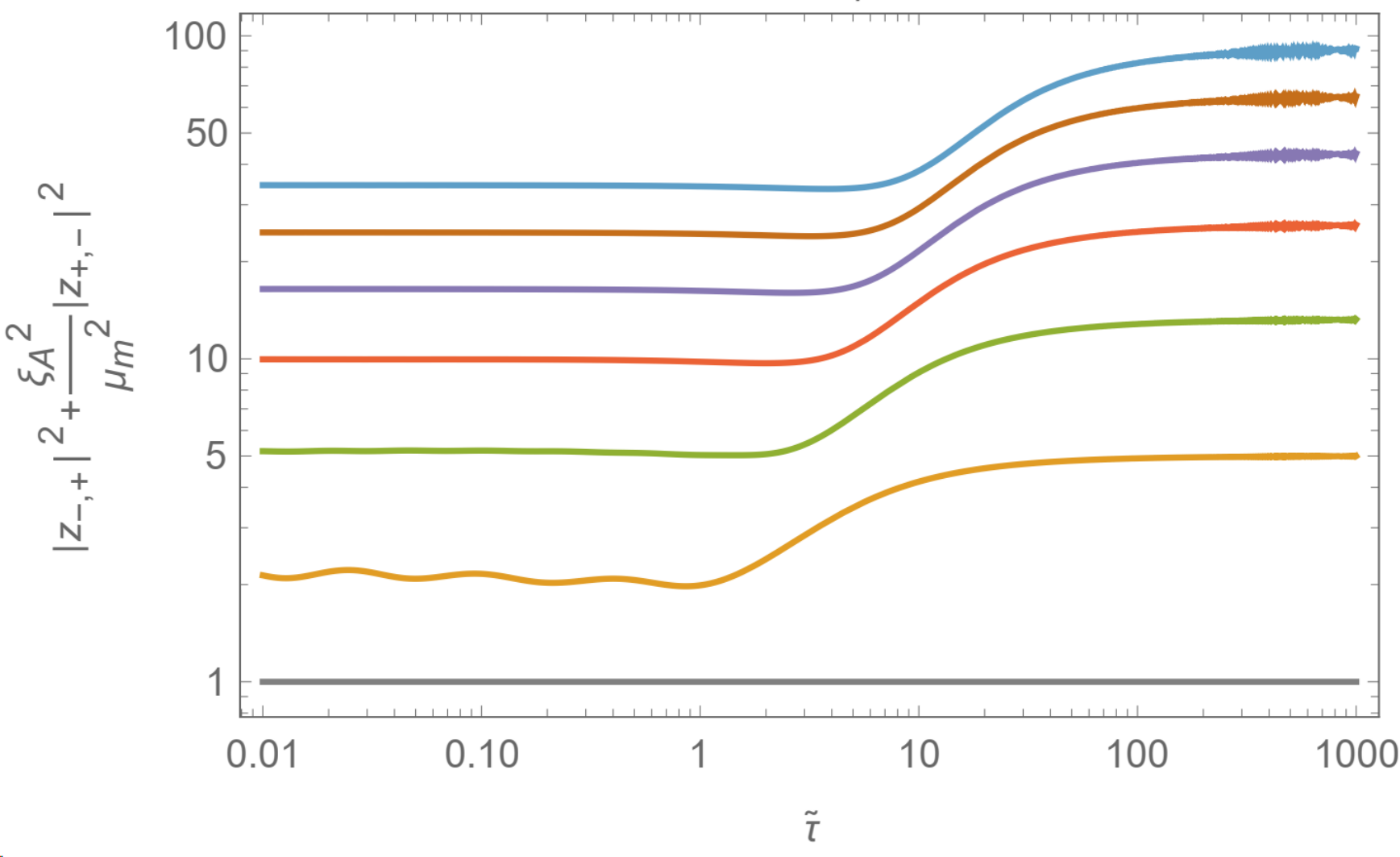}
\caption{$\lvert y_{s,+}\rvert^2+\lvert y_{-s,-}\rvert^2$, and $\lvert Z_{s,+}\rvert^2+\frac{\xp^2}{\mu^2_{\rm{m}}} \lvert Z_{-s,-}\rvert^2$ for the same parameters as Fig. \ref{fig:ysp-zsp-1}. }
\label{fig:ysp-ysp-1}  
\end{figure*}

\begin{itemize}
\item{\textbf{Neutral fermions:} In this limit the fermion is not charged under the $SU(2)$ gauge field, and is only derivatively coupled to the axion.  In Eq.s \eqref{ansatz-YZ} which formulate the deviation of the minus spinors from the plus one, we have $y_{s,+}(\x)=z_{s,+}=1$, and $y_{s,-}(\x)=\frac{\xa}{\mu_{{\rm{m}}}}z_{s,-}(\x)=0$. For the fermions which are coupled to the gauge field through $\xp$, this limit is the same as going to $\xp=0$. See the $\xa=0$ (gray) curves in Fig.s \ref{fig:ysp-zsp-44}-\ref{fig:ysp-ysp-1}. Therefore, the minus subspace is simply another copy of the plus subspace.}
\item{\textbf{Large mass limit:} Two different regimes with large mass limit are presented in Fig.s \ref{fig:ysp-zsp-44}-\ref{fig:ysp-ysp-4}. Figures \ref{fig:ysp-zsp-44} and \ref{fig:ysp-ysp-44} show systems with $\mu_{\rm{m}}=25$ and $\xpi=\xa$, while Fig.s \ref{fig:ysp-zsp-4} and \ref{fig:ysp-ysp-4} present systems with $\mu_{\rm{m}}=10$ and $\xpi=8$ for different values of $\xp$. The generic features of the system at this limit is as follows. 
The $y_{s,+}(\x)$, and $z_{s,+}(\x)$ ($y_{s,-}(\x)$) start from one (zero) at asymptotic past and after a fast transition phase each reduces (increases) to another constant number less than one after horizon crossing. The $z_{s,-}(\x)$ starts from one at asymptotic past and after horizon crossing approach a constant value of the order of $\frac{\mu_{{\rm{m}}}}{\xa}$. All of the curves roughly follows a $\tanh$-type behavior. Furthermore, in the very massive fermion limit, the combinations $\lvert y_{s,+}\rvert^2+\lvert y_{-s,-}\rvert^2$, and $\lvert Z_{s,+}\rvert^2+\frac{\xp^2}{\mu^2_{\rm{m}}} \lvert Z_{-s,-}\rvert^2$ ($s=\pm$) start from one at asymptotic past and remain very close to it throughout the mode's evolution.}
\item{\textbf{Large $\boldsymbol{\xp}$ limit:} A system with large values of $\xp$ and relatively small values of $\mu_{\rm{m}}$ and $\xpi$, i.e. $\mu_{\rm{m}}=0.1$ and $\xpi=0.05$ is presented in Fig.s \ref{fig:ysp-zsp-5} and \ref{fig:ysp-ysp-5}.}
\item{\textbf{SU(2) Schwinger process:} The system with no axion coupling, $\xpi=0$, is presented in Fig.s \ref{fig:ysp-zsp-1} and \ref{fig:ysp-ysp-1}. This limit is simply the $SU(2)$ Schwinger process during a (quasi) de-Sitter expansion.}
\end{itemize}

\section{Fermionic currents: point-splitting regularization}\label{F-current}
Working out the fermion fields in the previous section, we are now ready to compute the induced currents. The three currents induced by our fermions are presented in Eq.s \eqref{axial} and \eqref{fermion-number}. However, those expressions for currents are in their classical forms. Given the Graussman nature of the fermion fields, i.e. $\tilde{\Psi}_{\alpha}^{\dag}$ and $\tilde{\Psi}_{\beta}$ do not commute, the quantum currents are given by the anti-symmetrization of the fermionic fields. Therefore, the physical fermion (vector) current is given as \cite{parker2009quantum}
\bea
\langle J^{\mu} \rangle = \frac1a \delta^{\mu}_{\alpha} ~ \lvac  [\bar{\tilde{\Psi}}, \I\otimes\gamma^{\alpha}\tilde{\Psi} ] \rvac.
\eea
It is more convenient to replace the commutation on operators with the antinormal ordering on the creation and annihilation operators, $\vdots$, defined as
\bea\label{anti-normal-}
\vdots b^{~}_{\bf{k}} b^{\dag}_{\bf{k}} \vdots \equiv b^{~}_{\bf{k}} b^{\dag}_{\bf{k}} \an \vdots b^{\dag}_{\bf{k}} b^{~}_{\bf{k}} \vdots \equiv -b^{~}_{\bf{k}} b^{\dag}_{\bf{k}}.
\eea
Therefore, the physical fermion current can be written as
\bea\label{physical-fermion}
\langle J^{\mu} \rangle = \frac1a \delta^{\mu}_{\alpha} ~  \lvac \vdots  \bar{\tilde{\Psi}} \I\otimes \gamma^{\alpha}\tilde{\Psi} \vdots \rvac.
\eea
Similarly, we can find the physical isospin current and chiral current from their classical forms in Eq.s \eqref{fermion-number} and \eqref{axial}.

\subsection{Fermion number}\label{subsec:J-nab-J5}

Here we compute the net fermion number, $J^0$. 
From Eq. \eqref{physical-fermion}, we find the contribution of the plus fermions to the net fermion number as
\bea
\langle J^{0+} \rangle = \frac{1}{a^4} \sum_{s=\pm} \int {\rm{d}}^3k \bigg[ (v_s^{\uparrow*}v_s^{\uparrow}+v_s^{\downarrow*}v_s^{\downarrow})-(u_s^{\uparrow*}u_s^{\uparrow}+u_s^{\downarrow*}u_s^{\downarrow})\bigg]=0,\,
\eea
where in the last equality we used Eq. \eqref{V-U+}.
Similarly, the contribution of the minus fermions to the net fermion number is 
\bea
\langle J^{0-} \rangle = \frac{1}{a^4} \sum_{s,s'=\pm} \int {\rm{d}}^3k \bigg[ (v_{s,s'}^{\uparrow*}v_{s,s'}^{\uparrow}+v_{s,s'}^{\downarrow*}v_{s,s'}^{\downarrow})-(u_{s,s'}^{\uparrow*}u_{s,s'}^{\uparrow}+u_{s,s'}^{\downarrow*}u_{s,s'}^{\downarrow})\bigg]=0.\,
\eea
As a result, the net fermion number vanishes
\bea\label{J0+-}
\langle J^{0} \rangle = \langle J^{0+} \rangle + \langle J^{0-} \rangle = 0,
\eea
which is a consequence of the $C$-symmetry in the fermionic sector.

\subsection{Isospin current and chiral charge}

At this part, we compute the isospin current, $J^{\mu a}$, and the chiral current, $J^{\mu}_{5}$. The corresponding momentum integrals are UV divergent and need to be renormalized for which we will use the point-splitting regularization skim. 
The gauge invariant isospin and chiral currents with symmetric point separation are respectively given as
\bea
\langle J^{\mu a}(x^{\nu};\varepsilon^{\nu}) \rangle \equiv {\rm{symm}\lim_{\varepsilon\rightarrow 0}}  \ga \lvac \vdots \bar{\tilde{\Psi}}(x_f^{\nu}) \mathcal{W}(x,x_f) {\bf{T}}^a \otimes\gamma^{\mu} \mathcal{W}(x_b,x) \tilde{\Psi}(x_b^{\nu}) \vdots \rvac,
\eea
and 
\bea\label{point-split-J5}
\langle J^{\mu }_5(x^{\nu};\varepsilon^{\nu}) \rangle \equiv {\rm{symm}\lim_{\varepsilon\rightarrow 0}}   \lvac \vdots \bar{\tilde{\Psi}}(x_f^{\nu}) \mathcal{W}(x_b,x_f) \I \otimes (\gamma^{\mu} \gamma_5) \tilde{\Psi}(x_b^{\nu}) \vdots \rvac,
\eea
in which $\mathcal{W}(y,z)$ is the Wilson line defined as
\bea
\mathcal{W}(y,z)\equiv \mathcal{P}\bigg[\exp(-i\ga \int^{z}_{y} A_{\lambda}{\rm{d}}x^{\lambda})\bigg],
\eea
which is required to maintain the gauge invariance and trace is over the fundamental representation of the gauge group, $\mathcal{P}$ stands for path ordering, and $x^{\mu}_f$ and $x^{\mu}_b$ are the \textit{forward} and \textit{backward} point-splitting points, corresponding to the point-splitting \textit{physical} points \footnote{It is noteworthy to mention that in the expanding universe, the point-splitting, i.e. $X^{\mu} \rightarrow \{X^{\mu}_{f},X^{\mu}_{b}\}$ with $X^{\mu}_{f}=X^{\mu}_{b}+\varepsilon^{\mu}$, should be done to the physical coordinates and not the comoving ones. Otherwise, the UV divergent terms would be powers of $\frac{\tau}{\varepsilon}$, not $\frac{1}{\varepsilon}$. Such a mistake leads to an unphysical term of the form $\ln(a)$ in the currents which fakes an IR divergence.}
\bea
X^{\mu}_f \equiv (t + \frac{\varepsilon}{2}, {\bf{X}}^i + \frac{{\bep}^i}{2}) \an X^{\mu}_b \equiv ( t - \frac{\varepsilon}{2}, {\bf{X}}^i - \frac{{\bep}^i}{2}),
\eea
respectively and $\varepsilon^{\mu}$ is a constant 4-vector which we will eventually take to zero. Using the slow-roll inflation relations, $Ha(\tau)\simeq-\frac{1}{\tau}$ and $a(\tau)\simeq e^{Ht}$, we can read off the forward and backward point-splitting conformal times as
\bea\label{eq-tau-a-r}
\tau_f = (1-\frac{\varepsilon}{2}) \tau \an \tau_b = (1+\frac{\varepsilon}{2}) \tau.
\eea
and the spatial comoving coordinates as 
\bea
\bx_f =  \bx + \frac{\bep}{2a} \an \bx_b = \bx- \frac{\bep}{2a}.
\eea

In terms of the comoving fields and using the plus and minus decomposition, the chiral charge in Eq. \eqref{Q5} can be written as
\bea
Q_5 = \langle a J^0_5 \rangle =
Q_{5}^+ + Q_{5}^-,
\eea
where $Q_{5}^+$ and $Q_{5}^-$ are the contribution of the plus and minus fermions to $Q_{5}$ respectively. Similarly, the fermionic backreaction to the gauge field's equation in Eq. \eqref{eq:CurlyJ} can be written as
\bea
\mathcal{J} \equiv \frac{a}{3} \delta^i_{a} \langle J^{ai} \rangle = \mathcal{J}^+ + \mathcal{J}^-,
\eea
where $\mathcal{J}^{\pm}$ are the contributions of the plus and minus spinors to $\mathcal{J}$.

Taking the $\varepsilon^{\mu}$ goes to zero limit and using Eq. \eqref{III-Gamma0}, $Q_{5}^{\pm}(x^{\nu};\varepsilon)$ can be written as
\bea\label{Qpm-p}
Q_{5}^{\pm}(x^{\nu};\varepsilon) = - \frac{H^3\xp^3}{(2\pi)^2} + 
\frac{a}{(a_f a_b)^{2}}  \lvac \vdots  \bar{\Uppsi}^{\pm}(x^{\nu}_f) \gamma^{0}\gamma^5 \Uppsi^{\pm}(x^{\nu}_b) \vdots \rvac,
\eea
where $a_f$ and $a_b$ are the forward and backward scale factors, i.e.
\bea
a_f(\tau) \equiv a(\tau_{f}) \an a_b(\tau) \equiv a(\tau_{b}).
\eea
Moreover, using Eq. \eqref{II-Gammaalpha}, $\mathcal{J}^{\pm}(x^{\nu};\varepsilon)$ is  \footnote{More precisely, the explicate form of $\mathcal{J}^-(x^{\nu};\varepsilon)$ is 
\bea
\mathcal{J}^-(x^{\nu};\varepsilon) = 
- {\rm{symm}\lim_{\varepsilon\rightarrow 0}}  ~  \frac{1}{2}\frac{a\ga}{(a_f a_b)^{2}}  \lvac \vdots  \bar{\Uppsi}^-(x^{\nu}_f) \bigg( \gamma^{0}\gamma^{5} - 2\gamma^1\bigg) \Uppsi^-(x^{\nu}_b ) \vdots \rvac.
\eea
However, from the combination of the charge conjugation Eq. \eqref{V-U}, as well as antinormal ordering in Eq. \eqref{anti-normal}, we find that ${\rm{symm}\lim_{\varepsilon\rightarrow 0}}  \lvac \vdots  \bar{\Uppsi}^-(x^{\nu}_f) \gamma^1 \Uppsi^-(x^{\nu}_b) \vdots \rvac = 0$. Therefore, up to a total sign difference, the expectation value of $\mathcal{J}^-(x^{\nu};\varepsilon)$ takes the form of $\mathcal{J}^+(x^{\nu};\varepsilon)$.}
\bea\label{Jpmep}
\mathcal{J}^{\pm}(x^{\nu};\varepsilon) = 
{\rm{symm}\lim_{\varepsilon\rightarrow 0}}  ~ \frac{\pm a\ga}{2(a_f a_b)^{2}}  \lvac \vdots  \bar{\Uppsi}^{\pm}(x^{\nu}_f) \gamma^{0}\gamma^{5}\Uppsi^{\pm}(x^{\nu}_b) \vdots \rvac.
\eea
It is noteworthy to mention that the first term in Eq. \eqref{Qpm-p} is the contribution of $\bep^i$ terms in the presence of the gauge field and the Wilson line.
This term is the zeroth element of the Chern-Simons current and the result of the Abelian Adler-Bell-Jackiw anomaly \cite{Bell:1969ts, Adler:1969gk} (See Eq. \ref{Abelian-CA}). In the momentum integrals on the right hand sides of Eq. \eqref{Qpm-p} and \eqref{Jpmep}, all the divergent terms are given in terms of the temporal separation. That is due to the spatial isotropy of our setup. Therefore, from now on, we set ${\bep}^i$ to zero and only keep the separation in time parameter, i.e. $\varepsilon$.

Interestingly, both the isospin backreactions, $\mathcal{J}^{\pm}$, and chiral charges, $Q_{5}^{\pm}$, are specified in term of the following dimensionless quantity
\bea\label{Kp--}
\mathcal{K}^{\pm}(x^{\mu}; \varepsilon) \equiv {\rm{symm}\lim_{\varepsilon\rightarrow 0}} ~ \frac{a}{2(a_f a_b)^{2}}   \frac{(2\pi)^2}{H^3} \lvac \vdots  \Uppsi^{\pm\dag}(x^{\nu}_f) \gamma^{5}  \Uppsi^{\pm}(x^{\nu}_b) \vdots \rvac.
\eea
More precisely, the currents can be written as
\bea\label{J-Q-K}
\mathcal{J}^{\pm}(x^{\mu}; \varepsilon)  = \pm \frac{\ga H^3}{(2\pi)^2} \mathcal{K}^{\pm}(x^{\mu}; \varepsilon)  \an Q_{5}^{\pm}(x^{\mu}; \varepsilon) =  \frac{2H^3}{(2\pi)^2} \big( - \frac12\xp^3 + \mathcal{K}^{\pm}(x^{\mu}; \varepsilon) \bigg).
\eea
Besides, the fermionic backreaction to the axion background in Eq. \eqref{axion-BR} can be written as
\bea
\mathcal{B}(x^{\mu}; \varepsilon) = \frac{\beta\lambda}{2f} \bigg(\p_t Q_5 +3 H Q_5\bigg).
\eea
In the following, we compute $\mathcal{K}^{+}(x^{\mu}; \varepsilon)$ and $\mathcal{K}^{-}(x^{\mu}; \varepsilon)$ which are the contributions of the plus and minus fermions respectively.

\subsection*{$\Uppsi^{+}$ fermions:}

In the plus subspace and after using Eq.s \eqref{V-U+} and \eqref{Kp--}, $\mathcal{K}^{+}$ can be written  as
\bea\label{K+tau}
\mathcal{K}^{+}(x^{\mu};\varepsilon) = - {\rm{symm}\lim_{\varepsilon\rightarrow 0}} ~  \sum_{s=\pm} s~  \frac{a(2\pi)^2/H^3}{2(a_f a_b)^{2}} \int {\rm{d}}^3k\bigg[ u^{\uparrow*}_{s}(k,\tau_f)u^{\downarrow}_{s}(k,\tau_b) + u^{\downarrow*}_{s}(k,\tau_f)u^{\uparrow}_{s}(k,\tau_b)\bigg].\nonumber\\
\eea
This momentum integral is UV divergent and we use the point-splitting skim to regularize it. Here, we present the final results and the details of the computations are presented in App. \ref{sec:current-comp}. We find that $\mathcal{K}^{+}(\tau,\varepsilon)$ has a log divergent term and can be written as 
\bea
\mathcal{K}^{+}(\tau;\varepsilon) = 4 \kappa_{_{+,I}} \mu^2_{{\rm{m}}}  \ln\big(\frac{\varepsilon}{2}\big) + \mathcal{K}^{+}_{\rm{reg}}(\tau),
\eea
where $\mathcal{K}^{+}_{\rm{reg}}(\tau;\varepsilon)$ is the regularized current, given as
\bea\label{fin-J+}
&& \mathcal{K}^{+}_{\rm{reg}} =  \nonumber\\
&&  \kappa_{I}^+ \bigg\{ \frac{2}{3} (1-2(\kappa_{I}^{+})^2) \bigg(1 - \frac{\lvert \mu^+\rvert}{\kappa_{I}^+} \frac{\sinh(2\kappa_{I}^+\pi)}{\sinh(2\lvert\mu^+\rvert\pi)}\bigg) + \mu^2_{{\rm{m}}} \bigg(
-4  \psi^{(0)}(1) + 2  - \frac{8\lvert \mu^+\rvert}{3\kappa_{I}^+ }\frac{\sinh(2\kappa_{I}^+\pi)}{\sinh(2\lvert\mu^+\rvert\pi)}\bigg) + \nonumber\\
&& \mu^2_{{\rm{m}}}  \sum_{s=\pm}  {\rm{Re}}\bigg[  
   \frac{e^{2\lvert\mu^+\rvert\pi}-e^{-2s\kappa_{I}^+\pi}}{\sinh(2\lvert\mu^+\rvert\pi)} \psi^{(0)}(-is\kappa_{I}^+ -i\lvert\mu^+\rvert)  - \frac{e^{-2\lvert\mu^+\rvert\pi}-e^{-2s\kappa_{I}^+\pi}}{\sinh(2\lvert\mu^+\rvert\pi)} \psi^{(0)}(i\lvert\mu^+\rvert-is\kappa_{I}^+)
 \bigg] \bigg\},\nonumber\\
\eea
in which $\psi^{(0)}(z)\equiv \frac{d\Gamma(z)}{dz}$ is the digamma function and $\kappa_{I}^+$ is the imaginary part of $\kappa^{+}$, i.e. $\kappa_{I}^+=2\xpi-\frac{1}{2}\xp$, and $\vert \mu^{+}\rvert = [\mu^2_{{\rm{m}}} + (\kappa_I^+)^2]^{\frac12}$. Notice that $\mathcal{K}^{+}_{\rm{reg}} $ is proportional to $\kappa_{I}^+$ and the quantity inside the curly brackets is even under parity.

\subsection*{$\Uppsi^{-}$ fermions:}

For the minus subspace and after using Eq. \eqref{V-U-}, we obtain $\mathcal{K}^{-}(\tau,\varepsilon)$ as
\bea
\mathcal{K}^{-}(\tau;\varepsilon) = -   {\rm{symm}\lim_{\varepsilon\rightarrow 0}} ~  \frac{a(2\pi)^2/H^3}{2(a_f a_b)^{2}}  \sum_{s,s'=\pm}  s s' \int {\rm{d}}^3k \bigg[ u^{\uparrow*}_{s,s'}(k,\tau_f)u^{\downarrow}_{s,s'}(k,\tau_b) + u^{\downarrow*}_{s,s'}(k,\tau_f)u^{\uparrow}_{s,s'}(k,\tau_b)\bigg],\nonumber\\
\eea
which by decomposition defined by Eq. \eqref{ansatz-YZ}, it can be simplified to
\bea
\mathcal{K}^{-} = -   {\rm{symm}\lim_{\varepsilon\rightarrow 0}} ~  \frac{a(2\pi)^2/H^2}{2(a_f a_b)^{\frac32}}  \sum_{s=\pm}  s \int \frac{{\rm{d}}^3k}{k} \bigg[ \big( \lvert y_{s,+}\rvert^2 + \lvert y_{-s,-}\rvert^2\big) \lvert Y_{s}\rvert^2 - \big( \lvert z_{s,+}\rvert^2 + \frac{\xp^2}{\mu^2_{\rm{m}}} \lvert z_{-s,-}\rvert^2\big) \lvert Z_{s}\rvert^2 \bigg],\nonumber\\
\eea
Since we only have $y_{s,s'}(\x)$ and $z_{s,s'}(\x)$ functions numerically, we can not work out the current analytically. However, as it is expected from 
Eq.s \eqref{eq:YZ1storderm-2} and is showed in Sec. \ref{sec:numerics}, in the limit $\xp/\mu_{\rm{m}}\ll1$, the combinations $\big( \lvert y_{s,+}(\x)\rvert^2 + \lvert y_{-s,-}(\x)\rvert^2\big)$, and $\big( \lvert z_{s,+}(\x)\rvert^2 + \frac{\xp^2}{\mu^2_{\rm{m}}} \lvert z_{-s,-}(\x)\rvert^2\big)$ are always close to one. As a result, we can approximate $ \mathcal{K}^{-}(\tau;\varepsilon)$ in this limit as
\bea
\mathcal{K}^{-}(\tau;\varepsilon) \simeq -   {\rm{symm}\lim_{\varepsilon\rightarrow 0}} ~  \frac{a(2\pi)^2/H^2}{2(a_f a_b)^{\frac32}}  \sum_{s=\pm}  s \int \frac{{\rm{d}}^3k}{k} \big(  \lvert Y_{s}\rvert^2 - \lvert Z_{s}\rvert^2 \big).
\eea
Following the corresponding analysis in the plus subspace, we find the regularized current in the minus subspace, $\mathcal{K}^{-}_{\rm{reg}}$, as
\bea\label{fin-J-}
&& \mathcal{K}^{-}_{\rm{reg}} \simeq  \nonumber\\
&& \kappa_{I}^- \bigg\{ \frac{2}{3} (1-2(\kappa_{I}^{-})^2) \bigg(1 - \frac{\lvert \mu^-\rvert}{\kappa_{I}^-} \frac{\sinh(2\kappa_{I}^-\pi)}{\sinh(2\lvert\mu^-\rvert\pi)}\bigg) + \mu^2_{{\rm{m}}} \bigg(
-4  \psi^{(0)}(1) + 2  - \frac{8\lvert \mu^-\rvert}{3\kappa_{I}^- }\frac{\sinh(2\kappa_{I}^-\pi)}{\sinh(2\lvert\mu^-\rvert\pi)} \bigg) + \nonumber\\
&&  \mu^2_{{\rm{m}}} \sum_{s=\pm}  {\rm{Re}}\bigg[  
   \frac{e^{2\lvert\mu^-\rvert\pi}-e^{-2s\kappa_{I}^-\pi}}{\sinh(2\lvert\mu^-\rvert\pi)} \psi^{(0)}(-is\kappa_{I}^- -i\lvert\mu^-\rvert)  - \frac{e^{-2\lvert\mu^-\rvert\pi}-e^{-2s\kappa_{I}^-\pi}}{\sinh(2\lvert\mu^-\rvert\pi)} \psi^{(0)}(i\lvert\mu^-\rvert-is\kappa_{I}^-)
 \bigg]  \bigg\},\nonumber\\
\eea 
where $\kappa_{I}^-$ is the imaginary part of $\kappa^{-}$, i.e. $\kappa_{I}^-=2\xpi+\frac{1}{2}\xp$, and $\vert \mu^{-}\rvert = [\mu^2_{{\rm{m}}} + (\kappa_I^-)^2 + \xp^2]^{\frac12}$. As we mentioned earlier, is only valid in the limit $\xp/\mu_{\rm{m}}\ll1$.

\vskip 5mm

To summarize, we worked out the induced currents analytically by using the point-splitting regularization skim. In the minus subspace, our analytical solution for $\mathcal{J}^-$ and $Q_5^-$ (given in terms of $\mathcal{K}^-$) are approximations which are valid in $\xp/\mu_{\rm{m}}\ll1$ limit. To find $\mathcal{K}^{-}_{\rm{reg}}(\tau)$ in other parts of the parameter space, one needs a full numerical analysis, e.g. \cite{Mirzagholi:2019jeb}. The two approaches are only in agreement when the source of the particle production, $\mathcal{\kappa}_I^{\pm}$ is large. The reason is in the use of a hard cut-off in the numerical computation of the momentum integrals carried out in \cite{Mirzagholi:2019jeb}. As a result, the numerical approach subtracts the divergences correctly but misses the finite terms which appear by the regularization process.

\section{Dirac fermions in (quasi) de Sitter}\label{fermi-de}
In the previous section, we worked out the analytic form of the fermionic currents sourced by an $SU(2)$ gauge field and possibly derivatively coupled to the axion field in de Sitter. In this section, we take a closer look at the results and make a detailed comparison with other cases of fermions generated in different setups in de Sitter, i.e., neutral fermions coupled to axion, and Abelian fermions. We also compare the non-Abelian fermion Schwinger effect with its scalar counterpart.

\subsection{Charged $SU(2)$ fermions}\label{gauge-field-fermion}

\begin{figure*}[!h]
\centering
\includegraphics[scale=0.56]{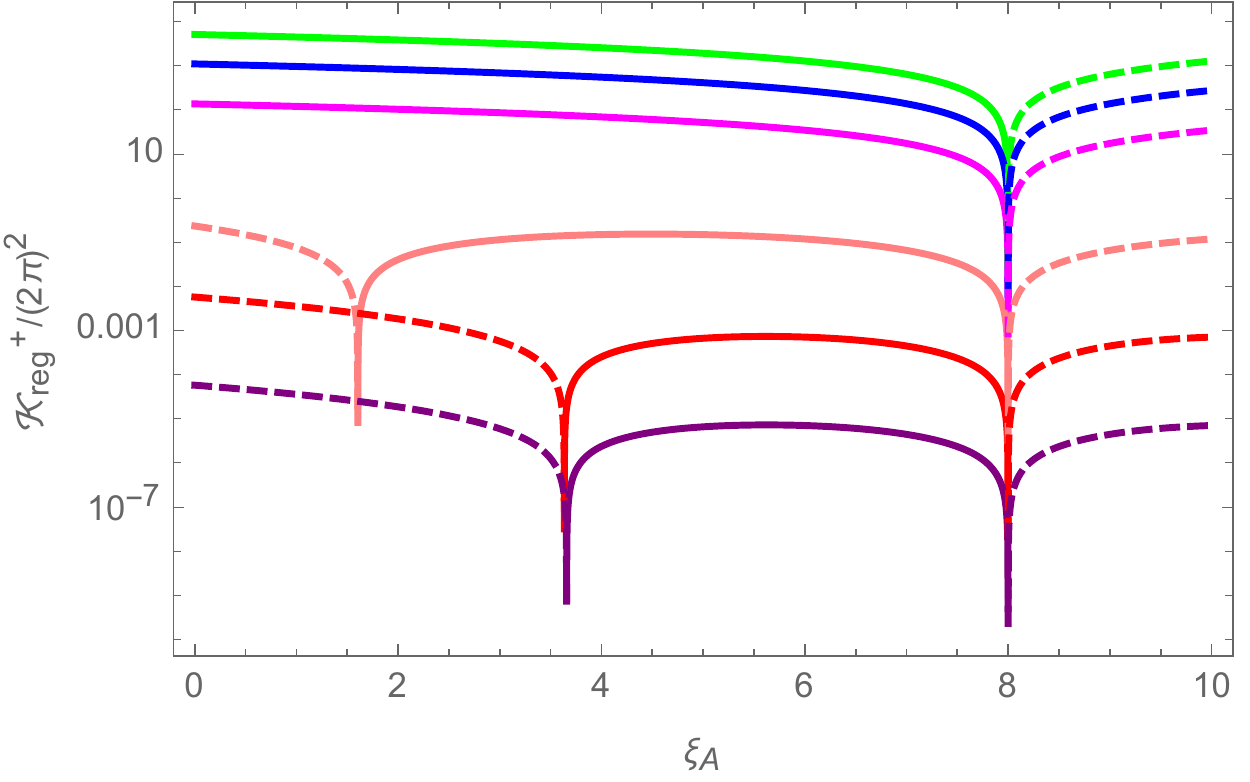} 
\includegraphics[scale=0.56]{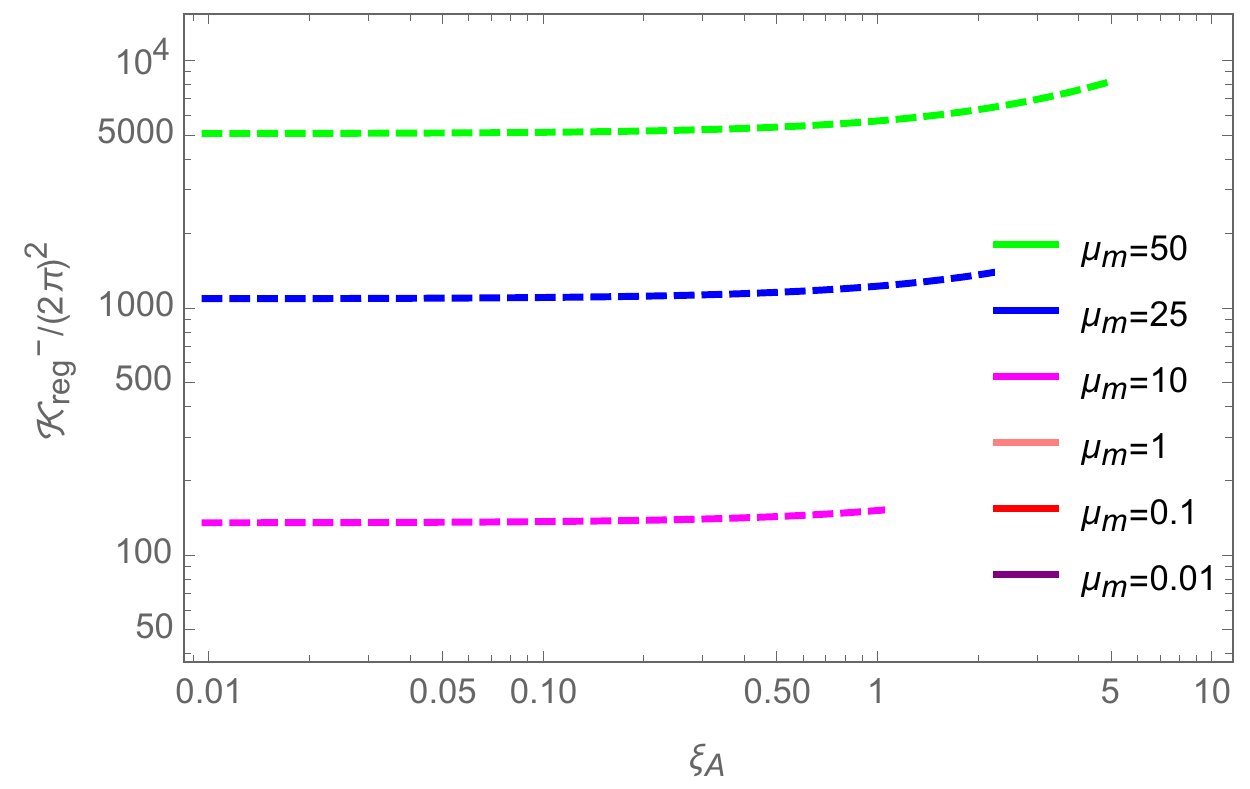}\\
\caption{The absolute values of $\mathcal{K}^{+}_{\rm{reg}}/(2\pi)^2$ (left panel) and $\mathcal{K}^{-}_{\rm{reg}}/(2\pi)^2$ (right panel) for Dirac fermions coupled to the $SU(2)$ gauge field with $\xpi=2$ in terms of $\xp$ and for different values of $\mu_{\rm{m}}$. The solid (dashed) lines denote positive (negative) values of current. The $\mathcal{K}^-_{\rm{reg}}$ lines are only plotted at the regime of the validity of our analytical approximation for this subspace, i.e. $\xp/\mu_{\rm{m}}<0.1$.}
\label{fig:K+--SU-axion}  
\end{figure*}

\begin{figure*}[!h]
\centering
\includegraphics[scale=0.56]{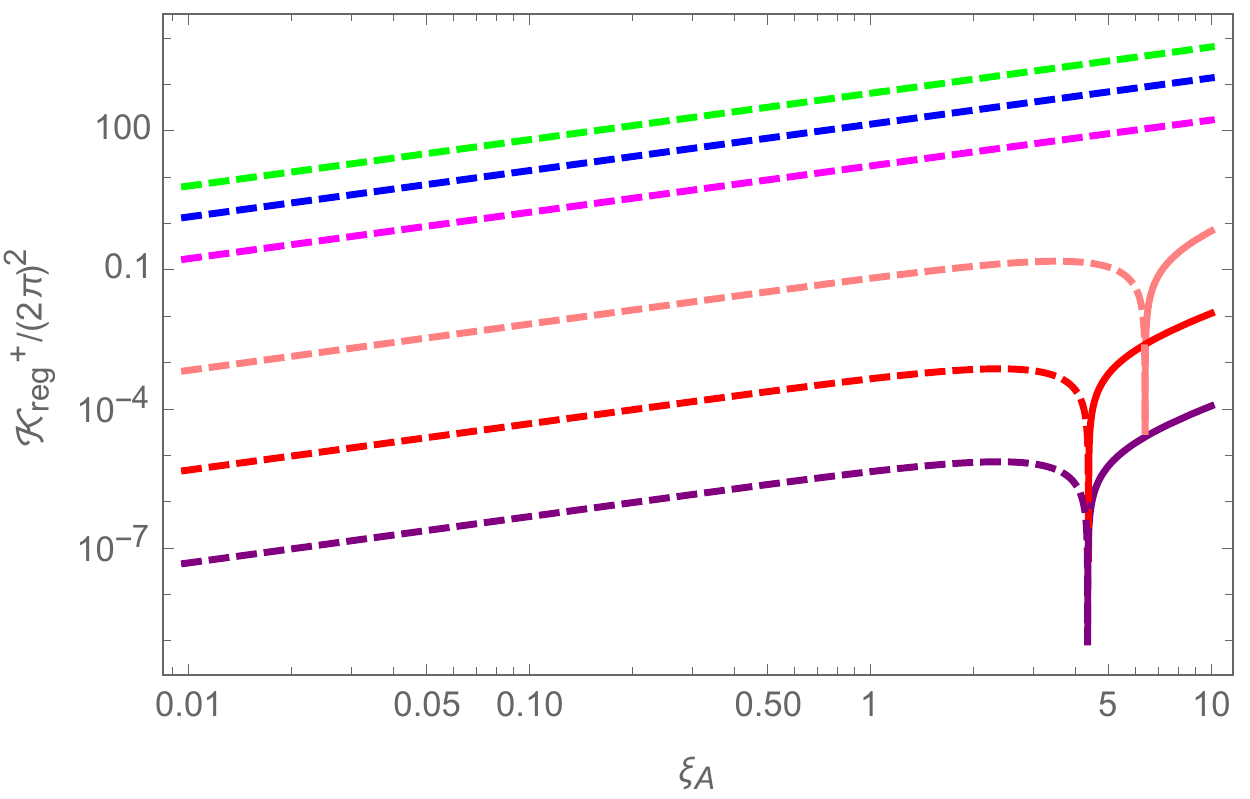} 
\includegraphics[scale=0.56]{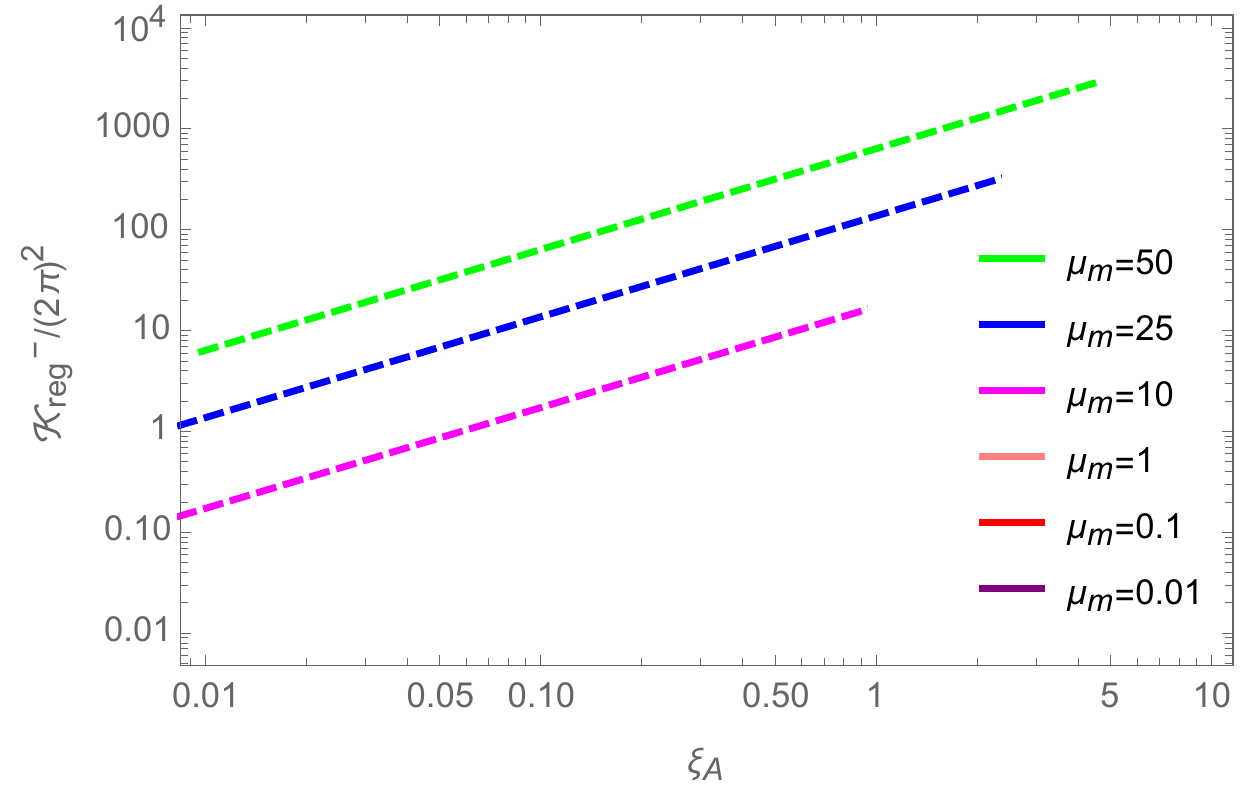}\\
\caption{$\mathcal{K}^{+}_{\rm{reg}}/(2\pi)^2$ (left panel) and $\mathcal{K}^{-}_{\rm{reg}}/(2\pi)^2$ (right panel) for Dirac fermions coupled to the $SU(2)$ gauge field but in the absence of the axion coupling, i.e. $\xpi=0$. The solid (dashed) lines denote positive (negative) values. Notice that the $\mathcal{K}^-_{\rm{reg}}$ are only plotted at the regime of the validity of our analytical approximation for this subspace.}
\label{fig:K+--SU}  
\end{figure*}

This setup of Dirac fermion doublet is reducible into two subspaces of Dirac fermions in the c-helicty frame (Sec. \ref{sec:setup})
\bea
\tilde\Psi_{\bf{k}} = \Psi^+_{\bf{k}} \oplus \Psi^-_{\bf{k}},\nonumber
\eea
where the supercript $+/-$ denotes the plus/minus subspace, and $\Psi^{\pm}_{\bf{k}}$ are two 4-spinors. In Sec. \ref{F-current}, we showed that our non-Abelian fermions have vanishing fermion number, i.e. 
$$\langle J^0\rangle=0.$$ 
as a consequence of $C$-symmetry in our setup. Then,
we showed that the isospin current backreacts to the (background) $SU(2)$ gauge field, $\mathcal{J} \equiv \frac{a}{3}\delta^i_a\langle J^{ai}\rangle =\mathcal{J}^{+} + \mathcal{J}^{-}$, and also has a non-zero chiral charge, $Q_{5} \equiv \langle aJ^0_5 \rangle =Q_{5}^++Q_{5}^-$. We regularized both of these quantities in Sec. \ref{sec:current-comp} and found
\bea\label{calJ-Q5}
\mathcal{J}_{\rm{reg}} = \pm \frac{\ga H^3}{(2\pi)^2} \big(\mathcal{K}^{+}_{\rm{reg}} - \mathcal{K}^{-}_{\rm{reg}}\big)  \an Q_{5{\rm{reg}}} = \frac{2H^3}{(2\pi)^2} \big( \mathcal{K}_{\rm{reg}}^{+} + \mathcal{K}_{\rm{reg}}^{-} - \xp^3 \big),
\eea
where $\mathcal{K}_{\rm{reg}}^{\pm}$ is a momentum integral which we computed analytically for the plus subspace in Eq.s \eqref{fin-J+} (and for minus subspace in $\xi_A/\mu_{_{\rm{m}}}\ll 1$ in Eq. \eqref{fin-J-}). During slow-roll, they are almost constant quantities in terms of the (slow-varying and $CP$-breaking) vacuum quantities $\xp$, $\xpi$ in Eq.s \eqref{def:xa-xphi}, as well as the mass of the fermion $\mu_{\rm{m}}=\frac{m}{H}$. Thus, the backreaction to the axion background, $\mathcal{B} \equiv \frac{\beta\lambda}{2f}\nabla_{\mu} J^{\mu}_5$, is given as
\bea
\mathcal{B}_{\rm{reg}} \simeq \frac{3\beta\lambda}{f} \frac{H^4}{(2\pi)^2} \big( \mathcal{K}^{+}_{\rm{reg}} + \mathcal{K}^{-}_{\rm{reg}} - \xp^3 \big).
\eea
The $\mathcal{K}^{\pm}_{\rm{reg}}$ are directly proportional to (and an odd function of) $\kappa^{\pm}_{I} = (2\xpi \mp \xp/2)$ which is the source of $CP$-violation and increase with the mass of the fermion, $\mu_{\rm{m}}$, which is the source of chiral symmetry breaking. Thus, $\mathcal{K}^{\pm}_{\rm{reg}}$ vanishes in the massless limit where the system has classical chiral symmetry. Moreover, in the absence of the $SU(2)$ gauge field and the axion interactions which are the sources of $CP$-breaking, the fermion fields are unpolarized and all the currents vanish. That is in agreement with our qualitative discussion based on symmetries in Sec. \ref{vacuum}. In the large and small mass limits $\mathcal{K}^{+}_{\rm{reg}}$ is 
\bea
\mathcal{K}^{+}_{\rm{reg}} = \begin{cases}
 4 \kappa_{I}^+ \mu^2_{{\rm{m}}}  \bigg[  \ln\mu_{{\rm{m}}} - \psi^{(0)}(1)  + \frac12 \bigg] +  \kappa_{I}^+ (1-\frac43(\kappa_{I}^+)^2) &\where  \mu_{\rm{m}}\gg 1 ,\\
  -\frac{4\pi}{3} \kappa_{I}^+ \mu_{\rm{m}}^2 \lvert\kappa_{I}^+ \rvert &\where  \mu_{\rm{m}} \ll 1.
\end{cases}
\eea
In the limit that $\kappa_{I}^+ \gg \mu_{\rm{m}}$, the $\mathcal{K}_{\rm{reg}}^+(\tau)$ takes the following asymptotic form
\bea
\mathcal{K}^+_{\rm{reg}}(\tau) \simeq - \frac{4(\kappa_{I}^+)^2 \lvert\kappa_{I}^+\rvert}{3} \big( 1 -  e^{-\frac{\mu^2_{\rm{m}}\pi}{\kappa_{I}^+}}\big)\simeq  -\frac{4\pi}{3}\mu_{\rm{m}}^2 \kappa_{I}^+\lvert\kappa_{I}^+\rvert.
\eea
 In Fig. \ref{fig:K+--SU-axion}, we plot $\mathcal{K}_{\rm{reg}}^{\pm}$ for fermion fields coupled both to the gauge field and axion which $\xpi=2$. Figure \ref{fig:K+--SU} shows $\mathcal{K}_{\rm{reg}}^{\pm}$ for fermions coupled to the isotropic $SU(2)$ gauge field but in the absence of the derivative coupling with the axion, i.e. $\beta=0$ and $\xpi=0$.

\subsubsection*{Light fermions}

In the massless limit, $\mathcal{K}^{\pm}_{\rm{reg}}=0$, and the chiral charge is
\bea
Q_{5\rm{reg}} = - \frac{2H^3 \xp^3}{(2\pi)^2},
\eea
and the fermion backreaction to the background axion field 
\bea
\mathcal{B}_{\rm{reg}} = - \frac{3\beta\lambda}{f}\frac{H^4 \xp^3}{(2\pi)^2}. 
\eea
Namely, the only non-zero effect is the Abelian chiral anomaly (See Sec. \ref{CSym}).

\subsubsection*{Heavy fermions}

The mass of the fermion and the energy scale of inflation cannot exceed the decay constant of the axion, $f$, which is the $UV$ cutoff of the theory. Apart from that, the size of the fermion backreactions on the background gauge field and axion put upper bounds on it. Note that unlike the non-perturbative tunneling effects in Minkowski space, our current analysis is perturbative and relays on assuming a classical background with infinite energy that sources the fermions. That approximation is broken once the fermion backreaction becomes large. 
 Following \cite{Maleknejad:2018nxz}, here we constrain the parameter space of the system based on the size of the fermion backreaction.

\textbf{I) Backreaction to the background $SU(2)$ gauge field:} The validity of the perturbation theory and slow-roll dynamics requires that the backreaction to the gauge field should utmost be as small as slow-roll suppressed terms. Demanding that condition, we find
\bea
\frac{\ga H^3}{(2\pi)^2}\mathcal{K}^{\pm}_{\rm{reg}}\ll H^2\psi,
\eea
which in the large mass limit, and requiring the RHS be smaller than $10^{-2} H^2\psi$, gives
\bea\label{BR-m}
m \lesssim \frac{\pi}{10\ga} \bigg(\frac{\xp}{\kappa_{I}^+}\bigg)^{\frac12} H,
\eea
which for typical values of $\kappa^+_{I}$ and $\xp$ of order one and $\ga\sim 10^{-2}-10^{-3}$ gives $m\lesssim 100 H$.

 \textbf{II) Backreaction to the background axion field:} Demanding that the femrion backreaction to the axion field equation should utmost be as small as slow-roll suppressed terms implies 
\bea
\frac{\lambda\beta}{f}\frac{H^4}{(2\pi)^2} \big( \mathcal{K}^{+}_{{\rm{reg}}} + \mathcal{K}^{-}_{{\rm{reg}}}+\xp^3 \big) \lesssim H\dot\varphi.
\eea
In $SU(2)$-axion models during slow-roll inflation, we have $\xp\sim \xi$ \cite{Maleknejad:2018nxz} which using in the above, it gives  
\bea
\xp \lesssim \frac{2\pi f}{H} \an m \lesssim 2\pi f.
\eea
For instance, for $f/H\sim 10^2$ we have $\xp \sim \frac{m}{H} < 600$.

\begin{figure*}[!h]
\centering
\includegraphics[scale=0.56]{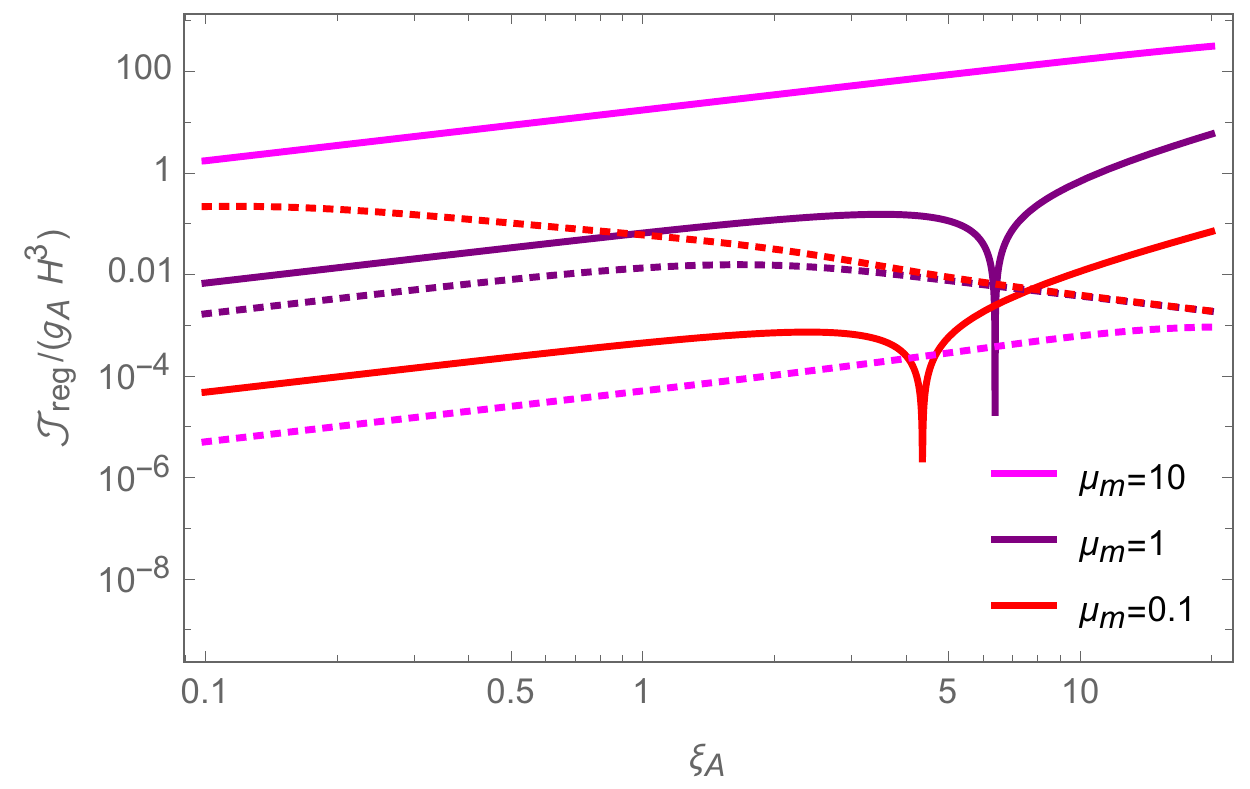} 
\caption{Comparing the scalar and fermion backreactions to the background $SU(2)$ gauge field. The solid lines denotes $\mathcal{J}_{\rm{reg}}^+$ for charged Dirac fermions (with $\xpi=0$) and the dotted lines represent the corresponding backreaction of the charged scalar fields worked out analytically in \cite{Lozanov:2018kpk}.}
\label{fig:J-scalar-fermion}  
\end{figure*}

\subsubsection{Charged scalars vs charged fermions}\label{scalar-fermion}

Here we compare the scalar and fermion backreactions to the background $SU(2)$ gauge field. As for the backreaction of the charged scalars, we use the analytical formula worked out in \cite{Lozanov:2018kpk}. In Fig. \ref{fig:J-scalar-fermion}, we show the regularized scalar and fermion backreactions. Interestingly, our vacuum produces more fermions than scalars in i) large $\xp$ and ii) large mass $\mu_{\rm{m}}>1$, i.e. $m>H$ limits. More precisely, the asymptotic form of $\mathcal{J}_{{\rm{reg}}}$ in the large $\xp$ for the scalars decreases like $1/\xp$ while for the Dirac fermions increase as $\xp^2$. That is because the $SU(2)$ gauge field VEV adds an additional mass term to the spin-0 field in such a way that the charged scalar has negligible particle production in the strong field limit \cite{Lozanov:2018kpk}. Moreover, by the increase of their mass, the scalar current decreases like $1/\mu_{\rm{m}}$ while the fermionic one increases like $\mu_{\rm{m}}^2$. As a result, the $SU(2)$-axion inflationary models can efficiently produce massive fermions.


\begin{figure*}[!h]
\centering
\includegraphics[scale=0.56]{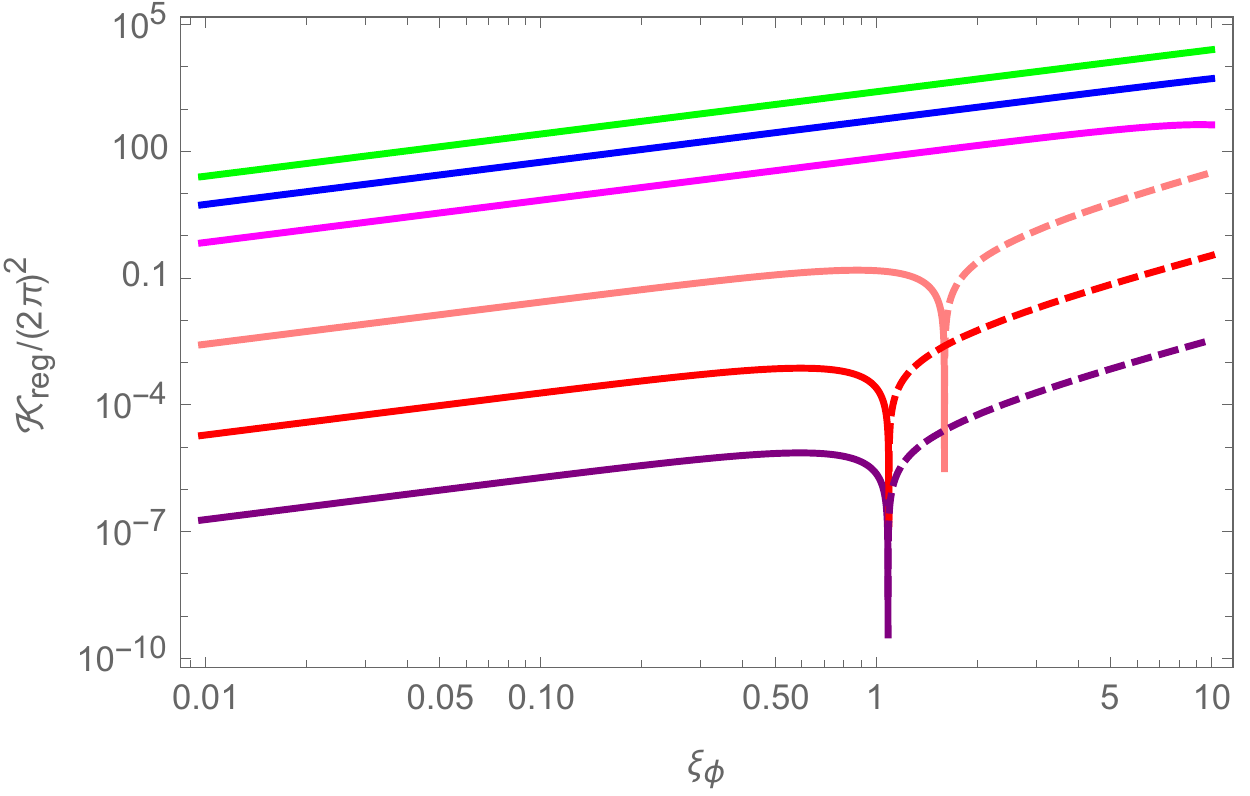} 
\caption{$\mathcal{K}_{\rm{reg}}/(2\pi)^2$ for uncharged Dirac fermions coupled to the axion in terms of $\xpi$ and for different values of $\mu_{\rm{m}}$. The colors correspond to the values similar to Fig. \ref{fig:K+--SU}. The solid (dashed) lines denote positive (negative) values.}
\label{fig:K+--axion}  
\end{figure*}





\subsection{Neutral fermions coupled to axion}

As a limit in our setup, here we focus on neutral Dirac fermions coupled derivative to the axion. \footnote{Neutral Dirac fermions derivatively coupled to the axion, its backreaction to the axion $\nabla_{\mu} J^{\mu}_5=-2im\bar{\Psi}\gamma_5\Psi$, and the fermionic effects on the curvature perturbations have been extensively studied in \cite{Adshead:2018oaa}. The $UV$ divergent momentum integrals are regularized using adiabatic regularization for fermions. The authors reported a singularity in the massless limit and used a change of basis to remove it. In this work, we did not find any singularities. In the massless limit in particular, we realize that the neutral fermions acquire conformal and chiral symmetry and all the observable currents vanish.} In the absence of the interaction with the gauge field, the minus subsystem is simply a second copy of the plus subsystem, hence $\mathcal{K}_{\rm{reg}}^{-}=\mathcal{K}_{\rm{reg}}^{+}=\mathcal{K}_{\rm{reg}}$. Thus from Eq. \eqref{calJ-Q5}, its backreaction to the gauge field vanishes, i.e. $$\mathcal{J}_{\rm{reg}}=\mathcal{J}_{\rm{reg}}^{+}+\mathcal{J}_{\rm{reg}}^{-}=0.$$
Moreover, the fermion number is zero, i.e. 
\bea
\langle J^0 \rangle =0.
\eea
However, it still has a non-zero chiral charge
 $$Q_{5\rm{reg}} = \frac{4H^3}{(2\pi)^2}\mathcal{K}_{\rm{reg}},$$
while the backreaction to the axion field equation is given as 
\bea
\mathcal{B} = \frac{6\beta\lambda}{f} \frac{H^4}{(2\pi)^2}\mathcal{K}_{\rm{reg}},
\eea 
where $\mathcal{K}_{\rm{reg}}$ is 
\bea\label{fin-J+---}
&& \mathcal{K}_{\rm{reg}} = \kappa_{I} \bigg\{ \frac{2}{3} (1-2\kappa_{I}^2) \bigg(1 - \frac{\lvert \mu\rvert}{\kappa_{I}} \frac{\sinh(2\kappa_{I}\pi)}{\sinh(2\lvert\mu\rvert\pi)}\bigg) + \mu^2_{{\rm{m}}} \bigg( 2
-4  \psi^{(0)}(1)  - \frac{8\lvert \mu\rvert}{3\kappa_{I} }\frac{\sinh(2\kappa_{I}\pi)}{\sinh(2\lvert\mu\rvert\pi)}\bigg)  \nonumber\\
&& + \mu^2_{{\rm{m}}}  \sum_{s=\pm}  {\rm{Re}}\bigg[  
   \frac{e^{2\lvert\mu\rvert\pi}-e^{-2s\kappa_{I}\pi}}{\sinh(2\lvert\mu\rvert\pi)} \psi^{(0)}(-is\kappa_{I} -i\lvert\mu\rvert)  - \frac{e^{-2\lvert\mu\rvert\pi}-e^{-2s\kappa_{I}\pi}}{\sinh(2\lvert\mu\rvert\pi)} \psi^{(0)}(i\lvert\mu\rvert-is\kappa_{I})
 \bigg] \bigg\},\nonumber
\eea
 in which $\kappa_I = 2\xpi$ and $\vert \mu\rvert = (\mu^2_{{\rm{m}}} + \kappa_I^2)^{\frac12}$. The chiral charge increases with the mass of the fermion like $\mu_{\rm{m}}^2$ and the axion parameter like $\xpi^2$. The $\mathcal{K}_{\rm{reg}}$ in this case is presented in Fig. \ref{fig:K+--axion}.

\subsection{Abelain fermions, small mass limit}

For the sake of completeness, let us now consider the case of a Dirac fermion coupled to an Abelian gauge field in the small mass limit in de Sitter. The aim is to explore the relation of the mass of the fermion and its backreaction in the $U(1)$ gauge field case qualitatively to explore if it also shows the increase with $m$ behavior which we found for the other cases. Consider a $U(1)$ gauge field in the temporal gauge, with a constant electric field as
\bea
A_{\mu} =(0,0,0, A_{3}(\tau)) \where A_{3}(\tau)= \frac{E_3}{H^2\tau},
\eea
and a Dirac fermion which is charged under it with the theory
\bea\label{U1f}
\sqrt{-g}\mathcal{L}_{\Uppsi} = \bar{\Uppsi} \big(i\gamma^{\alpha}\p_{\alpha} - e A_3\gamma^3 - m \big) \Uppsi,
\eea
where $\Uppsi$ is the canonical Dirac field (for which the spin connection disappears), and $m$ is the mass of the fermion.
In this case, the Noether current is
\bea
J^{\mu} = -\frac{e}{a^4} \delta^{\mu}_{\alpha}\langle \bar{\Uppsi} \gamma^{\alpha} \Uppsi\rangle = - \frac{e}{a^4} \bigg( \Uppsi^{\dag}_L \bar{\boldsymbol{\sigma}}^{\mu} \Uppsi_L + \Uppsi^{\dag}_R \boldsymbol{\sigma}^{\mu} \Uppsi_R \bigg),
\eea
and the fermion backreaction to the $U(1)$ gauge field is given as
\bea
\mathcal{J}_{\rm{reg}} = \langle J^3 \rangle =  \frac{e}{a^4} \bigg( \Uppsi^{\dag}_L \boldsymbol{\sigma}^{3} \Uppsi_L - \Uppsi^{\dag}_R \boldsymbol{\sigma}^{3} \Uppsi_R \bigg).
\eea
Moreover, in this case $\langle F\tilde F\rangle =0$, and hence both the vector and axial vector currents are conserved. Namely, the massless fermions have chiral symmetry, hence their backreaction to the gauge field should vanish. It implies that the backreaction should be directly related to the mass of the fermion which is the source of the chiral symmetry breaking \footnote{The fermion theory in Eq. \eqref{U1f} and its fermion production in de Sitter is studied analytically using adiabatic regularization skim in \cite{Hayashinaka:2016qqn}, and later with a different regularization condition in \cite{Hayashinaka:2018amz} which led to another result. However, the results of both of these studies disagree with our observation based on the symmetries of $\mathcal{J}$ in Eq. \eqref{cJ-U1}.}
\bea\label{cJ-U1}
\mathcal{J}_{\rm{reg}} \propto \frac{m}{H}.
\eea
As a result, our simple argument implies that even in the $U(1)$ gauge field case, the fermionic backreaction should increase with the mass of the fermion.

\section{Fermion dark matter, a quick view}\label{fermio-genesis}
Up to this point, we analytically studied the (dark) Dirac fermions charged under the $SU(2)$ VEV in a generic $SU(2)$-axion inflationary model. We realized that the spontaneous $CP$-violating vacuum structure of this class of models leads to a new and efficient mechanism to generate non-thermal massive dark fermions with a preferred helicity state, \textit{h-asymmetric}, during inflation. Moreover, the generated dark fermions can be very massive, which depending on the efficiency of their possible feebly interactions with the visible and lighter dark sectors, may decay to leptons and baryons or other dark particles after inflation. In other words, these Dirac fermions may replace the right-handed neutrinos in the standard thermal leptogenesis scenario and hence provide a novel natural scenario to explain the observed matter asymmetry in the Universe. However, this possibility, as well as the late time macroscopic properties of the produced dark fermions as a dark matter sector, are all highly model-dependent. They depend on the value of the gauge coupling, $\ga$, the possible mass of the gauge field at some symmetry breaking scale after inflation, and the details of the possible feebly interactions of the dark fermions with the visible and dark sectors which is beyond the scope of the current work. We leave these important questions for future exhaustive study. However, let us here roughly estimate the gist of the simplest possible scenario for fermio-genesis in this setup.

The three necessary Sakharov conditions to generate an imbalance of baryonic matter are i) C-symmetry and CP-symmetry violation, ii) interactions out of thermal equilibrium, and iii) baryon number, $B$, violation \cite{Sakharov:1967dj}. One can generalize it for any form of matter by replacing the third condition by the violation of its corresponding fermion number. The fermion doublets charged under our $SU(2)$ gauge field which we studied so far are C-symmetric. That, however, can be easily and naturally violated by the mass term, nonperturbative effects, and especially interactions with the SM fermions, which is beyond the scope of the current work. In the absence of these effects, the generated fermions are not asymmetric in the standard sense but fermions with h-asymmetry. 


For simplicity, we assume that the dark sector is made of one generation of massive dark fermion doublets, $\tilde{\Psi}$. We also assume that they are only gravitationally coupled to the visible sector. Thus, dark fermions are stable. The dark fermion has a constant energy density during inflation
\bea
\rho_{df}^{\rm{inf}} \sim m^2 H^2  \bigg(\kappa_I^{+~2} + \kappa_I^{-~2}\bigg),
\eea
where $H$ is the Hubble parameter during inflation. Notice that the contribution of the $\pm$ spinors into $\rho_{df}$ (which is positive definite) should be directly proportional and an even function of $\kappa_I^{\pm}$ which is the source of the particle production. After the end of inflation, the source of fermion production (gauge field and axion) disappears and therefore $\rho_{df}$ decays like ordinary matter, i.e. 
\bea
a^3 \rho_{df}(a) = a^3_{\rm{inf}}\rho^{\rm{inf}}_{df}.
\eea

\begin{figure*}[t] 
   \centering
     \includegraphics[width=2.5in]{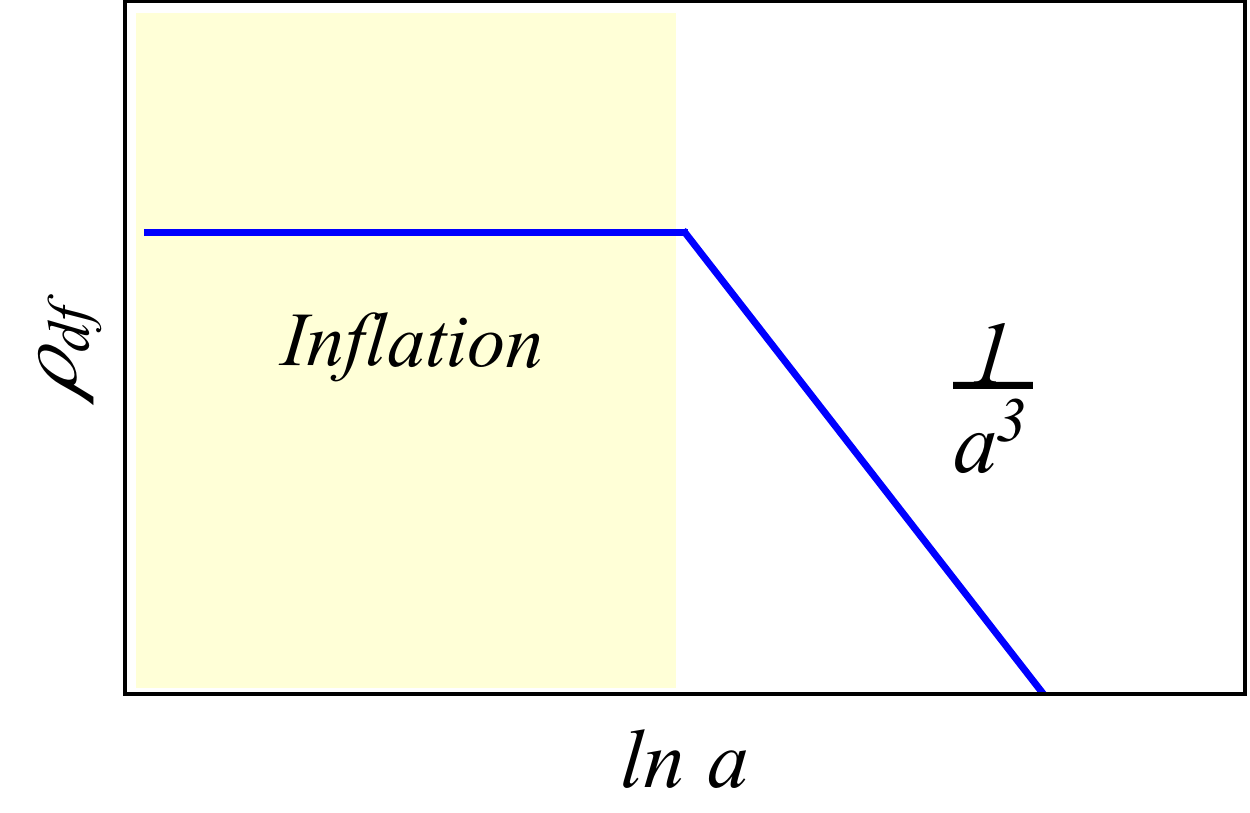} 
  \hspace{0.05in}
   \caption{The time evolution of the energy density of the dark fermions which are gravitationally coupled to the rest of the matter fields.}
   \label{fig:poles}   
\end{figure*}

Here, in order to estimate the energy density of the reheating, we consider the phenomenological reheating model below
\bea
\rho^{\rm{reh}} = \sigma \bigg(\frac{a_{\rm{inf}}}{a_{\rm{reh}}}\bigg)^4 \rho^{\rm{inf}} 
\eea
where $\sigma$ is the efficiency of the reheating process and relates $\rho^{\rm{reh}}$ and the energy density at the end of inflation, $\rho^{\rm{inf}}$. From that, one can read the number density of photons during reheating as
\bea
n_{\gamma}^{\rm{reh}} = \frac{6\sqrt{3}\zeta(3)}{\pi^2}\bigg(\frac{\sigma}{g_{\rm{eff}}}\bigg)^{\frac34} \bigg(\frac{a_{\rm{inf}}}{a_{\rm{reh}}}\bigg)^3 (H\mpl)^{\frac32},
\eea
where $g_{\rm{eff}}=427/4$ is the number of relativistic degrees of freedom at the time of reheating, and $\zeta(z)$ is the Riemann zeta function, $\zeta(3) = 1.2$. Now, we define the parameter $\theta_{df}$ as the ratio of the dark fermion energy density to the number density of photons as
\bea
\theta_{df}(a) \equiv \frac{\rho_{df}(a)}{n_{\gamma}(a)},
\eea
which has the dimension of mass and gives
\bea\label{theta0}
\theta_{df}^0 \simeq \theta_{df}^{\rm{reh}} \sim 24 \frac{m^2}{\mpl} \sigma^{-\frac34} \bigg( \frac{H}{\mpl}\bigg)^{\frac12} \bigg(\kappa_I^{+~2} + \kappa_I^{-~2}\bigg),
\eea
where we assume that the dark fermions are only gravitationally coupled to the visible (and other extra dark) sectors.

Given the fact that the dark matters energy fraction and the visible one are related as $\Omega_{D}\simeq 5\Omega_{B}$ and $\Omega_{df}<\Omega_{D}$ gives
\bea\label{et2}
\theta_{df}^0 \lesssim 5 m_p \eta^0_B,
\eea
in which $m_p\simeq 10^{-19}\mpl$ is the mass of proton and $\eta_B^0$ is the baryon asymmetry observed today, i.e. $\eta^0_B= 6 \times 10^{-10}$ \cite{Ade:2015xua}. From the combination of Eq.s \eqref{theta0} and \eqref{et2}, we find the upper value of the dark fermion as
\bea\label{mass}
\frac{m}{\mpl} \sim 10^{-1} \big( \frac{m_p}{\mpl}\eta_B^0 \big)^{\frac12} ~ \sigma^{\frac38} \bigg(\frac{H}{\mpl}\bigg)^{-\frac14}.
\eea
For a scale of inflation as high as GUT scales, $H\sim 10^{-6}\mpl$, and $\sigma$ of order one, that gives $m \sim 10~TeV$. Notice that it is much lower than the upper bound imposed by the size of the backreaction in Eq. \eqref{BR-m} and higher than the electroweak scale, $E_{EW}\sim 246~GeV$.

\section{Discussion}\label{sec:Discussion}
Models of inflation with an axion and $SU(2)$ gauge field have very rich phenomenology. In this class of models, the $SU(2)$ gauge field has a (slow-varying) homogeneous and isotropic VEV with an almost constant energy density during inflation. Both of the axion and effective VEV of the gauge field are pseudo-scalars (see Eq.s \eqref{GF-SU2}-\eqref{axion} and Fig. \ref{P-B}). As a result, both $P$ and $CP$ are spontaneously broken by the vacuum. Although the background evolution is insensitive to that symmetry breaking, cosmic perturbations with spin, e.g., fermions, and spin-2 fields notice that and become chiral (see Fig. \ref{the-vac}). In particular, the $P$-violation leads to (a short phase of tachyonic) production of the spin-2 field of the perturbed gauge field which produces chiral primordial gravitational waves. Fermions are even more sensitive to the structure of the vacuum as a consequence of the celebrated Atiyah–Singer index theorem. As first pointed out in \cite{Maleknejad:2014wsa, Noorbala:2012fh, Maleknejad:2016dci}, the spontaneous $CP$-violation and non-trivial topology of this vacuum with $\langle R\tilde R\rangle \neq 0$, provides a natural leptogenesis setting via the gravitational anomaly in SM. In this work, we unveiled yet another unique aspect of such $SU(2)$-axion vacuum for fermions which might open a new window toward particle cosmology.

In the context of $SU(2)$-axion inflation models, we considered a (dark) massive  Dirac fermion doublet charged under the $SU(2)$ gauge field and (for the sake of completeness, possibly) derivatively coupled to the axion. This setup has been recently introduced in \cite{Mirzagholi:2019jeb} in which the fermion backreaction to the gauge field background has been numerically computed for a small part of the parameter space, i.e., $m\sim H$. Here, we extensively and analytically studied the system and using the point-splitting regularization skim computed the induced currents (Eq.s \eqref{J-Q-K}, \eqref{fin-J+}, and \eqref{fin-J-}). The net fermion number vanishes due to the $C$-symmetry. However, the isospin current, $J^{\mu a} {\bf{T}}_a$, and the chiral current, $J^{\mu}_5$, are non-zero. We found that the dark fermion's backreaction, $\mathcal{J}=\frac{1}{3a}\delta^i_a\langle J^a_i\rangle $, and its chiral charge, $Q_5= \langle J^t_5\rangle$, are almost constant during inflation. They are both proportional to the scale of inflation as $H^3$, the two sources of $CP$-violation $\xp$ and $\xpi$ (corresponding to the VEV of the gauge field and axion), and both increases with the mass of the fermion which explicitly breaks the chiral symmetry in the fermionic sector (Sec. \ref{gauge-field-fermion}). In fact, the massless Dirac fermions enjoy conformal and chiral symmetries and have zero (tree-level) induced currents. The massless fermions only have a non-zero chiral charge and therefore backreacts to axion field according to the celebrated Adler-Bell-Jackiw anomaly. 
Besides, the currents vanish in the absence of the $CP$-breaking the vacuum. We compared the production of charged scalars with fermions of this setup in Sec. \ref{scalar-fermion}. Interestingly, the vacuum produces more fermions than scalars in two limits; i) when the mass of the corresponding field is higher than Hubble during inflation, i.e., $m>H$, and ii) in large $\xp$ limit. Apart from explicit computation for the $SU(2)$-axion vacuum in Sec. \ref{F-current}, based on symmetries in Sec.s \ref{vacuum} and \ref{fermi-de}, we explained why the fermion currents should increase with the mass regardless of the source, whether it is the VEV of a $SU(2)$, $U(1)$, or an axion. At first glance, this behavior seems counter-intuitive. However, the present and similar studies which were carried out in de Sitter, are perturbative computations based on assuming a classical background with infinite energy that can always source and support the particle production. However, once the size of backreaction becomes sizable, that approximation is not valid anymore.

We realized that the spontaneous $CP$-violation during inflation naturally leads to a new and efficient mechanism to generate non-thermal massive dark fermions during inflation. That is regardless of the details of the specific $SU(2)$-axion model.
The size of the fermionic backreaction as well as the present-day density fraction of dark matter put upper bounds on the fermion's mass respectively as $10^{2}~H$ (Eq. \eqref{BR-m}) and $(10^{10} \frac{\mpl}{H})^{\frac14}~GeV$ (Eq. \eqref{mass}). In deriving the latter, we assumed that the heavy fermions are only gravitationally coupled to the rest of the visible and dark sectors. For a GUT scale inflation, the dark fermions can be as heavy as $10~TeV$.

The generated dark fermions can be very massive, which depending on the efficiency of their possible feebly interactions with the visible sector, may decay to the visible and lighter dark sectors once the temperature drops below their mass. In other words, these Dirac fermions may replace the right-handed neutrinos in the standard thermal leptogenesis scenario and hence provide a novel natural scenario to explain the observed matter asymmetry in the Universe. However, this possibility, as well as the late time macroscopic properties of the produced (self-interacting) dark fermions as a dark matter sector, are all highly model-dependent. These promising possibilities depend on the value of the gauge coupling, $\ga$, the possible mass of the gauge field after inflation, and the details of the possible feebly interactions of the dark fermions with themselves and the visible sector which is beyond the scope of the current work and we leave for future exhaustive study.

\acknowledgments

Special thanks go to Eiichiro Komatsu for stimulating discussions over the years and bringing the importance of Schwinger effect in the early universe into my attention which led to a series of papers including the current work. I appreciate Kaloian Lozanov for his collaboration in the early stage of the project. I am grateful to David Curtin, Gia Dvali, Elisa Ferreira, Mohammad Khorrami, Yuki Watanabe, and Fabian Schmidt for valuable discussions.

\appendix

\section{Mathematical tools}\label{Math}
For the sake of self-sufficiency, this appendix reviews mathematical tools that we use throughout this work. This overview includes the definition of the direct sum and product, the spinor covariant derivative, and Whittaker functions, respectively. 

\subsection*{\textbf{A.1~~ Direct sum and product:}}

The vector space $\bf{V}$, is the direct sum of two subspaces, $\bf{U}_1$ and $\bf{U}_2$, as \bea
\bf{V}= \bf{U}_1\oplus \bf{U}_2,
\eea
 if and only if
$\bf{V}=\bf{U}_1+\bf{U}_2$, and $\bf{U}_1$ and $\bf{U}_2$ are independent. The Kronecker product of two matrices, ${\bf{A}}_{m\times n}$ and ${\bf{B}}_{q\times p}$, is defined as a $mp \times nq$ block matrix given by
\Beq\label{Kronecker}
{\bf{A}}\otimes {\bf{B}} = \begin{pmatrix}
A_{11} {\bf{B}} & \dots &  A_{1n}  {\bf{B}} \\
& \ddots & \\
A_{m1} {\bf{B}} & \dots & A_{mn}  {\bf{B}}
\end{pmatrix}.
\Eeq

\subsection*{\textbf{A.2~~ Spinor covariant derivative: }}

The 8-spinor covariant derivative in \eqref{slashD} is
\Beq
D_{\mu}\otimes\gamma^{\alpha}\tilde{\psi}\equiv\left(\I \nabla_{\mu}-ig_{\!A}{\bf{A}}_{\mu}\right) \otimes\gamma^{\alpha}\tilde{\psi}\,,
\Eeq
where the spin covariant derivative is
\Beq\label{Spin-dee}
\nabla_{\mu}\tilde{\psi} = [\II \partial_{\mu}+\omega_{\mu}] \tilde{\psi},
\Eeq
with $\omega_{\mu}$ being the spin-connections
\bea
\omega_{\mu}=-\frac{i}{2} \omega_{\mu}^{~\alpha\beta} \sigma_{\alpha\beta},
\eea
and $\sigma_{\alpha\beta} = \frac{i}{4}[\gamma_{\alpha},\gamma_{\beta}]$ being the spinor generators of the Lorentz algebra. 
The elements of the spin-connection $\omega_{\mu}^{~\alpha\beta}$ are given by
\bea
\omega_{\mu}^{~\alpha\beta} = {\bf{e}}^{~\alpha}_{\nu} \nabla_{\mu} {\bf{e}}^{\nu\beta}.
\eea
In FLRW spacetime using the conformal time, the verbeins are
\Beq
{\bf{e}}^{\mu}_{~\alpha}=a(\tau)^{-1}\delta^{\mu}_{\alpha},
\Eeq
and the only non-zero components of the spin connection coefficients are
\Beq\label{Spin-connect-FRW}
 \omega_{\mu}^{~i0}=-\omega_{\mu}^{~0i}= -\mathcal{H}\delta^i_{\mu}.
\Eeq

\subsection*{\textbf{A.3~~ Whittaker functions:}}

Whittaker functions $W_{\kappa,\mu}(z)$ and $M_{\kappa,\mu}(z)$ take the following aymptotic forms in the limit $\mid z\mid\rightarrow\infty$

\bea \label{WM-asymp}
W_{\kappa,\mu}(z) &\rightarrow & z^{\kappa}e^{-z/2}, \\   \label{WM-asymp-2}
 M_{\kappa,\mu}(z) &\rightarrow & \Gamma(2\mu+1)\bigg(\frac{i(-1)^{\mu-\kappa}z^{\kappa}e^{-z/2}}{\Gamma({-\kappa+\mu+\frac12})}+\frac{z^{-\kappa}e^{z/2}}{\Gamma({-\kappa+\mu+\frac12})}\bigg),
\eea
where $\mid \arg z\mid<\frac32\pi$.
Thus, for a complex $\kappa$, we have
\bea\label{general-BD}
\lim_{\tau\rightarrow -\infty}(2\x)^{-\kappa_{_{\rm{R}}}}e^{-\kappa_{_{\rm{I}}}\pi/2}W_{\kappa,\mu}(-2i\x) =  e^{-ik\tau},
\eea
where $\kappa_{_{\rm{R}}}$ and $\kappa_{_{\rm{I}}}$ are the real and imaginary parts of $\kappa$. 

The Mellin-Barnes integral representation of the Whittaker functions is \cite{Nist}
\bea\label{Mellin-Barnes}
W_{\kappa,\mu}(z) = \frac{e^{-\frac{z}{2}}}{2i\pi} \int_{\mathcal{C}_{\alpha}} \frac{\Gamma(\frac12+\mu+\alpha)\Gamma(\frac12-\mu+\alpha)\Gamma(-\kappa-\alpha)}{\Gamma(\frac12+\mu-\kappa)\Gamma(\frac12-\mu-\kappa)} z^{-\alpha}d\alpha  \where \lvert {\rm{arg}}(z)\rvert <\frac32\pi, \nonumber\\
\eea 
which holds when
\bea
\frac12\pm \mu -\kappa \neq 0, -1, -2 ,\dots,  \nonumber
\eea
and the contour of the integration, $\mathcal{C}_{\alpha}$, separates the poles of $\Gamma(\frac12+\mu+\alpha)
\Gamma(\frac12-\mu+\alpha)$ at
\bea\label{snpm}
\alpha_{n,\pm}=-\frac12-n\pm \mu \quad \forall n\in \mathbb{N},
\eea
and the poles of $\Gamma(-\kappa-\alpha)$ at
\bea\label{sn0}
\tilde{\alpha}_{n}=-\kappa+n \quad \forall n\in \mathbb{N},
\eea 
from each other as we show in Fig. \ref{fig:poles}. 
In our setup, $\mu^p=i\lvert \mu^p\rvert$ is pure imaginary and $\kappa^p_s$ is a complex number with a real part equal to $\lvert\rm{Re}\kappa\rvert= \frac12$. 
\begin{figure*}[t] 
   \centering
     \includegraphics[width=2.5in]{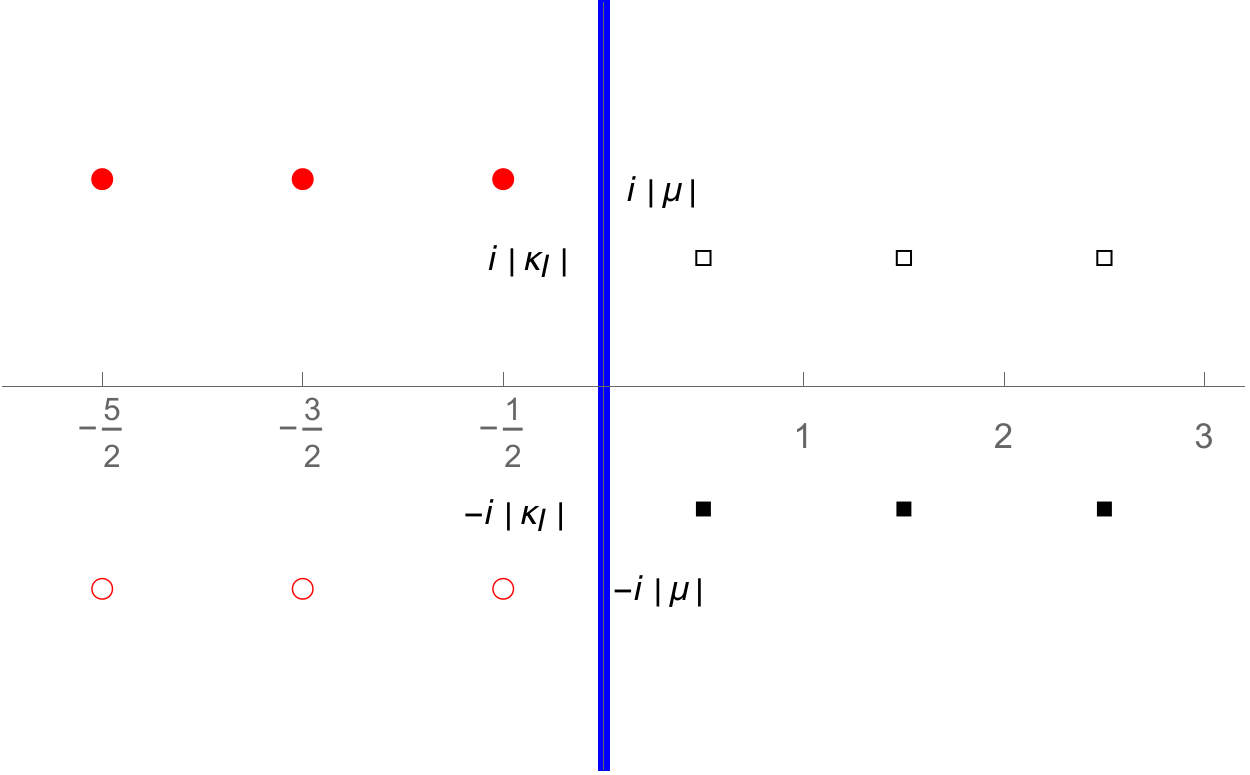} 
  \hspace{0.05in}
{\includegraphics[width=2.5in]{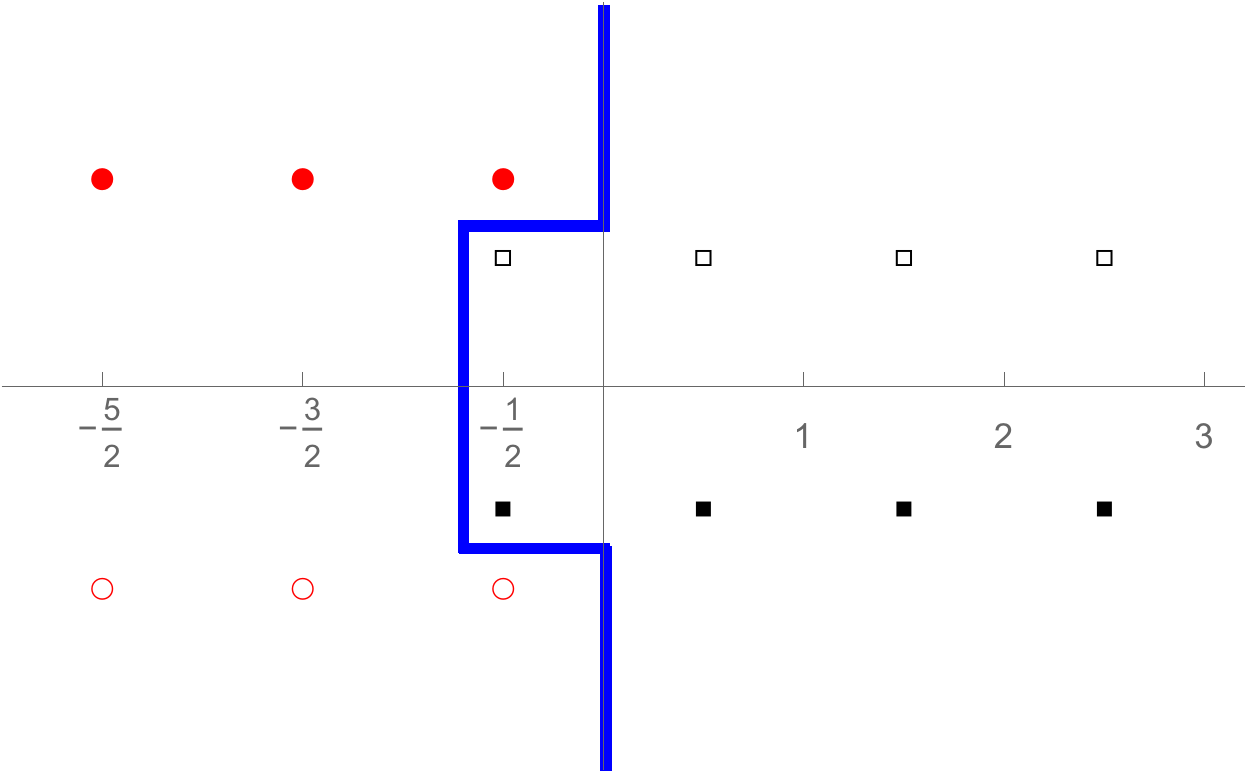}}
   \caption{Positions of poles $\alpha_{n,\pm}$ (circles) and $\tilde{\alpha}_{n,s}$ (squares) and the possible contouring for the validity of the Mellin-Barnes integral representation in \eqref{Mellin-Barnes} are shown where $\kappa_I$ is the imaginary part of $\kappa$ and $m\neq 0$. In the left panel, we show the locations of poles for $\rm{Re}\kappa=-\frac12$ and in the right panel we have the poles for $Re\kappa=\frac12$. The Mellin-Barnes integral is valid for any closed contour as far as it separates the circle poles from the square poles and it runs from $-i\infty$ to $+i\infty$.  }
   \label{fig:poles}   
\end{figure*}

\section{Color-spin helicity}\label{helicity}
This appendix gives a self-contained derivation of
the split of our setup as $\Uppsi^+\oplus\Uppsi^-$. It is a brief review of appendix B of \cite{Mirzagholi:2019jeb} by the author, which generalized the $2d$ helicity representation to the $4d$ color-spin helicity frame (c-helicity).

\subsection{$\Uppsi^+\oplus\Uppsi^-$}
\label{subsec:b1}

At this point, it is more insightful to go to the Fourier space and write the system in the Weyl frame. The 8-spinor in the isospin frame can be written in the Weyl frame as
$$\tilde T_1 \begin{pmatrix}
\Uppsi^1 \\
\Uppsi^2
\end{pmatrix} = \begin{pmatrix}
\Uppsi_L \\
\Uppsi_R
\end{pmatrix},$$
where $\tilde T_1$ is the following $8\times 8$ unitary matrix
\bea
\tilde T_1 = \frac{1}{\sqrt{2}}\begin{pmatrix}
\I & -\I & \bf{0} & \bf{0} \\
\bf{0} & \bf{0} & \I & -\I \\
\I & \I & \bf{0} & \bf{0} \\
\bf{0} & \bf{0} & \I & \I
\end{pmatrix}.
\eea
Moreover, the $\I\otimes \gamma^{\alpha}$ operators transform as
\bea
\tilde{T}_1.(\I\otimes \gamma^{\alpha}). \tilde{T}_1^{-1} = \begin{pmatrix}
\bf{0} & \bf{0} & \boldsymbol{\sigma}^{\alpha} & \bf{0} \\
\bf{0} & \bf{0} & \bf{0} & \boldsymbol{\sigma}^{\alpha} \\
 \bar{\boldsymbol{\sigma}}^{\alpha} & \bf{0} & \bf{0} & \bf{0} \\
 \bf{0} & \bar{\boldsymbol{\sigma}}^{\alpha} & \bf{0} & \bf{0} \\
\end{pmatrix},
\eea
where $\boldsymbol{\sigma}^{\alpha}$ and $\bar{\boldsymbol{\sigma}}^{\alpha}$ are 
\bea
\boldsymbol{\sigma}^{\alpha}= (\I, \boldsymbol{\sigma}^i)    \an \bar{\boldsymbol{\sigma}}^{\alpha}=(\I, -\boldsymbol{\sigma}^i),
\eea
and their indices are lowered with the 
Minkowski metric, i.e. $\boldsymbol{\sigma}_{\alpha}= \eta_{\alpha\beta} \boldsymbol{\sigma}^{\beta}$.
In the Weyl frame, the action \eqref{theory} takes the following form 
\bea\label{chiral-A}
S_{\Psi} = \int {\rm{d}}\tau {\rm{d}}k^3 (\Uppsi^{\dag}_{R,\bf{k}} \Uppsi^{\dag}_{L,\bf{k}}) ~ . ~ \tilde{\rm{L}}_{\bf{k}}(\tau )~.  \begin{pmatrix}
\Uppsi_{L,\bf{k}} \\
\Uppsi_{R,\bf{k}} 
\end{pmatrix},
\eea
where $\tilde{\rm{L}}_{\bf{k}}(\tau )$ is a $8\times 8$ operator given as
\bea\label{L-action}
\tilde{\rm{L}}_{\bf{k}}(\tau ) \equiv i \begin{pmatrix}
i\mu_{{\rm m}} \mH \II & \II \p_{\tau} + i \Sigma_4(\tau,\bk) \\
 \II \p_{\tau} - i\Sigma_4(\tau,\bk)  & i\mu_{{\rm m}} \mH  \II
\end{pmatrix},
\eea
and $\Sigma_4$ is the following $4\times 4$ operator
\bea\label{Sigma4-}
\Sigma_{_{4}}(\tau,\bk) =  \I\otimes k^i.\boldsymbol{\sigma}_i + \mH (2\xpi {\II} - \frac{\xp}{2} \boldsymbol{\tau}^i \otimes \boldsymbol{\sigma}_i).
\eea
As we see in Eq. \eqref{L-action}, in the absence of $\mu_{{\rm m}}$, massless case, the system in Eq. \eqref{chiral-A} is decomposed into two independent sub-sectors in terms of the L- and R-handed fields. However, in the massive case, we need to take one step further.

The ideal frame would be the one in which $\Sigma_4 $ is diagonalized. Such frame, if exists, must be made of the common eigenstates of the helicity operator, $  \I\otimes k^i.\boldsymbol{\sigma}_i $, and $ \boldsymbol{\tau}^i \otimes \boldsymbol{\sigma}_i$. However, these two operators have only two common eigenstates. Thus, $\Sigma_4$ is only block-diagonalizable and reducible up to two subspaces. 

As a mathematical tool, we define the color-spin helicity operator for any given 4-momentum $k^{\alpha}$, as
\bea
\mathfrak{h}(k^{\alpha}) \equiv \frac{\delta_{ia}}{k^2}~k^i.\boldsymbol{\tau}^a\otimes k^j.\boldsymbol{\sigma}_j,
\eea
in which the first Pauli matrix is a $su(2)$ generator while the second one is the spin operator.
The orthonormal eigenstates of this operator are
\Beq
E^{+}_{~+}(k^{\alpha})&=\frac{\check{k}^{\alpha}\bar{\boldsymbol{\tau}}_{\alpha}\otimes\check{k}^{\beta}\bar{\boldsymbol{\sigma}}_{\beta}}{2k(k+k^3)}\begin{pmatrix}1 \\ 0\\ 0 \\ 0\end{pmatrix}\,,\quad E^{+}_{~-}(k^{\alpha})=-\frac{\check{k}^{\alpha}\boldsymbol{\tau}_{\alpha}\otimes\check{k}^{\beta}\boldsymbol{\sigma}_{\beta}}{2k(k+k^3)}\begin{pmatrix}0 \\ 0\\ 0 \\ 1\end{pmatrix}\,,\\
E^{-}_{~+}(k^{\alpha})&=-\frac{\check{k}^{\alpha}\boldsymbol{\tau}_{\alpha}\otimes\check{k}^{\beta}\bar{\boldsymbol{\sigma}}_{\beta}}{2k(k+k^3)}\begin{pmatrix}0 \\ 0\\ 1 \\ 0\end{pmatrix}\, \quad E^{-}_{~-}(k^{\alpha})=-\frac{\check{k}^{\alpha}\bar{\boldsymbol{\tau}}_{\alpha}\otimes\check{k}^{\beta}\boldsymbol{\sigma}_{\beta}}{2k(k+k^3)}\begin{pmatrix}0 \\ 1\\ 0 \\ 0\end{pmatrix}\,,
\Eeq
where $\check{k}^{\alpha}$ is a four vector given as
\bea
\check{k}^{\alpha} \equiv (k,\bk) \where k=\sqrt{k^i . k^i }.
\eea
Thus, $\check{k}^{\alpha}$ is the four momentum of the massless field. 
The c-helicity $\mathfrak{h}(k^{\alpha})$, and helicity $\I\otimes k^i\boldsymbol{\sigma}_i$, have 4 eigenstates in common. However, $\mathfrak{h}(k^{\alpha})$ and $\boldsymbol{\tau}^i\otimes\boldsymbol{\sigma}_i$ have only two common eigenstates. More precisely,  $E^{p}_{~s}(k^{\alpha})$ with $p=\pm$ and $s=\pm$ satisfies the eigenstate equations
\Beq
 \I\otimes k^i.\boldsymbol{\sigma}_i  ~ E^{p}_{~s}(k^{\alpha}) = s ~ k E^{p}_{~s}(k^{\alpha})\,,
\Eeq
and the orthonormality condition
\bea
 \quad E^{p\dagger}_{~s}(k^{\alpha})\cdot E^{p'}_{~s'}(k^{\alpha})=\delta_{ss'}\delta^{pp'}\,.
\eea
However, only $p=+$ elements of $E^{p}_{~s}(k^{\alpha})$ are eigenstates of $\boldsymbol{\tau}^i\otimes\boldsymbol{\sigma}_i$ as well
\bea
\boldsymbol{\tau}^i\otimes ~ \boldsymbol{\sigma}_i E^{+}_{~s}(k^{\alpha}) = E^{+}_{~s}(k^{\alpha}).
\eea
The $E^p_{~s}(k^{\alpha})$ make an orthonormal basis  and hence the following unitary matrix
\bea\label{R-def}
R_{\bf{k}} = \big[ E^{+}_{~+}(k^{\alpha}) ~  E^{+}_{~-}(k^{\alpha}) ~  E^{-}_{~+}(k^{\alpha}) ~  E^{-}_{~-}(k^{\alpha}) \big] =\begin{pmatrix*}[c]e^{+}_{~+1} & e^{+}_{~-1} & e^{-}_{~+1} & e^{-}_{~-1}\\ e^{+}_{~+2} & e^{+}_{~-2} & e^{-}_{~+2} & e^{-}_{~-2}\\ e^{+}_{~+3} & e^{+}_{~-3} & e^{-}_{~+3} & e^{-}_{~-3}\\e^{+}_{~+4} & e^{+}_{~-4} & e^{-}_{~+4} & e^{-}_{~-4}\end{pmatrix*},
\eea
where $e^p_{~si}$ is the $i$th element of the $E^p_{~s}(k^{\alpha})$. For each given momentum $k^{\alpha}$, $R_{\bf{k}}$ takes the Weyl spinors to their c-helicity frame. Notice that the definition of $R_{\bf{k}}$ is a bit different than the corresponding matrix in \cite{Mirzagholi:2019jeb}. More precisely, comparing with \cite{Mirzagholi:2019jeb}, here we switched the orders of 3rd and 4th columns.

The 8-spinor in the Weyl representation can transformed to the c-helicity state as
\bea\label{Rspinor}
\tilde R_{\bf{k}} \begin{pmatrix}
\Uppsi_{L,\bf{k}} \\
\Uppsi_{R,\bf{k}} 
\end{pmatrix} = \begin{pmatrix}
\boldsymbol{\Uppsi}^+_{L,\bf{k}} \\
\boldsymbol{\Uppsi}^-_{L,\bf{k}} \\
\boldsymbol{\Uppsi}^+_{R,\bf{k}} \\
\boldsymbol{\Uppsi}^-_{R,\bf{k}}
\end{pmatrix},
\eea
where $\tilde R_{\bf{k}}$ is the following $8\times 8$ unitary operator 
\bea
\tilde R_{\bf{k}} \equiv \I \otimes R_{\bf{k}}^{-1}.
\eea
 This split would be clearer if we take another unitary transformation 
\bea
\tilde S =  \begin{pmatrix}
\I & 0 & 0 & 0\\
0 & 0 & \I & 0 \\
0 & \I & 0 & 0 \\
0 & 0 & 0 & \I
\end{pmatrix}.
\eea 
In fact, $\tilde S\tilde R_{\bf{k}}$ takes the 8-spinor transforms in Weyl frame to the c-helicity frame
\bea\label{pmspinor}
\tilde S\tilde R_{\bf{k}} \begin{pmatrix}
\Uppsi_{L,\bf{k}} \\
\Uppsi_{R,\bf{k}} 
\end{pmatrix} = \begin{pmatrix}
\Uppsi^+_{{\bf{k}},_{W}} \\
\Uppsi^-_{{\bf{k}},_{W}} 
\end{pmatrix} = \begin{pmatrix}
{\boldsymbol{\Uppsi}}^+_{L,\bf{k}} \\
{\boldsymbol{\Uppsi}}^+_{R,\bf{k}} \\
{\boldsymbol{\Uppsi}}^-_{L,\bf{k}} \\
{\boldsymbol{\Uppsi}}^-_{R,\bf{k}}
\end{pmatrix}.
\eea
As a result, the Lagrangian operator becomes the following block diagonal $8\times 8$ matrix
\bea
 L_{{\bf{k}},_{W}}(\tau) = \tilde S\tilde R_{\bf{k}} ~ . ~ \tilde L_{{\bf{k}},_{W}}(\tau) ~.~ (\tilde S\tilde R_{\bf{k}})^{-1} =  \begin{pmatrix}
L^+_{{\bf{k}},_{W}}(\tau) & 0 \\
0 & L^-_{{\bf{k}},_{W}}(\tau)
\end{pmatrix},
\eea
where $L^{\pm}_{{\bf{k}},_{W}}(\tau)$ are the following $4\times 4$ operations
\bea
L^{\pm}_{{\bf{k}},_{W}}(\tau) \equiv i \begin{pmatrix}
i\mu_{{\rm m}} \mH \I & \I \p_{\tau} + i \boldsymbol{\check{\Sigma}}^{\pm}(\tau,\bk) \\
 \I \p_{\tau} - i\boldsymbol{\check{\Sigma}}^{\pm}(\tau,\bk)  & i\mu_{{\rm m}} \mH  \I
\end{pmatrix}.
\eea
Here, $\boldsymbol{\check{\Sigma}}^{\pm}(\tau,\bk)$ are given as
\Beq
\label{eq:Sigmas}
\check{\boldsymbol{\Sigma}}^+ \equiv k\boldsymbol{\sigma}^3+ (2\xpi-\frac{\xi_A}{2})\mathcal{H}\I \, \an  \check{\boldsymbol{\Sigma}}^- \equiv k\boldsymbol{\sigma}^3 + (2\xpi + \frac{\xi_A}{2})\mathcal{H}\I  -\xi_A\mathcal{H}\boldsymbol{\sigma}^1\,.
\Eeq

 Therefore, the theory in Eq. \eqref{chiral-A} splits into two subsectors as
\bea
S_{\Psi}[\tilde\Uppsi] = S_+[\Uppsi^+_{_{W}}] + S_-[\Uppsi^-_{_{W}}],
\eea
where 
\bea\label{chiral-Ah}
S_{\pm} = \int \frac{{\rm{d}}\tau {\rm{d}}k^3}{(2\pi)^3} \bar{\Uppsi}^{\pm}_{{\bf{k}},_{W}} ~ . ~ \tilde{\rm{L}}^{\pm}_{{\bf{k}},_{W}}(\tau )~.  
\Uppsi^{\pm}_{{\bf{k}},_{W}}.
\eea
The $\tilde{\rm{L}}^{\pm}_{{\bf{k}},_{W}}(\tau )$ operators are given as
\bea\label{L+W}
\tilde{\rm{L}}^{+}_{{\bf{k}},_{W}}(\tau ) & \equiv & \left[i\gamma^0_{_{\rm{W}}} \partial_{\tau} - k\gamma^3_{_{\rm{W}}} -\left(2\xpi -\frac{\xi_A}{2}\right)\mathcal{H}\gamma^0_{_{\rm{W}}}\gamma^5_{_{\rm{W}}} - \mu_{{\rm m}}\mH \II \right]\,,\\ \label{L-W}
\tilde{\rm{L}}^{-}_{{\bf{k}},_{W}}(\tau ) & \equiv & \left[i\gamma^0_{_{\rm{W}}} \partial_{\tau} - k \gamma^3_{_{\rm{W}}} +\gamma^1_{_{\rm{W}}}\xi_A\mathcal{H}-\left(2\xpi +\frac{\xi_A}{2}\right)\mathcal{H}\gamma^0_{_{\rm{W}}}\gamma^5_{_{\rm{W}}} -\mu_{{\rm m}} \mH \II \right],
\eea
where $\gamma^{\alpha}_{_{\rm{W}}}$s are the gamma matrices in the Weyl representation and notice that
\bea
\gamma^0_{_{\rm{W}}}\gamma^5_{_{\rm{W}}} = \gamma^0\gamma^5 = \begin{pmatrix}
0 & \I \\
-\I & 0
\end{pmatrix}.
\eea

Up to now, we split the 8-spinor system into two 4-spinor systems each in a Weyl representation (See Eq. \eqref{pmspinor}). For the quantization purposes, it is more convenient to write the 4-d fields in the Dirac frame. The Weyl 4-spinors can be transformed to their Dirac representation as
\bea
\Uppsi^{\pm}_{\rm{D}} = \mathcal{D} ~ \Uppsi^{\pm},
\eea
where
\bea
\mathcal{D} = \frac{1}{\sqrt{2}} \begin{pmatrix}
\I & \I \\
-\I & ~~\I
\end{pmatrix} \an \tilde{\mathcal{D}} = \I\otimes \mathcal{D}.
\eea
The gamma matrices in the Weyl representation, $\gamma^{\alpha}_{_{\rm{W}}}$, and Dirac representation, $\gamma^{\alpha}$, are related as $\gamma^{\alpha} = \mathcal{D} \gamma^{\alpha}_{_{\rm{W}}} \mathcal{D}^{-1}$. Therefore, the operator $\tilde T_{2,\bf{k}}$ defined as
\bea
\tilde T_{2,\bf{k}} \equiv \tilde{\mathcal{D}}\tilde S \tilde R_{\bf{k}},
\eea
transforms the 8-spinor in the Weyl frame to the final Dirac frame as
\bea
\tilde T_{2,{\bf{k}}}  \begin{pmatrix}
\Uppsi_{L,\bf{k}} \\
\Uppsi_{R, \bf{k}}
\end{pmatrix} =  \begin{pmatrix}
\Uppsi_{\bf{k}}^{+} \\
\Uppsi_{\bf{k}}^{-}
\end{pmatrix},
\eea
where now $\Uppsi^{\pm}_{\bf{k}}$ are each in a Dirac frame. The matrix operators from the isospin to this final extended helicity representation are transformed as
\bea\label{II-Gamma0}
&& \tilde T_{2,{\bf{k}}} ~ \I\otimes\gamma^0 ~ \tilde T_{2,{\bf{k}}}^{-1} = \I\otimes\gamma^0,\\ \label{III-Gamma0}
&& \tilde T_{2,{\bf{k}}} ~ \I\otimes\gamma^0\gamma^5 ~ \tilde T_{2,{\bf{k}}}^{-1} = \I\otimes\gamma^0\gamma^5,
\eea
and 
\bea\label{II-Gammaalpha}
\delta^{\alpha}_a \tilde T_{2,{\bf{k}}} ~ {\bf{T}}^a\otimes\gamma^{\alpha} ~ \tilde T_{2,{\bf{k}}}^{-1} = \begin{pmatrix} \frac12 \gamma^0\gamma^5 & 0 \\ 0 & -\frac12\gamma^0\gamma^5 + \gamma^1 \end{pmatrix}.
\eea
Finally, the theory in Eq. \eqref{chiral-A} splits into two subsectors in terms of the plus and minus spinors as
\bea
S_{\Psi}[\tilde\Uppsi] = S^+[\Uppsi^+] + S^-[\Uppsi^-],
\eea
where 
\bea\label{chiral-Ah}
S^{\pm} = \int \frac{{\rm{d}}\tau {\rm{d}}k^3}{(2\pi)^3} \bar{\Uppsi}^{\pm}_{{\bf{k}}} ~ . ~ \tilde{\rm{L}}^{\pm}_{{\bf{k}},}(\tau )~.  
\Uppsi^{\pm}_{{\bf{k}}}.
\eea
with $\tilde{\rm{L}}^{\pm}_{\bf{k}}(\tau)$ as
\bea\label{L+D}
\tilde{\rm{L}}^{+}_{\bf{k}}(\tau ) & \equiv & \left[i\gamma^0 \partial_{\tau} - k\gamma^3 -\left(2\xpi -\frac{\xi_A}{2}\right)\mathcal{H}\gamma^0\gamma^5 - \mu_{{\rm m}}\mH \II \right]\,,\\
\label{L-D}
\tilde{\rm{L}}^{-}_{\bf{k}}(\tau ) & \equiv & \left[i\gamma^0 \partial_{\tau} - k \gamma^3 -\left(2\xpi +\frac{\xi_A}{2}\right)\mathcal{H}\gamma^0\gamma^5 -\mu_{{\rm m}} \mH \II +\gamma^1\xi_A\mathcal{H} \right].
\eea
This completed the proof that our theory splits into two irreducible representations as
\bea
\tilde \Uppsi_{\bf{k}} = \Uppsi^+_{\bf{k}} \oplus \Uppsi^-_{\bf{k}}.
\eea

\section{Charge conjugation and Parity}
\label{CC}
In this appendix, we focus on the discrete symmetries, charge conjugation, and parity. Sec.s \ref{CC} and \ref{PP} investigate the action of the charge conjugation and parity on the theory, respectively.

\subsection{Charge conjugation}
\label{CC}

At this point, we work out the charge conjugation operator and will show it is a symmetry of the theory in Eq. \eqref{theory}, i.e.
\bea\label{theory-app}
\mathcal{L}_{\Psi} = i \bar{\tilde\Uppsi} \bigg[ \I  \p_{\tau} \otimes \gamma^0 +  \bigg( \I \p_i  - \frac{i}{2}  \xa \mH \boldsymbol{\tau}^i \bigg) \otimes \gamma_{i} + i  \mu_{{\rm{m}}} \mH \III  + 2i \xi_{\varphi} \mH \I \otimes (\gamma^0 \gamma^5) \bigg] \tilde\Uppsi.~~~
\eea
Notice that the action above is presented in the real space and isospin frame. We define the charge conjugated field as
\bea\label{Psic-Psi}
\tilde\Uppsi_C \equiv \tilde{C} ~ \bar{\tilde{\Uppsi}}^{T},
\eea
where $\tilde{C}$ is the charge conjugation operator, and $T$ superscript stands for transpose with respect to Lorentz indices. Under the action of the charge conjugation the fermion charge $\ga$ will go to $g_{A,C} =-\ga$ which implies
\bea
\xi_{A,C} = \frac{g_{A,C} \psi}{H} = -\xp.
\eea
Moreover, $\tilde{C}$ is a $8\times 8$ matrix which satisfies 
\bea
\tilde{C}^{-1} \I\otimes\gamma^{\alpha} \tilde{C} = -\I\otimes\gamma^{\alpha T} .
\eea 
The $\tilde{C}$ matrix can be written as
\bea
\tilde{C} = \I \otimes C \where C = i\gamma^2\gamma^0,
\eea
and $\tilde{C}^{-1} = \I \otimes C^{-1}$. 
Then it is straightforward to check the following equalities
\bea
\tilde{C}^{-1} \boldsymbol{\tau}^i \otimes\gamma_{i} \tilde{C} &=& -  \boldsymbol{\tau}^i \otimes\gamma_{i}^T,\\
\tilde{C}^{-1} \I \otimes (\gamma^{0}\gamma^{5}) \tilde{C} &=&  \I \otimes (\gamma^{0}\gamma^{5})^{T}.
\eea

Now lets $\tilde{C}$ acts on action in Eq. \eqref{theory-app}. Here for the $\tilde\Uppsi_C$ field with charge $-\ga$, the action is
\bea\label{theory-app-2}
\mathcal{L}_{\Psi,C} = i \bar{\tilde\Uppsi}_C \bigg[ \I  \p_{\tau} \otimes \gamma^0 +  \bigg( \I \p_i  + \frac{i}{2}  \xa \mH \boldsymbol{\tau}^i \bigg) \otimes \gamma_{i} + i  \mu_{{\rm{m}}} \mH \III  + 2i \xi_{\varphi} \mH \I \otimes (\gamma^0 \gamma^5) \bigg] \tilde\Uppsi_C,\nonumber
\eea
which can be written in terms of $\tilde{\Uppsi}$ as
\bea\label{Psi-Psi-c}
\mathcal{L}_{\Psi,C} &=& - i \tilde{\Uppsi}^{T}\tilde{C}^{-1} \bigg[ \I  \p_{\tau} \otimes \gamma^0 +  \bigg( \I \p_i  + \frac{i}{2}  \xa \mH \boldsymbol{\tau}^i \bigg) \otimes \gamma_{i} + i  \mu_{{\rm{m}}} \mH \III  + 2i \xi_{\varphi} \mH \I \otimes (\gamma^0 \gamma^5) \bigg] \tilde{C}\bar{\tilde{\Uppsi}}^{T} \nonumber\\
&=&
 i \tilde{\Uppsi}^{T} \bigg[ \I  \p_{\tau} \otimes \gamma^{0T} +  \bigg( \I \p_i  + \frac{i}{2}  \xa \mH \boldsymbol{\tau}^i \bigg) \otimes \gamma_i^{T} - i  \mu_{{\rm{m}}} \mH \III  - 2i \xi_{\varphi} \mH \I \otimes (\gamma^0 \gamma^5)^{T} \bigg] \bar{\tilde{\Uppsi}}^{T}\nonumber\\
 &=&
- i \bar{\tilde{\Uppsi}} \bigg[ \I  \cev{\p}_{\tau} \otimes \gamma^{0} +  \bigg( \I \cev{\p}_i  + \frac{i}{2}  \xa \mH \boldsymbol{\tau}^i \bigg) \otimes \gamma_i - i  \mu_{{\rm{m}}} \mH \III  - 2i \xi_{\varphi} \mH \I \otimes (\gamma^0 \gamma^5) \bigg] \tilde{\Uppsi}\nonumber\\
&=& i \bar{\tilde{\Uppsi}} \bigg[ \I  \p_{\tau} \otimes \gamma^{0} +  \bigg( \I \p_i  - \frac{i}{2}  \xa \mH \boldsymbol{\tau}^i \bigg) \otimes \gamma_i + i  \mu_{{\rm{m}}} \mH \III  + 2i \xi_{\varphi} \mH \I \otimes (\gamma^0 \gamma^5) \bigg] \tilde{\Uppsi},
\eea
where the second to third line has passed by  transposition and using the fact that $\Uppsi$ and $\bar{\Uppsi}$ anti-commute and in the fourth line we dropped the total derivative terms. By an abuse of notation, it implies 
\bea
\mathcal{L}_{\Psi, C} = \mathcal{L}_{\Psi}.
\eea
That completed our proof that our action is invariant under the charge conjugation and therefore C-symmetric.

Equation \eqref{Psic-Psi} implies that the Fourier modes of the spinor and its charge conjugated field are related as
\bea
\tilde\Uppsi_{C{\bf{k}}} = i \I\otimes \gamma^2 ~ \tilde{\Uppsi}^*_{-{\bf{k}}}.
\eea
Going to the c-helicity frame, the above reduces to the following form
\bea
\Uppsi_{C{\bf{k}}}^{\pm} = i \gamma^2 ~ \Uppsi^{\pm*}_{-{\bf{k}}}.
\eea

\subsection{Parity}\label{PP}

Now lets explore the action of parity on the setup, which takes $x^{\mu}=(\tau,{\bf{x}})$ to $x^{\mu}_P=(\tau,-{\bf{x}})$, and hence 
\bea
\p_i \rightarrow \p_{i,P}\equiv -\p_i.
\eea
Besides, the gamma matrices under $P$ transform as
\bea
P \gamma^0 P^{-1} &=& \gamma^0 \an P \gamma^i P^{-1} = -\gamma^i.
\eea
The VEV of the axion and $SU(2)$ fields transform as
\bea
P \varphi(t) P^{-1} &=& -\varphi(t),\\
P A^a_{i}(t) P^{-1} &=& -A^a_{i}(t),
\eea
which implies that $\xi_{A,P}=-\xp$, and $\xi_{\varphi,P}=-\xpi$. Under the action of parity, the spinor transforms as
\bea
\tilde{\Uppsi}_{P}(\tau,{\bf{x}}) \equiv \tilde{P}\tilde{\Uppsi}(\tau,{\bf{x}}) = \tilde{\Uppsi}(\tau,-{\bf{x}}),
\eea
where $\tilde{P}$ is the $8\times 8$ parity matrix given as
\bea
\tilde{P} = \I \otimes P,
\eea

The theory in Eq. \eqref{theory-app} can be written in terms of the parity transformed field as
\bea\label{theory-P1}
 \mathcal{L}_{\Psi}(\tau,{\bf{x}}) &=& i \bar{\tilde\Uppsi}_{P} \tilde{P} \bigg[ \I  \p_{\tau} \otimes \gamma^0 +  \bigg( \I \p_i  - \frac{i}{2}  \xa \mH \boldsymbol{\tau}^i \bigg) \otimes \gamma_{i} + i  \mu_{{\rm{m}}} \mH \III  + 2i \xi_{\varphi} \mH \I \otimes (\gamma^0 \gamma^5) \bigg] \tilde{P}^{-1} \tilde\Uppsi_{P}\nonumber\\
&=& i \bar{\tilde\Uppsi}_{P} \bigg[ \I \p_{\tau} \otimes \gamma^0  + \bigg(   \I \p_{i,P} - \frac{i}{2}  \xa \mH \boldsymbol{\tau}^i \bigg) \otimes \gamma_{i} + i  \mu_{{\rm{m}}} \mH \III  + 2i \xi_{\varphi} \mH \I \otimes (\gamma^0 \gamma^5) \bigg] \tilde\Uppsi_{P}\nonumber\\
&=& i \bar{\tilde\Uppsi}_{P} \bigg[ \I \p_{\tau} \otimes \gamma^0  + \bigg( \I \p_{i,P} + \frac{i}{2}  \xi_{A,P} \mH \boldsymbol{\tau}^i \bigg) \otimes \gamma_{i} + i  \mu_{{\rm{m}}} \mH \III  - 2i \xi_{\varphi,P} \mH \I \otimes (\gamma^0 \gamma^5) \bigg] \tilde\Uppsi_{P}, \nonumber\\
\eea
which implies that
\bea
 \mathcal{L}_{\Psi}(\tau, {\bf{x}}) &\neq & \mathcal{L}_{\Psi}(\tau,-{\bf{x}}).
\eea 
Therefore, the axion-$SU(2)$ gauge field vacuum spontaneously broke the parity in the fermionic sector

\section{Current: point-splitting regularization}\label{sec:current-comp}
In this appendix, we present the details of computation of the current integral, $\mathcal{K}^{+}(\tau)$, in Eq. \eqref{K+tau}. Here, we use the point-splitting technique to renormalize $\mathcal{K}^{+}(\tau;\varepsilon)$ as
\bea
\mathcal{K}^{+}(\tau;\varepsilon) = - {\rm{symm}\lim_{\varepsilon\rightarrow 0}} ~  \sum_{s=\pm}  \frac{a~s~(2\pi)^2/H^3}{2(a_f a_b)^{2}} \int {\rm{d}}^3k\bigg[ u^{\uparrow*}_{s}(k,\tau_f)u^{\downarrow}_{s}(k,\tau_b) + u^{\downarrow*}_{s}(k,\tau_f)u^{\uparrow}_{s}(k,\tau_b)\bigg] ,\nonumber
\eea
where from Eq. \eqref{eq-tau-a-r} we have $$\tau_{f}=( 1 -\frac{\varepsilon}{2})\tau \an \tau_{b}=( 1 + \frac{\varepsilon}{2})\tau,$$
while $a_f\equiv a(\tau_f)$ and $a_b\equiv a(\tau_b)$ are the forward and backward scale factors, and the symmetrization is done under $\varepsilon\rightarrow -\varepsilon$. Using Eq.s \eqref{eq:uupdowngen} and \eqref{Ys-Zs}, we can write $\mathcal{K}^{+}$ as 
\bea\label{K++++}
\mathcal{K}^{+}(\tau;\varepsilon) = ~  \frac{1}{(2\pi)^2} \sum_{s=\pm}   is  \bigg[ e^{i\kappa_{s}\pi} \mathcal{I}_{\kappa_s,\mu^+} + \mu_{\rm{m}}^2 e^{i\tilde{\kappa}_{s}\pi}  \mathcal{I}_{\tilde\kappa_s,\mu^+} \bigg],
\eea
where for a given $\kappa$ and $\mu$, we define 
$\mathcal{I}_{\kappa,\mu}(\tau;\varepsilon) $ as
\bea\label{math-I-}
\mathcal{I}_{\kappa,\mu}(\tau;\varepsilon) \equiv (2\pi)^2 \big((1+\frac{\varepsilon}{2})(1-\frac{\varepsilon}{2})\big)^{\frac32}  {\rm{symm}\lim_{\varepsilon\rightarrow 0}} \int^{\infty}_0 \tau^2 k{\rm{d}}k \bigg[ W^*_{\kappa,\mu}(-2ik\tau_f)W_{\kappa,\mu}(-2ik\tau_b)  \bigg].\nonumber\\
\eea
Here we recall that $\mu^+=i\lvert \mu^+\rvert$, $\kappa^{+}_s = \frac12 + is\kappa_I^{+}$ and $\tilde\kappa^{+}_s = -\frac12 + is\kappa_I^{+}$. 
For simplicity, until otherwise stated, we present both $\kappa_s$ and $\tilde{\kappa}_s$ with $\kappa$ and $\mu^{+}$ by $\mu$. Moreover, we drop the $ {\rm{symm}\lim_{\varepsilon\rightarrow 0}}$ while we will keep in mind that all the terms should be symmetrized with respect to the sign of $\varepsilon$, and eventually we take the $\varepsilon$ goes to zero limit.

Doing the momentum integral, we arrive at
\bea\label{I-kappa-mu}
\mathcal{I}_{\kappa,\mu}(\tau;\varepsilon) =  \frac{ \big[(1+\frac{\varepsilon}{2})(1-\frac{\varepsilon}{2})\big]^{\frac32}\mathcal{G}_{\kappa,\mu}(\tau;\varepsilon)  }{\Gamma(\frac12+\mu-\kappa)\Gamma(\frac12-\mu-\kappa)\Gamma^*(\frac12+\mu-\kappa)\Gamma^*(\frac12-\mu-\kappa)},
\eea
where $\mathcal{G}_{\kappa,\mu}(\tau;\varepsilon)$ is
\bea \label{curl-G}
\mathcal{G}_{\kappa,\mu}(\tau;\varepsilon) \equiv  \frac14  \int_{\mathcal{C}_{\alpha}} {\rm{d}}\alpha  \bigg(-\frac{1}{1-\frac{\varepsilon}{2}}\bigg)^{\alpha} \Gamma(\frac12+\mu+\alpha)\Gamma(\frac12-\mu+\alpha)\Gamma(-\kappa-\alpha) G(\alpha),
\eea
and $G(\alpha)$ is defined as
\bea
G(\alpha) \equiv  \int_{\mathcal{C}_{\beta}} {\rm{d}}\beta  \big(1+\frac{\varepsilon}{2}\big)^{-\beta}\Gamma(-\kappa^*-\beta) \Gamma(\frac12-\mu+\beta)\Gamma(\frac12+\mu+\beta) \bigg(\frac{\varepsilon}{2}\bigg)^{\alpha+\beta-2} \Gamma(2-\alpha-\beta). \nonumber
\eea
The contours $\mathcal{C}_{\alpha}$ and $\mathcal{C}_{\beta}$ can be closed in either the left or right half-plane as far as they separate the poles of $\Gamma(\frac12+\beta\pm\mu)$ and $\Gamma(-\kappa-\beta)$. (See Fig. \ref{fig:poles})

In doing the complex integral in $G(\alpha)$ and closing the integral on the right-half plane, we have two types of poles
\bea\label{beta1T}
\beta_{1n} &\equiv & -\kappa^* +n,\\ \label{beta2T}
\beta_{2n} &\equiv & 2 - \alpha + n ~~ \where 2 + n - {\rm{Re}}\alpha >0,
\eea
where $n\in \mathbb{N}$.
Therefore, we can write $\mathcal{G}_{\kappa,\mu}$ as
\bea\label{Gmukappa}
&& \mathcal{G}_{\kappa,\mu}(\tau;\varepsilon) = -\frac{2i\pi}{4} \bigg(\frac{\varepsilon}{2}\bigg)^{-2} \sum_{n=0}^{\infty} \frac{1}{n!} \bigg(\frac{-\varepsilon/2}{1+\frac{\varepsilon}{2}}\bigg)^{n} \int_{\mathcal{C}_{\alpha}} {\rm{d}}\alpha  \bigg(\frac{-\varepsilon/2}{1-\frac{\varepsilon}{2}}\bigg)^{\alpha} \Gamma(\frac12+\mu+\alpha)\Gamma(\frac12-\mu+\alpha) \nonumber\\
&&  \Gamma(-\kappa-\alpha) \bigg[ \bigg(\frac{\varepsilon/2}{1+\frac{\varepsilon}{2}}\bigg)^{2-\alpha}  \Gamma(-\kappa^*-2-n+\alpha) \Gamma(\frac12-\mu+2+n-\alpha)\Gamma(\frac12+\mu+2+n-\alpha)
  \nonumber\\
&&+  \bigg(\frac{\varepsilon/2}{1+\frac{\varepsilon}{2}}\bigg)^{-\kappa^*}  \Gamma(2-\alpha+\kappa^*-n) \Gamma(\frac12-\mu-\kappa^*+n)\Gamma(\frac12+\mu-\kappa^*+n)\bigg].
\eea
For technical convenience we decompose $\mathcal{G}_{\kappa,\mu}(\tau;\varepsilon)$ as
$$\mathcal{G}_{\kappa,\mu}(\tau;\varepsilon) \equiv \mathcal{G}_1(\tau;\varepsilon) + \mathcal{G}_2(\tau;\varepsilon),$$
where $\mathcal{G}_i(\tau;\varepsilon)$ are the contributions of the $i$th term inside the square bracket to the $\mathcal{G}(\tau;\varepsilon)$. In the following we compute $\mathcal{G}_1(\tau;\varepsilon)$, and $\mathcal{G}_2(\tau;\varepsilon)$ respectively.

$\bullet$  $\boldsymbol{\mathcal{G}_1(\tau;\varepsilon)}:$ \hspace{0.3cm} Taking $\varepsilon \rightarrow 0$ and closing the contour on the left-half plane, the first term inside the square-bracket in Eq. \eqref{Gmukappa} gives 
\bea
&& \mathcal{G}_1 = -\frac{2i\pi}{4}  \int_{\mathcal{C}_{\alpha}} {\rm{d}}\alpha  (-1)^{\alpha} \Gamma(\frac12+\mu+\alpha) \Gamma(\frac12-\mu-\alpha)  \Gamma(\frac12-\mu+\alpha) \Gamma(\frac12+\mu-\alpha) \nonumber\\
&& \Gamma(-\kappa-\alpha) \Gamma(1+\kappa+\alpha)  \frac{ \Gamma(-\kappa^*+\alpha-2)}{\Gamma(1+\kappa+\alpha)} \big((\frac32-\alpha)^2-\mu^2\big)\big((\frac12-\alpha)^2-\mu^2\big), ~~~~~
\eea
which has an infinite number of type $\alpha_{\pm,q}=-q-\frac12\mp \mu$, and $\alpha_{0,q}=-q-\kappa-1$ with $q\in \mathbb{N}$. Therefore, it is more convenient to write the above as
\bea\label{G1-1}
&& \mathcal{G}_1 = -\frac{2i\pi}{4}  \int_{\mathcal{C}_{\alpha}} {\rm{d}}\alpha  (-1)^{\alpha} \Gamma(\frac12+\mu+\alpha) \Gamma(\frac12-\mu-\alpha)  \Gamma(\frac12-\mu+\alpha) \Gamma(\frac12+\mu-\alpha) \nonumber\\
&& \Gamma(-\kappa-\alpha) \Gamma(1+\kappa+\alpha) \bigg[w(\alpha)-w(\alpha-1)+\frac{\mathcal{A}_{\kappa,\mu}}{\alpha+\kappa}\bigg], ~~~~~
\eea
where $w(\alpha)$ and $\mathcal{A}_{\kappa,\mu}$ for each of $\reK=\pm\frac12$ and in terms of $f(\alpha)$
\bea
f(\alpha) =  \big((\frac32-\alpha)^2-\mu^2\big)\big((\frac12-\alpha)^2-\mu^2\big),
\eea
 are respectively as follows. 
  In case of $\reK=\frac12$, we have 
\bea\label{w-reK-1/2}
&& w(\alpha) = -\frac{\frac16 f(\alpha)+2(1-2\kappa)}{\alpha+\kappa} + \frac{ \frac12 f(\alpha+1) - \frac16 f(\alpha+2)}{\alpha+\kappa-1} - \frac{\frac16 f(\alpha+1)}{\alpha+\kappa-2} + 4 \alpha,~~~~~~\\
&& \mathcal{A}_{\kappa,\mu} =  2(1-2\kappa),
\eea
For $\reK=-\frac12$, $ w(\alpha)$ and $\mathcal{A}_{\kappa,\mu}$ are
\bea\label{w-reK--1/2}
&& w(\alpha) =  - \frac{f(\alpha+1)}{\alpha+\kappa} + \frac43 \alpha \bigg(\alpha^2 - \frac34(1+2\kappa)\alpha + 3 (\kappa^{2}-\mu^2 + \kappa + \frac16) \bigg) ,~~~~~~~~~\\ \label{cA-reK--1/2}
&& \mathcal{A}_{\kappa,\mu} =  - \frac12( 1+2\kappa)\bigg( 1  + 4 \kappa + 4 \kappa^{2} - 4 \mu^2 \bigg).
\eea

We can further simply Eq. \eqref{G1-1} as
$$\mathcal{G}_1 \equiv \mathcal{G}'_1 + \mathcal{G}''_1,$$
where $\mathcal{G}'_1$ is
\bea\label{G1-'}
&& \mathcal{G}'_1 = - \frac{2i\pi}{4}  \bigg(\int_{\mathcal{C}_{\alpha}}  {\rm{d}}\alpha - \int_{\mathcal{C}_{\alpha}-1}  {\rm{d}}\alpha \bigg)   (-1)^{\alpha} w(\alpha) \Gamma(\frac12+\mu+\alpha) \Gamma(\frac12-\mu-\alpha)  \Gamma(\frac12-\mu+\alpha) \nonumber\\
&& \Gamma(\frac12+\mu-\alpha) \Gamma(-\kappa-\alpha) \Gamma(1+\kappa+\alpha), 
\eea
and $\mathcal{G}''_1$ is
\bea\label{G1-''}
&& \mathcal{G}''_1 =  - \frac{2i\pi\mathcal{A}_{\kappa,\mu}}{4}  \int_{\mathcal{C}_{\alpha}} {\rm{d}}\alpha  \frac{(-1)^{\alpha}\Gamma(-\alpha-\kappa)\Gamma(\alpha+\kappa+1)}{(\alpha+\kappa)} \Gamma(\frac12+\mu+\alpha) \Gamma(\frac12-\mu-\alpha)   \nonumber\\
&& \Gamma(\frac12-\mu+\alpha)   \Gamma(\frac12+\mu-\alpha).~~~~~
\eea
Doing the integral in Eq. \eqref{G1-'} and closing the contour on the left-half plane, we have only three poles, $\alpha_{\pm}=-\frac12\pm \mu$ and $\alpha_0=-1-\kappa$, which lead to 
\bea\label{G1'-}
&& \frac{e^{i\kappa\pi}\mathcal{G}'_1/(2\pi)^2}{\Gamma(\frac12-\kappa+\mu) \Gamma(\frac12+\kappa-\mu) \Gamma(\frac12-\kappa-\mu) \Gamma(\frac12+\kappa+\mu)} = \nonumber\\
&& \frac{1}{4}   \bigg[ w(-1-\kappa) + \frac{i}{2} \bigg( w(-\frac12+\mu)  \frac{e^{2i\pi\kappa}+e^{2i\pi\mu}}{\sin(2\mu\pi)}  -   w(-\frac12-\mu)  \frac{e^{2i\pi\kappa}+e^{-2i\pi\mu}}{\sin(2\mu\pi)} \bigg)\bigg]. ~~~~~
\eea
However, $\mathcal{G}_1''$ includes an infinite number of simple poles
$$\alpha_{n}^{\pm} = -\frac12 \pm \mu -n \an \alpha^0_n= -1-n - \kappa,$$
where $n\in \mathbb{N}$. The contribution of $\alpha_n^{0}$-poles in $\mathcal{G}_1''$ is 
\bea\label{G1''1st}
 \frac{ e^{i\kappa\pi}/(2\pi)^2 ~ \mathcal{G}''_{1}\lvert_{\alpha^0}  }{\Gamma(\frac12+\mu+\kappa) \Gamma(\frac12-\mu-\kappa) \Gamma(\frac12-\mu+\kappa) \Gamma(\frac12+\mu-\kappa)} 
  =   \frac{ A_{\kappa,\mu}}{4} \psi^{(0)}(1).\nonumber\\
\eea
 Besides, the contribution of $\alpha^{\pm}_n$-poles gives 
\bea\label{G1''2nd}
&& \frac{e^{i\kappa\pi}/(2\pi)^2 ~ \mathcal{G}_1''\lvert_{\alpha^{\pm}} }{ \Gamma(\frac12+\mu+\kappa) \Gamma(\frac12-\mu-\kappa) \Gamma(\frac12-\mu+\kappa) \Gamma(\frac12+\mu-\kappa)} = 
 \nonumber\\
&&  \frac{i\mathcal{A}_{\kappa,\mu} }{8}    \sum_{n=0}^{\infty} \bigg( \frac{(e^{2i\kappa\pi}+e^{2i\mu\pi})}{\sin(2\mu\pi)}\frac{1}{(\mu + \kappa -\frac12 -n)} -
\frac{(e^{2i\kappa\pi}+e^{-2i\mu\pi})}{\sin(2\mu\pi)}\frac{1}{(-\mu + \kappa -\frac12 -n)} \bigg).~~~~~~~~~
\eea
Finally, from the combination of Eq.s \eqref{G1'-}-\eqref{G1''2nd}, we have 
\bea\label{G1---}
&& \frac{e^{i\kappa\pi}\mathcal{G}_1/(2\pi)^2}{\Gamma(\frac12-\kappa+\mu) \Gamma(\frac12+\kappa-\mu) \Gamma(\frac12-\kappa-\mu) \Gamma(\frac12+\kappa+\mu)} = \nonumber\\
&& \frac{1}{4}   \bigg[ w(-1-\kappa) + \frac{i}{2} \bigg( w(-\frac12+\mu)  \frac{e^{2i\pi\kappa}+e^{2i\pi\mu}}{\sin(2\mu\pi)}  -   w(-\frac12-\mu)  \frac{e^{2i\pi\kappa}+e^{-2i\pi\mu}}{\sin(2\mu\pi)} \bigg)  +  \mathcal{A}_{\kappa,\mu} \bigg\{ \psi^{(0)}(1) \nonumber\\ 
&& + \frac{i}{2\sin(2\mu\pi)} \bigg( (e^{2i\kappa\pi}+e^{2i\mu\pi}) \psi^{(0)}(\frac12-\kappa-\mu)  - (e^{2i\kappa\pi}+e^{-2i\mu\pi}) \psi^{(0)}(\frac12-\kappa+\mu)\bigg) \bigg\} \bigg]. \nonumber\\
\eea
We computed $\mathcal{G}_1$ and now we turn to compute $\mathcal{G}_2$.

$\bullet\bullet$  $\boldsymbol{\mathcal{G}_2(\tau;\varepsilon):}$ \hspace{0.3cm} Now we turn to compute the contribution of the second term inside the square-brackets of $\mathcal{G}_{\kappa,\mu}$ in Eq. \eqref{Gmukappa}, i.e. $\mathcal{G}_2$ as
\bea\label{G2}
&& \mathcal{G}_2 \equiv  -\frac{2i\pi}{4} \bigg(\frac{\varepsilon}{2}\bigg)^{-2} \bigg(\frac{\frac{\varepsilon}{2}}{1+\frac{\varepsilon}{2}}\bigg)^{-\kappa^*}  
\sum_{n=0}^{\infty} \frac{1}{n!} \bigg(-\frac{\frac{\varepsilon}{2}}{1+\frac{\varepsilon}{2}}\bigg)^{n} \Gamma(\frac12-\mu-\kappa^*+n)\Gamma(\frac12+\mu-\kappa^*+n)
 \nonumber\\
&&
\int_{\mathcal{C}_{\alpha}} {\rm{d}}\alpha  \bigg(-\frac{\frac{\varepsilon}{2}}{1-\frac{\varepsilon}{2}}\bigg)^{\alpha} \Gamma(\frac12+\mu+\alpha)\Gamma(\frac12-\mu+\alpha)   \Gamma(-\kappa-\alpha)  \Gamma(2-\alpha+\kappa^*-n).
\eea 
Closing the contour in the right-half plane, the above complex integral, lets call $\mathcal{G}_2$, can be decomposed as 
$$\mathcal{G}_{2} \equiv \mathcal{G}'_{2}  + \mathcal{G}''_{2},$$
in which $\mathcal{G}_2'$ includes the contribution of simple poles of $\Gamma(-\kappa-\alpha)$, and  $\mathcal{G}_2''$ which includes the second order poles. More precisely, for $m\in \mathbb{N}$, the poles are
\bea
&&\textit{Simple poles:} \qquad \qquad ~\alpha_m=-\kappa+m \quad  \where 0\leq m< 2+2\reK-n,\nonumber\\
&&\textit{Second-order poles:} \quad \alpha_m=2+2\reK-\kappa-n+m \where n\leq 2+2\reK+m.\nonumber
\eea

The contribution of simple poles to $\mathcal{G}_2$ gives $\mathcal{G}'_2$ as
\bea
&& \mathcal{G}'_2 = -\frac{(2\pi)^2 (-1)^{-\kappa}}{4} \bigg(\frac{\varepsilon}{2}\bigg)^{-2-2\reK} \bigg(\frac{1-\frac{\varepsilon}{2}}{1+\frac{\varepsilon}{2}}\bigg)^{\kappa} (1+\frac{\varepsilon}{2})^{2\reK} 
\sum_{n=0}^{1+2\reK} \frac{1}{n!} \bigg(-\frac{\frac{\varepsilon}{2}}{1+\frac{\varepsilon}{2}}\bigg)^{n} \nonumber\\
&& \Gamma(\frac12-\mu-\kappa^*+n)\Gamma(\frac12+\mu-\kappa^*+n)
\sum_{m=0}^{1+2\reK-n} \frac{1}{m!} \bigg(\frac{\frac{\varepsilon}{2}}{1-\frac{\varepsilon}{2}}\bigg)^{m}  \Gamma(2+2\reK-n-m) \nonumber\\
&& \Gamma(\frac12+\mu-\kappa+m)\Gamma(\frac12-\mu-\kappa+m).
\eea
Doing the series in the above, we find $\mathcal{G}'_2$ as
\bea\label{G1'-final}
&& \lim_{\varepsilon\rightarrow 0} \frac{e^{i\kappa\pi} \big[(1+\frac{\varepsilon}{2})(1-\frac{\varepsilon}{2})\big]^{\frac32}\mathcal{G}'_2/(2\pi)^2}{\Gamma(\frac12+\mu-\kappa)\Gamma(\frac12-\mu-\kappa)\Gamma(\frac12-\mu-\kappa^*)\Gamma(\frac12+\mu-\kappa^*) } = \frac{\mathcal{K}'}{4},
\eea
where $\mathcal{K}'$ takes the following forms for $\reK=\pm\frac12$ respectively.
In case of $\reK=\frac12$ and after symmetrization, we find
\bea\label{K'1st}
&& \mathcal{K}' =  \frac{(2\kappa-1)}{2} (1-4\kappa+4\kappa^2-4\mu^2).
\eea
For $\reK=-\frac12$, $\mathcal{K}' $ vanishes
\bea\label{K'2nd}
\mathcal{K}' = 0.
\eea
As we see in Eq.s \eqref{K'1st} and \eqref{K'2nd}, $\mathcal{G}'_2$ only includes finite terms.
 
 Now we turn to compute the contribution of second order poles in $\mathcal{G}_2$ which specifies $\mathcal{G}''_2$ as 
\bea\label{G2''}
&& \mathcal{G}''_2 \equiv  \frac{(2\pi)^2}{4} \bigg(\frac{\varepsilon}{2}\bigg)^{-2}   
\sum_{q=0}^{\infty} \sum_{n=0}^{2+2\reK+q}  \frac{(-1)^n}{n!} \bigg(\frac{\frac{\varepsilon}{2}}{1+\frac{\varepsilon}{2}}\bigg)^{n-\kappa^*} \Gamma(\frac12-\mu-\kappa^*+n)\Gamma(\frac12+\mu-\kappa^*+n)
 \nonumber\\
&&
\frac{d}{{\rm{d}}\alpha} \bigg[ (\alpha-\alpha_{0,q})^2 \bigg(-\frac{\frac{\varepsilon}{2}}{1-\frac{\varepsilon}{2}}\bigg)^{\alpha} \Gamma(\frac12+\mu+\alpha)\Gamma(\frac12-\mu+\alpha)   \Gamma(-\kappa-\alpha)  \Gamma(2-\alpha+\kappa^*-n)\bigg]_{\alpha=\alpha_{0,q}},\nonumber
\eea 
where $\alpha_{0,q}=2+\kappa^*-n+q$.
After taking the $\varepsilon$ goes to zero limit, we can write $\mathcal{G}''_2$ as 
\bea\label{G2''-final}
&& \frac{e^{i\kappa\pi} \lim_{\varepsilon\rightarrow 0} \mathcal{G}''_2/(2\pi)^2}{ \Gamma(\frac12-\mu-\kappa)\Gamma(\frac12+\mu-\kappa)\Gamma(\frac12+\mu+\kappa)\Gamma(\frac12-\mu+\kappa)} = \nonumber\\
&& 
\frac{1}{4}    
\bigg( \mathcal{A}_{\kappa,\mu}  \bigg[  \ln\big( \frac{\varepsilon}{2}\big) + i \pi
 + \psi^{(0)}(\frac12+\mu-\kappa) + \psi^{(0)}(\frac12-\mu-\kappa) - 2\psi^{(0)}(1) \bigg] \nonumber\\
 && -w(-1-\kappa) + \mathcal{K}''  \bigg), 
\eea
where $\mathcal{A}_{\kappa,\mu}$ and $w(-1-\kappa)$ for $\reK=\pm \frac12$ are given in Eq.s \eqref{w-reK-1/2}-\eqref{cA-reK--1/2} and $\mathcal{K}''$ is as follows.
In case of $\reK=\frac12$, $\mathcal{K}''$ is
\bea
\mathcal{K}'' = \frac43 \frac{(1-2\kappa)(7-14\kappa+8\kappa^2-4\mu^2)}{(1-4\kappa+4\kappa^2-4\mu^2)},
\eea
and in case of $\reK=-\frac12$, it is
\bea
\mathcal{K}'' = -\frac{1}{4}(1+2\kappa)(9+\frac{56}{3}\kappa+\frac{20}{3}\kappa^2 +  4\kappa^2-4\mu^2).
\eea
From the combination of Eq.s \eqref{G1'-final} and 
\eqref{G2''-final}, we have $ \mathcal{G}_2$ as 
\bea\label{G2-final}
&& \frac{e^{i\kappa\pi} [(1+\frac{\varepsilon}{2})(1-\frac{\varepsilon}{2})]^{\frac32} \mathcal{G}_2/(2\pi)^2}{\Gamma(\frac12+\mu-\kappa)\Gamma(\frac12-\mu-\kappa)\Gamma(\frac12-\mu+\kappa)\Gamma(\frac12+\mu+\kappa) } = \frac{1}{4}\bigg[ \frac{\Gamma(\frac12-\mu-\kappa^*)\Gamma(\frac12+\mu-\kappa^*) }{\Gamma(\frac12-\mu+\kappa)\Gamma(\frac12+\mu+\kappa) } \mathcal{K}' \nonumber\\
&& +  \mathcal{A}_{\kappa,\mu}  \bigg( \ln\big( \frac{\varepsilon}{2}\big)  + i \pi
 + \psi^{(0)}(\frac12+\mu-\kappa) + \psi^{(0)}(\frac12-\mu-\kappa) - 2\psi^{(0)}(1) \bigg) + \mathcal{K}'' -w(-1-\kappa)\bigg]. \nonumber\\
\eea

\subsection*{$\clubsuit$ $\boldsymbol{\mathcal{G}_{\kappa,\mu}(\tau;\varepsilon):} $} 
Working out both $\mathcal{G}_1$ and $\mathcal{G}_2$, we are now ready to compute $\mathcal{G}_{\kappa,\mu}(\tau;\varepsilon)$.
  From the combination of Eq.s \eqref{G1---} and \eqref{G2-final}, we find $\mathcal{G}_{\kappa,\mu}(\tau;\varepsilon)$ as
\bea\label{G---}
&& \frac{4 e^{i\kappa\pi} \big[(1+\frac{\varepsilon}{2})(1-\frac{\varepsilon}{2})\big]^{\frac32} \mathcal{G}_{\kappa,\mu}(\tau;\varepsilon)/(2\pi)^2}{\Gamma(\frac12-\kappa+\mu) \Gamma(\frac12-\kappa-\mu) \Gamma^*(\frac12-\kappa-\mu) \Gamma^*(\frac12-\kappa+\mu)} =  \nonumber\\ 
&&   \mathcal{K}' +  \frac{\Gamma(\frac12-\mu+\kappa)\Gamma(\frac12+\mu+\kappa) }{\Gamma(\frac12-\mu-\kappa^*)\Gamma(\frac12+\mu-\kappa^*) } \bigg\{ \mathcal{A}_{\kappa,\mu} \ln\big(\frac{\varepsilon}{2}\big)  +  \mathcal{K}''  - \Delta_1 +  \big( \frac{e^{2i\pi\kappa}+ \cosh2\pi\lvert\mu\rvert}{\sinh(2\lvert\mu\rvert\pi)} \big) \Delta_2 \nonumber\\
&& + \mathcal{A}_{\kappa,\mu}  \bigg[  i \pi - \psi^{(0)}(1)  +  \frac{e^{2i\kappa\pi}+e^{-2i\mu\pi}}{2\sinh(2\lvert\mu\rvert\pi)} \psi^{(0)}(\frac12-\kappa-\mu)  - \frac{e^{2i\kappa\pi}+e^{2i\mu\pi}}{2\sinh(2\lvert\mu\rvert\pi)} \psi^{(0)}(\frac12-\kappa+\mu)\bigg] \bigg\}, \nonumber\\
 \eea
in which we used $\mu=i\lvert \mu \rvert$ and 
$ \Delta_1$ and $\Delta_2$ are
\bea
&& \Delta_1 \equiv \frac12 \bigg[ w(-\frac12+\mu) + w(-\frac12-\mu) \bigg],\\
&& \Delta_2 \equiv \frac12 \bigg[ w(-\frac12+\mu) - w(-\frac12-\mu)\bigg].
\eea

\subsection*{$\clubsuit\clubsuit$ $\boldsymbol{\mathcal{K}^{+}(\tau;\varepsilon)}:$ }

Upon computing Eq. \eqref{G---} for $\kappa=\kappa_s$, we can compute $\mathcal{I}_{\kappa_s,\mu}(\tau;\varepsilon)$ as
\bea\label{Imukappa---}
&& 4 \frac{e^{i\kappa_s\pi}}{(2\pi)^2} \mathcal{I}_{\kappa_s,\mu}(\tau;\varepsilon) =  \frac{(1-2\kappa_s)(1-4\kappa_s+4\kappa_s^2-4\mu^2)}{2}  \bigg[  \ln\big(\frac{\varepsilon}{2}\big)   + i \pi -\psi^{(0)}(1) \bigg] 
 + 4(\frac12-\kappa_s)  \nonumber\\
&& \times[(\frac12-\kappa_s)^2-\mu^2]  \bigg[  
   \frac{e^{2i\kappa_s\pi}+e^{-2i\mu\pi}}{2\sinh(2\lvert\mu\rvert\pi)} \psi^{(0)}(\frac12-\kappa_s-\mu)  - \frac{e^{2i\kappa_s\pi}+e^{2i\mu\pi}}{2\sinh(2\lvert\mu\rvert\pi)} \psi^{(0)}(\frac12-\kappa_s+\mu) \bigg]
 \nonumber\\
&& +  1 - \frac{11}{3}\kappa_s + 4\kappa_s^2 - \frac43\kappa_s^3
+ \frac23\mu \big(4-9\kappa_s+6\kappa_s^2-4\mu^2\big) \bigg( \frac{e^{2i\pi\kappa_s}+ \cosh2\pi\lvert\mu\rvert}{\sinh(2\lvert\mu\rvert\pi)} \bigg).
\eea
Recalling that $\kappa_s=\frac12+is\kappa_I$ and summing over $s$, we arrive at
\bea\label{Imukappa---+}
&& 4 \sum_{s=\pm} is \frac{e^{i\kappa_s\pi}}{(2\pi)^2} \mathcal{I}_{\kappa_s,\mu}(\tau;\varepsilon) = 8\kappa_I (\lvert \mu\rvert^2 - \kappa_I^2)  \bigg[  \ln\big(\frac{\varepsilon}{2}\big) - \psi^{(0)}(1) + i \pi \bigg] + \frac43 \kappa_I (1-2\kappa_I^2)  \nonumber\\
&& + 4\kappa_I (\lvert \mu\rvert^2 - \kappa_I^2) \sum_{s=\pm}  \bigg[  
   \frac{e^{2\lvert\mu\rvert\pi}-e^{-2s\kappa_I\pi}}{2\sinh(2\lvert\mu\rvert\pi)} \psi^{(0)}(-is\kappa_I-i\lvert\mu\rvert)  - \frac{e^{-2\lvert\mu\rvert\pi}-e^{-2s\kappa_I\pi}}{2\sinh(2\lvert\mu\rvert\pi)} \psi^{(0)}(-is\kappa_I+ i\lvert\mu\rvert) 
 \bigg] \nonumber\\
&& 
+ 4i\kappa_I\lvert \mu\rvert \bigg( \frac{ \cosh2\pi\lvert\mu\rvert  - \cosh(2\kappa_I\pi)}{\sinh(2\lvert\mu\rvert\pi)} \bigg) - \frac{4\lvert \mu\rvert}{3}\big( 1  - 6\kappa_I^2 - 4\lvert\mu\rvert^2 \big) \frac{\sinh(2\kappa_I\pi)}{\sinh(2\lvert\mu\rvert\pi)}.
\eea
The above quantity is complex and we find its imaginary part as
\bea\label{im1}
&& {\rm{Im}}\bigg[ 4 \sum_{s=\pm} is \frac{e^{i\kappa_s\pi}}{(2\pi)^2} \mathcal{I}_{\kappa_s,\mu}(\tau;\varepsilon)\bigg] =  4i\kappa_I\lvert \mu\rvert \bigg( \frac{ \cosh2\pi\lvert\mu\rvert  - \cosh(2\kappa_I\pi)}{\sinh(2\lvert\mu\rvert\pi)} \bigg) \nonumber\\
&& - \frac{2\kappa_I (\lvert\mu\rvert^2-\kappa_I^2)}{\sinh(2\lvert\mu\rvert\pi)} \sum_{s=\pm} \bigg[  
   \frac{(e^{2\lvert\mu\rvert\pi}-e^{-2s\kappa_I\pi})}{2(s\kappa_I+\lvert\mu\rvert)}  -  \frac{(e^{-2\lvert\mu\rvert\pi}-e^{-2s\kappa_I\pi})}{2(s\kappa_I-\lvert\mu\rvert)} \bigg].
\eea
Moreover, by computing Eq. \eqref{G---} for $\kappa=\tilde{\kappa}_s$, we can compute $\mathcal{I}_{\tilde{\kappa}_s,\mu}(\tau;\varepsilon)$ as
\bea\label{Imukappa---}
&& 4 \frac{e^{i\tilde{\kappa}_s\pi}}{(2\pi)^2} \mathcal{I}_{\tilde{\kappa}_s,\mu}(\tau;\varepsilon) =    
 - 2(1+2\tilde{\kappa}_s) \bigg[ \ln\big(\frac{\varepsilon}{2}\big) + 1 + \frac13 \bigg(\frac{6+10\tilde{\kappa}_s+4\tilde{\kappa}_s^2}{1+4\tilde{\kappa}_s+4\tilde{\kappa}_s^2-4\mu^2}\bigg) \bigg]  \nonumber\\
&& - 2(1+2\tilde{\kappa}_s)  \bigg[ 
i\pi - \psi^{(0)}(1)  +  \frac{e^{2i\tilde{\kappa}_s\pi}+e^{-2i\mu\pi}}{2\sinh(2\lvert\mu\rvert\pi)} \psi^{(0)}(\frac12-\tilde{\kappa}_s-\mu)  - \frac{e^{2i\tilde{\kappa}_s\pi}+e^{2i\mu\pi}}{2\sinh(2\lvert\mu\rvert\pi)} \psi^{(0)}(\frac12-\tilde{\kappa}_s+\mu)\bigg] \nonumber\\
&& + \frac{8\mu}{3} \bigg( 1 + \frac{3+5\tilde{\kappa}_s+2\tilde{\kappa}_s^2}{ 1+4\tilde{\kappa}_s +4\tilde{\kappa}_s^2-4\mu^2}\bigg)  \bigg( \frac{e^{2i\pi\tilde{\kappa}_s}+ \cosh2\pi\lvert\mu\rvert}{\sinh(2\lvert\mu\rvert\pi)} \bigg).
\eea
Using the fact that $\tilde{\kappa}_s=-\frac12+is\kappa_I$ and summing over $s$, we find
\bea\label{Imukappa---}
&& 4 \sum_{s=\pm} is \frac{e^{i\tilde{\kappa}_s\pi}}{(2\pi)^2} \mathcal{I}_{\tilde{\kappa}_s,\mu}(\tau;\varepsilon) =    
8\kappa_I \bigg[ \ln\big(\frac{\varepsilon}{2}\big)  + i \pi - \psi^{(0)}(1) + 1 + \frac{1}{6} \bigg(\frac{1-2\kappa^2_I}{\lvert\mu\rvert^2 -\kappa_I^2}\bigg) \bigg]  \nonumber\\
&&  + 4\kappa_I \sum_{s=\pm} \bigg[ 
  \frac{e^{2\lvert\mu\rvert\pi}-e^{-2s\kappa_I\pi}}{2\sinh(2\lvert\mu\rvert\pi)} \psi^{(0)}(1-is\kappa_I - i\lvert\mu\rvert)  - \frac{e^{-2\lvert\mu\rvert\pi}-e^{-2s\kappa_I\pi}}{2\sinh(2\lvert\mu\rvert\pi)} \psi^{(0)}(1-is\kappa_I + i\lvert\mu\rvert) \bigg] \nonumber\\
&&  - \frac{4i\kappa_I\lvert\mu\rvert}{(\lvert\mu\rvert^2-\kappa_I^2)} \bigg( \frac{\cosh2\pi\lvert\mu\rvert - \cosh(2\pi\kappa_I)}{\sinh(2\lvert\mu\rvert\pi)} \bigg)  - \frac{4\lvert\mu\rvert}{3} \bigg( \frac{1 -6\kappa_I^2+4\lvert\mu\rvert^2}{(\lvert\mu\rvert^2-\kappa_I^2)}\bigg)   \frac{ \sinh(2\kappa_I\pi)}{\sinh(2\lvert\mu\rvert\pi)} .
\eea
Now, for the plus subspace with $\mu_{\rm{m}}^2 = \lvert \mu^+ \rvert^2 - \kappa_I^2$, it can be further simplified to
\bea\label{f-ItK}
&& 4 \mu_{\rm{m}}^2 \sum_{s=\pm} is \frac{e^{i\tilde{\kappa}_s\pi}}{(2\pi)^2} \mathcal{I}_{\tilde{\kappa}_s,\mu}(\tau;\varepsilon) =    
8\kappa_I  \bigg[ \mu_{\rm{m}}^2 \bigg( \ln\big(\frac{\varepsilon}{2}\big) + i \pi - \psi^{(0)}(1) \bigg) + \mu_{\rm{m}}^2 +  \frac{1-2\kappa^2_I}{6} \bigg]  \nonumber\\
&&  + 4\kappa_I \mu_{\rm{m}}^2 \sum_{s=\pm} \bigg[ 
  \frac{e^{2\lvert\mu\rvert\pi}-e^{-2s\kappa_I\pi}}{2\sinh(2\lvert\mu\rvert\pi)} \psi^{(0)}(1-is\kappa_I - i\lvert\mu\rvert)  - \frac{e^{-2\lvert\mu\rvert\pi}-e^{-2s\kappa_I\pi}}{2\sinh(2\lvert\mu\rvert\pi)} \psi^{(0)}(1-is\kappa_I + i\lvert\mu\rvert) \bigg] \nonumber\\
&&  - 4i\kappa_I\lvert\mu\rvert \bigg( \frac{\cosh2\pi\lvert\mu\rvert - \cosh(2\pi\kappa_I)}{\sinh(2\lvert\mu\rvert\pi)} \bigg)  - \frac{4\lvert\mu\rvert}{3} \big(1 -6\kappa_I^2+4\lvert\mu\rvert^2\big)   \frac{ \sinh(2\kappa_I\pi)}{\sinh(2\lvert\mu\rvert\pi)} .
\eea
As we see, it is a complex quantity with an imaginary part given as
\bea\label{im2}
&& {\rm{Im}}\bigg[ 4 \mu_{{\rm{m}}}^2\sum_{s=\pm} is \frac{e^{i\tilde{\kappa}_s\pi}}{(2\pi)^2} \mathcal{I}_{\tilde{\kappa}_s,\mu}(\tau;\varepsilon)\bigg] =  -4i\kappa_I\lvert \mu\rvert \bigg( \frac{ \cosh2\pi\lvert\mu\rvert  - \cosh(2\kappa_I\pi)}{\sinh(2\lvert\mu\rvert\pi)} \bigg) \nonumber\\
&& + \frac{2\kappa_I (\lvert\mu\rvert^2-\kappa_I^2)}{\sinh(2\lvert\mu\rvert\pi)} \sum_{s=\pm} \bigg[  
   \frac{(e^{2\lvert\mu\rvert\pi}-e^{-2s\kappa_I\pi})}{2(s\kappa_I+\lvert\mu\rvert)}  -  \frac{(e^{-2\lvert\mu\rvert\pi}-e^{-2s\kappa_I\pi})}{2(s\kappa_I-\lvert\mu\rvert)} \bigg].
\eea
Notice that it is exactly cancels the imaginary part of $4 \sum_{s=\pm} is \frac{e^{i\kappa_s\pi}}{(2\pi)^2} \mathcal{I}_{\kappa_s,\mu}(\tau;\varepsilon)$ in Eq. \eqref{im1}. Therefore, our final form for the current function, $\mathcal{K}^+(\tau,\varepsilon)$, in Eq. \eqref{K++++} is real. More precisely, combining Eq.s \eqref{Imukappa---+}, \eqref{im1}, \eqref{f-ItK}, and \eqref{im2}, we find the desired quantity
\bea
&& \mathcal{K}^{+}(\tau;\varepsilon) = \nonumber\\
&& 4 \kappa_I \mu^2_{{\rm{m}}}  \bigg[  \ln\big(\frac{\varepsilon}{2}\big) - \psi^{(0)}(1)  \bigg] + \frac23 \kappa_I (1-2\kappa_I^2) + 2 \kappa_I \mu^2_{{\rm{m}}}  - \frac{2\lvert \mu\rvert}{3}\big( 1  - 2\kappa_I^2 + 4 \mu^2_{{\rm{m}}} \big) \frac{\sinh(2\kappa_I\pi)}{\sinh(2\lvert\mu\rvert\pi)} \nonumber\\
&& +  \kappa_I \mu^2_{{\rm{m}}}  \sum_{s=\pm}  \bigg[  
   \frac{e^{2\lvert\mu\rvert\pi}-e^{-2s\kappa_I\pi}}{\sinh(2\lvert\mu\rvert\pi)} {\rm{Re}}\big[\psi^{(0)}(-is\kappa_I-i\lvert\mu\rvert)\big]  - \frac{e^{-2\lvert\mu\rvert\pi}-e^{-2s\kappa_I\pi}}{\sinh(2\lvert\mu\rvert\pi)} {\rm{Re}}\big[\psi^{(0)}(-is\kappa_I+ i\lvert\mu\rvert)\big] 
 \bigg]. \nonumber\\
\eea

\bibliographystyle{JHEP}
\bibliography{mybib}

\providecommand{\href}[2]{#2}\begingroup\raggedright\begin{thebibliography}{10}

\bibitem{Guth:1980zm}
A.~H. Guth, \emph{{The Inflationary Universe: A Possible Solution to the
  Horizon and Flatness Problems}},
  \href{https://doi.org/10.1103/PhysRevD.23.347}{\emph{Phys. Rev.} {\bfseries
  D23} (1981) 347}.

\bibitem{Sato:1980yn}
K.~Sato, \emph{{First Order Phase Transition of a Vacuum and Expansion of the
  Universe}}, {\emph{Mon. Not. Roy. Astron. Soc.} {\bfseries 195} (1981) 467}.

\bibitem{Linde:1981mu}
A.~D. Linde, \emph{{A New Inflationary Universe Scenario: A Possible Solution
  of the Horizon, Flatness, Homogeneity, Isotropy and Primordial Monopole
  Problems}}, \href{https://doi.org/10.1016/0370-2693(82)91219-9}{\emph{Phys.
  Lett.} {\bfseries 108B} (1982) 389}.

\bibitem{Albrecht:1982wi}
A.~Albrecht and P.~J. Steinhardt, \emph{{Cosmology for Grand Unified Theories
  with Radiatively Induced Symmetry Breaking}},
  \href{https://doi.org/10.1103/PhysRevLett.48.1220}{\emph{Phys. Rev. Lett.}
  {\bfseries 48} (1982) 1220}.

\bibitem{Ade:2015lrj}
{\scshape Planck} collaboration, P.~A.~R. Ade et~al., \emph{{Planck 2015
  results. XX. Constraints on inflation}},
  \href{https://doi.org/10.1051/0004-6361/201525898}{\emph{Astron. Astrophys.}
  {\bfseries 594} (2016) A20}
  [\href{https://arxiv.org/abs/1502.02114}{{\ttfamily 1502.02114}}].

\bibitem{Matsumura:2013aja}
T.~Matsumura et~al., \emph{{Mission design of LiteBIRD}},
  \href{https://arxiv.org/abs/1311.2847}{{\ttfamily 1311.2847}}.

\bibitem{Hazumi:2019lys}
M.~Hazumi et~al., \emph{{LiteBIRD: A Satellite for the Studies of B-Mode
  Polarization and Inflation from Cosmic Background Radiation Detection}},
  \href{https://doi.org/10.1007/s10909-019-02150-5}{\emph{J. Low. Temp. Phys.}
  {\bfseries 194} (2019) 443}.

\bibitem{Abazajian:2019eic}
K.~Abazajian et~al., \emph{{CMB-S4 Science Case, Reference Design, and Project
  Plan}},  \href{https://arxiv.org/abs/1907.04473}{{\ttfamily 1907.04473}}.

\bibitem{Chen:2009zp}
X.~Chen and Y.~Wang, \emph{{Quasi-Single Field Inflation and
  Non-Gaussianities}},
  \href{https://doi.org/10.1088/1475-7516/2010/04/027}{\emph{JCAP} {\bfseries
  1004} (2010) 027} [\href{https://arxiv.org/abs/0911.3380}{{\ttfamily
  0911.3380}}].

\bibitem{Baumann:2011nk}
D.~Baumann and D.~Green, \emph{{Signatures of Supersymmetry from the Early
  Universe}}, \href{https://doi.org/10.1103/PhysRevD.85.103520}{\emph{Phys.
  Rev.} {\bfseries D85} (2012) 103520}
  [\href{https://arxiv.org/abs/1109.0292}{{\ttfamily 1109.0292}}].

\bibitem{Noumi:2012vr}
T.~Noumi, M.~Yamaguchi and D.~Yokoyama, \emph{{Effective field theory approach
  to quasi-single field inflation and effects of heavy fields}},
  \href{https://doi.org/10.1007/JHEP06(2013)051}{\emph{JHEP} {\bfseries 06}
  (2013) 051} [\href{https://arxiv.org/abs/1211.1624}{{\ttfamily 1211.1624}}].

\bibitem{Arkani-Hamed:2015bza}
N.~Arkani-Hamed and J.~Maldacena, \emph{{Cosmological Collider Physics}},
  \href{https://arxiv.org/abs/1503.08043}{{\ttfamily 1503.08043}}.

\bibitem{Lee:2016vti}
H.~Lee, D.~Baumann and G.~L. Pimentel, \emph{{Non-Gaussianity as a Particle
  Detector}}, \href{https://doi.org/10.1007/JHEP12(2016)040}{\emph{JHEP}
  {\bfseries 12} (2016) 040}
  [\href{https://arxiv.org/abs/1607.03735}{{\ttfamily 1607.03735}}].

\bibitem{Kehagias:2017cym}
A.~Kehagias and A.~Riotto, \emph{{On the Inflationary Perturbations of Massive
  Higher-Spin Fields}},
  \href{https://doi.org/10.1088/1475-7516/2017/07/046}{\emph{JCAP} {\bfseries
  1707} (2017) 046} [\href{https://arxiv.org/abs/1705.05834}{{\ttfamily
  1705.05834}}].

\bibitem{Arkani-Hamed:2018kmz}
N.~Arkani-Hamed, D.~Baumann, H.~Lee and G.~L. Pimentel, \emph{{The Cosmological
  Bootstrap: Inflationary Correlators from Symmetries and Singularities}},
  \href{https://arxiv.org/abs/1811.00024}{{\ttfamily 1811.00024}}.

\bibitem{Liu:2019fag}
T.~Liu, X.~Tong, Y.~Wang and Z.-Z. Xianyu, \emph{{Probing P and CP Violations
  on the Cosmological Collider}},
  \href{https://arxiv.org/abs/1909.01819}{{\ttfamily 1909.01819}}.

\bibitem{Freese:1990rb}
K.~Freese, J.~A. Frieman and A.~V. Olinto, \emph{{Natural inflation with pseudo
  - Nambu-Goldstone bosons}},
  \href{https://doi.org/10.1103/PhysRevLett.65.3233}{\emph{Phys. Rev. Lett.}
  {\bfseries 65} (1990) 3233}.

\bibitem{Easther:2013kla}
R.~Easther and R.~Flauger, \emph{{Planck Constraints on Monodromy Inflation}},
  \href{https://doi.org/10.1088/1475-7516/2014/02/037}{\emph{JCAP} {\bfseries
  1402} (2014) 037} [\href{https://arxiv.org/abs/1308.3736}{{\ttfamily
  1308.3736}}].

\bibitem{McAllister:2014mpa}
L.~McAllister, E.~Silverstein, A.~Westphal and T.~Wrase, \emph{{The Powers of
  Monodromy}}, \href{https://doi.org/10.1007/JHEP09(2014)123}{\emph{JHEP}
  {\bfseries 09} (2014) 123} [\href{https://arxiv.org/abs/1405.3652}{{\ttfamily
  1405.3652}}].

\bibitem{Pajer:2013fsa}
E.~Pajer and M.~Peloso, \emph{{A review of Axion Inflation in the era of
  Planck}}, \href{https://doi.org/10.1088/0264-9381/30/21/214002}{\emph{Class.
  Quant. Grav.} {\bfseries 30} (2013) 214002}
  [\href{https://arxiv.org/abs/1305.3557}{{\ttfamily 1305.3557}}].

\bibitem{Marsh:2015xka}
D.~J.~E. Marsh, \emph{{Axion Cosmology}},
  \href{https://doi.org/10.1016/j.physrep.2016.06.005}{\emph{Phys. Rept.}
  {\bfseries 643} (2016) 1} [\href{https://arxiv.org/abs/1510.07633}{{\ttfamily
  1510.07633}}].

\bibitem{Obied:2018sgi}
G.~Obied, H.~Ooguri, L.~Spodyneiko and C.~Vafa, \emph{{De Sitter Space and the
  Swampland}},  \href{https://arxiv.org/abs/1806.08362}{{\ttfamily
  1806.08362}}.

\bibitem{Dvali:2018dce}
G.~Dvali, C.~Gomez and S.~Zell, \emph{{A Proof of the Axion?}},
  \href{https://arxiv.org/abs/1811.03079}{{\ttfamily 1811.03079}}.

\bibitem{Maleknejad:2014wsa}
A.~Maleknejad, \emph{{Chiral Gravity Waves and Leptogenesis in Inflationary
  Models with non-Abelian Gauge Fields}},
  \href{https://doi.org/10.1103/PhysRevD.90.023542}{\emph{Phys. Rev.}
  {\bfseries D90} (2014) 023542}
  [\href{https://arxiv.org/abs/1401.7628}{{\ttfamily 1401.7628}}].

\bibitem{Zeldovich:1974uw}
{\relax Ya}.~B. Zeldovich, I.~{\relax Yu}. Kobzarev and L.~B. Okun,
  \emph{{Cosmological Consequences of the Spontaneous Breakdown of Discrete
  Symmetry}}, {\emph{Zh. Eksp. Teor. Fiz.} {\bfseries 67} (1974) 3}.

\bibitem{Dvali:2018txx}
G.~Dvali, C.~Gomez and S.~Zell, \emph{{Discrete Symmetries Excluded by Quantum
  Breaking}},  \href{https://arxiv.org/abs/1811.03077}{{\ttfamily 1811.03077}}.

\bibitem{Maleknejad:2011sq}
A.~Maleknejad and M.~M. Sheikh-Jabbari, \emph{{Non-Abelian Gauge Field
  Inflation}}, \href{https://doi.org/10.1103/PhysRevD.84.043515}{\emph{Phys.
  Rev.} {\bfseries D84} (2011) 043515}
  [\href{https://arxiv.org/abs/1102.1932}{{\ttfamily 1102.1932}}].

\bibitem{Maleknejad:2011jw}
A.~Maleknejad and M.~M. Sheikh-Jabbari, \emph{{Gauge-flation: Inflation From
  Non-Abelian Gauge Fields}},
  \href{https://doi.org/10.1016/j.physletb.2013.05.001}{\emph{Phys. Lett.}
  {\bfseries B723} (2013) 224}
  [\href{https://arxiv.org/abs/1102.1513}{{\ttfamily 1102.1513}}].

\bibitem{Adshead:2013nka}
P.~Adshead, E.~Martinec and M.~Wyman, \emph{{Perturbations in Chromo-Natural
  Inflation}}, \href{https://doi.org/10.1007/JHEP09(2013)087}{\emph{JHEP}
  {\bfseries 09} (2013) 087} [\href{https://arxiv.org/abs/1305.2930}{{\ttfamily
  1305.2930}}].

\bibitem{Maleknejad:2012fw}
A.~Maleknejad, M.~M. Sheikh-Jabbari and J.~Soda, \emph{{Gauge Fields and
  Inflation}}, \href{https://doi.org/10.1016/j.physrep.2013.03.003}{\emph{Phys.
  Rept.} {\bfseries 528} (2013) 161}
  [\href{https://arxiv.org/abs/1212.2921}{{\ttfamily 1212.2921}}].

\bibitem{Maleknejad:2018nxz}
A.~Maleknejad and E.~Komatsu, \emph{{Production and Backreaction of Spin-2
  Particles of $SU(2)$ Gauge Field during Inflation}},
  \href{https://arxiv.org/abs/1808.09076}{{\ttfamily 1808.09076}}.

\bibitem{Dimastrogiovanni:2012ew}
E.~Dimastrogiovanni and M.~Peloso, \emph{{Stability analysis of chromo-natural
  inflation and possible evasion of Lyth's bound}},
  \href{https://doi.org/10.1103/PhysRevD.87.103501}{\emph{Phys. Rev.}
  {\bfseries D87} (2013) 103501}
  [\href{https://arxiv.org/abs/1212.5184}{{\ttfamily 1212.5184}}].

\bibitem{Adshead:2013qp}
P.~Adshead, E.~Martinec and M.~Wyman, \emph{{Gauge fields and inflation: Chiral
  gravitational waves, fluctuations, and the Lyth bound}},
  \href{https://doi.org/10.1103/PhysRevD.88.021302}{\emph{Phys. Rev.}
  {\bfseries D88} (2013) 021302}
  [\href{https://arxiv.org/abs/1301.2598}{{\ttfamily 1301.2598}}].

\bibitem{Thorne:2017jft}
B.~Thorne, T.~Fujita, M.~Hazumi, N.~Katayama, E.~Komatsu and M.~Shiraishi,
  \emph{{Finding the chiral gravitational wave background of an axion-SU(2)
  inflationary model using CMB observations and laser interferometers}},
  \href{https://doi.org/10.1103/PhysRevD.97.043506}{\emph{Phys. Rev.}
  {\bfseries D97} (2018) 043506}
  [\href{https://arxiv.org/abs/1707.03240}{{\ttfamily 1707.03240}}].

\bibitem{Agrawal:2017awz}
A.~Agrawal, T.~Fujita and E.~Komatsu, \emph{{Large tensor non-Gaussianity from
  axion-gauge field dynamics}},
  \href{https://doi.org/10.1103/PhysRevD.97.103526}{\emph{Phys. Rev.}
  {\bfseries D97} (2018) 103526}
  [\href{https://arxiv.org/abs/1707.03023}{{\ttfamily 1707.03023}}].

\bibitem{Agrawal:2018mrg}
A.~Agrawal, T.~Fujita and E.~Komatsu, \emph{{Tensor Non-Gaussianity from
  Axion-Gauge-Fields Dynamics : Parameter Search}},
  \href{https://doi.org/10.1088/1475-7516/2018/06/027}{\emph{JCAP} {\bfseries
  1806} (2018) 027} [\href{https://arxiv.org/abs/1802.09284}{{\ttfamily
  1802.09284}}].

\bibitem{Papageorgiou:2018rfx}
A.~Papageorgiou, M.~Peloso and C.~Unal, \emph{{Nonlinear perturbations from the
  coupling of the inflaton to a non-Abelian gauge field, with a focus on
  Chromo-Natural Inflation}},
  \href{https://doi.org/10.1088/1475-7516/2018/09/030}{\emph{JCAP} {\bfseries
  1809} (2018) 030} [\href{https://arxiv.org/abs/1806.08313}{{\ttfamily
  1806.08313}}].

\bibitem{Maleknejad:2016dci}
A.~Maleknejad, \emph{{Gravitational leptogenesis in axion inflation with SU(2)
  gauge field}},
  \href{https://doi.org/10.1088/1475-7516/2016/12/027}{\emph{JCAP} {\bfseries
  1612} (2016) 027} [\href{https://arxiv.org/abs/1604.06520}{{\ttfamily
  1604.06520}}].

\bibitem{Noorbala:2012fh}
A.~Maleknejad, M.~Noorbala and M.~M. Sheikh-Jabbari, \emph{{Leptogenesis in
  inflationary models with non-Abelian gauge fields}},
  \href{https://doi.org/10.1007/s10714-018-2435-8}{\emph{Gen. Rel. Grav.}
  {\bfseries 50} (2018) 110} [\href{https://arxiv.org/abs/1208.2807}{{\ttfamily
  1208.2807}}].

\bibitem{Alexander:2004us}
S.~H.-S. Alexander, M.~E. Peskin and M.~M. Sheikh-Jabbari, \emph{{Leptogenesis
  from gravity waves in models of inflation}},
  \href{https://doi.org/10.1103/PhysRevLett.96.081301}{\emph{Phys. Rev. Lett.}
  {\bfseries 96} (2006) 081301}
  [\href{https://arxiv.org/abs/hep-th/0403069}{{\ttfamily hep-th/0403069}}].

\bibitem{Lozanov:2018kpk}
K.~D. Lozanov, A.~Maleknejad and E.~Komatsu, \emph{{Schwinger Effect by an
  $SU(2)$ Gauge Field during Inflation}},
  \href{https://doi.org/10.1007/JHEP02(2019)041}{\emph{JHEP} {\bfseries 02}
  (2019) 041} [\href{https://arxiv.org/abs/1805.09318}{{\ttfamily
  1805.09318}}].

\bibitem{Chung:2011ck}
D.~J.~H. Chung, L.~L. Everett, H.~Yoo and P.~Zhou, \emph{{Gravitational Fermion
  Production in Inflationary Cosmology}},
  \href{https://doi.org/10.1016/j.physletb.2012.04.066}{\emph{Phys. Lett.}
  {\bfseries B712} (2012) 147}
  [\href{https://arxiv.org/abs/1109.2524}{{\ttfamily 1109.2524}}].

\bibitem{Hayashinaka:2016qqn}
T.~Hayashinaka, T.~Fujita and J.~Yokoyama, \emph{{Fermionic Schwinger effect
  and induced current in de Sitter space}},
  \href{https://doi.org/10.1088/1475-7516/2016/07/010}{\emph{JCAP} {\bfseries
  1607} (2016) 010} [\href{https://arxiv.org/abs/1603.04165}{{\ttfamily
  1603.04165}}].

\bibitem{Adshead:2015kza}
P.~Adshead and E.~I. Sfakianakis, \emph{{Fermion production during and after
  axion inflation}},
  \href{https://doi.org/10.1088/1475-7516/2015/11/021}{\emph{JCAP} {\bfseries
  1511} (2015) 021} [\href{https://arxiv.org/abs/1508.00891}{{\ttfamily
  1508.00891}}].

\bibitem{Adshead:2018oaa}
P.~Adshead, L.~Pearce, M.~Peloso, M.~A. Roberts and L.~Sorbo,
  \emph{{Phenomenology of fermion production during axion inflation}},
  \href{https://doi.org/10.1088/1475-7516/2018/06/020}{\emph{JCAP} {\bfseries
  1806} (2018) 020} [\href{https://arxiv.org/abs/1803.04501}{{\ttfamily
  1803.04501}}].

\bibitem{Adshead:2019aac}
P.~Adshead, L.~Pearce, M.~Peloso, M.~A. Roberts and L.~Sorbo,
  \emph{{Gravitational waves from fermion production during axion inflation}},
  \href{https://arxiv.org/abs/1904.10483}{{\ttfamily 1904.10483}}.

\bibitem{Adshead:2017znw}
P.~Adshead, A.~J. Long and E.~I. Sfakianakis, \emph{{Gravitational
  Leptogenesis, Reheating, and Models of Neutrino Mass}},
  \href{https://doi.org/10.1103/PhysRevD.97.043511}{\emph{Phys. Rev.}
  {\bfseries D97} (2018) 043511}
  [\href{https://arxiv.org/abs/1711.04800}{{\ttfamily 1711.04800}}].

\bibitem{Papageorgiou:2017yup}
A.~Papageorgiou and M.~Peloso, \emph{{Gravitational leptogenesis in Natural
  Inflation}}, \href{https://doi.org/10.1088/1475-7516/2017/12/007}{\emph{JCAP}
  {\bfseries 1712} (2017) 007}
  [\href{https://arxiv.org/abs/1708.08007}{{\ttfamily 1708.08007}}].

\bibitem{Alexander:2018fjp}
S.~Alexander, E.~McDonough and D.~N. Spergel, \emph{{Chiral Gravitational Waves
  and Baryon Superfluid Dark Matter}},
  \href{https://doi.org/10.1088/1475-7516/2018/05/003}{\emph{JCAP} {\bfseries
  1805} (2018) 003} [\href{https://arxiv.org/abs/1801.07255}{{\ttfamily
  1801.07255}}].

\bibitem{Domcke:2018gfr}
V.~Domcke, Y.~Ema, K.~Mukaida and R.~Sato, \emph{{Chiral Anomaly and Schwinger
  Effect in Non-Abelian Gauge Theories}},
  \href{https://doi.org/10.1007/JHEP03(2019)111}{\emph{JHEP} {\bfseries 03}
  (2019) 111} [\href{https://arxiv.org/abs/1812.08021}{{\ttfamily
  1812.08021}}].

\bibitem{Mirzagholi:2019jeb}
L.~Mirzagholi, A.~Maleknejad and K.~D. Lozanov, \emph{{Production and
  Backreaction of Fermions from Axion-$SU(2)$ Gauge Fields during Inflation}},
  \href{https://arxiv.org/abs/1905.09258}{{\ttfamily 1905.09258}}.

\bibitem{Maleknejad:2011jr}
A.~Maleknejad, M.~M. Sheikh-Jabbari and J.~Soda, \emph{{Gauge-flation and
  Cosmic No-Hair Conjecture}},
  \href{https://doi.org/10.1088/1475-7516/2012/01/016}{\emph{JCAP} {\bfseries
  1201} (2012) 016} [\href{https://arxiv.org/abs/1109.5573}{{\ttfamily
  1109.5573}}].

\bibitem{Maleknejad:2013npa}
A.~Maleknejad and E.~Erfani, \emph{{Chromo-Natural Model in Anisotropic
  Background}},
  \href{https://doi.org/10.1088/1475-7516/2014/03/016}{\emph{JCAP} {\bfseries
  1403} (2014) 016} [\href{https://arxiv.org/abs/1311.3361}{{\ttfamily
  1311.3361}}].

\bibitem{Maleknejad:2012as}
A.~Maleknejad and M.~M. Sheikh-Jabbari, \emph{{Revisiting Cosmic No-Hair
  Theorem for Inflationary Settings}},
  \href{https://doi.org/10.1103/PhysRevD.85.123508}{\emph{Phys. Rev.}
  {\bfseries D85} (2012) 123508}
  [\href{https://arxiv.org/abs/1203.0219}{{\ttfamily 1203.0219}}].

\bibitem{Watanabe:2009ct}
M.-a. Watanabe, S.~Kanno and J.~Soda, \emph{{Inflationary Universe with
  Anisotropic Hair}},
  \href{https://doi.org/10.1103/PhysRevLett.102.191302}{\emph{Phys. Rev. Lett.}
  {\bfseries 102} (2009) 191302}
  [\href{https://arxiv.org/abs/0902.2833}{{\ttfamily 0902.2833}}].

\bibitem{Adshead:2018emn}
P.~Adshead and A.~Liu, \emph{{Anisotropic Massive Gauge-flation}},
  \href{https://doi.org/10.1088/1475-7516/2018/07/052}{\emph{JCAP} {\bfseries
  1807} (2018) 052} [\href{https://arxiv.org/abs/1803.07168}{{\ttfamily
  1803.07168}}].

\bibitem{Adler:1969gk}
S.~L. Adler, \emph{{Axial vector vertex in spinor electrodynamics}},
  \href{https://doi.org/10.1103/PhysRev.177.2426}{\emph{Phys. Rev.} {\bfseries
  177} (1969) 2426}.

\bibitem{Bell:1969ts}
J.~S. Bell and R.~Jackiw, \emph{{A PCAC puzzle: $\pi^0 \to \gamma \gamma$ in
  the $\sigma$ model}}, \href{https://doi.org/10.1007/BF02823296}{\emph{Nuovo
  Cim.} {\bfseries A60} (1969) 47}.

\bibitem{Peskin:1995ev}
M.~E. Peskin and D.~V. Schroeder, \emph{{An Introduction to quantum field
  theory}}. Addison-Wesley, Reading, USA, 1995.

\bibitem{parker2009quantum}
L.~Parker and D.~Toms, \emph{Quantum Field Theory in Curved Spacetime:
  Quantized Fields and Gravity}, Cambridge Monographs on Mathematical Physics.
  Cambridge University Press, 2009.

\bibitem{Hayashinaka:2018amz}
T.~Hayashinaka and S.-S. Xue, \emph{{Physical renormalization condition for de
  Sitter QED}}, \href{https://doi.org/10.1103/PhysRevD.97.105010}{\emph{Phys.
  Rev.} {\bfseries D97} (2018) 105010}
  [\href{https://arxiv.org/abs/1802.03686}{{\ttfamily 1802.03686}}].

\bibitem{Sakharov:1967dj}
A.~D. Sakharov, \emph{{Violation of CP Invariance, C asymmetry, and baryon
  asymmetry of the universe}},
  \href{https://doi.org/10.1070/PU1991v034n05ABEH002497}{\emph{Pisma Zh. Eksp.
  Teor. Fiz.} {\bfseries 5} (1967) 32}.

\bibitem{Ade:2015xua}
{\scshape Planck} collaboration, P.~A.~R. Ade et~al., \emph{{Planck 2015
  results. XIII. Cosmological parameters}},
  \href{https://doi.org/10.1051/0004-6361/201525830}{\emph{Astron. Astrophys.}
  {\bfseries 594} (2016) A13}
  [\href{https://arxiv.org/abs/1502.01589}{{\ttfamily 1502.01589}}].

\bibitem{Nist}
F.~W. Olver, D.~W. Lozier, R.~F. Boisvert and C.~W. Clark, \emph{NIST Handbook
  of Mathematical Functions}. Cambridge University Press, New York, NY, USA,
  1st~ed., 2010.

\end{thebibliography}\endgroup

\end{document}